\pdfoutput=1
\documentclass[
  12pt          %
  ,letterpaper  %
  ,center       %
  ,noupper      %
  ,english,notranslate,fleqn]{uconnthesis}
\ThesisLineSpacing{2}
\usepackage[fleqn]{amsmath}
\usepackage{amssymb}
\usepackage{amsfonts}
\usepackage{amsthm}

\newcommand{\changefunction}[1]{%
  \expandafter\renewcommand\csname#1\endcsname[1][]%
    {\qopname\relax o{#1}\ifx\relax##1\relax\else^{##1}\fi}}
\changefunction{sin}
\changefunction{cos}
\changefunction{tan}

\renewcommand{\arcsin}{\sin[-1]}

\usepackage{dsfont}
\newcommand {\ic}[0]{i}
\usepackage{nccmath}
\usepackage{bm}

\usepackage{relsize}
\usepackage{indentfirst}

\usepackage[utf8]{inputenc}
\usepackage[T2A,T1]{fontenc}
\DeclareTextSymbolDefault{\cyrd}{T2A}
\DeclareTextSymbolDefault{\cyru}{T2A}
\DeclareTextSymbolDefault{\cyrv}{T2A}
\DeclareTextSymbolDefault{\cyra}{T2A}
\DeclareTextSymbolDefault{\cyrn}{T2A}
\DeclareTextSymbolDefault{\cyri}{T2A}
\DeclareTextSymbolDefault{\cyrr}{T2A}
\DeclareTextSymbolDefault{\cyre}{T2A}
\DeclareTextSymbolDefault{\cyrl}{T2A}
\DeclareTextSymbolDefault{\cyrsftsn}{T2A}
\DeclareTextSymbolDefault{\cyrs}{T2A}
\DeclareTextSymbolDefault{\cyro}{T2A}
\DeclareTextSymbolDefault{\CYRC}{T2A}
\usepackage{lmodern}
\usepackage[english]{babel}
\usepackage{microtype}

\let\stdsection\section
\renewcommand\section{\newpage\stdsection}

\usepackage{savesym}
\savesymbol{singlespacing}
\usepackage{setspace}
\restoresymbol{setspace}{singlespacing}

\usepackage[square,comma,numbers,nonamebreak,sort&compress,merge]{natbib}

\usepackage[usenames,dvipsnames]{color}
\definecolor{ultramarine}{RGB}{63, 0, 255}
\definecolor{medblue}{RGB}{0, 0, 100}
\definecolor{panblue}{RGB}{0,24,150}
\definecolor{carmine}{RGB}{150, 0, 24}
\usepackage{hyperref}
\hypersetup{colorlinks,
linkcolor=medblue,
citecolor=carmine,
urlcolor=panblue,
anchorcolor=OliveGreen}
\usepackage[toc,section=chapter,nopostdot,nonumberlist,translate=false]{glossaries}
\makeglossaries
\glstoctrue
\makeatletter
\let\oldmakefirstuc\makefirstuc
\renewcommand*{\makefirstuc}[1]{%
  \def\gls@add@space{}%
  \mfu@capitalisewords#1 \@nil\mfu@endcap
}
\def\mfu@capitalisewords#1 #2\mfu@endcap{%
  \def\mfu@cap@first{#1}%
  \def\mfu@cap@second{#2}%
  \gls@add@space
  \oldmakefirstuc{#1}%
  \def\gls@add@space{ }%
  \ifx\mfu@cap@second\@nnil
    \let\next@mfu@cap\mfu@noop
  \else
    \let\next@mfu@cap\mfu@capitalisewords
  \fi
  \next@mfu@cap#2\mfu@endcap
}
\makeatother

\newglossaryentry{condprobspace}
{
name={Conditional Probability Space},
text={conditional probability space},
description={The abstract space of possible experimental statistical parameters, also known as statistical space. A given experimental parameter, for example, could be the probability that Alice and Bob obtain correlated measurement outputs \emph{given} that Alice uses measurement apparatus $A1$ and Bob using apparatus $B2$. A fixed experimental setup yields a single points in conditional probability space in the limit of infinite experimental trials. (A Finite trial count results in an uncertainty of the coordinates of the experiment in statistical space, the study of which is generally subsumed in Martingale analysis \cite{YanbaoThesis,PhysRevA.81.032117,PhysRevA.84.062118,PhysRevA.88.052119}.) \emph{See} \textbf{\Gls{polytope}}.}
}
\newglossaryentry{clique}
{
name={Clique},
text={clique},
description={In graph theory, a clique is a set of vertices each of which is adjacent to all the others in that set. In the context of Orthogonality Graphs, a clique refers to a set of measurement outcomes (non-necessarily of the same measurement tuple) which are mutually exclusive, or contradictory. An example of a clique of incompatible measurement outcomes is given by the ten terms in the inequality \noeq{eq:RafaelClique}. See Ref. \cite{EPOriginal} for a quick introduction to the relevant concept. }
}
\newglossaryentry{QB}
{
name={QB},
description={An initialism of ``quantum bound''. In this thesis we make reference primarily to ${\mbox{QB}_3}^{(8)}$ featured in Ref. \cite{WolfeQB} and reproduced in \tab{tab:bounds}. The subscript $3$ indicates that it belongs to the third class of quantum bounds in the table, whereas the superscript $8$ indicates that ${\mbox{QB}_3}^{(8)}$ represents a boundary in the full $8$-dimensional conditional probability space of the (2,2,2) scenario \cite{WolfeQB}. ${\mbox{QB}_3}^{(8)}$ is a function-valued quantum bound, in that it describes a contiguous collection (or envelope) of linear quantum bounds. Linear quantum bounds receive the special designation ``Tsirelson Bounds'' \cite{Tsirelson1980}, or ``TB'', in this thesis. }
}
\newglossaryentry{NPA}
{
name={NPA},
description={An acronym for the three authors of the Navascués-Pironio-Acín (NPA) hierarchy \cite{NPA2007Short,NPA2008Long,NPAReview}. The hierarchy converges asymptotically to the genuine set of quantum nonlocal correlations. Finite levels of the hierarchy, however, permit additional correlations which are not achievable in a quantum nonlocality scenario. Of special interests in the $1+AB$ level of the NPA hierarchy, as this corresponds the the correlations achievable in a single-partite quantum contextuality scenario \cite{FritzCombinatorialLong,CabelloMultigraph} as well the set of correlations recovered by the Consistent Histories theory of quantum gravity \cite{ConsistentHistoriesContextuality}, the convex and closed-under-wirings variant of Consistent Exclusivity [\citealp{EPYan} and \citealp[Sec. 7.7]{FritzCombinatorialLong}], as well as various other principles \cite{AlmostQuantum2}.}
}
\newglossaryentry{polytope}
{
name={Polytope},
text={polytope},
description={A polytope is a higher-dimensional analog of a polyhedron. Of special interest are statistical polytopes in conditional probability space. For any nonlocality scenario, both the set of no-signalling conditions and the set of Bell inequalities form polytopes in conditional probability space which bound the set correlations consistent with general probabilistic theories and with local hidden variable models, respectively \cite{PROriginal,PRUnit,Brunner2013Bell,PopescuReviewNatureComm,ScaraniNotes,*ScaraniNotes2}. Polytopes are described by either the convex hull of extremal points, or equivalently, by linear equalities. The quantum boundary in conditional probability space is nonlinear, and therefore quantum correlations are described by a convex elliptope instead of a polytope.  \emph{See} \textbf{\Gls{condprobspace}}.}
}
\newglossaryentry{elliptope}
{
name={Elliptope},
text={elliptope},
description={An elliptope is a higher-dimensional analog of a convex curved ellipsoid. Of special interest is the quantum elliptope in conditional probability space. The quantum boundary in conditional probability space is known to be nonlinear, but does not presently have a closed form characterization \cite{MasanesFourD,HowMuchLarger,NPA2007Short}. One can use linear Tsirelson inequalities to generate a psuedo-quantum polytope which contains the genuine quantum elliptope within it; in the infinite limit the region bounded by the linear equalities converge to the quantum elliptope per the supporting hyperplane theorem \cite{WolfeQB}. The quantum elliptope can, alternatively, be inner characterized by the convex hull of extremal quantum-achievable conditional probability distributions. \emph{See} \textbf{\Gls{polytope}}.}
}
\newglossaryentry{ep}
{
name={Exclusivity Principle},
text={Exclusivity Principle},
description={A graph-theoretic principle which states that any exclusivity graph must have vertex probabilities less than or equal to one in every orthogonality clique \cite{EPOriginal,EPSpecker,EPExampleGraphs,EPHenson,EPYan,EP2013,EPTwoCities,EPTsirelson}. Quantum mechanics innately obeys this principle, because if a set of quantum projectors are all pairwise orthogonal, than any string of such projectors, or any length, is equal to the null vector. One can, however, imagine probabilistic theories which violate the Exclusivity Principle \cite{SpekkensSeer}; such correlations are not possible in a quantum reality, however. The Exclusivity Principle differs from Local Orthogonality in that it applies to contextuality scenarios as well as nonlocality scenarios. The Exclusivity Principles differs from Consistent Exclusivity in that it considers all possible extensions of the target scenario \cite{EPHenson,EPTsirelson}. \emph{See} \textbf{\Gls{clique}}.}
}
\newglossaryentry{ce}
{
name={Consistent Exclusivity},
text={Consistent Exclusivity},
description={Consistent Exclusivity is a special case of the Exclusivity Principle when inferences on the correlations of a contextuality scenario are made by considering the possibility that independent copies of the scenario could exist simultaneously [\citealp{EPHenson} and \citealp[Sec. 7.1]{FritzCombinatorialLong}]. Consistent Exclusivity is characterized by various integer levels, corresponding the number of independent copies from which inferences are being drawn. For correlations to be unambiguously consistently exclusive, they must be compatible with Consistent Exclusivity in the limit of infinite simultaneous copies. Consistent Exclusivity is directly related to the Shannon Capacity of nonorthogonality graphs \citep[Sec. 7.3]{FritzCombinatorialLong}. Extended Consistent Exclusivity requires additionally that the set of allowed correlations be convex and closed under wirings. Extended Consistent Exclusivity is known to perfectly recover quantum contextual - but not quantum nonlocal - correlations, in that it is equivalent to the $1+AB$ level of the NPA hierarchy \citep[Sec. 7.7]{FritzCombinatorialLong}. \emph{See} \textbf{\gls{ep}}, \textbf{\gls{NPA}}.}
}
\newglossaryentry{point}
{
name={Point},
text={point},
description={In this thesis, we reserve the use of ``point'' to indicate a particulate set of coordinates in  \textbf{\Gls{condprobspace}}. Thus, a point refers to a particular set of statistics which might characterize a nonlocality or contextuality scenario. Completely equivalent is the notion of a \textbf{behavior}, or a \text{probabilistic model}, and a bipartite \textbf{box} such as the famous Popescu and Rohlich - PR - box \cite{PROriginal,PRUnit,Brunner2013Bell,PopescuReviewNatureComm,ScaraniNotes,*ScaraniNotes2}. In this thesis we consider composite boxes, that is to say, a normalized mixture of boxes where the the weights are variable parameters. Such variable boxes defines curves or surfaces in  \textbf{\gls{condprobspace}}.}
}
\newglossaryentry{box}
{
name={Box},
text={box},
description={\emph{See} \textbf{\Gls{point}}.}
}
\newglossaryentry{lo}
{
name={Local Orthogonality},
text={Local Orthogonality},
description={A first principle which partially recovers quantum correlations. Nearly identitical to \textbf{\Gls{ce}}, Local Orthogonality applies only to multipartite nonlocality scenarios. It was developed by \citet{LONatureComm}, independently from Cabello's \textbf{\Gls{ep}} \cite{EPOriginal}. See Refs. \cite{LONatureComm,FritzCombinatorialLong,LONewShort,LOHardy,LOExploring}. Local Orthogonality constructs composite scenarios through a hypergraph formalism \cite{FritzCombinatorialLong}, but then reduces to standard orthogonality graphs when its primary principle is applied. \emph{See} \textbf{\gls{ep}}, \textbf{\gls{ce}}.}
}

\definecolor{purple}{RGB}{128,0,128}
\newcommand{\purp}[1]{{\color{purple}{#1}\color{black}}}

\usepackage{mathrsfs}
\usepackage{graphicx}
\usepackage[export]{adjustbox}
\usepackage{placeins}
\usepackage[format=plain]{caption}

\usepackage{verbatim}
\usepackage{enumitem}

\usepackage{booktabs}
\usepackage{geometry}
\usepackage{pdflscape}
\usepackage{rotating}
\usepackage{tabularx}
\newcolumntype{R}{>{\raggedleft\arraybackslash}X}
\newcolumntype{C}{>{\centering\arraybackslash}X}
\newcolumntype{L}{>{\raggedright\arraybackslash}X}

\usepackage{tabulary}

\usepackage{units}

\usepackage{braket}

\newcommand{\brackets}[1]{\lbrace{#1\rbrace}}
\newcommand {\id}[0]{\ensuremath{\mathds{1}}}

\newcommand{\LeftEqns}[1]{\begin{fleqn}[\leftmargini minus \leftmargini]\begin{align}#1\end{align}\end{fleqn}}

\newcommand{\LeftEqn}[1]{\LeftEqns{\begin{split}#1\end{split}}}
\newcommand{\FlushEqn}[1]{\LeftEqns{\begin{split}#1\end{split}}}

\newcommand{\acases}[1]{{\renewcommand{\arraystretch}{0.75}\left\lbrace\begin{array}{ll} #1\end{array}\right.}}

\newcommand{\ceq}[1]{Eq.\,(\ref{#1})}

\newcommand{\noeq}[1]{(\ref{#1})}
\newcommand{\fig}[1]{Fig.\,\ref{#1}}
\newcommand{\tab}[1]{Table\,\ref{#1}}

\newcommand{\ox}[2]{{\operatorname{X}\limits_{#2}^{}}}
\newcommand{\ochi}[2]{{\operatorname{\chi}\limits_{#2}^{}}}

\newcommand*{\dyad}[2]{ {\ket{#1}\hspace{-0.5ex}\bra{#2}}}
\newcommand*{\pdyad}[2]{ {\left(\ket{#1}\hspace{-0.5ex}\bra{#2}\right)}}
\newcommand{\brho}[0]{\bm{\mbox{\large \(\varrho\)}}}
\newcommand{\ot}[0]{\mathrel{\hspace{-0.3ex}\scriptstyle{\otimes}\hspace{-0.3ex}}}
\newcommand{\mindyad}[2]{\mathrel{\scriptstyle{\dyad{#1}{#2}}}}
\newcommand{\da}[0]{\mindyad{0}{0}}
\newcommand{\db}[0]{\mindyad{0}{1}}
\newcommand{\dc}[0]{\mindyad{1}{0}}
\newcommand{\dd}[0]{\mindyad{1}{1}}
\newcommand{\op}[4]{#1 \ot #2 \ot #3 \ot #4}

\newcommand {\jbar}[0]{\bar{J}}
\newcommand{\bp}{^\backprime}
\newcommand{\bpp}{^{\backprime\backprime}}

\newcommand{\nmap}[1]{{\operatorname{Map}_N{\mathlarger{\mathlarger{\left(\mathsmaller{\mathsmaller{ #1}}\right)}}}}}
\newcommand{\nmapalt}[0]{{\operatorname{Map}_N}}

\begin{document}

\abstract{This work develops analytic methods to quantitatively demarcate quantum reality from its subset of classical phenomenon, as well as from the superset of general probabilistic theories. Regarding quantum nonlocality, we discuss how to determine the quantum limit of Bell-type linear inequalities. In contrast to semidefinite programming approaches, our method allows for the consideration of inequalities with abstract weights, by means of leveraging the Hermiticity of quantum states. Recognizing that classical correlations correspond to measurements made on separable states, we also introduce a practical method for obtaining sufficient separability criteria. We specifically vet the candidacy of driven and undriven superradiance as schema for entanglement generation. We conclude by reviewing current approaches to quantum contextuality, emphasizing the operational distinction between nonlocal and contextual quantum statistics. We utilize our abstractly-weighted linear quantum bounds to explicitly demonstrate a set of conditional probability distributions which are simultaneously compatible with quantum contextuality while being incompatible with quantum nonlocality. It is noted that this novel statistical regime implies an experimentally-testable target for the Consistent Histories theory of quantum gravity.}

\title{Quantum Apices: Identifying Limits of Entanglement, Nonlocality, \& Contextuality}
\author{Elie Wolfe}
\authorspreviousdegreelong{
  B.A., Yeshiva University, 2006\\
  M.S., University of Connecticut, 2010
  }

\MajorAdvisor{Susanne Yelin}
\AssociateAdvisorA{Robin C\^ot\'e}
\AssociateAdvisorB{Alexander Russell}
\date{2014}
\dedication{
  to Noam and Eitan, \\ sweet and strong.
  }

\acknowledgements{
    \indent Without Dr. Susanne Yelin, this work could not have gotten off the ground. The stark truth is, that I might not have pursued any of this research, or any research at all really, were it not for Dr. Yelin's faith and support. I had one foot out the door already when Dr. Yelin intervened by welcoming me into her research group. Nothing I can write can adequately convey the gratitude I feel for the privilege of working with Dr. Yelin. She provided me with an enduring paradigm for excellence in advisorship.\\
    \indent Dr. Yelin opened up her own projects to me, and at the same time gave me complete freedom to pursue my own interests. She was able to be attentive and encouraging while also granting me flexibility and understanding to balance my family life. She was \emph{always} available for questions, be they about research,  writing, or tact. Every topic that I have looked into, every conference I have attended, every opportunity I have been afforded, are due to Dr. Yelin's consistent support. Would that I might but emulate her in advising students of my own, then I'd be so very proud.\\
    \indent I also extend my heartfelt appreciation to Dr. Robin Côté and Dr. Alex Russell for their positive encouragement and selfless dedication as members of my thesis committee, and to Dr. Philip Mannheim for opening door to the UConn physics department for me.\\
    \indent To my family, and especially my wife Eli, who somehow accommodated my excessive and unpredictable working hours: 
    You surely do not know the extent of your gift: You kept me going when I was tired, you realigned me when I were distracted, and you steadily reminded me what was \emph{truly} important. I love you.\\
    \indent I am grateful to Dr. Ad{\'a}n Cabello, Dr. Robert Spekkens, and Dr. Tobias Fritz for invaluable and inspiring conversations. They responded to even inane queries with infinite patience and genuine consideration. I'm moreover deeply indebted for ongoing and upcoming opportunities they have seen fit to lobby for on my behalf.
    \indent And a final special thanks to my (lifelong) copy editor, Vera Schwarcz.}
\maketitle

\frontmatter

\begin{spacing}{1.7}
\tableofcontents

\end{spacing}

\mainmatter

\chapter{Nonlocality}
\vspace*{\fill}
\begin{spacing}{1.5}
\begin{tabularx}{400pt}{m{1cm}Lm{1cm}}
& \noindent\textit{[Quantum statistical limits are] something strange, neither mathematics nor physics, of very little interest.}
\footnote{\textsc{Anonymous Reviewer}, 1980\\ \textit{Letter to Boris Tsirelson} rejecting to publish Tsirelson's derivation of his historic inequalities.
\\\citetext{quoted by Tsirelson in private communication}\\} & \\ & & \\
& \noindent\textit{Nonlocality is the most characteristic feature of quantum mechanics.}
\footnote{\textsc{Sandu Popescu}, 2014\\ \textit{Nature Physics Insight, Quantum Foundations}, summarizing the modern perspective \cite{PopescuReviewNatureComm}.} &
\end{tabularx}
\end{spacing}
\vspace*{\fill}

\section{From Classical to General No-Signalling}\label{sec:intro}
    It is now well understood that quantum measurements are contextual in the sense that the outcome of a measurement on a given subsystem can depend on the {\em context} of global measurements \cite{SpekkensSeer,Pentagrams,ExperimentalContextuality,ks18,CSWNew}. A {\em context} here means a tuple which uniquely maps all the subsystems to some measurements choices. A {\em context} can be {\em nonlocal} if the respective subsystems are spatially separated and the measurement choices for each {\em party}'s subsystem are selected nearly simultaneously. In such a scenario there is not enough time for a signal to propagate across the entire system to inform the subsystems of which global context has been selected. Nonetheless, the measurement outcomes manifest a striking awareness of the global context by yielding noticeably different statistics depending on the particular global context in play \cite{WernerWolfe2001,GisinFramework2012,BellAndNonContextualInequalities,ContextualityLeggettGarg,ContextualityWithSignalling,QMysteriesAravid,FritzCorrelationScenarios,GriffithsLocality,BellCausalityReview,BellCausalityArXiv,BellAssumptionsWiseman,Brunner2013Bell}.
   
    It is inevitably surprising to first learn that our universe does indeed exhibit such quantum nonlocality. How can such a thing be, if we accept that no signal can traverse any distance faster than light could? The answer is that the {\em marginal} statistics are independent of the global context. If we split the global system in two subsystems of any size, say subsystems $A$ and $B$, then the probability of seeing some set out measurement outcomes in $A$ upon selecting some choice of measurements for subsystem $A$, is guaranteed to be identical no matter what choice of measurements are selected for the space-like separated subsystem $B$. Thus, even Bayes' Theorem cannot offer any insight into which measurement choices were selected in subsystem $B$ given only the information accessible to system $A$. As such, the law of {\em No Signalling} is respected, even by quantum nonlocality. 
   
    So, we live in a nonlocal world: A world in which the systems we investigate demonstrate a sensitivity to global contexts but which nonetheless give no hint of what the global context actually is to the local investigators. Indeed, the investigators must collaborate through classical communication in order for them to even conclude that nonlocality has happened, as the evidence is concealed in the {\em correlation} (or anticorrelation) statistics. The specific classical assumptions which must be discarded upon witnessing such nonlocal correlations are discussed in Refs. \cite{FritzCorrelationScenarios,GriffithsLocality,BellCausalityReview,BellCausalityArXiv,BellAssumptionsWiseman}.
    
    If one accepts that the universe is nonlocal, one might presume that all forms of nonlocality which respect the no-signalling principle should be permitted. This, however, is not the case.
    
    A careful analysis of the mathematical formalism underpinning quantum theory reveals that there are conceivable nonlocal no-signalling statistics which are nevertheless completely inaccessible through quantum mechanics \cite{Tsirelson1980,PROriginal,PRUnit,BellSimple,roberts_thesis,HowMuchLarger,DiscreteEntanglementNonlocality,Brunner2013Bell,PopescuReviewNatureComm}. This no-go theorem is independent of the specific physical quantum system which might be utilized, or even the dimension of the Hilbert space: quantum mechanics excludes certain no-signalling statistics. And thus, the stage is set. The world is nonlocal, but not maximally nonlocal. One can define the set of all local statistics \cite{GeneralNoSignalling,WringingBellInequalities}, and one can define the set of all nonlocal no-signalling statistics \cite{GisinFramework2012,BellInequalitiesReview}, but the quantum set is intermediate.
    
    The first chapter of this thesis is dedicated to advancing and improving our understanding of the quantum set \cite{MixedStateQB,PeresBellMaximization99,WernerWolfQB,MinMaxQB,WehnerQB,WolfeQB,LatestTsirelson}. Indeed, while closed-form linear inequalities are known which tightly define local and nonlocal statistical \glspl{polytope} \cite{PROriginal,PRUnit,Brunner2013Bell,PopescuReviewNatureComm,ScaraniNotes,*ScaraniNotes2}, no closed-form description has been developed for the quantum set. The quantum boundary is certainly nonlinear \cite{MasanesFourD,HowMuchLarger,NPA2007Short}, and there is presently a conjecture that it cannot even be defined by any guaranteed-terminating algorithm  \citep[Conjecture (8.3.3)]{FritzCombinatorialLong}. Our contributions to this effort, reviewed herein, include the development of new inequalities which inscribe the quantum set \cite{WolfeQB,WolfeQBExperimental}; more broadly, this work supplies {\em formal justification of the method} used to derive the inequalities, correcting and explaining the pioneering work of \citet{Tsirelson1980}\footnote{``Cirel'son'' is the Romanization from the Cyrillic ``Цирельсон'' employed in articles published before 1983. ``Tsirel'son'' was the form used this author from 1983 through 1991. Since Tsirelson emigrated to Israel from Russia in 1991 he has used only the form ``Tsirelson'', which is the form we use in this thesis {\em except when explicitly citing} an early work. See \href[pdfnewwindow ]{http://www.math.tau.ac.il/~tsirel/faq1.html}{http://www.math.tau.ac.il/~tsirel/faq1.html} and Mathematical Reviews' 1982 author index, appendix C, page C1.}, which was published without proof. Entirely unpublished prior to this thesis is our discussion explaining the role of Hermitian polynomials in justifying Tsirelson's method, as well as the explicit contrast of our analytic quantum bounds with limits inferred from the principles of Information Causality \cite{ICOriginal,ICRecovery,ICGrayWyner,InfoCausArXiv}, Macroscopic Locality \cite{MacroscopicLocalityOriginal,ScaraniML,ICFailure}, and Local Orthogonality \cite{LONatureComm,FritzCombinatorialLong,LONewShort,LOHardy,LOExploring}.

\section{Qubits are Sufficient for Binary and Dichotomic}\label{sec:qubitssufficient}
    A nonlocality scenario is described by the number of parties who are spatially separated, the number of measurement choices accessible to each party, and the (discrete) dimension of the possible measurement outcomes associated with each measurement. The set of quantum-compatible statistics are those which can be reproduced by associating every measurement with a set of quantum projectors, assigning some global quantum state to be shared among the parties, and ensuring that {\em order} of measurements is irrelevant by imposing commutation for any projector pair belonging to two distinct parties. In practice it is convenient to implement this commutation structure by simply assigning each party to their own Hilbert space. It is not presently known if perhaps some generality is lost by using distinct Hilbert spaces for the parties, but to date there is no evidence for such a loss of generality. This question is known as Tsirelson's Problem, see Refs. \cite{TsirelsonProblemArXiv,ConnesEmbedding,PhysicalTsirelson,KirchbergConjecture}.
   
    So, our goal is to decide if a given statistical ``point'' is quantum-compatible for some nonlocality scenario. A statistical point, here, refers to a tuple of statistics defining some experimental results. This is equivalently referred to as a ``behavior'' \cite{NPA2008Long} or a ``probabilistic model'' \cite{FritzCombinatorialLong}.  There is no simple test to verify the quantumness of a statistical point, although the NPA hierarchy \cite{NPA2007Short,NPA2008Long,DohertyNPA,NPAReview,FritzCombinatorialLong} provides an asymptotically converging algorithm which becomes computationally intractable very quickly even for the most elementary of nonlocality scenarios. 
    
    If one restricts consideration to a particular shared quantum state, then it becomes relatively straightforward to determine all the nonlocal statistics (if any) that can be accessed by measurements on that state \cite{GivenQB.NQubit,GivenQB.Cabello,GivenQB.TwoQubit,GivenQB.arXiv}. One very important result, for example, is that the shared quantum state must be entangled in order to exhibit any nonlocality \cite{NonlocalityWithoutEntanglement,WernerWolfQB,NeedEntanglement,CryptoPRA,MultipartiteEntanglement}, and every entangled state can exhibit some form of nonlocality \cite{HiddenNonlocality,UnifiedCorrelations,PPTNonlocal,AllEntangledNonlocality,AllEntangledNonlocal,CloserConnections}. The goal in this thesis, however, is not to be tied down to any particular quantum state. Rather, we seek to quantify the set of nonlocal statistics accessible through quantum measurements upon {\em any} shared entangled state. 
    
    Although much progress has been made toward this goal \cite{MixedStateQB,WernerWolfQB,MinMaxQB,WehnerQB,VerstraeteBellMonogamy}, nevertheless, we find that increasing the dimension of the Hilbert spaces unlocks monotonically more nonlocal statistics. Thus, infinite dimensionality is required for true generality. One can even infer the dimension of the Hilbert spaces from the extent of some observed nonlocality \cite{Dimension2008,Dimension2010,Dimension2013,Dimension2013Cabello,Dimension2014Cabello}. 
    
    There is an important exception \cite{MasanesQubits}, which can be leveraged to great effect \cite{WolfeQB}. What \citet{MasanesQubits} proves (which \citet{Tsirelson1980} presumed) is that if the parties have access to two measurement choices each (``dichotomic''), and each measurement yields one of only two possible outcomes (``binary''), then shared qubits are completely sufficient to unlock all possible quantum nonlocal statistics; Using higher dimension Hilbert spaces would not provide any advantage. This means that no matter how multipartite the scenario, it can be studied without loss of generality by considering nothing more than shared qubits, as long as it is is binary and dichotomic. 
    
    
    For dichotomic and binary scenarios it is possible in principle to determine if some candidate statistics are quantum compatible by assessing whether or not they can be achieved using distributed qubits. To mathematically formalize what we mean, let us introduce a conventional notation for discussing binary and dichotomic multipartite scenarios. Firstly, it is convenient to think of the outcomes as equal to $\pm 1$ so that we can easily define both marginal and correlation expectation values,
\LeftEqn{\label{eq:expectoprobs}
    \Braket{A_x} &\equiv Prob\left(A_x=1\right)-Prob\left(A_x \neq 1 \right) \\
    \Braket{A_x B_y} &\equiv Prob\left(A_x \times B_y = 1\right)-Prob\left(A_x \times B_y \neq 1\right)  \\
    \Braket{A_x B_y C_z} &\equiv Prob\left(A_x\times B_y \times C_z = 1\right)-Prob\left(A_x\times B_y \times C_z \neq 1\right) 
}
where the subscript indexes each party's measurement choice, $x,y,z\dotsc\in\brackets{0,1}$, such that the total number of independent statistical parameters in given binary and dichotomic no-signalling nonlocality scenario is equal to $3^k-1$ where $k$ is the number of parties, consistent with more general parameter counts derived by \citet[Eqn. (9)]{PironioDimension} and \citet[Eqn. (8)]{Brunner2013Bell}.

Suppose we have some linear function of observables, such as a Bell inequality \cite{BellOriginal,GisinFramework2012,VariousCHSHBellInequalities,BellInequalitiesReview,WringingBellInequalities}, and we are interested in determining the maximum of this linear function consistent with multipartite quantum mechanics. Quantum generalizations of Bell inequalities are known as Tsirelson inequalities \cite{Tsirelson1980,WolfeQB}. A Tsirelson inequality is a general statement about quantum-compatible statistics; a Tsirelson inequality can be used to reject a statical point. If a point lies within a given Tsirelson inequality, however, this does not mean it is necessarily quantum compatible. For a point to be quantum compatible it must be within {\em every} possible Tsirelson inequality. Thus, the task of determining such quantum inequalities is a proxy task for assessing the quantum compatibility of a given point.\footnote{To quote Ref. \cite{ConsistentHistoriesContextuality}: ``A research program of `characterising quantum non-locality' has arisen with two closely related goals: to provide a method of determining whether a given experimental behaviour could have been produced by an ordinary quantum model, and to discover physical or information-theoretic principles that result in constraints on possible behaviours.'' Tsirelson inequalities are a tool for achieving the former goal.}

As noted by \citet[Theorem 1]{Tsirelson1980}, the maximum value of a quantum measurement is equal to its largest eigenvalue. Since the measurement operator is independent of the state, we can determine its maximal eigenvalue by constructing the most general quantum measurement without having to also express a general quantum state.\footnote{The state-independent approach to determining quantum maxima only applies if the target function is being maximized over \emph{all} quantum compatible points. If, on the other hand, some probabilistic degrees of freedom are pre-specified, then a quantum state must be introduced to perform constrained optimization. The best known example of quantum optimization over partially specified probability distributions is Hardy's nonlocality \cite{Hardy92,Hardy93,Hardy97,CabelloHardy,HardyCHSH}. Note that qubits are still sufficient to map out all binary and dichtomic Hardy nonlocality [\citealp{HardyQubits06,HardyQubits11,HardyQubits12,HardyVarientDIversion} and \citealp[Sec. IV.E]{SpekkensSeer}].} Following \citet{WolfeQB} we note that it is efficient to use a reflection symmetry to parameterize the two measurements of each party. Therefore, the general quantum measurement operator lives in the ${\mathbb{C}^2}^{\otimes k}$ Hilbert space, and has only one free variable per party. Our task is to determine the largest possible value this operator's largest eigenvalue can take. The Hermiticity of the quantum measurement is a distinct advantage in this process, as we now explain.

\section{The Advantage of Hermitian Polynomials}
A Hermitian matrix has only real eigenvalues. This has an important consequence for semidefinite programming. Suppose we have a matrix where some of the matrix elements are variables, and we would like to either maximize the largest eigenvalue or minimize the smallest eigenvalue over these variables. At first, this appears to be a problem of simultaneous optimization. The problem, however, can be recast more efficiently if we are guaranteed that the eigenvalues are real.

A degree $N$ polynomial has $N$ roots, by the fundamental theorem of algebra, however multiple roots can coincide if a root is also a critical point of the polynomial. Pursuant to the intermediate value theorem, we know there must be at least one critical point between any two {\em real} roots of a polynomial, as it must have changed direction in order to intersect the x-axis again. Note that this intermediacy of a critical point generalizes even to the case of a degenerate root, in the sense that we can define the critical point to be intermediate to two of the degenerate roots, effectively sandwiching it between two real roots separated by zero value. In other words,
\LeftEqn{\label{eq:polydef}
&\text{Let } p^{(0)}(x) \text{ be a polynomial in $x$}\text{, with }p^{(q)}(x) \equiv \frac{\partial^q p^{(0)}(x)}{\partial x^q}
\\&\text{where }\forall_{q\geq 0,i< j} :\: p^{(q)}\left(r^{(q)}_i\right) = 0 \text{ and } r^{(q)}_i \in \mathbb{R}  \text{ and } r^{(q)}_i \leq r^{(q)}_j
\\&\text{then }\quad \exists_k:\text{ such that }r^{(0)}_i \leq r^{(1)}_k \leq r^{(0)}_j \,,.
}
or, alternatively,
\LeftEqn{
\text{LEMMA:  }&\text{A polynomial with $N$ real roots (not necessarily distinct)}
\\&\text{has {\em at least} $N-1$ critical points (not necessarily distinct)}
\\&\text{{\em in the region} spanned by the real roots.}
}
If the polynomial does have complex roots, then it is possible for there to be critical points smaller than the smallest real root or larger than the largest real root, for example $p^{(0)}(x)=x^6-5x^4+7x^2-1$. When the polynomial is the characteristic polynomial of a Hermitian matrix, however, {\em all} the critical points must be within the range of the real roots; the fundamental theorem of algebra dictates that there are more roots of a polynomial than roots of its derivative. The intermediate value theorem demands that every pair of real roots sandwich some critical point.
\LeftEqn{
&\text{Let } p^{(0)}(x) \text{ be a {\em Hermitian} polynomial in $x$}
\\&\text{then }\quad \forall_{k} :\: \exists_{i<j}\text{ such that }r^{(0)}_i \leq r^{(1)}_k \leq r^{(0)}_j
}
Moreover because {\em every} root of the derivative now must be real in order to be intermediate to real roots, we can inductively realize the stronger statement
\LeftEqn{
&\text{Let } p^{(0)}(x) \text{ be a {\em Hermitian} polynomial in $x$}
\\&\text{then }\quad \forall_{q\geq 1,k} :\: \exists_{i<j}\text{ such that }r^{(q+1)}_i \leq r^{(q)}_k \leq r^{(q+1)}_j\,.
}
In other words,
\LeftEqn{\label{lemma:RootsAllInSpan}
\text{LEMMA:  }&\text{All the roots of a Hermitian polynomial's $N$'th derivative}
\\&\text{are real, and lie between the $(N-1)$'th derivative's roots.}
}
If we define {\em change} as any reversal of the sign of the polynomial or the sign of any of its derivatives, then Lemma \noeq{lemma:RootsAllInSpan} tells us that a Hermitian polynomial\footnote{Polynomials with exclusively real roots are also known as \emph{hyperbolic} polynomials \cite{Hyperbolic97,Hyperbolic01}.} has {\em no change} whatsoever in the region larger than its largest root or smaller than its smallest root. 

This allows us to give a somewhat unconventional definition for the largest real root or the smallest real root of a Hermitian polynomial in terms of infimum and supremum:
\LeftEqn{
&\text{Let } p^{(0)}(x) \text{ be a {\em Hermitian} polynomial in $x$, then}
\\&\quad r^{(0)}_{\max{}} \equiv \inf_{z\in \mathbb{R}}z\, { \text{, such that } \forall_{q\geq 0}:\: \operatorname{sgn}{\left[p^{(q)}(z)\right]} = \operatorname{sgn}{\left[\lim_{x\to \infty}{ p^{(q)}(x)}\right]}}
\\&\quad r^{(0)}_{\min{}} \equiv \sup_{z\in \mathbb{R}}z\, { \text{, such that } \forall_{q\geq 0}:\: \operatorname{sgn}{\left[p^{(q)}(z)\right]} = \operatorname{sgn}{\left[\lim_{x\to -\infty}{ p^{(q)}(x)}\right]}}\,.
}
The limits at $\pm \infty$ depend only on the leading coefficient of the polynomial, so the relevant concept can be simplified.
\LeftEqn{
&\text{Let } p^{(0)}(x) \text{ be a degree-$N$ {\em Hermitian} polynomial in $x$}
\\&\text{with leading term }c \times x^N,\text{ then taking }\tilde{p}^{(0)}(x)\equiv p^{(0)}(x)/c\,,
\\&\quad r^{(0)}_{\max{}} \equiv \inf_{z\in \mathbb{R}}z\, { \text{, such that } \forall_{q\geq 0}:\: \tilde{p}^{(q)}(z) > 0}
\\&\quad r^{(0)}_{\min{}} \equiv \sup_{z\in \mathbb{R}}z\, { \text{, such that } \forall_{q\geq 0}:\: (-1)^{N-q}\times \tilde{p}^{(q)}(z) > 0}\,.
}

This transforms the nature of the task of maximizing the largest eigenvalue or minimizing the smallest eigenvalue of a rank $N$ matrix. Instead of simultaneous optimization over multiple variables, the problem is recast now into optimization over the single dummy variable of the characteristic polynomial, but subject to $N$ ``for all''-type constraints which involve the matrix's genuine variable elements. For example,
\LeftEqn{\label{eq:minmax}
&\text{Let } p^{(0)}(x) \text{ be the characteristic polynomial of a Hermitian matrix}
\\&\text{with leading term $c \times x^N$ and with ``free'' variables $\{\mbox{vars}\}$, then}
\\&\quad \max{\left\{r^{(0)}_{\max{}}\right\}} \equiv \inf_{z\in \mathbb{R}}z\,  \text{, such that } \forall_{q\geq 0} \forall_{\{\mbox{vars}\}}:\: \tilde{p}^{(q)}(z) > 0
\\&\quad \min{\left\{r^{(0)}_{\max{}}\right\}} \equiv \inf_{z\in \mathbb{R}}z\, { \text{, such that } \forall_{q\geq 0}\exists_{\{\mbox{vars}\}}:\: \tilde{p}^{(q)}(z) > 0}
\\&\quad \min{\left\{r^{(0)}_{\min{}}\right\}} \equiv \sup_{z\in \mathbb{R}}z\, { \text{, such that } \forall_{q\geq 0}\forall_{\{\mbox{vars}\}}:\: (-1)^{N-q}\tilde{p}^{(q)}(z) > 0}
\\&\quad \max{\left\{r^{(0)}_{\min{}}\right\}} \equiv \sup_{z\in \mathbb{R}}z\, { \text{, such that } \forall_{q\geq 0}\exists{\{\mbox{vars}\}}:\: (-1)^{N-q}\tilde{p}^{(q)}(z) > 0}\,.
}
Note that the final condition of \ceq{eq:minmax} allows one to asses if a variable Hermitian matrix can - or cannot - be made positive semidefinite. The assessment can be rephrased as a question: ``Can the smallest eigenvalue be made larger than or equal to zero?'' Answering such a question requires only single-variate optimization, albeit constrained by a finite set of ``exists'' statements regarding the matrix's free variables. 

Note that for raw semidefinite programming the Hermitian matrix must be entirely numeric aside for the free variable. Our approach, however, allows one, in principle, to consider also non-numeric parameters in the Hermitian matrix aside from the free variable. We can now analytically compute the positive semidefinite domain of these parameters. That is, the condition $\max{\left\{r^{(0)}_{\min{}}\right\}}\geq 0$ effectively induces a restriction on the parameters.

We note that the Navascués-Pironio-Acín (NPA) hierarchy \cite{NPA2007Short,NPA2008Long,NPAReview} hinges on exactly this type of assessment. Their $\Gamma$ matrix, sometimes referred to as the {\em certificate}, is a Hermitian matrix parameterized by probabilities, namely both the marginal and correlation expectation values. The $\Gamma$ matrix also has extensive free variables, presumably corresponding to the expectation values of non-projective quantum operators, which are inherently unobservable. For probabilities consistent with a quantum bipartite experiment, $Gamma$ must be Hermitian and positive semidefinite. We can impose hermiticity on $\Gamma$ by construction, but the real test is demanding that $\Gamma$ be positive semidefinite. Do there exist free variable capable of making $Gamma$ positive semidefinite given the proposed experimental probabilities? Using \ceq{eq:minmax} we can compute the largest possible value of the smallest eigenvalue. If this value is less than zero, then the given probabilities are exposed as incompatible with the outcome of a quantum bipartite experiment.

Because \ceq{eq:minmax} permits non-numeric parameterization, this means that in principle one can obtain a restriction on the probabilities to be consistent with quantum from $\max{\left\{r^{(0)}_{\min{}}\right\}}\geq 0$. Indeed, the first level of the hierarchy has been already been used to define such a restriction [\citealp[Eq. (11)]{NPA2007Short}, see also \citealp[Eq. (11)]{ScaraniML}, \citealp[Eq. (3)]{ICRecovery}, and \citealp[Eq. (9)]{WolfeQB}]. It would be of great interest to derive the stronger restriction corresponding to the hierarchy level $Q^{1+AB}$ which has received extensive interest recently, as it is essential in Refs. \cite{LONatureComm,FritzCombinatorialLong,ConsistentHistoriesContextuality,AlmostQuantum2}. This special level of the hierarchy is discussed at length in Sec. \ref{sec:q1ab} of this thesis.

\section{Derivation of Linear Quantum Bounds}\label{sec:boundderiv}

We have established that the calculation of a Tsirelson inequality \cite{MixedStateQB,PeresBellMaximization99,WernerWolfQB,MinMaxQB,WehnerQB,WolfeQB,LatestTsirelson} is equivalent to a Hermitian matrix eigenvalue maximization problem, and that moreover, for binary and dichotomic scenarios, we have complete generality by considering each party as though in possession of only a single qubit. To demonstrate the general principle, we describe the calculation of a linear quantum bound for $k$ parties. 

We begin by defining a linear measurement operator to correspond to the general Tsirelson inequality with $3^k-1$ terms. For compactness, we illustrate explicitly only terms appearing in the bipartite scenario, which the multipartite generalization being fairly obvious. If the inequality is defined by seeking the quantum value of $\gamma$, where
\LeftEqn{\label{eq:simple}\underbrace{\sum\limits_{x=0}^{1}{c^A_x\Braket{A_x}}+\sum\limits_{y=0}^{1}{c^B_y\Braket{B_y}}+_{...}}_{\binom{k}{1} \text{ sums}}+\underbrace{\sum\limits_{x=0}^{1}{\sum\limits_{y=0}^{1}{c^{AB}_{xy}\Braket{A_x B_y}}}+_{...}}_{\binom{k}{2} \text{ sums}}+\cdots \,\leq\,\gamma
}
then the corresponding quantum linear operator can be defined as 
\LeftEqn{\label{eq:expec}
&Z  \equiv \sum\limits_{x=0}^{1}{c^A_x \widehat{A_x}\otimes \id\otimes_{...}}+\sum\limits_{y=0}^{1}{c^B_y \id\otimes\widehat{B_y}\otimes_{...}}+\sum\limits_{x,y=0}^{1}{{c^{AB}_{xy}\widehat{A_x}\otimes \widehat{B_y}\otimes_{...}}}+\dots
}
where without loss of generality we define each party's two operators as measurements in the x-y plane, with the second operator being the y-reflection of the first, such that
\LeftEqn{\label{eq:setup}
\renewcommand*{\arraystretch}{0.5}
\widehat{K_\alpha} \equiv v_K \widehat{\sigma_x}+(-1)^\alpha \sqrt{1-v_K} \widehat{\sigma_y} \,= \begin{pmatrix}
 0 & e^{-i (-1)^i v_K} \\
 e^{i (-1)^i v_K} & 0 
\end{pmatrix}
}
where $v_A$ and $v_B$ are the free variables in our matrix, they are arbitrary real numbers between zero and one. The operator $Z$ is therefore a $2^k\times 2^k$ Hermitian matrix. By using $\sigma_x$ and $\sigma_y$ as the spanning operators for each party, we are able to construct $Z$ such that it has exclusively zeroes along its diagonal. Let the characteristic polynomial of $Z$ be defined as $p_Z(m)$. The leading coefficient of this polynomial is equal to one, by virtue of the diagonal zeroes. 

The Tsirelson bound $\gamma$ is equal to the largest possible value of the largest root of $p_Z(m)$. Using the first definition in \ceq{eq:minmax} we can express this quite formally now as
\LeftEqn{\label{eq:gammasol}
\gamma \equiv \inf_{z\in \mathbb{R}}z\, \text{, such that }\, \forall_{q\in \mathbb{N},0\leq<q\leq 2^k-1}\: \forall_{0\leq v_A,v_B \leq 1}:\: 0< {\left. \frac{\partial^q p_Z(m)}{\partial m^q} \right\rvert}_{m=z}
}
or, since the last relevant derivative is always equal to $2^k! \times m$,
\LeftEqn{\label{eq:gammasol2}
\gamma = \inf_{z\geq 0}z\, \text{, such that }\, \forall_{q\in \mathbb{N},0\leq<q\leq 2^k-2}\: \forall_{0\leq v_A,v_B \leq 1}:\: 0< {\left. \frac{\partial^q p_Z(m)}{\partial m^q} \right\rvert}_{m=z}
}

From here on out let us specialize further and consider only the bipartite case of a binary and dichotomic scenario; we refer to this bipartite binary and dichotomic scenario as the (2,2,2) scenario \cite{GisinFramework2012}, although it is also often referred to as the CHSH scenario \cite{CHSHOriginal}. It is the most fundamental scenario in which quantum nonlocality is observed. The explicit conditions to determine any Tsirelson inequality in the (2,2,2) scenario are
\LeftEqn{\label{eq:gammasol222}
\gamma = \inf_{z\geq 0}z\, \text{, such that }\, \forall_{0\leq v_A,v_B \leq 1}:\: 0< \min\brackets{p_Z(m),p_Z^{\backprime}(m),p_Z^{\backprime\backprime}(m)}\,.
}

We calculated various novel marginal-involving Tsirelson inequality according to this procedure in \citet{WolfeQB}. Our results are summarized here in \tab{tab:bounds}.

\newgeometry{left=1.00in,right=1.0in,top=1.4in,bottom=1.4in}
\begin{landscape}
\renewcommand*{\arraystretch}{2.5}
\begin{table*}[hbtp]
\caption{\label{tab:bounds}Linear Bounds (Number-Valued and Function-Valued): Note that \({\mbox{TB}}^{(4)}\) is the well-known Tsirelson's bound, and \({\mbox{QB}_1}^{(4)}\) is its function-valued generalization. The bounds tabulated here should be read as $\sum\limits_{x=0}^{1}{c^A_x\Braket{A_x}}~+~\sum\limits_{y=0}^{1}{c^B_y\Braket{B_y}}~+~\sum\limits_{x=0}^{1}{\sum\limits_{y=0}^{1}{c^{AB}_{xy}\Braket{A_x B_y}}}~\leq~(?)\,$, where the upper bound to the inequally can be determined for each class of general probabilistic theory, making the inequality a Bell Inequality, a Tsirelson Inequality, or a convex-hull description of the No-Signalling \gls{polytope}.  \({\mbox{QB}_3}^{(8)}\) is invoked repeatedly in this thesis to construct constraints on the quantum region of various slices of the no-signalling polytope. See \ceq{eq:qb3forall} and \ceq{eq:qb3beta} in particular.}
\begin{tabularx}{\linewidth}{l|cccccccL|c|c|c}
 Name & $c^A_0$ & $c^A_1$ & $c^B_0$ & $c^B_1$ & $c^{\hspace{-0.5ex}AB}_{00}$ & $c^{\hspace{-0.5ex}AB}_{10}$ & $c^{\hspace{-0.5ex}AB}_{01}$ & $c^{\hspace{-0.5ex}AB}_{11}$ & \(\text{LHVM}_\text{Max}\) & \(\text{QM}_\text{Max}\) & \(\text{NOSIG}_\text{Max}\) \\
\midrule 
\({\mbox{TB}}^{(4)}\) & 0 & 0 & 0 & 0 & 1 & 1 & 1 & -1 & \({2}\) & \({2\sqrt{2}}\) & \({4}\) \\
\({\mbox{QB}_1}^{(4)}\) &  0 & 0 & 0 & 0 & 1 & 1 & 1 & x & \(|x+1|+2\) & \({\begin{cases}
  x+3 & \forall\; x\geq -\frac{1}{3} \\
  \sqrt{(x-1)^3/x} & \forall\; x\leq -\frac{1}{3}
 \end{cases}}\) & \(|x|+3\) \\
\midrule 
 & 1 & 0 & 0 & 0 & 1 & 1 & 1 & -1 & 3 & \(\sqrt{10}\) & 4 \\
 \({\mbox{QB}_2}^{(8)}\) & x & 0 & 0 & 0 & 1 & 1 & 1 & -1 & \(|x|+2\) & \({\begin{cases}
  |x|+2 & \forall\; |x|\geq 2 \\
  \sqrt{2x^2+8} & \forall\; |x|\leq 2
 \end{cases}}\) & \(\begin{cases}
  |x|+2 & \forall\; |x|\geq 2 \\
  4 & \forall\; |x|\leq 2
 \end{cases}\)\\
\midrule 
 & 1 & 1 & -1 & 0 & 1 & 1 & 1 & -1 & 3 & \(3\) & 4 \\
 \({\mbox{QB}_3}^{(8)}\) & x & x & -x & 0 & 1 & 1 & 1 & -1 & \(\begin{cases}
  3|x|-2 & \text{for } |x|\geq 2 \\
  |x|+2 & \text{for } |x|\leq 2
 \end{cases}\) & \({\begin{cases}
  3|x|-2 & \forall\; |x|\geq 2 \\
  |x|+2 & \forall\; 1\leq |x|\leq 2 \\
  \frac{x^2-\sqrt{(2-x^2)(4-3 x^2)}}{x^2-1} & \forall\; |x|\leq 1
 \end{cases}}\)  & \(\begin{cases}
  3|x|-2 & \forall\; |x|\geq 2 \\
  4 & \forall\; |x|\leq 2
 \end{cases}\)
\end{tabularx}
\end{table*}
\end{landscape}
\restoregeometry

\section{Comparison of Quantum Bounds}\label{sec:boundcomp}
Let us return to our primary task of trying to characterize the genuinely quantum \gls{elliptope} of statistical points. The fundamental (2,2,2) general No-Signalling \gls{polytope} \cite{PROriginal,PRUnit,Brunner2013Bell,PopescuReviewNatureComm,ScaraniNotes,*ScaraniNotes2,scarani2012device} is the convex hull of the 16 local deterministic points and the 8 PR-box maximally nonlocal points. The local hidden variable polytope is the convex hull of only the 16 local boxes. The facets of the classical polytope are the 8 Bell inequalities. Every Bell inequality is saturated by six local points. At each facet, the quantum elliptope must touch the relevant 6 local points but extend ``upward'', approaching (but not reaching) the maximally nonlocal point which lies above the facet. 


We would like to map a quantum curve in this region. To do so we consider a plane spanned by three boxes: The PR box, the all-one fully-deterministic box, and the ``origin''. In this thesis the concept of a probabilistic box is completely equivalent to a point in statistical space, or a  ``behavior'' \cite{NPA2008Long} or a ``probabilistic model'' \cite{FritzCombinatorialLong}. A box is nothing more than some multipartite conditional probability distribution. The plane spanned by these three points is exactly the ``slice'' studied in both Refs. \citep[Fig. 3]{ICRecovery} and \citep[Fig. 4]{LONatureComm}. 

The PR box \cite{PROriginal,PRUnit,Brunner2013Bell,PopescuReviewNatureComm,ScaraniNotes,*ScaraniNotes2} corresponds to the extremal nonlocal probability point which perfectly correlates Alice and Bob for three out of four of their possible pairwise measurement choices, but perfectly anticorrelates them should they happen to both choose measurements with index 1. It is named after \citet{PROriginal}, and can be expressed as
\LeftEqn{\label{eq:prboxdef}
P_{\mbox{PR}}(ab|xy) = \frac{\mathlarger{\delta_{\left(a\oplus b  = x y\right)}}}{2}
} 
where the $\oplus$ in \ceq{eq:prboxdef} indicates addition module two, ie. the bitwise XOR function.

The all-one fully-deterministic box is an extremal local point which not only saturates the conventional Bell inequality \cite{roberts_thesis} but also has no dependence whatsoever on the choice of Alice's or Bob's measurement choices. 
We indicate such measurement-invariant local boxes, even if they are {\em not} deterministic, by two numbers
\LeftEqn{\label{eq:plocalinvariant}
\renewcommand*{\arraystretch}{0.5}
P_{m,n}(ab|xy)=\left(\left\lbrace\begin{array}{rl} {m,}& {a=1} \\  {1-m,}& {a=0}\end{array}\right.\right)\times\left(\left\lbrace\begin{array}{rl} {n,}& {b=1} \\  {1-n,}& {b=0}\end{array}\right.\right)
}
noting that special cases include the fully-deterministic all-one box
\LeftEqns{\label{eq:pallone}
\renewcommand*{\arraystretch}{0.5}
&\text{All-one box:} & P_{\mbox{FD}} \equiv\,\, & P_{1,1}(ab|xy)=\left\lbrace\begin{array}{rl} {1,}& {a=b=1} \\  {0,}& {\text{otherwise}}\end{array}\right.  
&}
 the semi-deterministic Bob-random box\LeftEqns{\label{eq:pbobrandom}
&\text{Bob-random box:} & P_{\mbox{SD}} \equiv\,\, & P_{1,\frac{1}{2}}(ab|xy)=\left\lbrace\begin{array}{rl} {1/2,}& {a=1} \\  {0,}& {a=0}\end{array}\right. 
&}
and the maximally random ``white noise'' box
\LeftEqns{\label{eq:pallrandom} 
&\text{All-random box:} & P_{\mathlarger{\operatorname{{\emptyset}}}} \equiv\,\, & P_{\frac{1}{2},\frac{1}{2}}(ab|xy)=\frac{1}{4} &
}
The ``origin'' in statistical space refers to precisely the maximally random box, with no positive or negative bias along any expectation value, joint or marginal. We denote it here as $P_{\mathlarger{\operatorname{{\emptyset}}}}$, as per \ceq{eq:pallrandom}. 

We introduce a mixed box spanned by $P_{\mbox{PR}},\:P_{\mbox{FD}},\text{ and }P_{\mathlarger{\operatorname{{\emptyset}}}}$ such that
\LeftEqn{\label{eq:boxmix}
&P_{\xi\gamma}(ab|xy)=\xi P_{\mbox{PR}}(ab|xy)+\gamma P_{\mbox{FD}}(ab|xy)+\left(1-\gamma-\xi\right)P_{\mathlarger{\operatorname{{\emptyset}}}}(ab|xy)
}
and proceed to assess the constraints on $\gamma$ and $\xi$ implied by presuming that the experimental outcomes are mediated through fundamentally quantum mechanisms. The box $P_{\xi\gamma}$ defines the 2-dimensional statistical region
\LeftEqn{
&\Braket{A_0}=\Braket{A_1}=\Braket{B_0}=\Braket{B_1}=\gamma\,,
\\&  \Braket{A_0 B_0}=\Braket{A_1 B_0}=\Braket{A_0 B_1} = \gamma+\xi \,,\quad \Braket{A_1 B_1} = \gamma-\xi\,.
}
A well-known but relatively weak quantum bound is Uffink's bound \cite{Uffink,HowMuchLarger}, which also corresponds to the principle of Information Causality \cite{ICOriginal,ICRecovery,InfoCausArXiv}. 
\LeftEqns{\label{eq:uffink}
\left(\Braket{A_0 B_0}-\Braket{A_1 B_1}\right)^2+\left(\Braket{A_0 B_1}+\Braket{A_1 B_0}\right)^2\leq 4
}
For $P_{\xi\gamma}$ Uffink's bound yields the restriction
\LeftEqns
{
\xi\leq\frac{\sqrt{2-\gamma ^2}-\gamma}{2}
}

A much stronger quantum bound is that corresponding to the first level of the \GLS{NPA} hierarchy \cite{NPA2007Short,NPA2008Long}, which also corresponds to the principle of Macroscopic Locality \cite{MacroscopicLocalityOriginal,ScaraniML,ICFailure}. It is the most commonly used bound in modern quantum foundations comparisons, because it nicely balances explicit analytical accessibility with strong restrictive power. We denote this criterion by $\mbox{NPA}^1$, and it states that
\LeftEqn{\label{eq:NPA1}\sum\limits_{x,y=0}^{1}{{(-1)}^{x y}\arcsin{\left(\frac{\Braket{A_x B_y}-\Braket{A_x}\Braket{B_y}}{\sqrt{\left(1-\Braket{A_x}^2\right)\left(1-\Braket{B_y}^2\right)}}\right)}}\leq \pi
}
For $P_{\xi\gamma}$ the $\mbox{NPA}^1$ criterion it takes the explicit form
\LeftEqn{\label{eq:npaboxsum}\sum\limits_{x,y=0}^{1}{\arcsin{\left(\frac{{(-1)}^{x y}\gamma\left(\gamma-1\right)-\xi}{\gamma^2-1}\right)}}\leq \pi
}
One can invoke the identity
\LeftEqns{
\sin{\left(n\arcsin{\left(z\right)}\right)}=\sum\limits_{k=0}^{\infty}{\frac{n z^{(2k+1)}(2k-1+n)!!(2k-1-n)!!}{(2k+1)!(-1+n)!!(-1-n)!!}}
}
which, due to the $2k-1-n$ double factorial, terminates cleanly for odd $n$. Using $n=3$ to consolidate like terms from the sum in \ceq{eq:npaboxsum} we obtain
\LeftEqns{
\frac{(\gamma  (\gamma -1)-\xi )}{\gamma ^2-1} \left(4 \left(\frac{\gamma  (\gamma -1)-\xi }{\gamma ^2-1}\right)^2-3\right)\leq\frac{(\gamma -1) \gamma +\xi }{\gamma ^2-1}
}
which in terms of a bound on $\xi$ equivalent to
\LeftEqn{
\xi \leq &\gamma(\gamma-1)+\frac{\left(\gamma^2-1\right)^{2}}{\operatorname{h}{\left(\gamma\right)}}+\frac{\operatorname{h}{\left(\gamma\right)}}{6} \quad\text{with}
\\& \operatorname{h}{\left(\gamma\right)}\equiv \sqrt[3]{6 (1-\gamma )^3 (\gamma +1)^2 \left(9 \gamma +\sqrt{3 \gamma  (25 \gamma -4)-6}\right)}
}

We also would like to consider $\mbox{QB}_3^{(8)}$ from our earlier work seeking out new Tsirelson inequalities \cite{WolfeQB}, reproduced here in \tab{tab:bounds}. We specifically select $\mbox{QB}_3^{(8)}$ in order to illustrate its superior restrictive power relative to $\mbox{NPA}^1$ on this slice of the no signalling \gls{polytope}. It reads
\LeftEqns{\label{eq:qb3reproduced}
\forall_{c^2 \leq 1}\mathlarger{:}\, \begin{matrix}c\left( \left\langle A_0\right\rangle +\left\langle A_1\right\rangle -\left\langle B_0\right\rangle\right) -\left\langle A_1 B_1\right\rangle
\\ +\left\langle A_0 B_0\right\rangle +\left\langle A_1 B_0\right\rangle +\left\langle A_0 B_1\right\rangle \end{matrix} \leq \frac{c^2-\sqrt{\left(3 c^2-4\right) \left(c^2-2\right)}}{c^2-1}
}
For $P_{\xi\gamma}$ the left hand side of $\mbox{QB}_3^{(8)}$ is $(2+c)\gamma+4\xi$ such that we are effectively bounding $\xi$ by 
\LeftEqns{\label{eq:qb3forall}
&\xi\leq \frac{\min\limits_{c^2 \leq 1}{\bigg\lbrace{\frac{c^2-\sqrt{\left(3 c^2-4\right) \left(c^2-2\right)}}{c^2-1}-c \gamma\bigg\rbrace}}-2\gamma}{4}
\\\nonumber&\text{or, in Mathematica{\it\tiny\texttrademark},}\,\;\mbox{
\texttt{$\xi\leq \frac{\text{MinValue}\left[\left\{\frac{c^2-\sqrt{\left(3 c^2-4\right) \left(c^2-2\right)}}{c^2-1}-c \gamma ,0<c<1\right\},\{c\}\right]-2\gamma}{4}$}
}
}
which resolves to the explicit envelope defining the entire set of linear bounds\footnote{We have found \texttt{MinValue} to be the most efficient approach in Mathematica{\it\tiny\texttrademark}, but the envelope can also be derived using \texttt{Refine} and \texttt{ForAll}.}, namely via the minimal root of an order 8 (non-Hermitian) polynomial, 
\LeftEqns{\label{eq:envelope}
&\xi \leq \frac{\operatorname{\lambda_{\min}}{\big(\gamma\big)}-2\gamma}{4}\,\,\,,\,\,\,\,\text{ where}
\\\nonumber&\operatorname{\lambda_{min}}{\big(\gamma\big)} \equiv \min\limits_{\lambda_1\cdots\lambda_8}\text{ such that }\quad\Bigg(3 \left(\lambda ^2-8\right) \left(\lambda ^2+2 \lambda -2\right)^3+
\\\nonumber&
8 \gamma ^8-2 \gamma ^6 \left(13 \lambda ^2-56 \lambda +220\right)+\gamma ^4 \left(31 \lambda ^4-104 \lambda ^3-470 \lambda ^2+960 \lambda +4080\right)
\\\nonumber&-\gamma ^2 \left(16 \lambda ^6+10 \lambda ^5-449 \lambda ^4-1060 \lambda ^3+1360 \lambda ^2+7264 \lambda +8032\right)\Bigg)=0
}

We additionally consider a quantum bound due to Local Orthogonality \cite{LONatureComm,FritzCombinatorialLong,LONewShort,LOHardy,LOExploring}. In particular, we invoke the 10-term $\mbox{LO}^2$ inequality\footnote{This \gls{clique} was conveyed in private communication from the authors of Refs. \cite{LONatureComm}.}
\LeftEqn{\label{eq:RafaelClique}
&P(1111|1100)+P(1110|1100)+P(1101|1001)+P(1100|0001)
\\+&P(1011|1101P+P(1001|1001)+P(0111|1100)
\\+&P(0011|1111)+P(0010|0111)+P(0000|1010)\;\quad\leq\; 1
}
where the four-partite boxes in \ceq{eq:RafaelClique} are to be understood as a \emph{wiring} of two bipartite boxes \cite{LOExploring}, such that 
\LeftEqns{
P(abcd|xyzw)=P(ab|xy)\times P(cd|zw)\,.
}
If both the bipartite boxes that comprise $P(abcd|xyzw)$ are identical copies of $P_{\xi\gamma}$ then 
\LeftEqns{
&\frac{\xi  \left(3 \xi -2 \left(\gamma -1\right)\right)-3 \left(\gamma -2\right) \gamma +5}{8}\leq 1 \\
&\text{or, as a bound on }\,\xi:\quad \xi \leq \frac{\sqrt{10}-1}{3}\left(1-\gamma\right)
}
which is precisely what is plotted in Figure 4 of Ref. \cite{LONatureComm}.

\begin{figure}[ht]
\centering
\begin{minipage}[t]{1\textwidth}
    \includegraphics[width=1.0\linewidth]{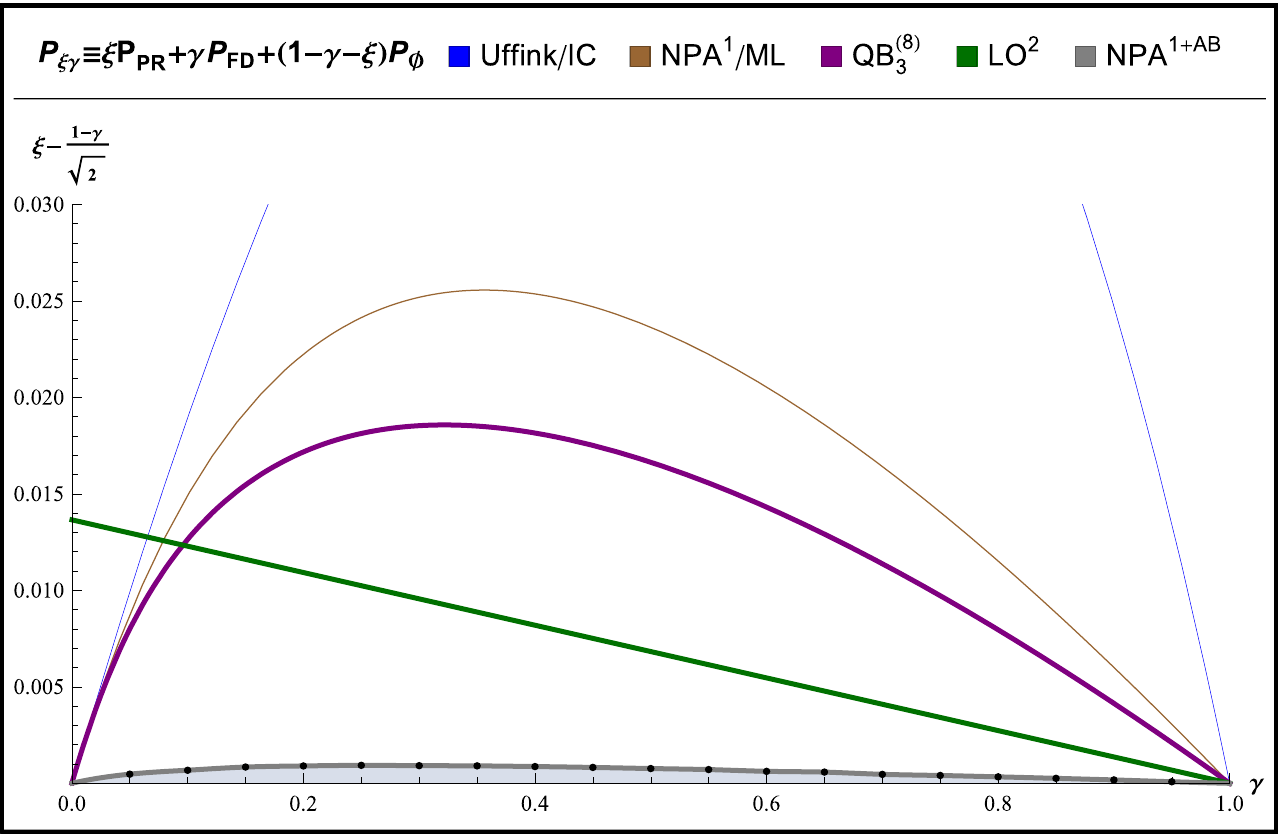}
    \caption{(Color online) A comparison of various quantum bounds and their implications for the quantum \gls{elliptope}. Prior to the determination of the $\mbox{LO}^2$ bounds first visualized in Ref. \cite{LONatureComm} in 2013, our own $\mbox{QB}_3^{(8)}$ was the best-known analytic boundary in this slice of the no-signalling \gls{polytope}. The grey region bounded numerically by $\mbox{NPA}^{1+AB}$ corresponds to quantum contextuality, which is known to be a relaxation to quantum nonlocality even for the (2,2,2) scenario \cite{AlmostQuantum2}. The black dots are the numeric points for which we calculated the maximal $\xi$ per $\mbox{NPA}^{1+AB}$.}\label{fig:boundcomptight}
\end{minipage}
\end{figure}

We know that the quantum \gls{elliptope} includes the deterministic box as well as the so-called Tsirelson box. By convexity, all points below the line connected these two points must be with the quantum elliptope. We therefore are particularly interested in the quantum region which lies outside this plane, if any. This line corresponds to boundary
\LeftEqns{
\xi \leq \frac{1}{\sqrt{2}}\left(1-\gamma\right)
}
which we subtract from all of the relevant nontrivial quantum bounds to obtain the plot of \fig{fig:boundcomptight}.

To illustrate that even all these quantum bounds are still inadequate, we have numerically calculated the maximum possible $\xi$ permitted by the $1+AB$ level of the NPA hierarchy, $\mbox{NPA}^{1+AB}$. The $\mbox{NPA}^{1}$ criterion is a relaxation of $\mbox{NPA}^{1+AB}$, which itself is a relaxation of the genuine bipartite quantum boundary. This is discussed further in Sec. \ref{sec:q1ab} of this thesis.

We note here that on this slice of the polytope the condition of $\mbox{QB}_3^{(8)}$ is grossly inadequate at characterizing the true quantum elliptope, as evidenced by both $\mbox{LO}^2$ as well as $\mbox{NPA}^{1+AB}$. This inadequacy was not known at the time Ref. \cite{WolfeQB} was published. In that reference a volume analysis was presented in an effort to compare the relative sizes of the statistical region consistent with local, quantum, and no-signalling models. The assumption was that the genuine quantum boundary was well approximated by the collection of linear quantum bounds dominated by $\mbox{QB}_3^{(8)}$. This assumption was in retrospect not valid, and therefore we do not reproduce the results of that volume analysis in this thesis. 

We conclude this chapter by noting that the study of characterizing quantum nonlocality has become an entire burgeoning field in quantum information theory. Quantum nonlocality has enjoyed a meteoric rise to prominence, as indicated by this chapter's opening epigraphs, and is presently at the forefront of many current active research areas. Nonlocality is the device-independent resource which powers many novel applications in quantum information theory, such as the device-independent variants of quantum cryptography \cite{CryptoPRA,CryptoPRL,masanesDIQKD,CryptoDIQKD,RenatoQKDNature,DIRandomness,scarani2012device,CryptoDIQKDarXiv2006,CryptoDIQKDarXiv,CryptoDIQKDEarly,CryptoDIQKD2010,CryptoDIQKDNJP} and state tomography \cite{DITomagraphy2013,DITomagraphy2014}, among other extraordinary uses.

\chapter{Entanglement}
\vspace*{\fill}
\begin{spacing}{1.5}
\begin{tabularx}{400pt}{m{1cm}Lm{1cm}}
& \noindent\textit{Entanglement is a trick that quantum magicians use to produce phenomena that cannot be imitated
by classical magicians.}
\footnote{\textsc{Asher Peres}\\ \textit{Quoted by Dagmar Bruß}, providing a favored definition of entanglement \citet{characterizingentanglement}.}  &
\end{tabularx}
\end{spacing}
\vspace*{\fill}

\section{Entanglement as a Resource}
Entanglement is well-understood to be the fundamental resource requisite for quantum nonlocality. While there are possible indications that entanglement is not fundamental for quantum computational speedup \cite{DQC1Original,Discord2001,DQC1,Discord2010,Discord2011,NegativityPostDiscord,MinEntangComputing}, and that the amount of entanglement in a system might not even directly translate into the strength of nonlocality a system can exhibit \cite{UPBOriginal,EntanglementSurprise2005,EntanglementSurprise2005b,EntanglementSurprise2007,EntanglementSurprise2011,EntanglementSurprise2014,RandomnessFromNonlocality,SteeringNonlocalityBrunner,SteeringNonlocalityArXiv,EntanglementInequivalent2014}, there is no doubt that entanglement represents a firm qualitative demarcation in the nature of correlations that a quantum system can be utilized to generate. Firstly, separable states give rise only to classical correlations \cite{UnifiedCorrelations}. The converse, ie. the statement that every entangled state violates a Bell inequality, is known as Gisin's Theorem \cite{Gisin1991,GisinPure2012,EnhancedGisinSteering,shimony2009quantum}. Gisin's theorem was proven true for all pure states in 2012  \cite{GisinPure2012}, two decades after its conjecture. Long believed to hold for mixed states as well \cite{acin2001bound,DiscreteEntanglementNonlocality}, the general proof extending Gisin's theorem to mixed states appears to have been established very recently as well, albeit with some qualifications \cite{AllEntangledNonlocality,SteeringEllipsoids,EnhancedGisinSteering,AllEntangledNonlocal,CloserConnections}. Thus, determining if a state is separable or entangled is fundamentally critical to our understanding of quantum nonlocality. 

There is extensive interest in schema for generating entanglement \cite{DownConversion,ResonantFluorescence,DickePrepration,SpinEntanglementMorrison,TothDickePRL} due to the plethora of nonclassical tasks which entanglement makes possible. Examples include Quantum Key Distribution \cite{CryptoPRA,CryptoPRL,CryptoDIQKD,RenatoQKDNature,DickeNetworking}, Quantum Computation \cite{mosca_algorithms_revisited,MBQC,MBQC2,QuantOptBook,Schumacher2010Quantum,Nielsen2011Quantum,Wilde2013Quantum}, and Precision Measurement \cite{SpinSqueezing2001,TothSpinSqueezing,NoriSpinSqueezing,Gross2010Nature,PrecisionToth,Gross2012IOP,FisherPrecisionMeasurement}. Various measures of entanglement \cite{TothDickePRL,DickeEntanglement,OneNegativeEigenvalue,NegativityIntro,NilpotentReview,GeomQuantStatesBook,PlenioEntanglementMeasures,multireview,entang.review.toth,SymmetricEquivalents,MultipartiteConc,ConvVsNeg,ConcVsGeom,Geometric3Qubits,GeometricMultiQubit,SteeringEntanglement} have been developed to quantify the resource value of entangled systems for the different tasks. These various entanglement measures reveal the presence of different classes of entanglement \cite{3Qubits2Ways,ThreeQubitClassification04,ThreeQubitClassification06,ClassificationHyperdeterminants,NQubitEquivalences,characterizingentanglement}, but they do not, however, tightly characterize the set of entangled states. Entanglement measures are effectively {\em necessary} separability criteria, in that for a quantum state to be separable it must have a measure of zero in the relevant entanglement metric, but they are not {\em sufficient}. A mixed quantum state may yield zero on some entanglement measure and yet nevertheless not be a separable state. All readily-evaluatable separability criteria are of the necessary-but-not-sufficient variety \cite{PPTAsher,PPTHorodecki,PPTNonlocal,ppt4qubit,pptNqubit}. Such is the essential nature of entanglement witnesses \cite{UPBOriginal,UnifiedCorrelations,EntangWitnessExtend}. It is worth noting that an \emph{asymptotically} sufficient separability criterion has been developed \cite{EntanglementHierarchy,EntanglementHierarchyMultipartite} in a manner that parallels the semidefinite hierarchy for characterizing quantum correlations \cite{NPAReview}.

To certify that a system is separable, one must show that the density matrix can be decomposed as a convex mixture of separable states. There is no general rule for how to go about seeking such a decomposition, even if one is convinced that such a decomposition exists.  \citet{KrausCiracKarnasLewenstein2000,KarnasLewenstein2001} describe a limited algorithm to certify biseparability \cite{Szalay2012Partial}, but these presume that the decomposition can be represented as a finite mixture, which is not always the case \cite{SuperradSeparable}. \citet{CiracSpinSqueezing} delineate an in-principle approach suitable for symmetric states, but it is not amenable for practical application. Refs. \cite{BipartiteMUBEntanglement,BipartiteSICEntanglement} provide sufficient criteria for \(\mathbb{C}^d\otimes \mathbb{C}^d\) mixed states based on mutually unbiased bases \cite{MUBConstruction,MUBReview,MUBSICPOVM} and symmetric informationally-complete POVMs \cite{SICPOVM2002,SICPOVM2004}, but these criteria are limited to system with a particular bipartite structure. There is, therefore, an resolved desire for a general method capable of practically certifying separability; such a method is needed, for example, to delineate truly entangled phenomenon as distinct from merely superficially cooperative-behaving systems \cite{eckert2002quantum,SuperradSeparable,MacroscropicQuantumness,MacroscopicArXiv}.

As established in Sec. \ref{sec:qubitssufficient}, for multipartite scenarios which are binary and dichotomic, the full range of quantum nonlocality can be achieved using qubits \cite{MasanesQubits}. Additionally, we note that no single party holds a privileged position with regards to nonlocality. That is, the quantum nonlocality \gls{elliptope} in statistical space must be invariant with respect to permutations of the parties, maintaining this natural symmetry of the local and no-signalling \glspl{polytope} \cite{GisinFramework2012,roberts_thesis,WernerWolfQB,WehnerQB,ScaraniNotes,*ScaraniNotes2}. 

This suggests that we should prioritize permutationally-invariant multi-qubit states in our study of entanglement and separability certification, because the quantum statistical boundary for binary and dichotomic scenarios corresponds physically to measurements on such quantum states. This direct correspondence between nonlocality and entanglement only holds for permutationally-invariant multi-qubit states. See Refs. \cite{WernerWolfQB,GivenQB.NQubit,AcinDetecting} for a survey of recent works advancing the quantification of nonlocality in multi-qubit states. It is especially interesting that all entangled permutationally-invariant multi-qubit states are necessarily genuinely multipartite entangled \cite{pptNqubit}, but it should also be noted that genuine multipartite entanglement does not necessarily imply genuinely multipartite nonlocality \cite{EntanglementInequivalent2014}. 

The second chapter of this thesis is dedicated to studying entanglement in permutationally-invariant multi-qubit states. Our contributions include the development of an explicit method for certifying separability for diagonally-symmetric multi-qubit mixed states \cite{SuperradSeparable}, the counter-intuitive finding that uniquely-quantum effects in Dicke model superradiance take place absent any entanglement \cite{SuperradSeparable}, and the analytic validation of \emph{driven} Dicke model superradiance as a scheme for generating spin-squeezed entangled states \cite{DrivenSuperrad}. Entirely unpublished prior to this thesis is our idea to utilize the structure of independently evolving systems in order to develop sufficient separability criteria for cooperatively evolving systems. The analysis of driven superradiance in terms of \emph{normalized} Dicke states, and the closed form expressions for the $N$-particle SDS jacobian and diagonally symmetric PPT conditions,  are also unique to this thesis.

\section{Separability of Diagonally Symmetric States}

Any permutationally invariant pure state can be expressed in terms of symmetric\footnote{Our use of the term ``symmetric'' here is equivalent to permutationally invariant, which is to say, a bosonic symmetry of indistinguishable particles. This is in contrast to the more reserved use of the terminology ``symmetric'' such as is considered by \citet{SymmetricEquivalents}.} Dicke basis states. The Dicke states are the superposition of equal-energy states; each is essentially a normalized sum-over-all-permutations of
some (separable) computational-basis state, such as
\begin{align}\label{eq:exampleket}
\ket{D^4_{1}}=\frac{\ket{0001}+\ket{0010}+\ket{0100}+\ket{1000}}{\sqrt{4}}.
\end{align}
or in general
\begin{align}\label{eq:firstdef}
\renewcommand{\arraystretch}{0.5}
   \ket{D^N_{n_1}} &=\sqrt{\frac{n_0!n_1!}{N!}}\sum_{\begin{subarray}{c}\scriptstyle{\text{perms.}}\\ \scriptstyle{\brackets{\ket{0},\ket{1}}}\end{subarray}}{\ket{\underbrace{0...0}_{n_0},\underbrace{1...1}_{n_1}}}
\end{align}
where $n_i$ is number of instances of $\ket{i}$ in the computational-basis state being permuted, and we have introduced the shorthand $n_0=N-n_1$, where $N$ the total number of qubits. Indexing Dicke states by $N$ and $n_1$ is the convention used in Refs. \cite{BestSeparableQubits,AcinDetecting}, whereas in our earlier work in Ref. \cite{SuperradSeparable} we used $n_0$ and $n_1$.

We begin by studying the separability properties of mixed states which are diagonal in the Dicke basis. To indicate such states we use the subscript GDS for ``General Diagonally Symmetric''. The general mixed state which is diagonal in the Dicke basis can be parameterized as
\begin{align}\label{eq:gds}
    \rho_{\text{\tiny GDS}}=\sum\limits_{n_1=0}^{N}{\chi_{n_1} \ket{D^N_{n_1}}\bra{D^N_{n_1}}}
\end{align}
where the $\chi_{n_1}$ represent the eigenvalues in the eigendecomposition of $\rho_{\text{\tiny GDS}}$, which, in the convention of quantum optics, we refer to as the populations of $\rho_{\text{\tiny GDS}}$.

Certifying separability amounts proving the existence of a decomposition of target mixed state into some convex combination of separable states; determining the existence of such a decomposition is ``hard''. We show that it is effective to instead ask if the target mixed state ``fits'' some preconstructed separable form. For GDS states, we take as our preconstructed separable form a given mixture of symmetric separable states. First, the fully-general separable multi-qubit pure state which is permutationally invariant is just the $N$-fold tensor product of a fully-general two-level pure state, that is,
\begin{align}
    &\hspace{-\mathindent}\ket{\psi\left[y,\phi \right]}=\left(\sqrt{y} \ket{0}+\sqrt{1-y} e^{\ic\phi}\ket{1}\right)^{\otimes N}\text{ or, in operator form }\rho^N\equiv\ket{\psi}\bra{\psi}
\\\nonumber&    \hspace{-\mathindent}\text{ie. }\rho^N\left[y,\phi \right]=\Big(y \dyad{0}{0}+(1-y)\dyad{1}{1}+\sqrt{y(1-y)}\left(e^{-\ic\phi}\dyad{0}{1}+e^{\ic\phi}\dyad{1}{0}\right)\Big)^{\otimes N}
\end{align}
which we would like to express more explicitly. By multinomial expansion we can expand $\rho^N\left[y,\phi \right]$ in terms of a sum over four exponents,  $\gamma_{00},\gamma_{10},\gamma_{01},\gamma_{11}$, which are to be understood as ranging over nonnegative integers $\bm{\gamma}\in \mathbb{Z}^+$ in such a manner that the sum of the exponents total $N$, $\gamma_{00}+\gamma_{10}+\gamma_{01}+\gamma_{11}=N$. In this manner we are also taking tensor-exponents of each of the four operator-basis product states, summing over all permutations of the operators as well. That is, $\rho^N\left[y,\phi \right]=$
\LeftEqn{
\renewcommand*{\arraystretch}{0.5}
    &\hspace{-\mathindent}=\sum_{\lbrace{\text{all }\gamma \rbrace}}^N{
    y^{(\gamma_{00}+\gamma_{01}/2)}(1-y)^{(\gamma_{11}+\gamma_{10}/2)}e^{\ic\phi(\gamma_{10}-\gamma_{01})}\sum\limits_{\begin{subarray}{l} \text{operator permutations} \\
\brackets{\dyad{0}{0},\dyad{0}{1},\dyad{1}{0},\dyad{1}{1}} \end{subarray}}{
    \renewcommand*{\arraystretch}{0.5}\rho\begin{bmatrix}\gamma_{00} & \gamma_{01} \\ \gamma_{10} & \gamma_{11} \end{bmatrix}}}
\\&\hspace{-\mathindent}\text{where }\;\renewcommand*{\arraystretch}{0.5}
\rho\begin{bmatrix}\gamma_{00} & \gamma_{01} \\ \gamma_{10} & \gamma_{11} \end{bmatrix}\equiv \pdyad{0}{0}^{\otimes \gamma_{00}}\pdyad{1}{1}^{\otimes \gamma_{11}}\pdyad{0}{1}^{\otimes \gamma_{01}}\pdyad{1}{0}^{\otimes \gamma_{10}}
}
is the natural generalization of computational basis states to product states. Note that the sum over operator permutations is intentionally {\em not} normalized as each permutation of each has equal weight in the expansion of $\rho^1\left[y,\phi \right]^{\otimes N}$. 

We elect to consider only such states which are mixed uniformly over all $\phi$, namely $\rho^N\left[y\right]\equiv \left(2\pi\right)^{-1}\int_{0}^{2\pi}{\rho^N\left[y,\phi\right]\;\mbox{d}\phi}\,$. While this does induce a loss of generality, this is desirable in order that $\rho^N\left[y\right] \in \brho_{\text{\tiny GDS}}$. Note that 
\LeftEqn{
\renewcommand*{\arraystretch}{0.5}
\int_{0}^{2\pi}{e^{\ic \phi\left(\gamma_{10}-\gamma_{01}\right)}\;\mbox{d}\phi}=\left\lbrace\begin{array}{rl} 1,& {\gamma_{10} = \gamma_{01}}\\ 0,& {\gamma_{10} \neq \gamma_{01}}\end{array}\right.
}
which allows us to perform a four-to-three change-of-variable, such that $\gamma_{10}= \gamma_{01}\rightarrow \kappa,\;\; \gamma_{00} \rightarrow n_0-\kappa,\;\; \gamma_{11} \rightarrow n_1-\kappa$. Note that in these three new variables, the earlier implicit condition $\gamma_{00}+\gamma_{10}+\gamma_{01}+\gamma_{11}=N$ is {\em automatically satisfied} once we define $n_0=N-n_1$, but to preserve the positivity of both $\gamma_{00}$ and $\gamma_{11}$ we must be careful to upper bound $\kappa \leq \operatorname{min}[n_0,n_1]$. Thus the uniform mixing over all phases results in simply
\begin{align}\label{eq:rhony}
\renewcommand*{\arraystretch}{0.5}
\hspace{-\mathindent}\rho^N\left[y\right]&= \sum\limits_{n_1=0}^{N}{\sum\limits_{\kappa=0}^{\operatorname{min}[n_0,n_1]}{{y}^{n_0}{(1-y)}^{n_1}\sum\limits_{\begin{subarray}{l} \text{operator permutations} \\
\brackets{\dyad{0}{0},\dyad{0}{1},\dyad{1}{0},\dyad{1}{1}} \end{subarray}}{
   \renewcommand*{\arraystretch}{0.5} \rho\begin{bmatrix}\left(n_0 - \kappa\right) & \kappa \\ \kappa & \left(n_1 -\kappa\right) \end{bmatrix}}}} 
\end{align}

To translate explicitly into the Dicke basis states we rearrange the order of summation and make use of a binomial argument, namely
\begin{align}\label{eq:counting}
\renewcommand*{\arraystretch}{0.5}
\hspace{-\mathindent}\sum\limits_{\begin{subarray}{c} \text{perms.} \\
\brackets{4\scriptstyle{\times}\; \dyad{j}{k}} \end{subarray}}{\sum\limits_{\kappa}^{}{   \renewcommand*{\arraystretch}{0.5} \rho\begin{bmatrix}\left(n_0 - \kappa\right) & \kappa \\ \kappa & \left(n_1 -\kappa\right) \end{bmatrix}}} 
     = \begin{pmatrix}\mathlarger{\sum}\limits_{\begin{subarray}{c}\scriptstyle{\text{perms.}}\\ \scriptstyle{\brackets{\ket{0},\ket{1}}}\end{subarray}}{\ket{\underbrace{0...0}_{n_0}\underbrace{1...1}_{n_1}}}\end{pmatrix}\hspace{-1ex}\begin{pmatrix}\mathlarger{\sum}\limits_{\begin{subarray}{c}\scriptstyle{\text{perms.}}\\ \scriptstyle{\brackets{\bra{0},\bra{1}}}\end{subarray}}{\bra{\underbrace{0...0}_{n_0}\underbrace{1...1}_{n_1}}}\end{pmatrix}
\end{align}
The left hand side of \ceq{eq:counting} is a double sum, over permutations of the four operators as well as over all possible partition schemes indexed by $k$. This is equivalent to the right hand side of \ceq{eq:counting}, namely taking the product of unpaired permutation summations. This counting scheme follows from $\sum_{\kappa}{\frac{N!}{\kappa!(n_0-\kappa)!(n_1-\kappa)!\kappa!}}={\left(\frac{N!}{n_0!n_1!}\right)}^2\,$. 

To clarify what is meant in \ceq{eq:counting} let us consider the explicit example where $n_0=3$  and $n_1=1$ as in \ceq{eq:exampleket}. We'll start with the right hand side of \ceq{eq:exampleket} and work backward to to the left hand side.
\LeftEqns{
\renewcommand*{\arraystretch}{0.5}
&\hspace{-\mathindent}\begin{pmatrix}\mathlarger{\sum}\limits_{\text{perms.}}{\ket{\underbrace{0...0}_{3}\underbrace{1...1}_{1}}}\end{pmatrix}\begin{pmatrix}\mathlarger{\sum}\limits_{\text{perms.}}{\bra{\underbrace{0...0}_{3}\underbrace{1...1}_{1}}}\end{pmatrix} \\\nonumber
&\hspace{-\mathindent}=\Big(\ket{0001}+\ket{0010}+\ket{0100}+\ket{1000}\Big) \Big(\bra{0001}+\bra{0010}+\bra{0100}+\bra{1000}\Big) \\\nonumber
&\hspace{-\mathindent}=\Big(\op{\da}{\da}{\da}{\dd}+\op{\da}{\da}{\dd}{\da}+\op{\da}{\dd}{\da}{\da}+\op{\dd}{\da}{\da}{\da}\Big)\\\nonumber
&\hspace{-\mathindent}\hphantom{=}+\Big(\op{\da}{\da}{\db}{\dc}+\op{\da}{\da}{\dc}{\db}+\op{\da}{\db}{\da}{\dc}+\op{\da}{\dc}{\da}{\db}\\\nonumber
&\hspace{-\mathindent}\hphantom{=}+\hphantom{\Big(}\op{\da}{\db}{\dc}{\da}+\op{\da}{\dc}{\db}{\da}+\op{\db}{\da}{\da}{\dc}+\op{\dc}{\da}{\da}{\db}\\\nonumber
&\hspace{-\mathindent}\hphantom{=}+\hphantom{\Big(}\op{\db}{\da}{\dc}{\da}+\op{\dc}{\da}{\db}{\da}+\op{\db}{\dc}{\da}{\da}+\op{\dc}{\db}{\da}{\da}\Big)\\\nonumber
&\hspace{-\mathindent}\renewcommand*{\arraystretch}{0.5}=\sum\limits_{\text{perms.}}{\rho\begin{bmatrix} 3 & 0 \\ 0 & 1 \end{bmatrix}}+\sum\limits_{\text{perms.}}{\rho\begin{bmatrix} 2 & 1 \\ 1 & 0 \end{bmatrix}}\renewcommand*{\arraystretch}{0.5}=\sum\limits_{\text{perms.}}{\sum\limits_{\kappa=0}^{\operatorname{min}[3,1]}{\rho\begin{bmatrix}\left(3 - \kappa\right) & \kappa \\ \kappa & \left(1 -\kappa\right) \end{bmatrix}}}
}
Of course by inspection of \ceq{eq:firstdef} it is clear that $\sum\limits_{\text{perms.}}{\ket{\underbrace{0...0}_{n_1}\underbrace{1...1}_{n_1}}}=\sqrt{\frac{N!}{n_0! n_1!}}\ket{D^N_{n_1}}$ such that we can conveniently now express \ceq{eq:rhony} in a manner which makes it clear that $\rho^N\left[y\right] \in \brho_{\text{\tiny GDS}}\,$, namely
\begin{align}\label{eq:rhosdsy}
\rho^N\left[y\right]&= \sum\limits_{n_1=0}^{N}{N!\frac{{y}^{n_0}{(1-y)}^{n_1}}{n_0! n_1!}\ket{D^N_{n_1}}\bra{D^N_{n_1}}}\,.
\end{align}
This suggests a suitably-generic parameterization of separable diagonally symmetric states, namely
\LeftEqns{\label{eq:qubitSDS}
\rho_{\text{\tiny SDS}}&= \sum\limits_{n_0=0}^{N}{\sum\limits_{j=1}^{j_\text{max}}{N!x_j\frac{{y_j}^{n_0}{(1-y_j)}^{n_1}}{n_0! n_1!}\ket{D^N_{n_1}}\bra{D^N_{n_1}}}}
}
which implies a sufficient criterion for separability of diagonally symmetric states: Does the state $\rho_{\text{\tiny GDS}}$ fit for the form of $\rho_{\text{\tiny GDS}}$? Formally, a comparison of Eqs. (\ref{eq:gds}) and (\ref{eq:qubitSDS}) indicates that 
\LeftEqn{\label{eq:sepcrit}
&\rho_{\text{\tiny GDS}}\in\brho_{\text{\tiny SDS}}\;\text{ iff }\;\exists\;{x_1\dots x_{j_\text{max}}, y_1 \dots y_{j_\text{max}}}\\
 &\text{satisfying }\;\forall_{n_1} \quad {\chi_{n_1}} \,=\, N!\sum_{j=1}^{j_{\text{max}}}{\frac{x_j {y_j}^{n_0}\left(1-{y_j}\right)^{n_1}}{n_0!n_1!}}
    \\&\text{such that }\;\forall_{j}:\; 0\le \,x_j\,,\,y_j\,\le 1\,.
}

\section{Volume Analysis of Separability Criteria}

It is possible to show that not only is the separability criterion implied by conditions \noeq{eq:sepcrit} sufficient to certify separability, but that it is furthermore also apparently necessary! That is to say, we can show that {\em every} fully separable symmetric $N$-qubit state is of the form of $\rho_{\text{\tiny GDS}}$. We establish the universality of $\rho_{\text{\tiny GDS}}$ by showing that the volume of states which it parameterizes is equal to the volume of diagonally-symmetric states which satisfy the necessary separability criterion of positivity under partial transpose (PPT) \cite{PPTAsher,PPTHorodecki}.

The property of PPT is generally necessary but insufficient for separability \cite{UPBOriginal,acin2001bound,Szalay2012Partial,PPTNonlocal}, although for symmetric states it known to be {\em is} sufficient for $N=2,3$, but still insufficient for $N\geq 4$ \cite{eckert2002quantum,ppt4qubit,pptNqubit,ConvVsNeg}. Our finding of $\operatorname{SDSVol}_{N}=\operatorname{PPTDSVol}_{N}$ implies that the PPT criterion apparently {\em} is sufficient for $N\geq 4$ for {\em diagonally} symmetric states. We provide a complete numeric proof for $N=4$, and additional numeric evidence to suggest that the correspondence is not broken for larger $N$.

To define a volume of a set of quantum mixes states we must establish a metric on the spaces of density matrices; the metric can be arbitrary but must be consistent. We choose the populations of $\rho_{\text{\tiny GDS}}$ as our integration coordinates, ie. $\chi_0\dots\chi_N\,$. Note that this is in contrast to the more conventional metrics established by Życzkowski \emph{et~al.} \cite{Zyczkowski98,Zyczkowski99,Zyczkowski01,ZyczkowskiHS,ZyczkowskiBures}.

When computing the volume of $\operatorname{SDSVol}_{N}$ it is natural to integrate not using the populations $\chi_{n_1}$ but rather the variables $x_j, y_j\,$. This requires that we introduce not only a volume element to compensate for the change of variable of the integration basis, but that we furthermore take care to establish a one-to-one mapping between $x_1\dots x_{j_\text{max}}, y_1 \dots y_{j_\text{max}}$ and $\chi_0\dots\chi_N\,$. 

This implies that $j_{\text{max}}$ must be chosen such that the central system of equations comprising criterion \noeq{eq:sepcrit} should be well behaved, i.e. that there should be exactly $N+1$ variables $x_j, y_j$ {\em total} appearing in the $N+1$ equations. Considering that $x_j$ and $y_j$ always come in pairs, this poses an obstacle for those instances when $N+1$ is odd. Our solution is to take $j_{\text{max}}=\lceil(N+1)/2\rceil$ and to mediate the extraneous variable by manually adjusting $y_{\left(N+2\right)/2}=0$. This ansatz is validated when we find that $\operatorname{SDSVol}_{N}=\operatorname{PPTDSVol}_{N}$ still holds when $N$ is an even number. We summarize by defining
\LeftEqn{\label{eq:jmaxdefs}
jx_{\text{max}}=\lceil\frac{N+1}{2}\rceil \\
jy_{\text{max}}=\lfloor\frac{N+1}{2}\rfloor \\
y_{jx_\text{max}} = 0 \,\text{ if }\,jx_\text{max} \neq jy_{\text{max}}
}

We must also break the symmetry of the decomposition inherent to conditions \noeq{eq:sepcrit}, in the sense that as it stands, any solution for $y_1$ in terms of $\chi_0\dots\chi_N$ can be exchanged with a solution for $y_2$ by also exchanging $x_1$ and $x_2$, implying a degeneracy of solutions. We can solve this by imposing the ordering $\forall_{1\leq i<j\leq j_\text{max}}\:y_i \geq y_j$ which fits nicely with the sometimes-relevant constraint $y_{\left(N+2\right)/2}=0$. In practice it is easiest to integrate without restriction, and then to compensate by dividing by $jy_{\text{max}}!$ to account for the degeneracy.

The volume element we mentioned is of course the absolute value of the determinant of Jacobian matrix for the change-of-variable. The $N+1$ columns of this matrix corresponds to the populations $\chi_{n_1}$. The rows of this matrix corresponds to differentiation of the populations with respect to the $N+1$ new variables $x_j, y_1 \dots y_j$, as the populations are with regards to these new variable in Eq. \noeq{eq:sepcrit}. One can show that this determinant has the form
\LeftEqn{\label{eq:jacform}
jac_N=Z_N \prod\limits_{k=0}^{\lfloor\frac{N+1}{2}\rfloor}{x_k\prod\limits_{j=0,j\neq k}^{\lceil\frac{N+1}{2}\rceil}{\left(y_j-y_k\right)^2}}\quad\text{where}\quad Z_N \equiv \prod\limits_{n_1=0}^N{\frac{N!}{n_0! n_1!}}
}
which is valuable as it allows us to readily factor the integral into one integral over the $x_j$ and another integral over the $y_j$. Note that the index of $x$ in \ceq{eq:jacform} goes up only to $jy_\text{max}$, whereas the index of $y$ goes up to $jx_\text{max}$. This is a result of $x_{jx_\text{max}}$ not appearing at all in the Jacobian if ${jx_\text{max}}\neq {jy_\text{max}}\,$. Recall also that when $N$ is even then $y_{jx_\text{max}}=0$ as per \ceq{eq:jmaxdefs}.

The integral over the $x_j$ is not entirely trivial, as we must include some condition of normalization when considering a volume of states, $1=\mbox{Tr}\left[\rho_{\text{\tiny GDS}}\right]=\sum_{n_1=0}^{N}{\chi_{n_1}}=\sum_{j=1}^{j_\text{max}}{x_j}\,$. We could define integration limits that are inherently normalized, but it is easier to simply introduce a Dirac delta function in the integrand to enforce normalization. One can verify that
\LeftEqn{
\int\limits_0^1 ... \int\limits_0^1 \delta{\left(1-\sum\limits_{j=1}^{\lceil\frac{N+1}{2}\rceil}{x_j}\right)}\prod\limits_{j=0}^{\lfloor\frac{N+1}{2}\rfloor}{x_k} \;\mbox{d}x_1...\mbox{d}x_{\lceil\frac{N+1}{2}\rceil}=\frac{1}{N!}
}
which means that
\LeftEqn{
\operatorname{SDSVol}_{N}&=\frac{Z_N}{\lfloor\frac{N+1}{2}\rfloor !N!}\int\limits_0^1 ... \int\limits_0^1 \prod\limits_{j=0,j\neq k}^{\lceil\frac{N+1}{2}\rceil}{\left(y_j-y_k\right)^2} \;\mbox{d}y_1...\mbox{d}y_{\lfloor\frac{N+1}{2}\rfloor}\\
&=\prod_{z=1}^{N}{z^{(z-1)}\frac{(z-1)!}{(2 z-1)!}} \quad\text{(proof not shown)}
}
which we explicitly give for $N=4$, namely
\LeftEqn{
\hspace{-\mathindent}\operatorname{SDSVol}_{4}&=\frac{Z_4}{\lfloor\frac{4+1}{2}\rfloor !4!}\int\limits_0^1 \!\!\int\limits_0^1  {y_1}^{2} {y_2}^{2} {(y_1-y_2)}^{4}\;\mbox{d}y_1\mbox{d}y_2 \\
&=\frac{96}{2\times 24}\times {525}^{-1} \approx \left(3809.5\right)\times {10}^{-6}\,.
}

To calculate the volume of the positive-under-partial-transpositions diagonally symmetric states, $\operatorname{PPTDSVol}_{N}$, we perform the integration directly in the basis of the populations, so we must introduce indicator functions to enforce that we only count PPT states. Here the PPT conditions mean that all eigenvalues are nonnegative {\em for all} bipartitions of the qubits for partial transposition. The permutation symmetry of $\rho_{\text{\tiny GDS}}$ means we need only consider $\lfloor N/2\rfloor$ bipartitions: partial transposition of the first qubit $\rho^{\operatorname{PT}_{1|N}}$ or of the first two qubits $\rho^{\operatorname{PT}_{2|N-1}}$, etc, akin to the considerations in Refs. \cite{ppt4qubit,symmetricMultiparticleEntanglement}. 

For diagonally symmetric states the positivity conditions associated with each bipartition has a clean form, namely
\LeftEqn{\label{eq:pptsummary}
\rho^{\operatorname{PT}_{q|N-q}} \geq 0 \quad \text{iff}\quad &\forall_{0\leq m\leq N-2q}\,:\:\operatorname{Det}[\tilde{\rho}_m^{q}] \geq 0
}
where $\tilde{\rho}_m^q$ is a $(q+1)\times(q+1)$ matrix with elements
\LeftEqn{
\tilde{\rho}_m^{q}[[x,y]]\equiv \chi_{m+x+y}\sqrt{\frac{\binom{q}{x}\binom{q}{y}\binom{N-q}{m+x}\binom{N-q}{m+y}}{{\binom{N}{m+x+y}}^2}}
}
which implies the special case PPT condition for one-qubit bipartitionas a result of the closed-form expression
\LeftEqn{
\operatorname{Det}[\tilde{\rho}_m^{1}]=\frac{(m+2)(N-m)\chi_m\chi_{m+2}-(m+1)(N-m-1)\chi_{m+1}}{N^2}
}
corresponding to the PPT condition used by \citet[Eq. (14)]{BestSeparableQubits}.

Per \ceq{eq:pptsummary}, we set up our integral for the volume of PPT diagonally-symmetric states in terms of Heaviside Theta functions, such that
\LeftEqn{
\hspace{-\mathindent}\operatorname{PPTDSVol}_{N}&=\int\limits_0^1 ... \int\limits_0^1 \delta{\left(1-\sum\limits_{n_1=0}^{N}{\chi_{n_1}}\right)} \prod_{q=1}^{\lfloor N/2\rfloor}{\prod_{m=0}^{N-2q}{\operatorname{\Theta}{\left(\operatorname{Det}[\tilde{\rho}_m^{q}]\right)}}}\;\mbox{d}\chi_0...\mbox{d}\chi_N
}
which is difficult to evaluate in general. For $N=4$ we have the explicit integral
\LeftEqn{
\hspace{-\mathindent}\operatorname{PPTDSVol}_{4}&=\int\limits_0^1 \!\!\!\int\limits_0^1 \!\!\!\int\limits_0^1 \!\!\!\int\limits_0^1 \!\!\!\int\limits_0^1  \:\Theta \left(8 \chi _0 \chi _2-3 \chi _1^2\right) \Theta \left(9 \chi _1 \chi _3-4 \chi _2^2\right) \Theta \left(8 \chi _2 \chi _4-3 \chi _3^2\right) \\
&\quad\times\Theta \left(9 \left(\chi _1 \chi _3+8 \chi _0 \chi _4\right) \chi _2 -2 \chi _2^3 -27 \left(\chi _4 \chi _1^2+\chi _0 \chi _3^2\right)\right) \\
&\quad\times \delta{\left(1-\chi_0-\chi_1-\chi_2-\chi_3-\chi_4\right)}\;\mbox{d}\chi_0 \mbox{d}\chi_1 \mbox{d}\chi_2 \mbox{d}\chi_3 \mbox{d}\chi_4 \\
&=\left(3808\pm 2\right)\times {10}^{-6}
}
evaluated numerically. Because we {\em must} have $\operatorname{PPTDSVol}\geq \operatorname{SDSVol}$ we are forced to revise $\operatorname{PPTDSVol}_{N=4}$ to the upper limit of its numerical uncertainty, which indicates convincingly that $\operatorname{PPTDSVol}_4=\operatorname{SDSVol}_4\,$.

For larger $N$ we have numerical evidence that $\operatorname{PPTDSVol}_N=\operatorname{SDSVol}_N$ by a Monte Carlo argument. We programmed a numerical survey of billions and billions of random states which were positive under all partial transpositions, and without exception, ever such state also satisfied the necessary separability criterion \noeq{eq:sepcrit}. This would not be expected if $\operatorname{PPTDSVol}_N > \operatorname{SDSVol}_N\,$ and is as such a proof by contraposition of apparent equivalence between $\operatorname{PPTDSVol}_N$ and $\operatorname{SDSVol}_N\,$.

\section{Separability of Superradiance}
An example of a diagonally symmetric system of physical consequence is the Dicke model of superradiance \cite{Dicke54,superrad.original,superrad.yelinPRA,superrad.yelinBook}. Superradiancece is a phenomenon in which excited atoms spaced together very closely radiate in an induced cascade. This can occur if the volume of the system is smaller than the wavelength of the emitted radiation. The Dicke model of superradiance is the maximally idealized phenomenological model. The idealization employed in the Dicke model is that of perfect indistinguishability of the particles, such that we treat the system as existing entirely in only highest symmetry of the Hilbert space. Experimentally it corresponds to the small-volume limit and an absence of dipole-dipole induced dephasing. A thorough treatment of the volume-dependent many-body effects not considered in the Dicke model can be found in Refs. \cite{superrad.yelinPRA,superrad.yelinBook}.

The Dicke model describe the spontaneous decay of the (open) system with decay rate $\Gamma$ by means of Lindblad operators. The Liouville master equation \cite{Breuer2007Theory,scully1997quantum,QuantOptBook,NonMarkovianity} which governs the time evolution of driven Dicke model superradiance is
\begin{align}\label{eq:lindblad}
&\hspace{-\mathindent}\frac{\partial \rho}{\partial t} =\Gamma\left(A^- \rho A^+ - \frac{A^+ A^- \rho + \rho A^+ A^-   }{2}\right)
\end{align}
where 
\begin{align}\label{eq:Dplusminus}
    A^+ &= \sum\limits_{n=1}^N{\underbrace{\id ... \id}_{n-1}\otimes \pdyad{1}{0} \otimes \underbrace{\id ... \id}_{N-n}} 
\end{align}
with the annihilation operator being the adjoint of the creation operator, $A^-=\left(A^+\right)^\dag$. The high symmetry of the Lindblad operators lead to a time evolution entirely within the manifold of the diagonally symmetric states, such that if initially diagonally symmetric then it remains diagonally symmetric, of the form $\rho_{\text{\tiny GDS}}$ per \ceq{eq:gds}. 

\begin{figure}[p]
\centering
    \includegraphics[width=1.0\linewidth]{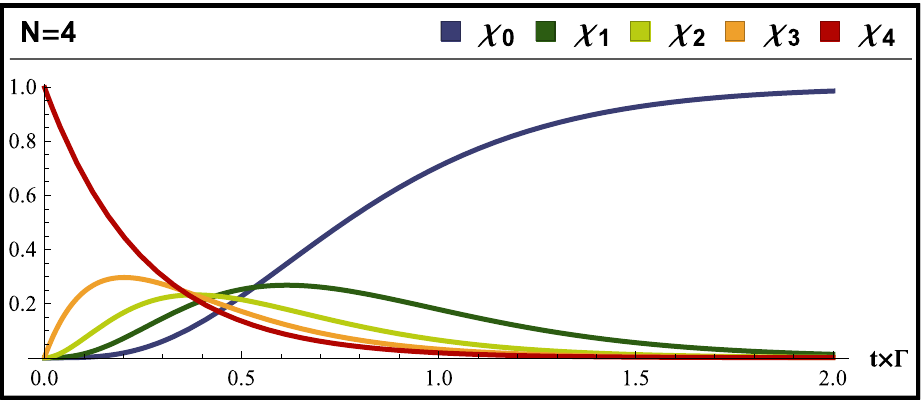}
    \caption{(Color online) Superradiance populations as a function of time, per \ceq{eq:superrad}. The system starts out entirely in the maximally excited state and evolves towards existing entirely in the ground state. At intermediate times the system is a mixture of populations.}\label{fig:n4superplot}
\vspace{31pt}
    \includegraphics[width=1.0\linewidth]{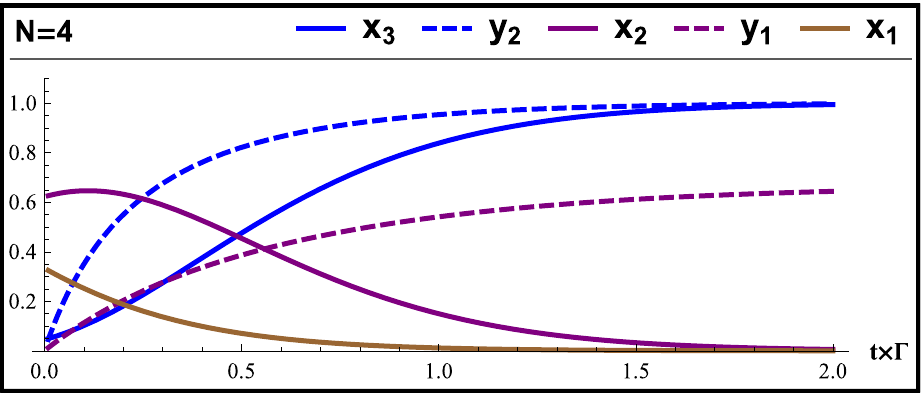}
    \caption{(Color online) This plot illustrates the permanent full separability of a superradiating system, for $N=4$. The various decomposition elements are plotted as a function of time; they have been solved-for from the system of polynomial equations defined in \ceq{eq:sepcrit}. For $N=4$ there are five decomposition elements, as $y_3$ has been manually set to zero in accordance with \ceq{eq:jmaxdefs}.}\label{fig:n4decompplot}
\end{figure}

Thus, the rate equation for Dicke model superradiance \citet{superrad.original} can be expressed in the form of a recursive relation between the populations, namely
\LeftEqn{\label{eq:superrad}
    \frac{\partial \ochi{n_0}{n_1}\left(t\right)}{\partial t}\;=\;-\Gamma\left(n_0+1\right)n_1 \ochi{n_0}{n_1}\left(t\right)+\Gamma n_0\left(n_1+1\right)\ochi{n_0-1}{n_1+1}\left(t\right)
}
as per \citet{superrad.original}. We choose to consider initially separable states and ask if superradiance time evolution leads to entanglement. The only diagonally symmetric separable state which superradiates is the maximally excited state, and thus we take as our initial conditions 
\begin{align}\label{eq:initconds}
	\ochi{n_0}{n_1} \left(t\rightarrow 0\right)=
	\renewcommand*{\arraystretch}{0.5}\left\lbrace\begin{array}{rl} 1,& {n_1=N,n_0=0}\\ 0,& {n_1<N,n_0>0}\end{array}\right.\,.
\end{align}
When we solve for the $\rho_{\text{\tiny SDS}}$ weights and amplitudes per \ceq{eq:sepcrit} we find that all the $x$ and $y$ are real number between zero and one throughout the entire time evolution, that is $\rho_{\text{\tiny SuperRad}}\in \brho_{\text{\tiny SDS}}\,$. Indeed, we numerically verified that for pure Dicke Model superradiance, conditions (\ref{eq:sepcrit}) are satisfied for all $\tau>0$, thereby certifying full separability throughout the time evolution, for $N\leq 8$. This is demonstrated graphically in \fig{fig:n4decompplot}.

\FloatBarrier

\section{Independently Radiating Model}

It is valuable to consider, by an apophatic contrasting approach, a system which has no quantum enhanced effects and radiates according to a constant-hazard decay principle, essentially the large volume limit. To understand statistical independence we must first recognize the relationship between populations and probability functions. For example, if we define a probability distribution $\operatorname{PDF}[t]$ giving the likelihood that a single two-level atom will decay at any instant in time, then the ground state occupancy of that atom over time is equivalent to the cumulative probability distribution, $\operatorname{CDF}[t]$. The excited states occupancy, being one minus the ground state occupancy, corresponds therefore to the survival function, also known as the reliability function, $\operatorname{REL}[t]$. The failure rate, ie. the emission intensity, is given by the statistical hazard function, $\operatorname{HAZ}[t]$. The hazard function is defined as 
\LeftEqn{
\operatorname{HAZ}[t]=\frac{\operatorname{PDF}[t]}{1-\operatorname{CDF}[t]}=-\frac{\operatorname{REL}'[t]}{\operatorname{REL}[t]}
}
such that if the atoms behave statistically independently then each has a constant hazard function $\operatorname{HAZ}[t]=\Gamma$ leading to the occupancy differential equation $\operatorname{REL}'[t]=-\Gamma \operatorname{REL}[t]$. 

Suppose now we have an infinite number of two-level atoms radiating according to constant hazard $\Gamma$, which we shall refer to as standardradiance as a foil to superradiance. Let us mentally group these infinite atoms into sets of size $N$. Any given set has $n_1$ excited atoms and $n_0$ ground-state atoms, where the ratio changes in time. At any intermediate time some sets will have more or less excited atoms than others. We {\em define} the populations to mean the {\em relative frequency} of having sets with $n_1$ excited atoms. A set of $n_1$ atoms has an $n_1$-fold hazard for the possibility that any one of its atoms might decay. On the other hand, the collection of $n_1$-excited sets is constantly being replenished by the decay of any atom from the sets with $n_1+1$ excited atoms, which have a hazard rate to transition equal to $\Gamma(n_1+1)$. Thus the independently-radiating rate equation for standardradiance populations is
\LeftEqn{\label{eq:standardrad}
    \frac{\partial \ox{n_0}{n_1}\left(t\right)}{\partial t}\;=\;-\Gamma n_1 \ox{n_0}{n_1}\left(t\right)+\Gamma\left(n_1+1\right)\ox{n_0-1}{n_1+1}\left(t\right)
}
where we use $\operatorname{X}$ for standardradiance to distinguish it from $\operatorname{\chi}$ used for superradiance in \ceq{eq:superrad}. 

\begin{figure}[p]
    \includegraphics[width=1.0\linewidth]{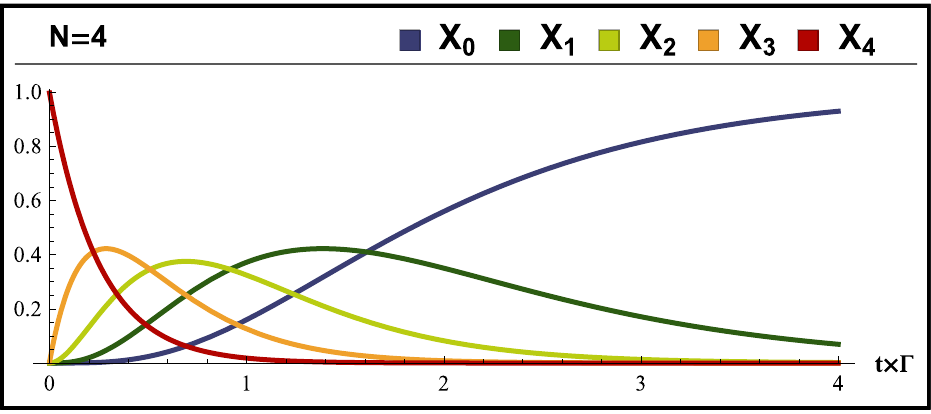}
    \caption{(Color online) Standardradiance populations as a function of time, per \ceq{eq:standardrad} and \ceq{eq:newsepform}. The system starts out entirely in the maximally excited state and evolves towards existing entirely in the ground state. At intermediate times the system is a mixture of populations. Note that the evolution is ``slower'' than that of superradiance; note the much shorter timescale plotted in \fig{fig:n4superplot}.}\label{fig:n4standardplot}
\vspace{49pt}
    \includegraphics[width=1.0\linewidth]{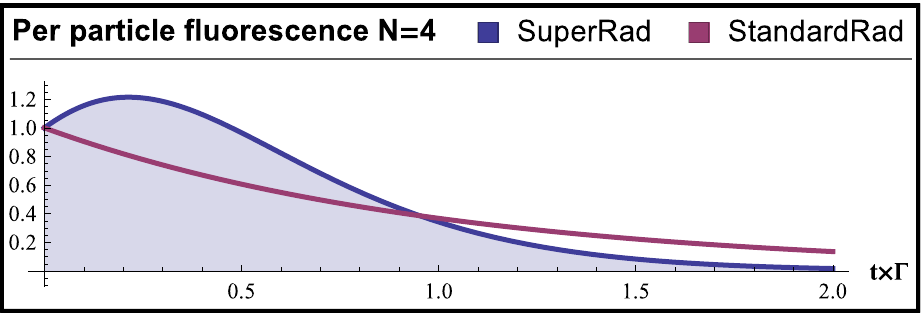}
    \caption{(Color online) This is a plot of the per-particle fluorescence rates for both superradiance and standardradiance, per \ceq{eq:frates}. See how the superradiance per-particle florescence rate has a higher peak emissivity (along with a narrower distribution width) from whence comes its name. }\label{fig:n4emissionplot}
\end{figure}

From a set of populations evolving in time it is possible to compute the per-particle florescence rate, ie. the per-particle probability density function to see a photon coming out of the system. For standardradiance the per-particle florescence rate is {\em presumed} to be an exponential distribution per the premise of constant hazard. By contrast, the superradiance per-particle florescence rate has a higher peak emissivity along with a narrower distribution width, from whence comes its name. The per-particle florescence rate for both forms of radiation is given simply by
\noindent\FlushEqn{\label{eq:frates}
\text{Superradiance PDF} &= \sum_{n_1}{\frac{n_1}{N} \frac{\partial \ochi{n_0}{n_1}\left(t\right)}{\partial t}} = \sum_{n_1}{\frac{\left(n_0+1\right)n_1}{N} \ochi{n_0}{n_1}\left(t\right)}
\\\text{Standardradiance PDF} &= \sum_{n_1}{\frac{n_1}{N} \frac{\partial \ox{n_0}{n_1}\left(t\right)}{\partial t}} = \sum_{n_1}{\frac{n_1}{N} \ochi{n_0}{n_1}\left[t\right]}=\Gamma e^{-\Gamma t}\,.
}
The relatives rates of florescence are  for  $N=4$ are contrasted in \fig{fig:n4emissionplot}.

Let us pause for a moment to prove what is already physically obvious, namely that standardradiance is always described by a separable system, ie. $\rho_{\text{\tiny StandardRad}}\in \brho_{\text{\tiny SDS}}\,$. As we proceed in this proof, we shall find that standardradiance corresponds to a very special case.

We seek a closed-form solution to the standardradiance populations. To this end we note that the independent radiation model is fully equivalent to the well-studied physical system of nuclear decay. We analogize the radiative evolution of set of $N$-atoms through various excitation levels with a decay chain consisting of $N$ nuclides. Specifically, we recognize that the concentration of a given nuclide in a sample parallels the relative frequency of finding $N$-atom sets containing a particular $n_1$ count of excited atoms. Decay chains in nuclear systems are described by Bateman equations \cite{Bateman2006}. Bateman equations give the concentration of the $k$'th nuclide at time t by
\LeftEqns{
\mbox{Concentration}_k \left[t\right] = \mathlarger{\sum}\limits_{j=1}^{k}{\begin{pmatrix}
\frac{\mathlarger{e^{-\lambda_j t} }}{\prod\limits_{p=1,p \neq j}^{k}{
\lambda_j-\lambda_p}}
\end{pmatrix}}\prod\limits_{p=1}^{k-1}{\lambda_p}
}
where $\lambda_j$ is the decay constant for nuclide $j$. For our model of independently radiating atoms, the decay constant is nothing more than the multiplicity of excited atoms in a given population\footnote{In principle Bateman's equations can be adapted to give a closed-form solution to superradaiance populations as well. The outer sum is not analytically amenable however. The best simplification we have been able to identify is $\ochi{n_0}{n_1}\left(t\right)=\sum\limits_{j=0}^{n_0} \frac{(-1)^{j} e^{-(j+1) t (N-j)} n_0! N! (N-2 j-1) \left(n_1-j-2\right)! }{j! n_1! \left(n_0-j\right)! (N-j-1)!}\,.$} times $\Gamma$. Note the Bateman equations index the initial nuclide by $k=1$, whereas for standardradiance, however, we index the initial state $\chi_{n_1=N}$ by $n_0 = 0$. Therefore to translate the Bateman equations for our purposes we set $k \rightarrow n_0 + 1$ and $\lambda_{n_0} \rightarrow \Gamma n_1$ or $\lambda_k \rightarrow \Gamma\left(N+1-k\right)$ such that we have
\LeftEqns{
\ox{n_0}{n_1}\left(t\right) = \mathlarger{\sum}\limits_{j=1}^{n_0+1}{\begin{pmatrix}
\frac{\mathlarger{e^{-\Gamma\left(N+1-j\right)t}} }{\prod\limits_{p=1,p\neq j}^{n_0+1}{\Gamma\left(j-p\right)}
}\end{pmatrix}}\prod\limits_{p=1}^{n_0}{\Gamma\left(N-p+1\right)}\,.
}
This variant of the Bateman equations, with integer-values decay constants, can be readily simplified, to yield
\LeftEqn{\label{eq:newsepform}
\ox{n_0}{n_1}\left(t\right) = \frac{N!}{n_0!n_1!}{\left(1-e^{-\Gamma t}\right)}^{n_0}{\left(e^{-\Gamma t}\right)}^{n_1}\,.
}

It is valuable at this point to introduce a dimensionless and rescaled time parameter, namely 
\LeftEqn{
y = 1-e^{-\Gamma t}\quad\text{ ie. }\quad t=-\frac{\ln{\left(1-y\right)}}{\Gamma}
}
so that $y$ ranging between zero and one maps to $t$ ranging from zero to infinity. With this rescaled time we see that 
\LeftEqn{
&\ox{n_0}{n_1}\left[y\right] = N!\frac{{y}^{n_0}{\left(1-y\right)}^{n_1}}{n_0!n_1!}\\
\hspace{-\mathindent}\text{ie.}\quad &\rho_{\text{\tiny StandardRad}}\left[y\right]=\sum\limits_{n_1=0}^{N}{N!\frac{{y}^{n_0}{(1-y)}^{n_1}}{n_0! n_1!}\ket{D^N_{n_1}}\bra{D^N_{n_1}}}=\rho^N\left[y\right]
}
and therefore not only $\rho_{\text{\tiny StandardRad}}\in \brho_{\text{\tiny SDS}}\,$, but furthermore the standarradiating systems are themselves the {\em basis} states of the SDS parameterization!

This has two important physical ramifications. Firstly, whereas the SDS states were originally constructed purely for their mathematical form, this correspondence with standardradiating system provides a physical interpretation of those states. Secondly, this tells us that we could have defined a sufficient separability criteria capable of certifying the full separability of superradiance {\em without requiring a priori separable states with similar form}. By {\em presuming} that the phenomenological model of independently radiating atoms must be absent entanglement, one can define a separability criteria by testing whether or not it is possible to express the superradiant state $\rho_{\text{\tiny SuperRad}}\left[t^* \right]$ in terms of a {\em mixture} of fundamentally-independently-radiating systems at various snapshots in time, $\rho_{\text{\tiny SuperRad}}\left[t_i \right]$. This amounts to the test of possible decomposition $\rho_{\text{\tiny SuperRad}}\left[t^* \right] \overset{\text{?}}{\rightleftharpoons} \sum\limits_{j=1}{x_j \rho_{\text{\tiny SuperRad}}\left[t_i\right]}\,$. Just as per \ceq{eq:sepcrit}, the possibility of decomposition into particular separable states implies a sufficient separability criterion directly in terms of the matrix elements, namely
\LeftEqn{\label{eq:indepsepcrit}
&\rho_{\text{\tiny SuperRad}}\left[t^*\right]\in\brho_{\text{\tiny StandardRad}}\;\text{ iff }\;\exists\;{x_1\dots x_{j_\text{max}}, t_1 \dots t_{j_\text{max}}}\\
 &\text{satisfying }\;\forall_{n_1} \quad {\chi_{n_1}} \,=\, \sum_{j=1}^{j_{\text{max}}}{x_j \ox{n_0}{n_1}}
    \\&\text{such that }\;\forall_{j}:\; 0\le \,x_j\,\le 1\,\text{ and }\,0\le \,t_j\,\le \infty
}
which, if we change times variables such that $t_j\to\frac{-\ln{\left(1-y_j\right)}}{\Gamma}$ and we substitute the closed form expression for the standardradiance populations given by \ceq{eq:newsepform}, then the criterion \noeq{eq:indepsepcrit} is transformed identically into the criterion \noeq{eq:sepcrit}.

This alternative method of separability certification, namely by forcing a comparison with a statistically-independent analog of the target model, is of particularly practically value. We hope that similar comparisons, and the sufficient separability criteria which follow, may be of use to other researchers attempting to certify separability as well.

\FloatBarrier

\section{Entanglment via Driven Superradiance}
Although we have established, perhaps counter to expectations that there is no entanglement in Dicke model superradiance, it is interesting to explore a variant of this model in which we can show that entanglement is indeed generated. The model which we consider now is that of driven Dicke model superradiance, which has been considered repeatedly \cite{Drummund1978,Drummond1980,Breuer2007Theory,superrad2010,DrivenSuperradAlmost,DrivenSuperradPorras}, and which \citet{DrivenSuperradPorras} have shown leads to a spin-squeezed steady state. Here we analytically affirm the numerical results of \citet{DrivenSuperradPorras}, however using a spin-squeezing parameter more sensitive at detecting entangled states. This section of the thesis summarizes the salient results discussed in our work in Ref. \cite{DrivenSuperrad}. Please not that in Ref. \cite{DrivenSuperrad} we parameterized the density matrix using unnormalized Dicke states in the $j,m$ spin notation. To be consistent with the body of this thesis, however, we have translated our results into the notation of established by the Dicke basis of \ceq{eq:firstdef}.

Driven superradiance is a generalization of Dicke model superradiance when the system is additionally driven by some external field; we take the external driving frequency in our model to be $\omega$ and use the Rotating Wave Approximation \cite{scully1997quantum,RWADerivation1,RWADerivation2}. Thus the Liouville master equation \cite{Breuer2007Theory,NonMarkovianity} which governs driven superradiance is 
\begin{align}\label{eq:drivenlindblad}
&\hspace{-\mathindent}\frac{\partial \rho}{\partial t} =-\ic\left[\frac{\omega}{2}\left(A^+ + A^- \right),\rho\right]+\Gamma\left(A^- \rho A^+ - \frac{A^+ A^- \rho + \rho A^+ A^-   }{2}\right)\,,
\end{align}
a simple extension of \ceq{eq:lindblad}, and where the raising and lowering operators are still defined by \ceq{eq:Dplusminus}.

To solve \ceq{eq:drivenlindblad} we need not consider a fully-general density matrix $\rho$. Firstly, the equation is symmetric with respect to permutation of the individual qubit Hilbert spaces, so we can take our density matrix to be symmetric, that is, expandable in symmetric basis states of the Dicke states of \ceq{eq:firstdef}, although no longer diagonal in that basis. Second, the raising and lowering nature of the driving potential allows us to infer which matrix elements must be real and which must be (entirely) imaginary, and therefore we can define a sufficiently-general $N$-particle density matrix
\LeftEqn{\label{eq:rhodef}
\rho_N&=\sum\limits_{n_a=0}^{N}{\sum\limits_{n_b=0}^{N}{\operatorname{X(N)}\limits_{n_a}^{n_b} {\ic}^{\left(n_a-n_b\right)}{\ket{D^N_{n_a}}\bra{D^N_{n_b}}}}}
\\&\text{ with real symmetric }\operatorname{X(N)}\limits_{n_a}^{n_b}=\operatorname{X(N)}\limits_{n_b}^{n_a} \in \mathbb{R} 
}
for shorthand, we shall refer to the various $\operatorname{X(N)}\limits_{n_a}^{n_b}$ as the matrix elements of $\rho_N$, although technically we ought to account for the phase as well. 

It is possible to infer a rate equation in terms of the matrix elements from \ceq{eq:rhodef} in the same manner that \ceq{eq:superrad} follows from \ceq{eq:lindblad}. To do so we need only recall the effects of the ladder operators on the Dicke states, namely
\LeftEqn{
\renewcommand*{\arraystretch}{0.5}
\hspace{-\mathindent}A^{+} \ket{D^N_{n}}&=\left\lbrace\begin{array}{rl} {\sqrt{\left(n+1\right)\left(N-n\right)}\ket{D^N_{n+ 1}}\equiv f_+(n)\ket{D^N_{n+ 1}}\,,}&{n+1\leq N }\\{ 0\,,}&{n+1> N}\end{array}\right.\quad\text{and}\\
\hspace{-\mathindent}A^{-} \ket{D^N_{n}}&=\left\lbrace\begin{array}{rl} \sqrt{\left(n\right)\left(N-n+1\right)}\ket{D^N_{n-1}}\equiv f_-(n)\ket{D^N_{n- 1}}\,,&n-1\geq 0 \\ 0\,,&n-1< 0\end{array}\right.
}
which allows us to apply the operators of \ceq{eq:drivenlindblad} inside the summations in \ceq{eq:rhodef}. If we then re-index the dummy variables of summation so as to have a common index in the Dicke basis ${\ket{D^N_{n_a}}\bra{D^N_{n_b}}}$, as opposed to a common index in $\operatorname{X(N)}\limits_{n_a}^{n_b}$, the we obtain a set of coupled first-order differential equations defined by 
\LeftEqns{\label{eq:drivendiffeq}
\frac{\partial\operatorname{X(N)}\limits_{n_a}^{n_b}(t)}{\partial t}&=\frac{\omega}{2}\left(
f_+(n_a)\operatorname{X(N)}\limits_{n_a+1}^{n_b}(t)
+f_+(n_b)\operatorname{X(N)}\limits_{n_a}^{n_b+1}(t) \right)\\\nonumber
&-\frac{\omega}{2}\left(f_-(n_a)\operatorname{X(N)}\limits_{n_a-1}^{n_b}(t)
+f_-(n_b)\operatorname{X(N)}\limits_{n_a}^{n_b-1}(t)\right) \\\nonumber
&+\Gamma\left(f_+(n_a)f_+(n_b)\operatorname{X(N)}\limits_{n_a+1}^{n_b+1}(t)+\frac{f_-(n_a)^2+f_-(n_b)^2}{2}\operatorname{X(N)}\limits_{n_a}^{n_b}(t)\right)
}
which have no imaginary elements, hence justifying the manual choice of phases in \ceq{eq:rhodef}. Setting the left hand side of \ceq{eq:drivendiffeq} to zero defines the steady state condition, along with
\begin{align}\label{eq:tracecondition}
\operatorname{tr}{\left[\rho_N\right]}=\sum\limits_{n=0}^{n}{\operatorname{X(N)}\limits_{n}^{n}} =1
\end{align}
to account for normalization.

To obtain, practically, the steady-state matrix elements from \ceq{eq:drivendiffeq} we need to iterate it over all possible $0\leq n_a,n_b \leq N$, amounting to $(N+1)^2$ equations. Without loss of generality we can invoke the symmetry of the matrix elements to consider only $0\leq n_a\leq n_b \leq N$, which reduces the set of equations by about a factor of two. Even leveraging the symmetry, however, the set of linear equations scales like  $\mathcal{O}{\left(N^2\right)}$, and thus has quadratic computational complexity.

Spin Squeezing provides a valuable metric of entanglement \cite{SpinSqueezing2001,NoriSpinSqueezing,TothSpinSqueezing,CiracSpinSqueezing,OptimalSpinSqueezingParameterEarly,OptimalSpinSqueezingParameter}, with extensive immediate application in precision metrology \cite{SpinSqueezing2001,TothSpinSqueezing,Gross2010Nature,NoriSpinSqueezing,PrecisionToth,Gross2012IOP,FisherPrecisionMeasurement}. We use the explicit form of the spin squeezing parameter of \citet[Eq. (57)]{NoriSpinSqueezing} and \citet[Eq. (45)]{SpinSqueeze2013}, as follows:
\begin{align}\label{eq:chi}
&\hspace{-\mathindent}{\xi^2}=\frac{\Braket{{\jbar_1}^2+{\jbar_2}^2}-\sqrt{\Braket{{\jbar_1}^2-{\jbar_2}^2}^2+\Braket{{\jbar_1}{\jbar_2}+{\jbar_2}{\jbar_1}}^2}}{2/N}
\end{align}
where
\begin{align}\begin{split}\label{eq:j12}
    \jbar_1 &= \jbar_y\cos{\phi}-\jbar_x\sin{\phi}\\
    \jbar_2 &= \jbar_x\cos{\theta}\cos{\phi}+\jbar_y\cos{\theta}\sin{\phi}-\jbar_z\sin{\theta} 
\end{split}\end{align}    
and
\begin{align}\begin{split}\label{eq:phitheta}
    \theta &= \cos^{-1}{\left(\nicefrac{\Braket{\jbar_z}}{\sqrt{{\Braket{\jbar_x}}^2+{\Braket{\jbar_y}}^2+{\Braket{\jbar_z}}^2}}\right)} \\
    \phi &= \tan^{-1}{\left(\nicefrac{\Braket{\jbar_y}}{\Braket{\jbar_x}}\right)\,\text{, sensitive to quadrant,}}
\end{split}\end{align}
and where
\begin{align}
    \jbar_{x/y/z} &= \frac{1}{N}\sum\limits_{n=1}^N{\underbrace{\id ... \id}_{n-1}\otimes \sigma_{x/y/z} \otimes \underbrace{\id ... \id}_{N-n}} \,.
\end{align}
This particular measure of spin squeezing is denoted with a subscript $S$ in Ref. \citep[Table 1]{NoriSpinSqueezing}, where it is credited to \citet{KitagawaUeda}.

The calculation of \({\xi}^2\) can be immensely simplified by recognizing that the entire system's spin is encoded in the $\rho$'s one or two particle reduced states. For states with real and imaginary parts {\`a} la \ceq{eq:rhodef} we show in the Supplementary Online Materials of Ref. \cite{DrivenSuperrad} that
\begin{align}\label{eq:assumedmin}
\xi^2_{_N}&=1+\left(N-1\right)\Braket{\sigma_x \otimes \sigma_x \;\rho_N^{(2)}}\,,
\end{align}
for driven superradiance, where subscript $N$ indicates this special-case form. We have also introduced here $\rho_N^{(d)}$ to indicate the reduced state of $d$ particles, and we elect to explicitly specify the reduced-state $\rho$ in the expectation value purely for pedagogical clarity. Note that \ceq{eq:assumedmin} is also derived for symmetric states in Ref. \citep[Eq. (7)]{SymmetricSpinSqueezing}.

Spin-squeezing is defined by ${\xi^2<1}$, which is also a sufficient criterion for the presence of entanglement. With no loss of generality we therefore have certification of nonzero entanglement \cite{entang.review.toth,multireview,characterizingentanglement,SymmetricEquivalents,eckert2002quantum} via
\begin{align}\label{eq:guarantee}
\forall_N:\,\,\rho_N\in \brackets{\bm{\mbox{\large \(\varrho\)}}_\text{entangled}}\text{ if }\Braket{\sigma_x \otimes \sigma_x \;\rho_N^{(2)}}<0\,
\end{align}
which, since $\forall_N:\,\,\Braket{\sigma_x \;\rho_N^{(1)}}=0$, means that \ceq{eq:guarantee} is just a special case of the the general entanglement criteria of \cite[Eq. (33)]{RamseySpinSqueezing,SymmetricSpinSqueezing,SymmetricEquivalentsPRL,OptimalSpinSqueezingParameter}, which recognizes that all separable symmetric states satisfy $\Braket{A \otimes A}-\Braket{A \otimes \id} \geq 0$ for all Hermitian operators A.

The essential contribution of our work in Ref. \cite{DrivenSuperrad} is to provide an explicit expression for $\Braket{\sigma_x \otimes \sigma_x \;\rho_N^{(2)}}$ directly in terms of the unreduced matrix elements $\operatorname{X(N)}\limits_{n_a+1}^{n_b}$ of $\rho_N$ per \ceq{eq:rhodef}. We found that $\rho_N^{(d)}$ can be expressed as linear map acting on $\rho_d$. we briefly summarize the argument there, translating it to the notation of this thesis. We recognize that the Dicke states can be expressed in an arbitrary bipartitioned basis through the Clebsch-Gordan coefficients \cite{Sharp1960,Schertler1996Combinatorial,Biedenharn1981,Coala2010,OHara2001,Guseinov2009,Louck2008,Ulfbeck2007,WolfeCG}, such that
\LeftEqn{
\hspace{-\mathindent}\rho_N=&\sum\limits_{n_a\bp=0}^{d}\sum\limits_{n_a\bpp=0}^{N-d}\sum\limits_{n_b\bp=0}^{d}\sum\limits_{n_b\bpp=0}^{N-d} \Bigg( \operatorname{X(N)}\limits_{n_a\bp+n_a\bpp}^{n_b\bp+n_b\bpp} {\ic}^{\left(n_a\bp+n_a\bpp-n_b\bp-n_b\bpp\right)}\times \operatorname{CG}(n_a\bp,n_a\bpp)\operatorname{CG}(n_b\bp,n_b\bpp)\\
&\times\ket{D^d_{n_a\bp}}\otimes\ket{D^{N-d}_{n_a\bpp}}\bra{D^d_{n_b\bp}}\otimes\bra{D^{N-d}_{n_b\bpp}} \Bigg)
}
where $\operatorname{CG}(n\bp,n\bpp)$ is shorthand for the full-spin Clebsch-Gordan coefficient. Explicitly,
\LeftEqn{
\operatorname{CG}(n\bp,n\bpp)\equiv\frac{\binom{d}{n\bp}\binom{N-d}{n\bpp}}{\binom{N}{n\bp+n\bpp}}=\sqrt{\frac{d!(N-d)!(n\bp+n\bpp)!(N-n\bp-n\bpp)!}{n\bp!(d-n\bp)!n\bpp!(N-d-n\bpp)!N!}}
}
per \citet[Eq. (7)]{WolfeCG} and \citet[Eq. (10)]{OHara2001}.
Now the definition of the reduced state requires tracing out the second ``half'' of the partitioned Hilbert space, namely
\LeftEqn{
\rho_N^{(d)}=\sum\limits_{k=0}^{N-d}{\left(\id\otimes\bra{D^{N-d}_{k}}\right)\rho_N\left(\id\otimes\ket{D^{N-d}_{k}}\right)}
}
which acts to isolate only those elements which are diagonal in the Dicke basis of that Hilbert space, such that we have
\LeftEqn{
\hspace{-\mathindent}\rho_N^{(d)}=&\sum\limits_{k=0}^{N-d}\sum\limits_{n_a\bp=0}^{d}\sum\limits_{n_b\bp=0}^{d} \Bigg(\operatorname{CG}(n_a\bp,k)\operatorname{CG}(n_b\bp,k) \operatorname{X(N)}\limits_{n_a\bp+k}^{n_b\bp+k} {\ic}^{\left(n_a\bp-n_b\bp\right)}\ket{D^d_{n_a\bp}}\bra{D^d_{n_b\bp}} \Bigg)
}
which we choose to express as a linear map acting on $\rho_d$, such that
\LeftEqn{\label{eq:sigmamap}
&\rho_N^{(d)}=\nmapalt{\big(\rho_d\big)}\,,\;\text{ where 
}\\
&\nmap{\operatorname{X(d)}\limits_{n_a\bp}^{n_b\bp}}\to \sum\limits_{k=0}^{N-d}{\operatorname{CG}(n_a\bp,k)\operatorname{CG}(n_b\bp,k) \operatorname{X(N)}\limits_{n_a\bp+k}^{n_b\bp+k}}\,.
}
Note that this mapping can readily be generalized for all symmetric states, not just those of the ansatz of \ceq{eq:rhodef}.

Now, since one can readily verify that
\LeftEqns{
\Braket{\sigma_x \otimes \sigma_x \;\rho_2}=&\sum\limits_{s=0}^{1}{(-1)^s\binom{2}{s}\operatorname{X(2)}\limits_{1-s}^{1+s}}\quad\text{and since}\\\label{eq:sigmavalue}
\Braket{\sigma_x \otimes \sigma_x \;\rho_N^{(2)}}&=\nmap{\Braket{\sigma_x \otimes \sigma_x \;\rho_2}}\\\nonumber
&=\sum\limits_{k=0}^{N-2}{\sum\limits_{s=0}^{1}{(-1)^s\binom{2}{s}\operatorname{CG}(1+s,k)\operatorname{CG}(1-s,k)\operatorname{X(N)}\limits_{1-s+k}^{1+s+k}}}
}
we are now able to explicitly define the spin squeezing parameter of \ceq{eq:assumedmin} directly in term of the matrix elements. Furthermore, we can simplify \ceq{eq:sigmavalue} by noting that, for the relevant parameter range,
\LeftEqn{
\binom{2}{s}&\operatorname{CG}(1+s\bp,k)\operatorname{CG}(1-s\bp,k)\\
&=\frac{2}{N(N-1)} \sqrt{(i+1)(i+1+s)(N-i-1)(N-i-1+s)}\,.
}

Finally, reindexing the summation and inserting our result into \ceq{eq:assumedmin},  we conclude that
\LeftEqn{\label{eq:altchi}
\xi^2_{_N}&=1+\frac{2}{N}\sum\limits_{q=1}^{N-1}{\sum\limits_{s=0}^{1}{(-1)^s \sqrt{(q)(q+s)(N-q)(N-q+s)}\operatorname{X(N)}\limits_{q-s}^{q+s}}}\,,
}
a simple expression that only draws upon the diagonal and the one-off-diagonal matrix elements of $\rho_N$.

\begin{figure}[ht]
\centering
\includegraphics[width=0.965\linewidth,center]{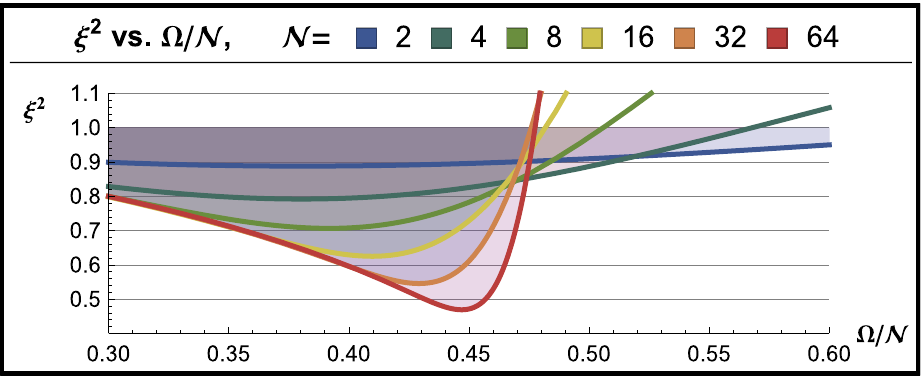}
\caption{(Color online) We graph  the spin-squeezing parameter $\xi^2$ for a steady state driven superradiant system as a function of $\Omega/N$ for various small $N$. Recall that $\Omega$ is the ratio of the driving frequency to relaxation frequency, $\Omega \equiv \omega/\Gamma$. The state is spin-squeezed whenever $\xi^2<1$; shown as shaded regions in this graph. The minima of the curves descends further with increasing $N$.}\label{fig:figXvO}
\vspace{2.8pt}
\includegraphics[width=0.965\linewidth,center]{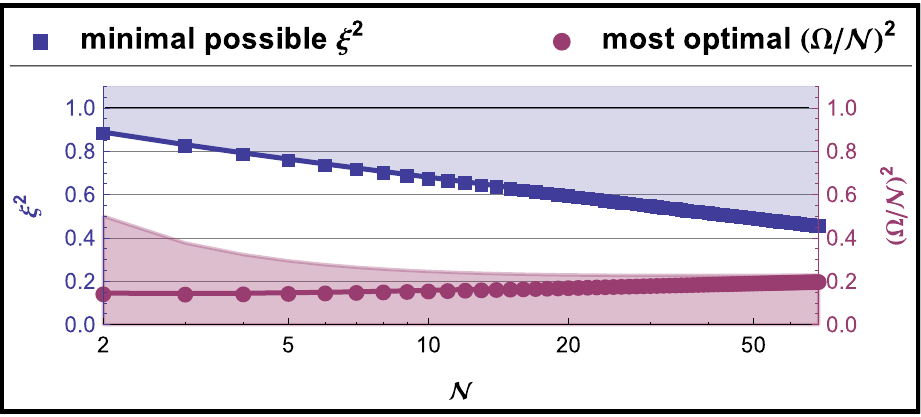}
\caption{(Color online) We plot the best-case scenario values for entanglement generation as a function of system size $N$. Note the dual meaning of the $Y$ axis: The upper curve indicates the minimal possible $\xi^2$, it is shaded upward to indicate that all larger values are also achievable. The lower curve indicates the optimal choice of $(\Omega/N)^2$ to achieve the corresponding minimal $\xi^2$. The lower shading indicates the complete parameter region where the steady state is spin squeezed. Note the logarithmic scaling of the $X$ axis.} 
\label{fig:figXvN}
\end{figure}

\begin{landscape}
\begin{figure}[p]
\includegraphics[width=1\linewidth,left]{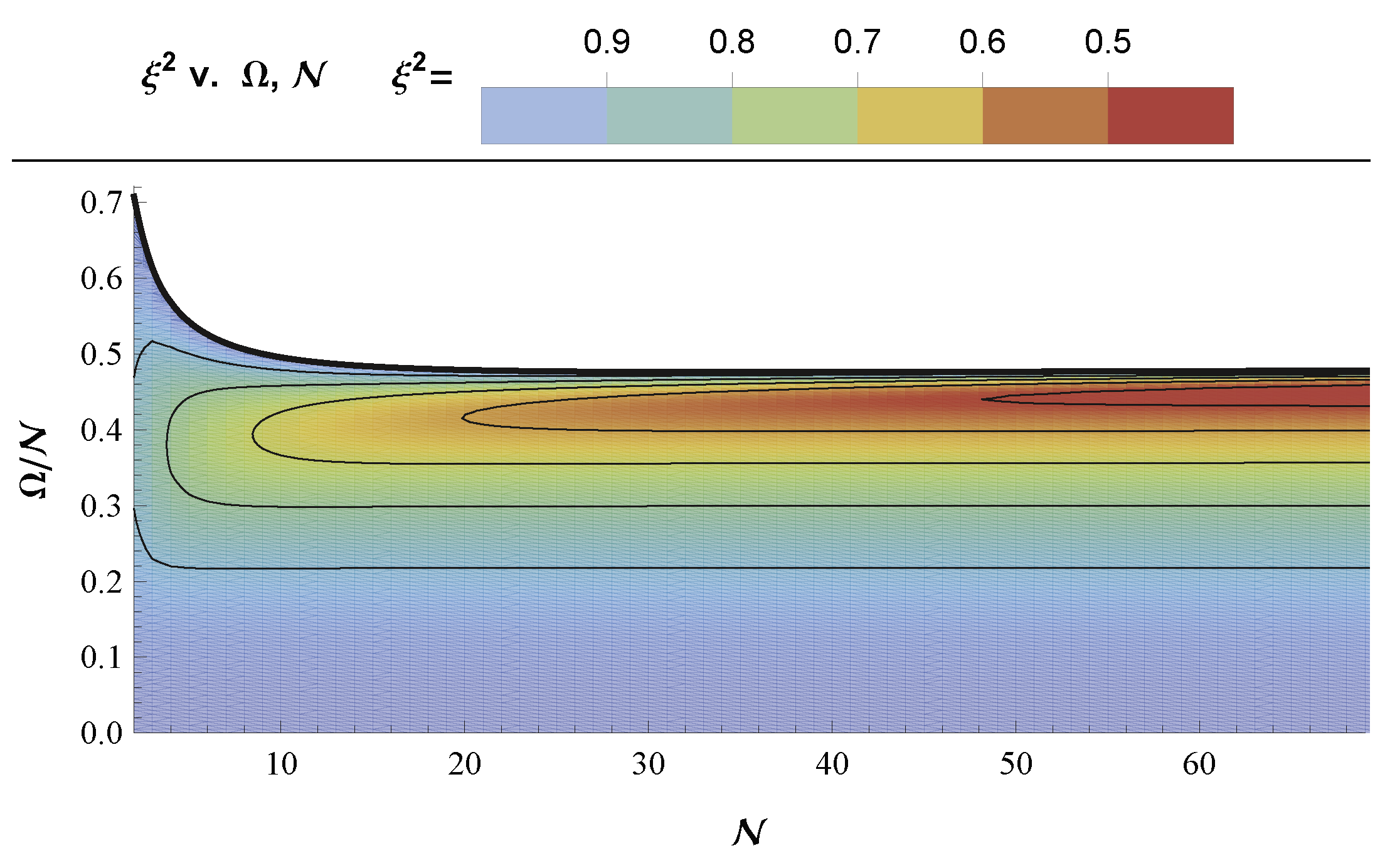}
\caption{(Color online) This contour plot shows the spin-squeezing parameter $\xi^2$ for a steady state driven superradiant system as a function of $\Omega = \omega/\Gamma$ over a dense set of $N$, among which are the discrete $N$ plotted in \fig{fig:figXvO}. 
We plot only the region where $\xi^2<1$. Red indicates strongest spin squeezing, ie. minimal $\xi^2<1$. Although hard to see, the $\xi^2=1$ boundary is {\em not} monotonically decreasing; rather, it's minimized at $\Omega/N=36$.}\label{fig:figOvN}
\end{figure}
\end{landscape}

Our question now is can we find some $\omega$ for a given $\Gamma$ such that we can drive the system into an entangled state characterized by $\xi^2<1$? Yes! We quantify the entanglement of the steady state in terms of $\Omega\equiv \omega/\Gamma\,$, defined as the ratio of the two experimental parameters. We find the steady state to be spin squeezed, ie. with measure $\xi^2<1$, for sufficiently small $\Omega$; see for example \fig{fig:figXvO}. To make a general statement, we note that for all $N$, when $ |\Omega| \lesssim 0.475 N$ the resulting steady state is always at least somewhat spin-squeezed state, see \fig{fig:figOvN}.

One would like to know how to tune $\Omega$ so as to maximize the entanglement in the resulting steady state. To this end, see \fig{fig:figXvN} where it appears that the optimal $\Omega$ scales like $(\Omega/N)^2 \sim a \ln{N}+b$ for large $N$. It is also desirable to quantify the maximal extent of the spin squeezing that can achieved in the model of driven Dicke superradiance. Per \fig{fig:figXvN}, the squeezing extent rapidly strengthens for large systems. Indeed, the value of the best-possible $\xi^2$ almost appears to drop off logarithmically as a function of $N$, descending below 0.5 at the right edge of \fig{fig:figXvN} with no sign yet of tapering off. This suggest that by increasing the size of the system, $\xi^2$ can perhaps be made arbitrarily small in the steady state of this model. With the usual caveats that genuine superradiance suffers from volume-dependent effect not accounted for in the Dicke model \cite{superrad.yelinPRA,superrad.yelinBook}, this result nevertheless further suggest that driven superradiance may be a viable scheme for generating large tightly squeezed states.

It is worth noting that the spin squeezing parameter is related to to the entanglement monotone {\em Negativity} \cite{OneNegativeEigenvalue,NegativityIntro,multireview,PlenioEntanglementMeasures}. The Negativity is equal to the combined magnitude of all negative eigenvalues in the partial transpose of $\rho$, ie.
\begin{align}\label{eq:negativity}
\operatorname{\mathscr{N}}{\hspace{-0.25ex}\Big(\rho\Big)} &\equiv \frac{\big(\sum_{i}{\left|\lambda_i\right|}\big)-1}{2}\,.
\end{align}
The Negativity is a common benchmark of a state's distillability and resource value for nonlocality \cite{GivenQB.NQubit,GivenQB.Cabello}. 

For a $2\times 2$ system, such as $\rho_N^{(2)}$, it is known that the partial transpose is always full rank and has at most one negative eigenvalue \cite{OneNegativeEigenvalue}, in which case the Negativity is the magnitude of that single negative eigenvalue. By direct computation we find that $\nicefrac{\Braket{\sigma_x \otimes \sigma_x \;\rho_2^{\vphantom{(2)}}}}{2}$ is one of the eigenvalues of $\rho_2^{PT}$, and thus via the mapping of \ceq{eq:sigmamap} we also have that $\nicefrac{\Braket{\sigma_x \otimes \sigma_x \;\rho_N^{(2)}}}{2}$ must be an eigenvalue of general ${\rho_N^{(2),PT}}$. What we see is that the spin-squeezing parameter is effectively a linear function of the reduced state Negativity, such that \ceq{eq:assumedmin} has the corollary 
\begin{align}\label{eq:negrelation}
\text{If }\xi^2_{_N}<0\,,\text{ then }\;\xi^2_{_N} & = 1-2\left(N-1\right)\operatorname{{\mathscr{N}}}{\hspace{-0.25ex}\Big(\rho_N^{(2)}\Big)}\,.
\end{align}
See Refs. \cite{OneNegativeEigenvalue,ConvVsNeg} for a translation between the Negativity and Concurrence entanglement monotones, as the Concurrence has in some sense become a conventional standard metric for multiparticle entanglement \cite{MultipartiteConc}, such as in Refs. \cite{SpinEntanglementMorrison,ConcVsGeom}. Spin squeezing is directly related to the two-particle Concurrence in Ref. \citep[Eq. (5)]{RamseySpinSqueezing} and to the CCNR criteria in Ref. \citep[Obs. 2]{SymmetricEquivalents}.

\newgeometry{left=1.00in,right=1.0in,top=1.4in,bottom=1.4in}
\restoregeometry

The bipartite entanglement detected by spin squeezing is an indication of the nonlocal capacity of the state, and vice versa. \citet{EntanglementOnlyTwoBodyToth} discuss how one cannot infer the degree of entanglement and the extent of spin squeezing from the observation of bipartite statistical correlations. \citet{AcinDetecting} discuss how general nonlocality of a quantum state can be inferred by Bell-testing it using exclusively bipartite measures such as that of \ceq{eq:guarantee}, see also Refs. \cite{BellTwoBodyOnlyArXiv,BellInequalityHomogenization}. \citet{PrecisionViaNonlocality} goes so far as to demonstrate that the well-known enhancement of precision measurement due to entangled states \cite{SpinSqueezing2001,TothSpinSqueezing,NoriSpinSqueezing,Gross2010Nature,PrecisionToth,Gross2012IOP,FisherPrecisionMeasurement} is actually more a consequence of the nonlocality capacity of the state than its entanglement.

We conclude that while entanglement may not be a direct proxy for nonlocal potential \cite{UPBOriginal,EntanglementSurprise2005,EntanglementSurprise2005b,EntanglementSurprise2007,EntanglementSurprise2011,EntanglementSurprise2014,RandomnessFromNonlocality,SteeringNonlocalityBrunner} it is deeply connected to nonlocality \cite{AllEntangledNonlocality,AllEntangledNonlocal,CloserConnections}, and is furthermore an extraordinarily valuable resource in that it allows for a plethora of non-classical phenomena \cite{CryptoPRA,CryptoPRL,CryptoDIQKD,RenatoQKDNature,DickeNetworking,mosca_algorithms_revisited,MBQC,MBQC2,QuantOptBook,Schumacher2010Quantum,Nielsen2011Quantum,Wilde2013Quantum,SpinSqueezing2001,TothSpinSqueezing,NoriSpinSqueezing,Gross2010Nature,PrecisionToth,Gross2012IOP,FisherPrecisionMeasurement}. 

Entangled qubits represent a uniquely valuable subtype of quantum resource in that they \emph{are} a direct proxy to multipartite binary and dichotomic quantum nonlocality \cite{MasanesQubits}, see also Refs. \cite{GivenQB.NQubit,GivenQB.Cabello,NQubitEquivalences}.  More broadly, symmetric quantum states are valuable in that they mirror the symmetry structure of multipartite no-signalling \glspl{polytope} \cite{GisinFramework2012,roberts_thesis,WernerWolfQB,WehnerQB,ScaraniNotes,*ScaraniNotes2}. The relationship between entanglement and nonlocality remains an open question; we hope that the techniques for assessing and generating entanglement that we have developed in this chapter may prove useful in furthering quantum information theory, both conceptually and practically.

\chapter{Contextuality}
\vspace*{\fill}
\begin{spacing}{1.5}
\begin{tabularx}{400pt}{m{1cm}Lm{1cm}}
& \noindent\textit{The fundamental theorem of quantum mechanics [is...] if you have several questions, and you can answer any two of them, then you can also answer all of them.}
\footnote{\textsc{Ernst Specker}, 2009\\ \textit{Recorded by Adán Cabello}, reflecting on the Kochen-Specker Theorem \cite{EPSpeckerVid}.} &
\end{tabularx}
\end{spacing}
\vspace*{\fill}

\section{The Graph-Theoretic Formalism}
As mentioned in Sec. \ref{sec:intro}, the outcomes of quantum measurements are contextual. Contextuality is broader than merely nonlocality, in the sense that quantum contextuality can be exhibited without requiring spatially separated subsystems. Contextuality, tautologically, is evident whenever measurement outcomes for a single-partite experiment are inconsistent with noncontextually (albeit probabilistically) assigning outcomes to the various measurements \cite{MerminContextuality,PreperationContextuality}. For explicit examples of quantum contextuality distinct from quantum nonlocality, see Refs. \cite{SpekkensSeer,Pentagrams,ExperimentalContextuality,ks18,CSWNew,BellAndNonContextualInequalities,KCBS,contextualitycircle}. It is precisely the inability to pre-assign probabilities to even a single-partite scenario (``noncomposite'' in the language of \citet{TwoQubitsSeparabilityContextuality}) scenario which makes contextuality more broadly applicable than nonlocality \cite{BellAndNonContextualInequalities,ContextualityLeggettGarg}.

Just as with nonlocality, contextuality is valued as a resource for information theoretic tasks \cite{naturecompeditorial,emersoncontextuality}. Further similarly, the extent of \emph{quantum} contextuality is \emph{intermediate} between absolute noncontextuality and maximal contextuality conceivable without violating probability normalization. The distinction between classes has recently been formalized with the aid of graph-theoretic approach due to \citet*{CSWOld}. This formalism is also reviewed in Refs. \cite{EPOriginal,FritzCombinatorialLong,CSWNew}. 

One can express the contextuality scenario as an exclusivity graph. This is done by assigning each vertex of the graph a label indicating some particular measurement outcome tuple (output) for some particular apparatus choice tuple (input). Noncontextuality is the notion that each individual output depends only on the individual input queried, whereas contextuality is when the output depends differently on the input depending on {\em which tuple} of inputs it is drawn from. To complete the exclusivity graph we draw edges between vertexes that are incompatible. By incompatible we mean that at least one input is common between both vertexes such that the output for that input is not the same at both vertexes. We refer to incompatible vertexes as orthogonal, adjacent, or exclusive. 

In this work, exclusivity between vertexes is represented through edges, thus in our notation exclusivity=orthogonality=adjacency. This is the notation used in Refs. \cite{CSWOld,EPOriginal,VariousCHSHBellInequalities,LONatureComm,CSWNew,LOExploring,CabelloMultigraph}. By contrast, in Refs. \cite{FritzCombinatorialLong,LONewShort} the authors work within the framework of nonorthogonality graphs, electing to represent exclusivity by non-adjacent vertices. \footnote{Note that different conventions for the graphical representation of exclusivity are used in different works even by the same authors. \citet{FritzCombinatorialLong} are also authors of Refs. \cite{LONatureComm,LOExploring}.} Our work accommodates both conventions by explicitly specifying which graphical representation is meant in each statement. 
To do this, we refer to graphs where exclusivity is represented by adjacency as orthogonality graphs - $\operatorname{O}{\left(G\right)}$. Graphs where exclusive vertexes are not adjacent are referred to as nonorthogonality graphs - $\operatorname{NO}{\left(G\right)}$.

Firstly, note that the total probability assigned to two adjacent vertexes in $\operatorname{O}{\left(G\right)}$ must not exceed 1. This is a simple consequence of probability normalization; summing the probability of seeing {\em all} various different outputs for a given constant input equals one, so any two different outputs for a given input have total probability no greater than one. We now proceed to assign probabilities to the various vertexes per various classes of statistical behaviors.

A deterministic behavior is one which assigns zero-or-one probabilities to the various vertexes. Note that the {\em total probability} of any deterministic behavior on the graph is bounded by the {\em Independence Number} - $\alpha\left(\operatorname{O}{\left(G\right)}\right)$, which is the vertex count of the largest possible set of vertexes in $G$ which do not mutually conflict, that is, there are no edges linking vertexes in an independent subset. A probabilistic noncontextual behavior essentially presumes that for every experimental query the experimenter is probabilistic exposed to some different possible deterministic underlying behavior, with the weights of the likelihood of the different behavior being normalized probabilities. Note that because the total probability of every deterministic behavior is bounded by $\alpha$, so to any convex combination of behaviors will also be bounded by $\alpha$. As such, having total probability less than or equal to the orthogonality graph's independence number is a universal feature of all noncontextual behaviors. Such behaviors are called NCHV, or noncontextual hidden variable models. 
\begin{align}\label{eq:nchvmax}
\text{If }\overrightarrow{p(v)}\in \text{ NCHV, then }\sum_v{p(v)}\leq {\alpha{\left(\operatorname{O}{\left(G\right)}\right)}}
\end{align}
Note that \ceq{eq:nchvmax} is a necessary condition for classicality, but not a sufficient one. A necessary and sufficient condition is given by \citet[Prop. 4.3.1]{FritzCombinatorialLong}.

A quantum contextual behavior requires modelling the probability with quantum measurement operators, ie. projectors, acting on a quantum state. As thoroughly discussed in Refs. \cite{CSWOld,EPOriginal,FritzCombinatorialLong,CSWNew,CabelloMultigraph}, the total probability of any quantum behavior on a graph is bounded by the graph's {\em Lov{\'a}sz Theta Number} - $\vartheta$. \footnote{There is a conflict in the definition of the Lov{\'a}sz theta number of a graph, depending on one's preference for representing orthogonality by adjacency or non-adjacency. We use the modern convention per \citet{knuth1993sandwich}. This is the convention used by both \citet{CSWNew} and \citet{FritzCombinatorialLong}.}. Following Refs. \cite{CSWOld,CSWNew} we have that
\LeftEqn{
&\vartheta(O(G)) = \max_{|\Psi\rangle,\:|\psi_v\rangle} \sum_{v\in V(G)} \:|\langle\Psi|\psi_v\rangle|^2
\\&\qquad\text{where ${|\Psi\rangle,\:|\psi_v\rangle}$ range over all unit vectors}
\\&\qquad\text{such that $u\perp v$ and $u\neq v$ implies $|\psi_u\rangle\perp|\psi_v\rangle$}.
}
where the orthogonal representation may be defined for $\operatorname{O}{\left(G\right)}$ if one insists that orthogonality ($u\perp v$) be represented by non-adjacency ($u\not\sim v$) such as in Refs.  \cite{knuth1993sandwich,FritzCombinatorialLong}. As such we have
\LeftEqn{\label{eq:qcontextual}
\text{If }\overrightarrow{p(v)}\in Q_{\mbox{contextual}},\text{ then }\sum_v{p(v)}\leq {\vartheta{\left(\operatorname{O}{\left(G\right)}\right)}}
}
which, analogously to \ceq{eq:nchvmax}, is a necessary condition for statistics to be quantum-contextual-achievable, but not a sufficient one. A necessary and sufficient condition for quantum contextuality in terms of the Lov{\'a}sz Theta Number is given by \citet[Prop. 6.3.2]{FritzCombinatorialLong}. In the language of graph theory, the quantum measurement projectors are called ``ribs'' and the quantum state being measured is called the ``handle'', after an archetypal five-vertex exclusivity graph with a three dimensional orthogonal  representation that looks like remarkably like an umbrella [\citealp[Theorem 2]{LovaszHandle}; see also \citealp{UPBOriginal,CabelloMultigraph}].

We note with some interest that we are not aware of a graph invariant which captures all general contextuality scenario beyond quantum contextuality. Such highly-general scenarios, however, are defined by {\em only} the restriction that adjacent vertexes have total probability less than one, see \citet{SpekkensSeer} for illustrative examples. Every probabilistic model of \citet{FritzCombinatorialLong}, by definition there, satisfies this general criterion.


Recently there has been intense interest in trying to derive quantum contextual statistics purely from a minimal set of informational principles \cite{MultiPrinciples2001,MultiPrinciples2011Masanes,MultiPrinciples2011Chiribella,MultiPrinciples2012}. Effort has shifted now to searching for a \emph{single} principle capable of recovering quantum nonlocality and contextuality. Landmark candidates included Macroscopic Locality and Information Causality \cite{MacroscopicLocalityOriginal,ScaraniML,ICOriginal,ICRecovery,ICFailure,InfoCausArXiv,ICGrayWyner}. The most recent and celebrated ideas, however, are the Exclusivity Principle \cite{EPOriginal,EPSpecker,EPExampleGraphs,EPHenson,EPYan,EP2013,EPTwoCities,EPTsirelson} and Local Orthogonality \cite{LONatureComm,FritzCombinatorialLong,LONewShort,LOHardy,LOExploring}.

Expressed in the graphical formalism, both these new principles posit imposing total probability less than one for any set of mutually exclusive vertices. Bounding \gls{clique} total probability is a more restrictive condition than merely bounding pairwise probability \cite{SpekkensSeer}. Indeed Refs. \cite{EPOriginal,LONatureComm,FritzCombinatorialLong,LONewShort,LOHardy,LOExploring,CSWNew,EP2013,EPTsirelson} have exploited this simple idea to tremendous effect.

A third approach has also been pioneered independently by \citet{SorkinOriginal}. Sorkin's Quantum Measure Theory was uniquely motivated by efforts to formulate a quantum theory of gravity, and serves as motivation for the  Consistent Histories interpretation of quantum mechanics \cite{Sorkin94quantummeasure,HistoriesReviewDowker}. Quantum Measure Theory operationally posits ``Sorkin's Sum Rule'' which forbids third-order interference phenomena. Recently it has been noted that Sorkin's Sum Rule recovers various properties of contextuality and nonlocality \cite{AltSorkingQ1a,AltSorkingQ1b}. In this thesis we treat Sorkin's principle as merely a special case of the Exclusivity Principle since \citet{SorkinQ1} has shown that Sorkin's Principle implies Consistent Exclusivity.\footnote{We note that the sum rule is effectively equivalent to requiring zero tripartite Interaction Information \citep[Eq. (6)]{BellInteractionInformation} if one understands third-order interference as merely a special physical case of tripartite Interaction Information. Interaction Information is the generalization of Mutual Information, which expresses the amount information (redundancy or synergy) bound up in a set of variables \emph{beyond} that which is present in any subset of those variables. Phrased as such, it becomes clear that Sorkin is concurring with Specker's sentiment [\citealp{KSOriginal,EPSpecker,EPSpeckerVid,SeevinckSpeckerTranslation}, and \citealp[Sec. 7.1]{FritzCombinatorialLong}] that ``A collection of propositions about a quantum mechanical system is precisely simultaneously decidable when they are pairwise simultaneously decidable.'' It is not clear to the authors why this correspondence between the principles of Sorkin and Specker has not been noted in earlier literature.}

There are two important limitations which we must keep in mind to modulate any expectations that principles such as the aforementioned might fully recover quantum nonlocality:

Firstly, the statistical constraints due to the Exclusivity Principle (EP) and Local Orthogonality (LO) have not been shown to recover quantum contextual statistics. \citet[Sec. 7.3]{FritzCombinatorialLong} has shown that `Consistent Exclusivity'' (CE) is related to the {\em Shannon Capacity} of nonorthogonality graphs, $\Theta{\left(\operatorname{NO}{\left(G\right)}\right)}$. CE is a \emph{gedankenexperiment} based on applying EP or LO to an infinite number of copies of a contextuality scenario, in which each copy is identical and independent. The existence of graphs for which the Shannon Capacity diverges from the Lov{\'a}sz Theta Number shows that CE fails, in general, to recover quantum contextual statistics \citep[Theorem 7.4.3]{FritzCombinatorialLong}. Admittedly, EP/LO may have ramifications beyond those which can be gleaned from the CE thought experiment, such as demonstrated recently by \citet{EPTsirelson}. It is therefore open that EP/LO might perhaps recover quantum contextual statistics. An extension of EP due to \citet{EPYan} is known to exactly correspond to quantum contextuality; see \citep[Theorem 7.7.2]{FritzCombinatorialLong}.


Secondly, and especially germane for this thesis, is that the statistical constraints due quantum contextuality fail to tightly define the constraints of quantum nonlocality. Although quantum nonlocality is innately a demonstration of quantum contextuality, the statistical boundaries of single-partite quantum contextuality per \ceq{eq:qcontextual} are {\em not sufficient} to define the set of nonlocal quantum statistics. As a consequence, EP/LO are {\em} not candidates for recovering quantum nonlocal statistics, only quantum contextual statistics.

This is not particular surprising, given that the  Lov{\'a}sz theta number is defined in terms of general quantum projective measurements without reference to any kind of product structure. In a genuine multipartite quantum experiment, although the shared state ${|\Psi\rangle}$ may be entangled, the measurement projectors ${|\psi_v\rangle}$ must represent tuples, ie. tensor products, of individual projectors for the different parties in different Hilbert spaces; at the very least the projectors must commute. It is the product structure (mandated commutativity) in the measurement projectors which gives rise to the ``gap'' between the quantum-contextual and quantum-nonlocal \glspl{elliptope} in conditional probability space. Contrast \citet{FritzCombinatorialLong}`s Proposition 5.2.1, where the projectors commute, with Corollary 6.4.2 there, where the projectors only \emph{appear} to commute, in that they are permutation invariant {\em with respect to a particular state}, ie. with respect to the handle associated with the orthogonal representation of the exclusivity graph. See also Refs. [\citealp{NonlocalityContextualityCommutativity}, and \citealp[Lemma 3]{AlmostQuantum2}]. 

The formal mathematical distinction between quantum nonlocality and quantum contextuality is the commutation of the parties' projectors which is ``imposed'' in quantum nonlocality. This is such a critical point that it bears repeating explicitly.

A nonlocality scenario can be ``relaxed'' to a contextuality scenario by maintaining the same exclusivity (hyper)graph, but generating the probabilities which live on the graph via noncommuting projectors. The measurements of Alice and Bob are now just considered different dimensions in a larger sample space; in general, nonlocality scenarios can readily be viewed as contextuality scenarios by identifying each party with a different degree of freedom of a single physical system. The fact that the temporal order of Alice and Bob's measurements is irrelevant can be implemented by imposing permutation invariance of the projectors with respect to whatever quantum state is being measured. 

As expressed by \citet{FritzCombinatorialLong} and \citet{AlmostQuantum2}, this idea can be extended to distinguish $N$-partite quantum nonlocality from a quantum contextuality scenario involving $N$ simultaneous measurements as follows: Let there be a normalized state \(\ket{\phi}\in {\cal H}\) and normalized projector operators \(\{E^{a,x}_k\}\subset B\left( {\cal H}\right)\) such that \(\sum_a E_k^{a,x}=\id\), for all $x$, $k$. This can be used to define the conditional probability distribution $P(a_1,...,a_N|x_1,...,x_N)=\bra{\phi}\bigotimes_{k=1}^n E^{a_k,x_k}_k\ket{\phi}$. Now the condition on the projectors which distinguishes nonlocality from contextuality is as follows:

\begin{description}
 \item[Contextuality:] The projectors commute with respect to the order of the simultaneous measurements: \(E^{a_1,x_1}_1...E^{a_N,x_N}_N\ket{\phi}=E^{a_{\pi(1)},x_{\pi(1)}}_{\pi(1)}...E^{a_{\pi(N)},x_{\pi(N)}}_{\pi(N)}\ket{\phi}\), where \(\pi\in S_N\) is an arbitrary permutation of the ``parties'' \(\{1,...,N\}\).
 \item[Nonlocality:]  The projectors for each party are assigned the own Hilbert spaces: \(\{E^{a,x}_k\}\subset B({\cal H}_k)\), implying that the global Hilbert space is effectively defined as \({\cal H} = \bigotimes_{k=1}^N {\cal H}_k\,\).
\end{description}
Note that projectors with satisfy the nonlocality condition inherently also satisfy the contextuality condition, but not vice versa. 

In this thesis we refer to this relaxation of nonlocality as precisely contextuality because it allows for the joint probability distribution to be generated by noncomposite measurements. Given a nonlocality scenario, one would naturally define the corresponding quantum contextual correlations as those which share the same exclusivity graph as the nonlocality scenario but which can be generated by noncomposite quantum measurements \cite[Proposition 6.3.1.d]{FritzCombinatorialLong}; indeed this is equivalent the the operational definition above in terms of ``almost'' commuting projectors \cite[Corollary 6.4.2]{FritzCombinatorialLong}. This perspective on quantum nonlocality is also expressed by \citet{CabelloMultigraph}, to whom we are grateful for sharing a preprint of Ref. \cite{CabelloMultigraph}.

The third chapter of this thesis is wholly dedicated to exploring the statistical discrepancy between correlations consistent with quantum nonlocality and those consistent with quantum contextuality, and the implications of this distinction. Our contribution to this effort is the novel identification of a \emph{region} in conditional probability space where the correlations which are impossible to implement in a quantum nonlocality scenario can nevertheless be achieved through a quantum contextuality scenario. Our results in this chapter have never before been published prior to this thesis.

\section{Quantum Contextuality is the NPA\textsuperscript{1+AB} Set}\label{sec:q1ab}
Quantum contextuality, as just defined, corresponds to what we refer to here as the $\mbox{NPA}^{1+AB}$ level of the NPA hierarchy \cite{FritzCombinatorialLong,NPA2007Short,NPA2008Long}, although it is also referred to as ${{Q}^{1+AB}}$ \cite{ConsistentHistoriesContextuality}, or  $\mbox{Q}_1$ \cite{FritzCombinatorialLong}, or $\tilde{\mbox{Q}}$ \cite{AlmostQuantum2}. The subtle distinction between quantum correlations in the sense of contextuality vs. nonlocality has received increasing attention in recent years [\citealp{FritzCombinatorialLong,AlmostQuantum2,CabelloMultigraph} and \citealp[Eq. (20)]{NonlocalityContextualityCommutativity}], although very few explicit statistical discrepancies have been identified. We note that other references rarely refer to such statistics as quantum contextuality statistics, even though an \emph{unlabelled} exclusivity graph, even if motivated by a multipartite scenario, tautologically defines an abstract contextuality scenario. The correspondence of the quantum correlations compatible with an unlabelled exclusivity graph and the $\mbox{NPA}^{1+AB}$ level of the NPA hierarchy applies only to \emph{complete} exclusivity graphs, representing Bell scenarios in the sense of \citet{FritzCombinatorialLong}, see Proposition 6.3.1.d there. An important example of a complete Bell scenario exclusivity graph for which $\mbox{NPA}^{1+AB}$ corresponds to quantum contextuality is the complete CHSH exclusivity graph illustrated in Figure 2 of Ref. \cite{LOExploring}. For exclusivity \emph{subgraphs}, the $\mbox{NPA}^{1+AB}$ level is more restrictive than full general noncomposite quantum correlations; see for example the pentagonal Bell inequalities considered in Ref. \cite{CabelloMultigraph}.

\begin{figure}[ht]
\centering
\begin{minipage}[t]{1\textwidth}
    \includegraphics[width=1.0\linewidth]{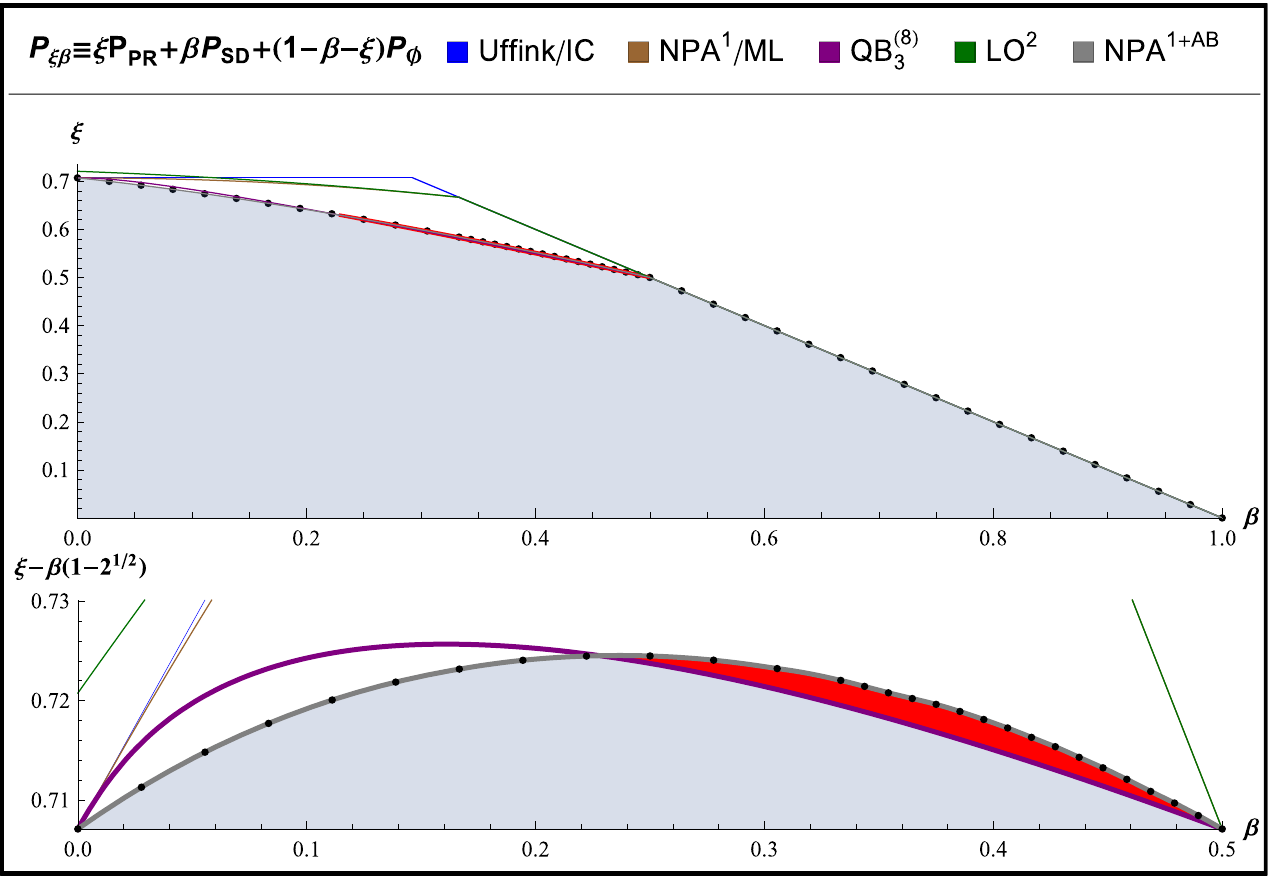}
    \caption{(Color online) A comparison of quantum bounds along a slice of the no-signalling \gls{polytope} which illustrates the inadequacy of $\mbox{NPA}^{1+AB}$ for characterizing multipartite quantum correlations. The grey shaded area represents the cumulative best known approximation of the genuine quantum boundary. Note that unlike in \fig{fig:boundcomptight}, the contextuality quantum bound corresponding to $\mbox{NPA}^{1+AB}$ is not tight along this boundary. Specifically, for $\nicefrac{1}{4}\lesssim \beta < \nicefrac{1}{2}$, it is inadequate. The region of the quantum contextuality \gls{elliptope} which is known to lie outside of the quantum nonlocality elliptope is shaded in red. The lower figure is a closeup view highlighting this region of interest, with a $\beta$-dependent rescaling of the y-axis for visual clarity. For $\beta > \nicefrac{1}{2}$ the quantum elliptope exactly coincides with the no-signalling polytope, and thus the quantum boundary is simply $\xi \leq 1-\beta$ for $\beta > \nicefrac{1}{2}$. It should be noted that the $\mbox{LO}^2$ and $\mbox{NPA}^1$ bounds bump into the no-signalling polytope earlier, at $\beta = \nicefrac{1}{3}$. The black dots are the numeric points for which we calculated the maximal $\xi$ per $\mbox{NPA}^{1+AB}$.}\label{fig:betatohalf}
\end{minipage}
\end{figure}

Some physicists have even suggested that perhaps quantum nonlocality is only an approximation to quantum contextuality, and that the universe might in fact permit $\mbox{NPA}^{1+AB}$-saturating correlations in a genuine bipartite scenario; \citet{PopescuReviewNatureComm} has questioned ``\textit{...whether nature is in fact more nonlocal than expected from quantum theory.}'' \citet{ConsistentHistoriesContextuality} have further explicitly shown that the existance of correlations consistent with $\mbox{NPA}^{1+AB}$ is the (strongest possible) consequence of the Consistent Histories interpretation due to Griffiths, Omnès, Gell-Mann, and Hartle, \citep[and references therein]{HistoriesShort,HistoriesReviewDowker,HistoriesReview}, specifically the convex and closed-under-wirings variant of the Strongly Positive Joint Quantum Measure formalism, $\mbox{SPJQM}_b$ \citep[Sec. 5-6]{ConsistentHistoriesContextuality}.\footnote{To quote \citet{ConsistentHistoriesContextuality}: ``Thus, if $Q^{1+AB}$ is strictly larger than $Q$, as indicated by the computational evidence, then non-local correlations beyond those achievable in ordinary quantum mechanics are achievable within strongly positive quantum measure theory.'' $Q^{1+AB}$ refers to the set of all points in conditional probability space consistent with the $\mbox{NPA}^{1+AB}$ criterion, and $Q$ refers to the set of nonlocal correlations achievable by assigning Alice and Bob to separate Hilbert spaces, see Ref. \cite{AlmostQuantum2}. The sentiment that $Q \neq Q^{1+AB}$ is merely ``indicated by the computational evidence'' is deprecated by the stronger analytical results graphed in \fig{fig:betatohalf} of this thesis.} Therefore, it is a matter of urgent importance to improve our understanding of the different experimental predictions of conventional multipartite quantum mechanics versus $\mbox{NPA}^{1+AB}$.


As mentioned, there are a few known instances where $\mbox{NPA}^{1+AB}$ gives a nontrivially weaker restriction of the allowed probabilities than does genuine bipartite nonlocality as implemented through quantum mechanics. The first examples were the I3322 scenario \cite{I3322Original,I3322NPA1,I3322NPA2} and the CGLMP scenario for $d=3$ \cite{CGLMP02,CGLMP08}, and now more recently a proof has been provided for the fundamental (2,2,2) scenario \cite{AlmostQuantum2}. While \citet{AlmostQuantum2} provide a singular counterexample point, we here demonstrate an entire region where $\mbox{NPA}^{1+AB}$ lies outside of the genuine bipartite quantum \gls{elliptope}.

To demonstrate the inadequacy of $\mbox{NPA}^{1+AB}$ we require another (stronger) quantum bound which we can show that $\mbox{NPA}^{1+AB}$ is relatively looser than. This is somewhat challenging as the default criterion of $\mbox{NPA}^1$ \ceq{eq:NPA1} is already a relaxation of  by $\mbox{NPA}^{1+AB}$. To get around this, we use our own $\mbox{QB}_3^{(8)}$ from Ref. \cite{WolfeQB}, reproduced here in \tab{tab:bounds} and \ceq{eq:qb3reproduced}, as our contrasting bound. Recall that the $\mbox{NPA}^{1+AB}$ certifies points as non-quantum if the corresponding $\Gamma$ matrix cannot be made positive semidefinite. We reproduce the $\Gamma$ matrix corresponding to binary and dichotomic $\mbox{NPA}^{1+AB}$ in the $9\times 9$ explicit matrix in \tab{tab:npa1}.

\newgeometry{left=1.00in,right=1.0in,top=1.4in,bottom=1.4in}
\begin{landscape}
\renewcommand*{\arraystretch}{2.5}
\begin{table*}[hbtp]\centering
\caption{\label{tab:npa1}We reproduce the $\Gamma$ matrix corresponding to $\mbox{NPA}^{1+AB}$ level of Navascués-Pironio-Acín hierarchy \cite{NPA2007Short,NPA2008Long} for the bipartite binary and dichotomic scenario. The coordinates in conditional probability space define the eight ``given'' parameters: $\Braket{A_0},\Braket{A_1},\Braket{B_0},\Braket{B_1},\Braket{A_0 B_0},\Braket{A_1 B_0},\Braket{A_0 B_1},\Braket{A_1 B_1}$. There are also eight ``free'' variables in the matrix which have no experimentally-determinable values, namely: $\Braket{A_0.A_1}, \Braket{B_0.B_1}, \Braket{A_0.A_1 B_0}, \Braket{A_0.A_1 B_1}, \Braket{A_0 B_0.B_1}, \Braket{A_1 B_0.B_1}, \Braket{A_0.A_1 B_0.B_1}, \Braket{A_0.A_1 B_1.B_0}$. In the matrix below, the free variables are colored purple and are not wrapped by BraKet. A conditional probability distribution is inside (outside) the $\mbox{NPA}^{1+AB}$ \gls{elliptope} iff values exist (do not exist) for the free variables such that the matrix $\Gamma$ can be made positive semidefinite.}

\begin{tabularx}{\linewidth}{|cCCCCCCCC|}

\toprule
\(\id \)&\( \Braket{A_0}  \)&\( \Braket{A_1}  \)&\( \Braket{B_0}  \)&\( \Braket{B_1}  \)&\( \Braket{A_0 B_0}  \)&\( \Braket{A_0 B_1}  \)&\( \Braket{A_1 B_0}  \)&\( \Braket{A_1 B_1}  \)\\\(
 \Braket{A_0}  \)&\( \id \)&\( \purp{A_0.A_1}  \)&\( \Braket{A_0 B_0}  \)&\( \Braket{A_0 B_1}  \)&\( \Braket{B_0}  \)&\( \Braket{B_1}  \)&\( \purp{A_0.A_1 B_0}  \)&\( \purp{A_0.A_1 B_1}  \)\\\(
 \Braket{A_1}  \)&\( \purp{A_0.A_1}  \)&\( \id \)&\( \Braket{A_1 B_0}  \)&\( \Braket{A_1 B_1}  \)&\( \purp{A_0.A_1 B_0}  \)&\( \purp{A_0.A_1 B_1}  \)&\( \Braket{B_0}  \)&\( \Braket{B_1}  \)\\\(
 \Braket{B_0}  \)&\( \Braket{A_0 B_0}  \)&\( \Braket{A_1 B_0}  \)&\( \id \)&\( \purp{B_0.B_1}  \)&\( \Braket{A_0}  \)&\( \purp{B_0.B_1 A_0}  \)&\( \Braket{A_1}  \)&\( \purp{B_0.B_1 A_1}  \)\\\(
 \Braket{B_1}  \)&\( \Braket{A_0 B_1}  \)&\( \Braket{A_1 B_1}  \)&\( \purp{B_0.B_1}  \)&\( \id \)&\( \purp{B_0.B_1 A_0}  \)&\( \Braket{A_0}  \)&\( \purp{B_0.B_1 A_1}  \)&\( \Braket{A_1}  \)\\\(
 \Braket{A_0 B_0}  \)&\( \Braket{B_0}  \)&\( \purp{A_0.A_1 B_0}  \)&\( \Braket{A_0}  \)&\( \purp{B_0.B_1 A_0}  \)&\( \id \)&\( \purp{B_0.B_1}  \)&\( \purp{A_0.A_1}  \)&\( \purp{A_0.A_1 B_0.B_1}  \)\\\(
 \Braket{A_0 B_1}  \)&\( \Braket{B_1}  \)&\( \purp{A_0.A_1 B_1}  \)&\( \purp{B_0.B_1 A_0}  \)&\( \Braket{A_0}  \)&\( \purp{B_0.B_1}  \)&\( \id \)&\( \purp{A_0.A_1 B_1.B_0}  \)&\( \purp{A_0.A_1}  \)\\\(
 \Braket{A_1 B_0}  \)&\( \purp{A_0.A_1 B_0}  \)&\( \Braket{B_0}  \)&\( \Braket{A_1}  \)&\( \purp{B_0.B_1 A_1}  \)&\( \purp{A_0.A_1}  \)&\( \purp{A_0.A_1 B_1.B_0}  \)&\( \id \)&\( \purp{B_0.B_1}  \)\\ \(
 \Braket{A_1 B_1}  \)&\( \purp{A_0.A_1 B_1}  \)&\( \Braket{B_1}  \)&\( \purp{B_0.B_1 A_1}  \)&\( \Braket{A_1}  \)&\( \purp{A_0.A_1 B_0.B_1}  \)&\( \purp{A_0.A_1}  \)&\( \purp{B_0.B_1}  \)&\( \id \)\\
 \bottomrule

\end{tabularx}
\end{table*}
\end{landscape}
\restoregeometry

In Sec. \ref{sec:boundcomp} we considered the parametric probabilistic box $P_{\xi\gamma}=\xi P_{\mbox{PR}}+\gamma P_{\mbox{FD}}+\left(1-\gamma-\xi\right)P_{\mathlarger{\operatorname{{\emptyset}}}}$ spanned by the PR box, the fully-deterministic box, and the totally random box, per \ceq{eq:boxmix}. We found that, on this facet, the criterion $\mbox{NPA}^{1+AB}$ was far and away the most restrictive quantum bound known. In other words, there did not appear to be any instance in which quantum nonlocality was distinct from quantum contextuality along that particular slice of the no-signalling \gls{polytope}.

We have found another slice, however, for which we can show that $\mbox{NPA}^{1+AB}$ is outside of, ie. not tight against, the \gls{elliptope} of bipartite quantum correlations. The three boxes which span this new slice are the PR box, the totally random box, and the semi-deterministic Bob-random box of \ceq{eq:pbobrandom}. Analogous to \ceq{eq:boxmix} we define here a box mixture which lives in this span, namely
\LeftEqn{\label{eq:newbox}
&P_{\xi \beta}(ab|xy)=\xi P_{\mbox{PR}}(ab|xy)+\beta P_{\mbox{SD}}(ab|xy)+\left(1-\beta-\xi\right)P_{\mathlarger{\operatorname{{\emptyset}}}}(ab|xy)
}
Recall that $P_{\mbox{SD}}$ is a local box which returns 1 for Alice (independent of her measurement choice), and is completely random for Bob (independent of his measurement choice). 

The 2-dimensional statistical region we are considering with this box span, is therefore
\LeftEqn{\label{eq:newboxparameters}
\Braket{A_0}=\Braket{A_1}=\beta\,,\quad \Braket{B_0}=\Braket{B_1}=0
\\  \Braket{A_0 B_0}=\Braket{A_1 B_0}=\Braket{A_0 B_1}=-\Braket{A_1 B_1} = \xi 
}

\FloatBarrier
Using $P_{\xi \beta}$ per \ceq{eq:newbox}, and recognizing that normalization of probabilities demands $\xi \leq 1-\beta$, it is easy to recalculate the familiar quantum bounds we first considered in Sec. \ref{sec:boundcomp}. We define the bounds piecewise along $\beta$, switching to $\xi\leq 1-\beta$ when that becomes the more restrictive condition on $\xi$. It is easy to see that Uffink's bound of \ceq{eq:uffink} becomes
\LeftEqn{\label{eq:uffinkbeta}
\xi \leq \acases{ 2^{-\nicefrac{1}{2}} & \text{for }0\leq \beta \leq 1-2^{-\nicefrac{1}{2}} \\ 1-\beta & \text{for }1-2^{-\nicefrac{1}{2}} \leq \beta \leq 1 }
}
and the bound corresponding to the first level of the NPA hierarchy per \ceq{eq:NPA1} is simply
\LeftEqn{
\xi \leq \acases{ 2^{-\nicefrac{1}{2}}\sqrt{1-\beta ^2} & \text{for }0\leq \beta \leq \nicefrac{1}{3} \\ 1-\beta & \text{for }\nicefrac{1}{3} \leq \beta \leq 1 }
}\,,
an obvious improvement over Uffink's bound in \ceq{eq:uffinkbeta}. 

Using the clique of mutually-exclusive events tallied in \ceq{eq:RafaelClique} we can show that $\mbox{LO}^2$ implies for the $P_{\xi \beta}$ box
\LeftEqn{
\xi \leq \acases{ \frac{\sqrt{2} \sqrt{(1-\beta ) (\beta +5)}+\beta-1}{3} & \text{for }0\leq \beta \leq \nicefrac{1}{3} \\ 1-\beta & \text{for }\nicefrac{1}{3} \leq \beta \leq 1 }\,.
}

Lastly, we can recycle the envelope of linear bounds given in \ceq{eq:envelope} in order to characterize the constraint implied by $\mbox{QB}_3^{(8)}$ of \ceq{eq:qb3reproduced} for the $P_{\xi \beta}$ box of \ceq{eq:newbox}. By direct substitution of \ceq{eq:newboxparameters} into the function-valued bound \ceq{eq:qb3reproduced}, we see that
\LeftEqns{\label{eq:qb3beta}
&\xi\leq \frac{\min\limits_{c^2 \leq 1}{\bigg\lbrace{\frac{c^2-\sqrt{\left(3 c^2-4\right) \left(c^2-2\right)}}{c^2-1}-c (2\beta)\bigg\rbrace}}}{4}\,.
}
Note that \ceq{eq:qb3beta} shares the same minimization term as \ceq{eq:qb3forall}, such that we can identify the envelope of bounds on $P_{\xi \beta}$ due to $\mbox{QB}_3^{(8)}$ as
\LeftEqn{
\xi \leq \acases{ \frac{\operatorname{\lambda_{\min}}{\big(2 \beta\big)}}{4} & \text{for }0\leq \beta \leq \nicefrac{1}{2} \\ 1-\beta & \text{for }\nicefrac{1}{2} \leq \beta \leq 1 }
}
where $\operatorname{\lambda_{\min}}{\big(\gamma\big)}$ is defined in \ceq{eq:envelope}.

\section{Concluding Thoughts}
When we plot the quantum boundary due to $\mbox{QB}_3^{(8)}$ per \ceq{eq:qb3beta} in \fig{fig:betatohalf} we observe a region in which this boundary is more restrictive than even $\mbox{NPA}^{1+AB}$. This affirms that finding that quantum contextuality allows for statistics not accessible through quantum nonlocality \cite{FritzCombinatorialLong,AlmostQuantum2}.

Note that any computational error in the numeric determination of the $\mbox{NPA}^{1+AB}$ boundary would result in an \emph{underestimation} of $\xi$ for every discrete $\beta$, such that the plotted $\mbox{NPA}^{1+AB}$ curve represents a \emph{lower bound}, and consequently there is no ambiguity that $\mbox{NPA}^{1+AB}$ is plainly more permissive than $\mbox{QB}_3^{(8)}$ in the region highlighted in \fig{fig:betatohalf}. The underestimation of $\xi$ results from the positivity constraint of the $\Gamma$ matrix of \tab{tab:npa1}; the computer may falsely reject a permitted $\xi$ if it is unable to instances of free variables to make $\Gamma$ positive semidefinite in a finite number of iterations.


Thus \fig{fig:betatohalf} reiterates the fundamental limitation inherent to both the Exclusivity Principle (EP) \cite{EPOriginal,EPSpecker,EPExampleGraphs,EPYan,EP2013,EPTwoCities,EPTsirelson} and Local Orthogonality (LO) \cite{LONatureComm,FritzCombinatorialLong,LONewShort,LOHardy,LOExploring} as means to recover quantum \emph{nonlocal} correlations. Since those principles are {\em expressed} in terms of the graph-theoretic formalism of quantum contextuality they are innately blind to the distinction between quantum contextual and quantum nonlocal correlations. 

Our identification of a broad region in condition probability space where $\mbox{NPA}^{1+AB}$ diverges from the predictions of quantum multipartite nonlocality opens the door to experimental testing of the true nature of quantum reality. A bipartite experiment which could trace out the $\mbox{NPA}^{1+AB}$ curve in \fig{fig:betatohalf}, such as a generalization of that performed recently by \citet{WolfeQBExperimental}, would violate the predictions of conventional quantum mechanics. It would, instead, provide long-sought-after experimental evidence for relativistic quantum theories, such as Consistent Histories \cite{ConsistentHistoriesContextuality}.


It is inspiring to recognize that the pursuit of a physical principle capable of recovering quantum nonlocality - as opposed to contextuality - is still a wide open question in quantum foundations. Nonlocality is the dominant feature exploited in the quantum information theory. Yet, it remains an unexplained consequence of the mathematics underpinning quantum mechanics. Although much is known about quantum nonlocality, its fundamental \textit{raison d'être} still escapes us. The quest for a more satisfying explanation will surely lead to numerous applications in the future. Once a strong conceptual foundation is established, who can fathom the potential of the ``second quantum revolution'' \cite{bell2004speakable,QuantumRevArXiv}?

\chapter*{References}
\addcontentsline{toc}{chapter}{References}

\bibliographystyle{apsrev4mod}
\begin{spacing}{1.25}
\bibliography{ThesisRefs}

\begin{thebibliography}{288}%
\makeatletter
\providecommand \@ifxundefined [1]{%
 \@ifx{#1\undefined}
}%
\providecommand \@ifnum [1]{%
 \ifnum #1\expandafter \@firstoftwo
 \else \expandafter \@secondoftwo
 \fi
}%
\providecommand \@ifx [1]{%
 \ifx #1\expandafter \@firstoftwo
 \else \expandafter \@secondoftwo
 \fi
}%
\providecommand \natexlab [1]{#1}%
\providecommand \enquote  [1]{``#1''}%
\providecommand \bibnamefont  [1]{#1}%
\providecommand \bibfnamefont [1]{#1}%
\providecommand \citenamefont [1]{#1}%
\providecommand \href@noop [0]{\@secondoftwo}%
\providecommand \href [0]{\begingroup \@sanitize@url \@href}%
\providecommand \@href[1]{\@@startlink{#1}\@@href}%
\providecommand \@@href[1]{\endgroup#1\@@endlink}%
\providecommand \@sanitize@url [0]{\catcode `\\12\catcode `\$12\catcode
  `\&12\catcode `\#12\catcode `\^12\catcode `\_12\catcode `\%12\relax}%
\providecommand \@@startlink[1]{}%
\providecommand \@@endlink[0]{}%
\providecommand \url  [0]{\begingroup\@sanitize@url \@url }%
\providecommand \@url [1]{\endgroup\@href {#1}{\urlprefix }}%
\providecommand \urlprefix  [0]{URL }%
\providecommand \Eprint [0]{\href }%
\providecommand \doibase [0]{http://dx.doi.org/}%
\providecommand \selectlanguage [0]{\@gobble}%
\providecommand \bibinfo  [0]{\@secondoftwo}%
\providecommand \bibfield  [0]{\@secondoftwo}%
\providecommand \translation [1]{[#1]}%
\providecommand \BibitemOpen [0]{}%
\providecommand \bibitemStop [0]{}%
\providecommand \bibitemNoStop [0]{.\EOS\space}%
\providecommand \EOS [0]{\spacefactor3000\relax}%
\providecommand \BibitemShut  [1]{\csname bibitem#1\endcsname}%
\let\auto@bib@innerbib\@empty
\bibitem [{\citenamefont {Popescu}(2014)}]{PopescuReviewNatureComm}%
  \BibitemOpen
  \bibfield  {author} {\bibinfo {author} {\bibfnamefont {S.}~\bibnamefont
  {Popescu}},\ }\bibfield  {title} {\enquote {\bibinfo {title} {{Nonlocality
  Beyond Quantum Mechanics}},}\ }\href {http://dx.doi.org/10.1038/nphys2916}
  {\bibfield  {journal} {\bibinfo  {journal} {Nature Physics}\ }\textbf
  {\bibinfo {volume} {10}},\ \bibinfo {pages} {264} (\bibinfo {year}
  {2014})}\BibitemShut {NoStop}%
\bibitem [{\citenamefont {Liang}\ \emph {et~al.}(2011)\citenamefont {Liang},
  \citenamefont {Spekkens},\ and\ \citenamefont {Wiseman}}]{SpekkensSeer}%
  \BibitemOpen
  \bibfield  {author} {\bibinfo {author} {\bibfnamefont {Y.-C.}\ \bibnamefont
  {Liang}}, \bibinfo {author} {\bibfnamefont {R.~W.}\ \bibnamefont {Spekkens}},
  \ and\ \bibinfo {author} {\bibfnamefont {H.~M.}\ \bibnamefont {Wiseman}},\
  }\bibfield  {title} {\enquote {\bibinfo {title} {{Specker’s Parable of the
  Overprotective Seer: A Road to Contextuality, Nonlocality and
  Complementarity}},}\ }\href
  {http://www.sciencedirect.com/science/article/pii/S0370157311001517}
  {\bibfield  {journal} {\bibinfo  {journal} {Physics Reports}\ }\textbf
  {\bibinfo {volume} {506}},\ \bibinfo {pages} {1} (\bibinfo {year}
  {2011})}\BibitemShut {NoStop}%
\bibitem [{\citenamefont {Badzi{\c{a}}g}\ \emph {et~al.}(2011)\citenamefont
  {Badzi{\c{a}}g}, \citenamefont {Bengtsson}, \citenamefont {Cabello},
  \citenamefont {Granstr{\"{o}}m},\ and\ \citenamefont {Larsson}}]{Pentagrams}%
  \BibitemOpen
  \bibfield  {author} {\bibinfo {author} {\bibfnamefont {P.}~\bibnamefont
  {Badzi{\c{a}}g}}, \bibinfo {author} {\bibfnamefont {I.}~\bibnamefont
  {Bengtsson}}, \bibinfo {author} {\bibfnamefont {A.}~\bibnamefont {Cabello}},
  \bibinfo {author} {\bibfnamefont {H.}~\bibnamefont {Granstr{\"{o}}m}}, \ and\
  \bibinfo {author} {\bibfnamefont {J.-{\AA}.}\ \bibnamefont {Larsson}},\
  }\bibfield  {title} {\enquote {\bibinfo {title} {{Pentagrams and
  Paradoxes}},}\ }\href {http://dx.doi.org/10.1007/s10701-010-9433-3}
  {\bibfield  {journal} {\bibinfo  {journal} {Foundations of Physics}\ }\textbf
  {\bibinfo {volume} {41}},\ \bibinfo {pages} {414} (\bibinfo {year}
  {2011})}\BibitemShut {NoStop}%
\bibitem [{\citenamefont {Cabello}(2008)}]{ExperimentalContextuality}%
  \BibitemOpen
  \bibfield  {author} {\bibinfo {author} {\bibfnamefont {A.}~\bibnamefont
  {Cabello}},\ }\bibfield  {title} {\enquote {\bibinfo {title} {{Experimentally
  Testable State-Independent Quantum Contextuality}},}\ }\href
  {http://link.aps.org/doi/10.1103/PhysRevLett.101.210401} {\bibfield
  {journal} {\bibinfo  {journal} {Physical Review Letters}\ }\textbf {\bibinfo
  {volume} {101}},\ \bibinfo {pages} {210401} (\bibinfo {year}
  {2008})}\BibitemShut {NoStop}%
\bibitem [{\citenamefont {Cabello}(1996)}]{ks18}%
  \BibitemOpen
  \bibfield  {author} {\bibinfo {author} {\bibfnamefont {A.}~\bibnamefont
  {Cabello}},\ }\bibfield  {title} {\enquote {\bibinfo {title}
  {{Bell-Kochen-Specker Theorem: A Proof with 18 Vectors}},}\ }\href
  {http://dx.doi.org/10.1016/0375-9601(96)00134-x} {\bibfield  {journal}
  {\bibinfo  {journal} {Physics Letters A}\ }\textbf {\bibinfo {volume}
  {212}},\ \bibinfo {pages} {183} (\bibinfo {year} {1996})}\BibitemShut
  {NoStop}%
\bibitem [{\citenamefont {Cabello}\ \emph {et~al.}(2014)\citenamefont
  {Cabello}, \citenamefont {Severini},\ and\ \citenamefont {Winter}}]{CSWNew}%
  \BibitemOpen
  \bibfield  {author} {\bibinfo {author} {\bibfnamefont {A.}~\bibnamefont
  {Cabello}}, \bibinfo {author} {\bibfnamefont {S.}~\bibnamefont {Severini}}, \
  and\ \bibinfo {author} {\bibfnamefont {A.}~\bibnamefont {Winter}},\
  }\bibfield  {title} {\enquote {\bibinfo {title} {{Graph-Theoretic Approach to
  Quantum Correlations}},}\ }\href
  {http://link.aps.org/doi/10.1103/PhysRevLett.112.040401} {\bibfield
  {journal} {\bibinfo  {journal} {Physical Review Letters}\ }\textbf {\bibinfo
  {volume} {112}},\ \bibinfo {pages} {040401} (\bibinfo {year}
  {2014})}\BibitemShut {NoStop}%
\bibitem [{\citenamefont {Werner}\ and\ \citenamefont
  {Wolf}(2001{\natexlab{a}})}]{WernerWolfe2001}%
  \BibitemOpen
  \bibfield  {author} {\bibinfo {author} {\bibfnamefont {R.~F.}\ \bibnamefont
  {Werner}}\ and\ \bibinfo {author} {\bibfnamefont {M.~M.}\ \bibnamefont
  {Wolf}},\ }\bibfield  {title} {\enquote {\bibinfo {title} {{Bell Inequalities
  and Entanglement}},}\ }\href {http://arxiv.org/abs/quant-ph/0107093}
  {\bibfield  {journal} {\bibinfo  {journal} {Quantum Information \&
  Computation}\ }\textbf {\bibinfo {volume} {1}},\ \bibinfo {pages} {1}
  (\bibinfo {year} {2001}{\natexlab{a}})}\BibitemShut {NoStop}%
\bibitem [{\citenamefont {Jean-Daniel}\ \emph {et~al.}(2012)\citenamefont
  {Jean-Daniel}, \citenamefont {Cyril}, \citenamefont {Nicolas}, \citenamefont
  {Nicolas},\ and\ \citenamefont {Yeong-Cherng}}]{GisinFramework2012}%
  \BibitemOpen
  \bibfield  {author} {\bibinfo {author} {\bibfnamefont {B.}~\bibnamefont
  {Jean-Daniel}}, \bibinfo {author} {\bibfnamefont {B.}~\bibnamefont {Cyril}},
  \bibinfo {author} {\bibfnamefont {B.}~\bibnamefont {Nicolas}}, \bibinfo
  {author} {\bibfnamefont {G.}~\bibnamefont {Nicolas}}, \ and\ \bibinfo
  {author} {\bibfnamefont {L.}~\bibnamefont {Yeong-Cherng}},\ }\bibfield
  {title} {\enquote {\bibinfo {title} {{A Framework for the Study of Symmetric
  Full-Correlation Bell-Like Inequalities}},}\ }\href
  {http://stacks.iop.org/1751-8121/45/i=12/a=125301} {\bibfield  {journal}
  {\bibinfo  {journal} {Journal of Physics A: Mathematical and Theoretical}\
  }\textbf {\bibinfo {volume} {45}},\ \bibinfo {pages} {125301} (\bibinfo
  {year} {2012})}\BibitemShut {NoStop}%
\bibitem [{\citenamefont {Abramsky}\ and\ \citenamefont
  {Hardy}(2012)}]{BellAndNonContextualInequalities}%
  \BibitemOpen
  \bibfield  {author} {\bibinfo {author} {\bibfnamefont {S.}~\bibnamefont
  {Abramsky}}\ and\ \bibinfo {author} {\bibfnamefont {L.}~\bibnamefont
  {Hardy}},\ }\bibfield  {title} {\enquote {\bibinfo {title} {{Logical Bell
  Inequalities}},}\ }\href {http://link.aps.org/doi/10.1103/PhysRevA.85.062114}
  {\bibfield  {journal} {\bibinfo  {journal} {Physical Review A}\ }\textbf
  {\bibinfo {volume} {85}},\ \bibinfo {pages} {062114} (\bibinfo {year}
  {2012})}\BibitemShut {NoStop}%
\bibitem [{\citenamefont {{Acacio de Barros}}\ \emph
  {et~al.}(2014)\citenamefont {{Acacio de Barros}}, \citenamefont
  {{Dzhafarov}}, \citenamefont {{Kujala}},\ and\ \citenamefont
  {{Oas}}}]{ContextualityLeggettGarg}%
  \BibitemOpen
  \bibfield  {author} {\bibinfo {author} {\bibfnamefont {J.}~\bibnamefont
  {{Acacio de Barros}}}, \bibinfo {author} {\bibfnamefont {E.~N.}\ \bibnamefont
  {{Dzhafarov}}}, \bibinfo {author} {\bibfnamefont {J.~V.}\ \bibnamefont
  {{Kujala}}}, \ and\ \bibinfo {author} {\bibfnamefont {G.}~\bibnamefont
  {{Oas}}},\ }\bibfield  {title} {\enquote {\bibinfo {title} {{Unifying Two
  Methods of Measuring Quantum Contextuality}},}\ }\href
  {http://arxiv.org/abs/1406.3088} {\bibfield  {journal} {\bibinfo  {journal}
  {arXiv:1406.3088}\ } (\bibinfo {year} {2014})}\BibitemShut {NoStop}%
\bibitem [{\citenamefont {{Dzhafarov}}\ and\ \citenamefont
  {{Kujala}}(2014)}]{ContextualityWithSignalling}%
  \BibitemOpen
  \bibfield  {author} {\bibinfo {author} {\bibfnamefont {E.~N.}\ \bibnamefont
  {{Dzhafarov}}}\ and\ \bibinfo {author} {\bibfnamefont {J.~V.}\ \bibnamefont
  {{Kujala}}},\ }\bibfield  {title} {\enquote {\bibinfo {title} {{Probabilistic
  Contextuality in EPR/Bohm-type Systems with Signaling Allowed}},}\ }\href
  {http://arxiv.org/abs/1406.0243} {\bibfield  {journal} {\bibinfo  {journal}
  {arXiv:1406.0243}\ } (\bibinfo {year} {2014})}\BibitemShut {NoStop}%
\bibitem [{\citenamefont {Aravind}(2004)}]{QMysteriesAravid}%
  \BibitemOpen
  \bibfield  {author} {\bibinfo {author} {\bibfnamefont {P.~K.}\ \bibnamefont
  {Aravind}},\ }\bibfield  {title} {\enquote {\bibinfo {title} {Quantum
  mysteries revisited again},}\ }\href
  {http://scitation.aip.org/content/aapt/journal/ajp/72/10/10.1119/1.1773173}
  {\bibfield  {journal} {\bibinfo  {journal} {American Journal of Physics}\
  }\textbf {\bibinfo {volume} {72}},\ \bibinfo {pages} {1303} (\bibinfo {year}
  {2004})}\BibitemShut {NoStop}%
\bibitem [{\citenamefont {Tobias}(2012)}]{FritzCorrelationScenarios}%
  \BibitemOpen
  \bibfield  {author} {\bibinfo {author} {\bibfnamefont {F.}~\bibnamefont
  {Tobias}},\ }\bibfield  {title} {\enquote {\bibinfo {title} {{Beyond Bell's
  Theorem: Correlation Scenarios}},}\ }\href
  {http://stacks.iop.org/1367-2630/14/i=10/a=103001} {\bibfield  {journal}
  {\bibinfo  {journal} {New Journal of Physics}\ }\textbf {\bibinfo {volume}
  {14}},\ \bibinfo {pages} {103001} (\bibinfo {year} {2012})}\BibitemShut
  {NoStop}%
\bibitem [{\citenamefont {Griffiths}(2011)}]{GriffithsLocality}%
  \BibitemOpen
  \bibfield  {author} {\bibinfo {author} {\bibfnamefont {R.}~\bibnamefont
  {Griffiths}},\ }\bibfield  {title} {\enquote {\bibinfo {title} {{Quantum
  Locality}},}\ }\href {http://dx.doi.org/10.1007/s10701-010-9512-5} {\bibfield
   {journal} {\bibinfo  {journal} {Foundations of Physics}\ }\textbf {\bibinfo
  {volume} {41}},\ \bibinfo {pages} {705} (\bibinfo {year} {2011})}\BibitemShut
  {NoStop}%
\bibitem [{\citenamefont {{Gill}}(2012)}]{BellCausalityReview}%
  \BibitemOpen
  \bibfield  {author} {\bibinfo {author} {\bibfnamefont {R.~D.}\ \bibnamefont
  {{Gill}}},\ }\bibfield  {title} {\enquote {\bibinfo {title} {{Statistics,
  Causality and Bell's Theorem}},}\ }\href {http://arxiv.org/abs/1207.5103}
  {\bibfield  {journal} {\bibinfo  {journal} {arXiv:1207.5103}\ } (\bibinfo
  {year} {2012})}\BibitemShut {NoStop}%
\bibitem [{\citenamefont {{Hofer-Szab{\'o}}}\ and\ \citenamefont
  {{Vecserny{\'e}s}}(2014)}]{BellCausalityArXiv}%
  \BibitemOpen
  \bibfield  {author} {\bibinfo {author} {\bibfnamefont {G.}~\bibnamefont
  {{Hofer-Szab{\'o}}}}\ and\ \bibinfo {author} {\bibfnamefont {P.}~\bibnamefont
  {{Vecserny{\'e}s}}},\ }\bibfield  {title} {\enquote {\bibinfo {title} {{On
  the concept of Bell's Local Causality in Local Classical and Quantum
  Theory}},}\ }\href {http://arxiv.org/abs/1407.3610} {\bibfield  {journal}
  {\bibinfo  {journal} {arXiv:1407.3610}\ } (\bibinfo {year}
  {2014})}\BibitemShut {NoStop}%
\bibitem [{\citenamefont {{Wiseman}}(2014)}]{BellAssumptionsWiseman}%
  \BibitemOpen
  \bibfield  {author} {\bibinfo {author} {\bibfnamefont {H.~M.}\ \bibnamefont
  {{Wiseman}}},\ }\bibfield  {title} {\enquote {\bibinfo {title} {{The Two
  Bell's Theorems of John Bell}},}\ }\href {http://arxiv.org/abs/1402.0351}
  {\bibfield  {journal} {\bibinfo  {journal} {arXiv:1402.0351}\ } (\bibinfo
  {year} {2014})}\BibitemShut {NoStop}%
\bibitem [{\citenamefont {Brunner}\ \emph {et~al.}(2014)\citenamefont
  {Brunner}, \citenamefont {Cavalcanti}, \citenamefont {Pironio}, \citenamefont
  {Scarani},\ and\ \citenamefont {Wehner}}]{Brunner2013Bell}%
  \BibitemOpen
  \bibfield  {author} {\bibinfo {author} {\bibfnamefont {N.}~\bibnamefont
  {Brunner}}, \bibinfo {author} {\bibfnamefont {D.}~\bibnamefont {Cavalcanti}},
  \bibinfo {author} {\bibfnamefont {S.}~\bibnamefont {Pironio}}, \bibinfo
  {author} {\bibfnamefont {V.}~\bibnamefont {Scarani}}, \ and\ \bibinfo
  {author} {\bibfnamefont {S.}~\bibnamefont {Wehner}},\ }\bibfield  {title}
  {\enquote {\bibinfo {title} {{Bell Nonlocality}},}\ }\href
  {http://link.aps.org/doi/10.1103/RevModPhys.86.419} {\bibfield  {journal}
  {\bibinfo  {journal} {Reviews of Modern Physics}\ }\textbf {\bibinfo {volume}
  {86}},\ \bibinfo {pages} {419} (\bibinfo {year} {2014})}\BibitemShut
  {NoStop}%
\bibitem [{\citenamefont {Cirel'son}(1980)}]{Tsirelson1980}%
  \BibitemOpen
  \bibfield  {author} {\bibinfo {author} {\bibfnamefont {B.~S.}\ \bibnamefont
  {Cirel'son}},\ }\bibfield  {title} {\enquote {\bibinfo {title} {{Quantum
  Generalizations of Bell's Inequality}},}\ }\href
  {http://dx.doi.org/10.1007/BF00417500} {\bibfield  {journal} {\bibinfo
  {journal} {Letters in Mathematical Physics}\ }\textbf {\bibinfo {volume}
  {4}},\ \bibinfo {pages} {93} (\bibinfo {year} {1980})}\BibitemShut {NoStop}%
\bibitem [{\citenamefont {Popescu}\ and\ \citenamefont
  {Rohrlich}(1994)}]{PROriginal}%
  \BibitemOpen
  \bibfield  {author} {\bibinfo {author} {\bibfnamefont {S.}~\bibnamefont
  {Popescu}}\ and\ \bibinfo {author} {\bibfnamefont {D.}~\bibnamefont
  {Rohrlich}},\ }\bibfield  {title} {\enquote {\bibinfo {title} {{Quantum
  Nonlocality as an Axiom}},}\ }\href {http://dx.doi.org/10.1007/BF02058098}
  {\bibfield  {journal} {\bibinfo  {journal} {Foundations of Physics}\ }\textbf
  {\bibinfo {volume} {24}},\ \bibinfo {pages} {379} (\bibinfo {year}
  {1994})}\BibitemShut {NoStop}%
\bibitem [{\citenamefont {Barrett}\ and\ \citenamefont
  {Pironio}(2005)}]{PRUnit}%
  \BibitemOpen
  \bibfield  {author} {\bibinfo {author} {\bibfnamefont {J.}~\bibnamefont
  {Barrett}}\ and\ \bibinfo {author} {\bibfnamefont {S.}~\bibnamefont
  {Pironio}},\ }\bibfield  {title} {\enquote {\bibinfo {title}
  {{Popescu-Rohrlich Correlations as a Unit of Nonlocality}},}\ }\href
  {http://link.aps.org/doi/10.1103/PhysRevLett.95.140401} {\bibfield  {journal}
  {\bibinfo  {journal} {Physical Review Letters}\ }\textbf {\bibinfo {volume}
  {95}},\ \bibinfo {pages} {140401} (\bibinfo {year} {2005})}\BibitemShut
  {NoStop}%
\bibitem [{\citenamefont {Maccone}(2013)}]{BellSimple}%
  \BibitemOpen
  \bibfield  {author} {\bibinfo {author} {\bibfnamefont {L.}~\bibnamefont
  {Maccone}},\ }\bibfield  {title} {\enquote {\bibinfo {title} {{A Simple Proof
  of Bell's Inequality}},}\ }\href
  {http://scitation.aip.org/content/aapt/journal/ajp/81/11/10.1119/1.4823600}
  {\bibfield  {journal} {\bibinfo  {journal} {American Journal of Physics}\
  }\textbf {\bibinfo {volume} {81}},\ \bibinfo {pages} {854} (\bibinfo {year}
  {2013})}\BibitemShut {NoStop}%
\bibitem [{\citenamefont {Roberts}(2004)}]{roberts_thesis}%
  \BibitemOpen
  \bibfield  {author} {\bibinfo {author} {\bibfnamefont {D.}~\bibnamefont
  {Roberts}},\ }\emph {\bibinfo {title} {{Aspects of Quantum Non-Locality}}},\
  \href
  {http://www.bris.ac.uk/physics/media/theory-theses/roberts-d-thesis.pdf}
  {Ph.D. thesis} (\bibinfo {year} {2004})\BibitemShut {NoStop}%
\bibitem [{\citenamefont {Cabello}(2005)}]{HowMuchLarger}%
  \BibitemOpen
  \bibfield  {author} {\bibinfo {author} {\bibfnamefont {A.}~\bibnamefont
  {Cabello}},\ }\bibfield  {title} {\enquote {\bibinfo {title} {{How much
  Larger Quantum Correlations are Than Classical Ones}},}\ }\href
  {http://link.aps.org/doi/10.1103/PhysRevA.72.012113} {\bibfield  {journal}
  {\bibinfo  {journal} {Physical Review A}\ }\textbf {\bibinfo {volume} {72}},\
  \bibinfo {pages} {012113} (\bibinfo {year} {2005})}\BibitemShut {NoStop}%
\bibitem [{\citenamefont {Spengler}\ \emph {et~al.}(2011)\citenamefont
  {Spengler}, \citenamefont {Huber},\ and\ \citenamefont
  {Hiesmayr}}]{DiscreteEntanglementNonlocality}%
  \BibitemOpen
  \bibfield  {author} {\bibinfo {author} {\bibfnamefont {C.}~\bibnamefont
  {Spengler}}, \bibinfo {author} {\bibfnamefont {M.}~\bibnamefont {Huber}}, \
  and\ \bibinfo {author} {\bibfnamefont {B.~C.}\ \bibnamefont {Hiesmayr}},\
  }\bibfield  {title} {\enquote {\bibinfo {title} {{A Geometric Comparison of
  Entanglement and Quantum Nonlocality in Discrete Systems}},}\ }\href
  {http://stacks.iop.org/1751-8121/44/i=6/a=065304} {\bibfield  {journal}
  {\bibinfo  {journal} {Journal of Physics A: Mathematical and Theoretical}\
  }\textbf {\bibinfo {volume} {44}},\ \bibinfo {pages} {065304} (\bibinfo
  {year} {2011})}\BibitemShut {NoStop}%
\bibitem [{\citenamefont {Masanes}\ \emph {et~al.}(2006)\citenamefont
  {Masanes}, \citenamefont {Ac{\'i}n},\ and\ \citenamefont
  {Gisin}}]{GeneralNoSignalling}%
  \BibitemOpen
  \bibfield  {author} {\bibinfo {author} {\bibfnamefont {L.}~\bibnamefont
  {Masanes}}, \bibinfo {author} {\bibfnamefont {A.}~\bibnamefont {Ac{\'i}n}}, \
  and\ \bibinfo {author} {\bibfnamefont {N.}~\bibnamefont {Gisin}},\ }\bibfield
   {title} {\enquote {\bibinfo {title} {{General Properties of Nonsignaling
  Theories}},}\ }\href {http://link.aps.org/doi/10.1103/PhysRevA.73.012112}
  {\bibfield  {journal} {\bibinfo  {journal} {Physical Review A}\ }\textbf
  {\bibinfo {volume} {73}},\ \bibinfo {pages} {012112} (\bibinfo {year}
  {2006})}\BibitemShut {NoStop}%
\bibitem [{\citenamefont {Braunstein}\ and\ \citenamefont
  {Caves}(1989)}]{WringingBellInequalities}%
  \BibitemOpen
  \bibfield  {author} {\bibinfo {author} {\bibfnamefont {S.~L.}\ \bibnamefont
  {Braunstein}}\ and\ \bibinfo {author} {\bibfnamefont {C.~M.}\ \bibnamefont
  {Caves}},\ }\bibfield  {title} {\enquote {\bibinfo {title} {{Wringing Out
  Better Bell Inequalities}},}\ }\href
  {http://www.sciencedirect.com/science/article/pii/0920563289904416}
  {\bibfield  {journal} {\bibinfo  {journal} {Nuclear Physics B - Proceedings
  Supplements}\ }\textbf {\bibinfo {volume} {6}},\ \bibinfo {pages} {211 }
  (\bibinfo {year} {1989})}\BibitemShut {NoStop}%
\bibitem [{\citenamefont {Rosset}\ \emph {et~al.}(2014)\citenamefont {Rosset},
  \citenamefont {Bancal},\ and\ \citenamefont
  {Gisin}}]{BellInequalitiesReview}%
  \BibitemOpen
  \bibfield  {author} {\bibinfo {author} {\bibfnamefont {D.}~\bibnamefont
  {Rosset}}, \bibinfo {author} {\bibfnamefont {J.-D.}\ \bibnamefont {Bancal}},
  \ and\ \bibinfo {author} {\bibfnamefont {N.}~\bibnamefont {Gisin}},\
  }\bibfield  {title} {\enquote {\bibinfo {title} {{Classifying 50 Years of
  Bell Inequalities}},}\ }\href {http://arXiv.org/abs/1404.1306} {\bibfield
  {journal} {\bibinfo  {journal} {arXiv:1404.1306}\ } (\bibinfo {year}
  {2014})}\BibitemShut {NoStop}%
\bibitem [{\citenamefont {Braunstein}\ \emph {et~al.}(1992)\citenamefont
  {Braunstein}, \citenamefont {Mann},\ and\ \citenamefont
  {Revzen}}]{MixedStateQB}%
  \BibitemOpen
  \bibfield  {author} {\bibinfo {author} {\bibfnamefont {S.~L.}\ \bibnamefont
  {Braunstein}}, \bibinfo {author} {\bibfnamefont {A.}~\bibnamefont {Mann}}, \
  and\ \bibinfo {author} {\bibfnamefont {M.}~\bibnamefont {Revzen}},\
  }\bibfield  {title} {\enquote {\bibinfo {title} {{Maximal Violation of Bell
  Inequalities for Mixed States}},}\ }\href
  {http://link.aps.org/doi/10.1103/PhysRevLett.68.3259} {\bibfield  {journal}
  {\bibinfo  {journal} {Physical Review Letters}\ }\textbf {\bibinfo {volume}
  {68}},\ \bibinfo {pages} {3259} (\bibinfo {year} {1992})}\BibitemShut
  {NoStop}%
\bibitem [{\citenamefont {Peres}(1999)}]{PeresBellMaximization99}%
  \BibitemOpen
  \bibfield  {author} {\bibinfo {author} {\bibfnamefont {A.}~\bibnamefont
  {Peres}},\ }\bibfield  {title} {\enquote {\bibinfo {title} {{All the Bell
  Inequalities}},}\ }\href {{http://dx.doi.org/10.1023/A%3A1018816310000}}
  {\bibfield  {journal} {\bibinfo  {journal} {Foundations of Physics}\ }\textbf
  {\bibinfo {volume} {29}},\ \bibinfo {pages} {589} (\bibinfo {year}
  {1999})}\BibitemShut {NoStop}%
\bibitem [{\citenamefont {Werner}\ and\ \citenamefont
  {Wolf}(2001{\natexlab{b}})}]{WernerWolfQB}%
  \BibitemOpen
  \bibfield  {author} {\bibinfo {author} {\bibfnamefont {R.~F.}\ \bibnamefont
  {Werner}}\ and\ \bibinfo {author} {\bibfnamefont {M.~M.}\ \bibnamefont
  {Wolf}},\ }\bibfield  {title} {\enquote {\bibinfo {title} {{All-Multipartite
  Bell-Correlation Inequalities for Two Dichotomic Observables per Site}},}\
  }\href {http://link.aps.org/doi/10.1103/PhysRevA.64.032112} {\bibfield
  {journal} {\bibinfo  {journal} {Physical Review A}\ }\textbf {\bibinfo
  {volume} {64}},\ \bibinfo {pages} {032112} (\bibinfo {year}
  {2001}{\natexlab{b}})}\BibitemShut {NoStop}%
\bibitem [{\citenamefont {Filipp}\ and\ \citenamefont
  {Svozil}(2004)}]{MinMaxQB}%
  \BibitemOpen
  \bibfield  {author} {\bibinfo {author} {\bibfnamefont {S.}~\bibnamefont
  {Filipp}}\ and\ \bibinfo {author} {\bibfnamefont {K.}~\bibnamefont
  {Svozil}},\ }\bibfield  {title} {\enquote {\bibinfo {title} {{Generalizing
  Tsirelson's Bound on Bell Inequalities Using a Min-Max Principle}},}\ }\href
  {http://link.aps.org/doi/10.1103/PhysRevLett.93.130407} {\bibfield  {journal}
  {\bibinfo  {journal} {Physical Review Letters}\ }\textbf {\bibinfo {volume}
  {93}},\ \bibinfo {pages} {130407} (\bibinfo {year} {2004})}\BibitemShut
  {NoStop}%
\bibitem [{\citenamefont {Wehner}(2006)}]{WehnerQB}%
  \BibitemOpen
  \bibfield  {author} {\bibinfo {author} {\bibfnamefont {S.}~\bibnamefont
  {Wehner}},\ }\bibfield  {title} {\enquote {\bibinfo {title} {{Tsirelson
  Bounds for Generalized Clauser-Horne-Shimony-Holt Inequalities}},}\ }\href
  {http://link.aps.org/doi/10.1103/PhysRevA.73.022110} {\bibfield  {journal}
  {\bibinfo  {journal} {Physical Review A}\ }\textbf {\bibinfo {volume} {73}},\
  \bibinfo {pages} {022110} (\bibinfo {year} {2006})}\BibitemShut {NoStop}%
\bibitem [{\citenamefont {Wolfe}\ and\ \citenamefont {Yelin}(2012)}]{WolfeQB}%
  \BibitemOpen
  \bibfield  {author} {\bibinfo {author} {\bibfnamefont {E.}~\bibnamefont
  {Wolfe}}\ and\ \bibinfo {author} {\bibfnamefont {S.~F.}\ \bibnamefont
  {Yelin}},\ }\bibfield  {title} {\enquote {\bibinfo {title} {{Quantum Bounds
  for Inequalities Involving Marginal Expectation Values}},}\ }\href
  {http://link.aps.org/doi/10.1103/PhysRevA.86.012123} {\bibfield  {journal}
  {\bibinfo  {journal} {Physical Review A}\ }\textbf {\bibinfo {volume} {86}},\
  \bibinfo {pages} {012123} (\bibinfo {year} {2012})}\BibitemShut {NoStop}%
\bibitem [{\citenamefont {Bae}\ and\ \citenamefont
  {Son}(2014)}]{LatestTsirelson}%
  \BibitemOpen
  \bibfield  {author} {\bibinfo {author} {\bibfnamefont {G.}~\bibnamefont
  {Bae}}\ and\ \bibinfo {author} {\bibfnamefont {W.}~\bibnamefont {Son}},\
  }\bibfield  {title} {\enquote {\bibinfo {title} {{Axiomatic Approach for the
  Functional Bound of Generic Bell's Inequality}},}\ }\href
  {http://arxiv.org/abs/1407.1030} {\bibfield  {journal} {\bibinfo  {journal}
  {arXiv:1407.1030}\ } (\bibinfo {year} {2014})}\BibitemShut {NoStop}%
\bibitem [{\citenamefont {Miniatura}\ \emph {et~al.}(2011)\citenamefont
  {Miniatura}, \citenamefont {Kwek}, \citenamefont {Ducloy}, \citenamefont
  {Gr{\'e}maud}, \citenamefont {Englert}, \citenamefont {Cugliandolo},\ and\
  \citenamefont {Ekert}}]{ScaraniNotes}%
  \BibitemOpen
  \bibfield  {author} {\bibinfo {author} {\bibfnamefont {C.}~\bibnamefont
  {Miniatura}}, \bibinfo {author} {\bibfnamefont {L.}~\bibnamefont {Kwek}},
  \bibinfo {author} {\bibfnamefont {M.}~\bibnamefont {Ducloy}}, \bibinfo
  {author} {\bibfnamefont {B.}~\bibnamefont {Gr{\'e}maud}}, \bibinfo {author}
  {\bibfnamefont {B.}~\bibnamefont {Englert}}, \bibinfo {author} {\bibfnamefont
  {L.}~\bibnamefont {Cugliandolo}}, \ and\ \bibinfo {author} {\bibfnamefont
  {A.}~\bibnamefont {Ekert}},\ }\href
  {http://books.google.com/books?id=CvVxzoYSjyoC} {\emph {\bibinfo {title}
  {{Ultracold Gases and Quantum Information: Lecture Notes of the Les Houches
  Summer School in Singapore: Volume 91, July 2009}}}}\ (\bibinfo  {publisher}
  {OUP Oxford},\ \bibinfo {year} {2011})\BibitemShut {NoStop}%
\bibitem [{\citenamefont {{Scarani}}(2009)}]{ScaraniNotes2}%
  \BibitemOpen
  \bibfield  {author} {\bibinfo {author} {\bibfnamefont {V.}~\bibnamefont
  {{Scarani}}},\ }\bibfield  {title} {\enquote {\bibinfo {title} {{Quantum
  Information: Primitive Notions and Quantum Correlations}},}\ }\href
  {http://arxiv.org/abs/0910.4222} {\bibfield  {journal} {\bibinfo  {journal}
  {arXiv:0910.4222}\ } (\bibinfo {year} {2009})}\BibitemShut {NoStop}%
\bibitem [{\citenamefont {Masanes}(2003)}]{MasanesFourD}%
  \BibitemOpen
  \bibfield  {author} {\bibinfo {author} {\bibfnamefont {L.}~\bibnamefont
  {Masanes}},\ }\bibfield  {title} {\enquote {\bibinfo {title} {{Necessary and
  Sufficient Condition for Quantum-Generated Correlations}},}\ }\href
  {http://arXiv.org/abs/quant-ph/0309137} {\bibfield  {journal} {\bibinfo
  {journal} {arXiv:quant-ph/0309137}\ } (\bibinfo {year} {2003})}\BibitemShut
  {NoStop}%
\bibitem [{\citenamefont {Navascu{\'e}s}\ \emph {et~al.}(2007)\citenamefont
  {Navascu{\'e}s}, \citenamefont {Pironio},\ and\ \citenamefont
  {Ac{\'i}n}}]{NPA2007Short}%
  \BibitemOpen
  \bibfield  {author} {\bibinfo {author} {\bibfnamefont {M.}~\bibnamefont
  {Navascu{\'e}s}}, \bibinfo {author} {\bibfnamefont {S.}~\bibnamefont
  {Pironio}}, \ and\ \bibinfo {author} {\bibfnamefont {A.}~\bibnamefont
  {Ac{\'i}n}},\ }\bibfield  {title} {\enquote {\bibinfo {title} {{Bounding the
  Set of Quantum Correlations}},}\ }\href
  {http://link.aps.org/doi/10.1103/PhysRevLett.98.010401} {\bibfield  {journal}
  {\bibinfo  {journal} {Physical Review Letters}\ }\textbf {\bibinfo {volume}
  {98}},\ \bibinfo {pages} {010401} (\bibinfo {year} {2007})}\BibitemShut
  {NoStop}%
\bibitem [{\citenamefont {Fritz}\ \emph {et~al.}(2012)\citenamefont {Fritz},
  \citenamefont {Leverrier},\ and\ \citenamefont
  {Sainz}}]{FritzCombinatorialLong}%
  \BibitemOpen
  \bibfield  {author} {\bibinfo {author} {\bibfnamefont {T.}~\bibnamefont
  {Fritz}}, \bibinfo {author} {\bibfnamefont {A.}~\bibnamefont {Leverrier}}, \
  and\ \bibinfo {author} {\bibfnamefont {A.~B.}\ \bibnamefont {Sainz}},\
  }\bibfield  {title} {\enquote {\bibinfo {title} {{A Combinatorial Approach to
  Nonlocality and Contextuality}},}\ }\href {http://arXiv.org/abs/1212.4084v2}
  {\bibfield  {journal} {\bibinfo  {journal} {arXiv:1212.4084v2}\ } (\bibinfo
  {year} {2012})}\BibitemShut {NoStop}%
\bibitem [{\citenamefont {Vermeyden}\ \emph {et~al.}(2013)\citenamefont
  {Vermeyden}, \citenamefont {Bonsma}, \citenamefont {Noel}, \citenamefont
  {Donohue}, \citenamefont {Wolfe},\ and\ \citenamefont
  {Resch}}]{WolfeQBExperimental}%
  \BibitemOpen
  \bibfield  {author} {\bibinfo {author} {\bibfnamefont {L.}~\bibnamefont
  {Vermeyden}}, \bibinfo {author} {\bibfnamefont {M.}~\bibnamefont {Bonsma}},
  \bibinfo {author} {\bibfnamefont {C.}~\bibnamefont {Noel}}, \bibinfo {author}
  {\bibfnamefont {J.~M.}\ \bibnamefont {Donohue}}, \bibinfo {author}
  {\bibfnamefont {E.}~\bibnamefont {Wolfe}}, \ and\ \bibinfo {author}
  {\bibfnamefont {K.~J.}\ \bibnamefont {Resch}},\ }\bibfield  {title} {\enquote
  {\bibinfo {title} {{Experimental Violation of Three Families of Bell's
  Inequalities}},}\ }\href {http://link.aps.org/doi/10.1103/PhysRevA.87.032105}
  {\bibfield  {journal} {\bibinfo  {journal} {Physical Review A}\ }\textbf
  {\bibinfo {volume} {87}},\ \bibinfo {pages} {032105} (\bibinfo {year}
  {2013})}\BibitemShut {NoStop}%
\bibitem [{\citenamefont {Pawłowski}\ \emph {et~al.}(2009)\citenamefont
  {Pawłowski}, \citenamefont {Paterek}, \citenamefont {Kaszlikowski},
  \citenamefont {Scarani}, \citenamefont {Winter},\ and\ \citenamefont
  {Zukowski}}]{ICOriginal}%
  \BibitemOpen
  \bibfield  {author} {\bibinfo {author} {\bibfnamefont {M.}~\bibnamefont
  {Pawłowski}}, \bibinfo {author} {\bibfnamefont {T.}~\bibnamefont {Paterek}},
  \bibinfo {author} {\bibfnamefont {D.}~\bibnamefont {Kaszlikowski}}, \bibinfo
  {author} {\bibfnamefont {V.}~\bibnamefont {Scarani}}, \bibinfo {author}
  {\bibfnamefont {A.}~\bibnamefont {Winter}}, \ and\ \bibinfo {author}
  {\bibfnamefont {M.}~\bibnamefont {Zukowski}},\ }\bibfield  {title} {\enquote
  {\bibinfo {title} {{Information Causality as a Physical Principle}},}\ }\href
  {http://dx.doi.org/10.1038/nature08400} {\bibfield  {journal} {\bibinfo
  {journal} {Nature}\ }\textbf {\bibinfo {volume} {461}},\ \bibinfo {pages}
  {1101} (\bibinfo {year} {2009})}\BibitemShut {NoStop}%
\bibitem [{\citenamefont {Allcock}\ \emph {et~al.}(2009)\citenamefont
  {Allcock}, \citenamefont {Brunner}, \citenamefont {Pawłowski},\ and\
  \citenamefont {Scarani}}]{ICRecovery}%
  \BibitemOpen
  \bibfield  {author} {\bibinfo {author} {\bibfnamefont {J.}~\bibnamefont
  {Allcock}}, \bibinfo {author} {\bibfnamefont {N.}~\bibnamefont {Brunner}},
  \bibinfo {author} {\bibfnamefont {M.}~\bibnamefont {Pawłowski}}, \ and\
  \bibinfo {author} {\bibfnamefont {V.}~\bibnamefont {Scarani}},\ }\bibfield
  {title} {\enquote {\bibinfo {title} {{Recovering Part of the Boundary between
  Quantum and Nonquantum Correlations from Information Causality}},}\ }\href
  {http://link.aps.org/doi/10.1103/PhysRevA.80.040103} {\bibfield  {journal}
  {\bibinfo  {journal} {Physical Review A}\ }\textbf {\bibinfo {volume} {80}},\
  \bibinfo {pages} {040103} (\bibinfo {year} {2009})}\BibitemShut {NoStop}%
\bibitem [{\citenamefont {{Beigi}}\ and\ \citenamefont
  {{Gohari}}(2011)}]{ICGrayWyner}%
  \BibitemOpen
  \bibfield  {author} {\bibinfo {author} {\bibfnamefont {S.}~\bibnamefont
  {{Beigi}}}\ and\ \bibinfo {author} {\bibfnamefont {A.}~\bibnamefont
  {{Gohari}}},\ }\bibfield  {title} {\enquote {\bibinfo {title} {{Information
  Causality is a Special Point in the Dual of the Gray-Wyner Region}},}\ }\href
  {http://arxiv.org/abs/1111.3151} {\bibfield  {journal} {\bibinfo  {journal}
  {arXiv:1111.3151}\ } (\bibinfo {year} {2011})}\BibitemShut {NoStop}%
\bibitem [{\citenamefont {{Pawłowski}}\ and\ \citenamefont
  {{Scarani}}(2011)}]{InfoCausArXiv}%
  \BibitemOpen
  \bibfield  {author} {\bibinfo {author} {\bibfnamefont {M.}~\bibnamefont
  {{Pawłowski}}}\ and\ \bibinfo {author} {\bibfnamefont {V.}~\bibnamefont
  {{Scarani}}},\ }\bibfield  {title} {\enquote {\bibinfo {title} {{Information
  Causality}},}\ }\href {http://arxiv.org/abs/1112.1142} {\bibfield  {journal}
  {\bibinfo  {journal} {arXiv:1112.1142}\ } (\bibinfo {year}
  {2011})}\BibitemShut {NoStop}%
\bibitem [{\citenamefont {Navascu\'{e}s}\ and\ \citenamefont
  {Wunderlich}(2010)}]{MacroscopicLocalityOriginal}%
  \BibitemOpen
  \bibfield  {author} {\bibinfo {author} {\bibfnamefont {M.}~\bibnamefont
  {Navascu\'{e}s}}\ and\ \bibinfo {author} {\bibfnamefont {H.}~\bibnamefont
  {Wunderlich}},\ }\bibfield  {title} {\enquote {\bibinfo {title} {{A Glance
  Beyond the Quantum Model}},}\ }\href
  {http://dx.doi.org/10.1098/rspa.2009.0453} {\bibfield  {journal} {\bibinfo
  {journal} {Proceedings of the Royal Society A: Mathematical, Physical and
  Engineering Science}\ }\textbf {\bibinfo {volume} {466}},\ \bibinfo {pages}
  {881} (\bibinfo {year} {2010})}\BibitemShut {NoStop}%
\bibitem [{\citenamefont {Yang}\ \emph {et~al.}(2011)\citenamefont {Yang},
  \citenamefont {Navascu{\'e}s}, \citenamefont {Sheridan},\ and\ \citenamefont
  {Scarani}}]{ScaraniML}%
  \BibitemOpen
  \bibfield  {author} {\bibinfo {author} {\bibfnamefont {T.~H.}\ \bibnamefont
  {Yang}}, \bibinfo {author} {\bibfnamefont {M.}~\bibnamefont {Navascu{\'e}s}},
  \bibinfo {author} {\bibfnamefont {L.}~\bibnamefont {Sheridan}}, \ and\
  \bibinfo {author} {\bibfnamefont {V.}~\bibnamefont {Scarani}},\ }\bibfield
  {title} {\enquote {\bibinfo {title} {{Quantum Bell Inequalities from
  Macroscopic Locality}},}\ }\href
  {http://link.aps.org/doi/10.1103/PhysRevA.83.022105} {\bibfield  {journal}
  {\bibinfo  {journal} {Physcal Review A}\ }\textbf {\bibinfo {volume} {83}},\
  \bibinfo {pages} {022105} (\bibinfo {year} {2011})}\BibitemShut {NoStop}%
\bibitem [{\citenamefont {Cavalcanti}\ \emph {et~al.}(2010)\citenamefont
  {Cavalcanti}, \citenamefont {Salles},\ and\ \citenamefont
  {Scarani}}]{ICFailure}%
  \BibitemOpen
  \bibfield  {author} {\bibinfo {author} {\bibfnamefont {D.}~\bibnamefont
  {Cavalcanti}}, \bibinfo {author} {\bibfnamefont {A.}~\bibnamefont {Salles}},
  \ and\ \bibinfo {author} {\bibfnamefont {V.}~\bibnamefont {Scarani}},\
  }\bibfield  {title} {\enquote {\bibinfo {title} {{Macroscopically Local
  Correlations can Violate Information Causality}},}\ }\href
  {http://dx.doi.org/10.1038/Ncomms1138} {\bibfield  {journal} {\bibinfo
  {journal} {Nature Communications}\ }\textbf {\bibinfo {volume} {1}},\
  \bibinfo {pages} {136} (\bibinfo {year} {2010})}\BibitemShut {NoStop}%
\bibitem [{\citenamefont {Fritz}\ \emph
  {et~al.}(2013{\natexlab{a}})\citenamefont {Fritz}, \citenamefont {Sainz},
  \citenamefont {Augusiak}, \citenamefont {Brask}, \citenamefont {Chaves},
  \citenamefont {Leverrier},\ and\ \citenamefont {Acin}}]{LONatureComm}%
  \BibitemOpen
  \bibfield  {author} {\bibinfo {author} {\bibfnamefont {T.}~\bibnamefont
  {Fritz}}, \bibinfo {author} {\bibfnamefont {A.~B.}\ \bibnamefont {Sainz}},
  \bibinfo {author} {\bibfnamefont {R.}~\bibnamefont {Augusiak}}, \bibinfo
  {author} {\bibfnamefont {J.~B.}\ \bibnamefont {Brask}}, \bibinfo {author}
  {\bibfnamefont {R.}~\bibnamefont {Chaves}}, \bibinfo {author} {\bibfnamefont
  {A.}~\bibnamefont {Leverrier}}, \ and\ \bibinfo {author} {\bibfnamefont
  {A.}~\bibnamefont {Acin}},\ }\bibfield  {title} {\enquote {\bibinfo {title}
  {{Local Orthogonality as a Multipartite Principle for Quantum
  Correlations}},}\ }\href {http://dx.doi.org/10.1038/ncomms3263} {\bibfield
  {journal} {\bibinfo  {journal} {Nature Communications}\ }\textbf {\bibinfo
  {volume} {4}},\ \bibinfo {pages} {2263} (\bibinfo {year}
  {2013}{\natexlab{a}})}\BibitemShut {NoStop}%
\bibitem [{\citenamefont {Fritz}\ \emph
  {et~al.}(2013{\natexlab{b}})\citenamefont {Fritz}, \citenamefont
  {Leverrier},\ and\ \citenamefont {Sainz}}]{LONewShort}%
  \BibitemOpen
  \bibfield  {author} {\bibinfo {author} {\bibfnamefont {T.}~\bibnamefont
  {Fritz}}, \bibinfo {author} {\bibfnamefont {A.}~\bibnamefont {Leverrier}}, \
  and\ \bibinfo {author} {\bibfnamefont {A.~B.}\ \bibnamefont {Sainz}},\
  }\bibfield  {title} {\enquote {\bibinfo {title} {{Probabilistic Models on
  Contextuality Scenarios}},}\ }\href {http://arXiv.org/abs/1307.0145}
  {\bibfield  {journal} {\bibinfo  {journal} {arXiv:1307.0145}\ } (\bibinfo
  {year} {2013}{\natexlab{b}})}\BibitemShut {NoStop}%
\bibitem [{\citenamefont {Das}\ \emph {et~al.}(2013)\citenamefont {Das},
  \citenamefont {Banik}, \citenamefont {Gazi}, \citenamefont {Rai},\ and\
  \citenamefont {Kunkri}}]{LOHardy}%
  \BibitemOpen
  \bibfield  {author} {\bibinfo {author} {\bibfnamefont {S.}~\bibnamefont
  {Das}}, \bibinfo {author} {\bibfnamefont {M.}~\bibnamefont {Banik}}, \bibinfo
  {author} {\bibfnamefont {M.~R.}\ \bibnamefont {Gazi}}, \bibinfo {author}
  {\bibfnamefont {A.}~\bibnamefont {Rai}}, \ and\ \bibinfo {author}
  {\bibfnamefont {S.}~\bibnamefont {Kunkri}},\ }\bibfield  {title} {\enquote
  {\bibinfo {title} {{Local Orthogonality Provides a Tight Upper Bound for
  Hardy's Nonlocality}},}\ }\href
  {http://link.aps.org/doi/10.1103/PhysRevA.88.062101} {\bibfield  {journal}
  {\bibinfo  {journal} {Physical Review A}\ }\textbf {\bibinfo {volume} {88}},\
  \bibinfo {pages} {062101} (\bibinfo {year} {2013})}\BibitemShut {NoStop}%
\bibitem [{\citenamefont {Sainz}\ \emph {et~al.}(2014)\citenamefont {Sainz},
  \citenamefont {Fritz}, \citenamefont {Augusiak}, \citenamefont {Brask},
  \citenamefont {Chaves}, \citenamefont {Leverrier},\ and\ \citenamefont
  {Acín}}]{LOExploring}%
  \BibitemOpen
  \bibfield  {author} {\bibinfo {author} {\bibfnamefont {A.~B.}\ \bibnamefont
  {Sainz}}, \bibinfo {author} {\bibfnamefont {T.}~\bibnamefont {Fritz}},
  \bibinfo {author} {\bibfnamefont {R.}~\bibnamefont {Augusiak}}, \bibinfo
  {author} {\bibfnamefont {J.~B.}\ \bibnamefont {Brask}}, \bibinfo {author}
  {\bibfnamefont {R.}~\bibnamefont {Chaves}}, \bibinfo {author} {\bibfnamefont
  {A.}~\bibnamefont {Leverrier}}, \ and\ \bibinfo {author} {\bibfnamefont
  {A.}~\bibnamefont {Acín}},\ }\bibfield  {title} {\enquote {\bibinfo {title}
  {{Exploring the Local Orthogonality Principle}},}\ }\href
  {http://link.aps.org/doi/10.1103/PhysRevA.89.032117} {\bibfield  {journal}
  {\bibinfo  {journal} {Physical Review A}\ }\textbf {\bibinfo {volume} {89}},\
  \bibinfo {pages} {032117} (\bibinfo {year} {2014})}\BibitemShut {NoStop}%
\bibitem [{\citenamefont {{Scholz}}\ and\ \citenamefont
  {{Werner}}(2008)}]{TsirelsonProblemArXiv}%
  \BibitemOpen
  \bibfield  {author} {\bibinfo {author} {\bibfnamefont {V.~B.}\ \bibnamefont
  {{Scholz}}}\ and\ \bibinfo {author} {\bibfnamefont {R.~F.}\ \bibnamefont
  {{Werner}}},\ }\bibfield  {title} {\enquote {\bibinfo {title} {{Tsirelson's
  Problem}},}\ }\href {http://arxiv.org/abs/0812.4305} {\bibfield  {journal}
  {\bibinfo  {journal} {arXiv:0812.4305}\ } (\bibinfo {year}
  {2008})}\BibitemShut {NoStop}%
\bibitem [{\citenamefont {Junge}\ \emph {et~al.}(2011)\citenamefont {Junge},
  \citenamefont {Navascués}, \citenamefont {Palazuelos}, \citenamefont
  {Perez-Garcia}, \citenamefont {Scholz},\ and\ \citenamefont
  {Werner}}]{ConnesEmbedding}%
  \BibitemOpen
  \bibfield  {author} {\bibinfo {author} {\bibfnamefont {M.}~\bibnamefont
  {Junge}}, \bibinfo {author} {\bibfnamefont {M.}~\bibnamefont {Navascués}},
  \bibinfo {author} {\bibfnamefont {C.}~\bibnamefont {Palazuelos}}, \bibinfo
  {author} {\bibfnamefont {D.}~\bibnamefont {Perez-Garcia}}, \bibinfo {author}
  {\bibfnamefont {V.~B.}\ \bibnamefont {Scholz}}, \ and\ \bibinfo {author}
  {\bibfnamefont {R.~F.}\ \bibnamefont {Werner}},\ }\bibfield  {title}
  {\enquote {\bibinfo {title} {{Connes's Embedding Problem and Tsirelson's
  Problem}},}\ }\href
  {http://scitation.aip.org/content/aip/journal/jmp/52/1/10.1063/1.3514538}
  {\bibfield  {journal} {\bibinfo  {journal} {Journal of Mathematical Physics}\
  }\textbf {\bibinfo {volume} {52}},\ \bibinfo {eid} {012102} (\bibinfo {year}
  {2011})}\BibitemShut {NoStop}%
\bibitem [{\citenamefont {Navascués}\ \emph {et~al.}(2012)\citenamefont
  {Navascués}, \citenamefont {Cooney}, \citenamefont {Pérez-García},\ and\
  \citenamefont {Villanueva}}]{PhysicalTsirelson}%
  \BibitemOpen
  \bibfield  {author} {\bibinfo {author} {\bibfnamefont {M.}~\bibnamefont
  {Navascués}}, \bibinfo {author} {\bibfnamefont {T.}~\bibnamefont {Cooney}},
  \bibinfo {author} {\bibfnamefont {D.}~\bibnamefont {Pérez-García}}, \ and\
  \bibinfo {author} {\bibfnamefont {N.}~\bibnamefont {Villanueva}},\ }\bibfield
   {title} {\enquote {\bibinfo {title} {{A Physical Approach to Tsirelson’s
  Problem}},}\ }\href {http://dx.doi.org/10.1007/s10701-012-9641-0} {\bibfield
  {journal} {\bibinfo  {journal} {Foundations of Physics}\ }\textbf {\bibinfo
  {volume} {42}},\ \bibinfo {pages} {985} (\bibinfo {year} {2012})}\BibitemShut
  {NoStop}%
\bibitem [{\citenamefont {Fritz}(2012)}]{KirchbergConjecture}%
  \BibitemOpen
  \bibfield  {author} {\bibinfo {author} {\bibfnamefont {T.}~\bibnamefont
  {Fritz}},\ }\bibfield  {title} {\enquote {\bibinfo {title} {{Tsirelson's
  Problem and Kirchberg's Conjecture}},}\ }\href
  {http://www.worldscientific.com/doi/abs/10.1142/S0129055X12500122} {\bibfield
   {journal} {\bibinfo  {journal} {Reviews in Mathematical Physics}\ }\textbf
  {\bibinfo {volume} {24}},\ \bibinfo {pages} {1250012} (\bibinfo {year}
  {2012})}\BibitemShut {NoStop}%
\bibitem [{\citenamefont {Miguel}\ \emph {et~al.}(2008)\citenamefont {Miguel},
  \citenamefont {Stefano},\ and\ \citenamefont {Antonio}}]{NPA2008Long}%
  \BibitemOpen
  \bibfield  {author} {\bibinfo {author} {\bibfnamefont {N.}~\bibnamefont
  {Miguel}}, \bibinfo {author} {\bibfnamefont {P.}~\bibnamefont {Stefano}}, \
  and\ \bibinfo {author} {\bibfnamefont {A.}~\bibnamefont {Antonio}},\
  }\bibfield  {title} {\enquote {\bibinfo {title} {{A Convergent Hierarchy of
  Semidefinite Programs Characterizing the Set of Quantum Correlations}},}\
  }\href {http://stacks.iop.org/1367-2630/10/i=7/a=073013} {\bibfield
  {journal} {\bibinfo  {journal} {New Journal of Physics}\ }\textbf {\bibinfo
  {volume} {10}},\ \bibinfo {pages} {073013} (\bibinfo {year}
  {2008})}\BibitemShut {NoStop}%
\bibitem [{\citenamefont {Doherty}\ \emph {et~al.}(2008)\citenamefont
  {Doherty}, \citenamefont {cherng Liang}, \citenamefont {Toner},\ and\
  \citenamefont {Wehner}}]{DohertyNPA}%
  \BibitemOpen
  \bibfield  {author} {\bibinfo {author} {\bibfnamefont {A.~C.}\ \bibnamefont
  {Doherty}}, \bibinfo {author} {\bibfnamefont {Y.}~\bibnamefont {cherng
  Liang}}, \bibinfo {author} {\bibfnamefont {B.}~\bibnamefont {Toner}}, \ and\
  \bibinfo {author} {\bibfnamefont {S.}~\bibnamefont {Wehner}},\ }\bibfield
  {title} {\enquote {\bibinfo {title} {{The Quantum Moment Problem and Bounds
  on Entangled Multi-Prover Games}},}\ }in\ \href
  {http://arxiv.org/abs/0803.4373} {\emph {\bibinfo {booktitle} {{In
  Proceedings of 23rd IEEE Conference on Computational Complexity}}}}\
  (\bibinfo {year} {2008})\BibitemShut {NoStop}%
\bibitem [{\citenamefont {Pironio}\ \emph {et~al.}(2010)\citenamefont
  {Pironio}, \citenamefont {Navascués},\ and\ \citenamefont
  {Acín}}]{NPAReview}%
  \BibitemOpen
  \bibfield  {author} {\bibinfo {author} {\bibfnamefont {S.}~\bibnamefont
  {Pironio}}, \bibinfo {author} {\bibfnamefont {M.}~\bibnamefont {Navascués}},
  \ and\ \bibinfo {author} {\bibfnamefont {A.}~\bibnamefont {Acín}},\
  }\bibfield  {title} {\enquote {\bibinfo {title} {{Convergent Relaxations of
  Polynomial Optimization Problems with Noncommuting Variables}},}\ }\href
  {http://dx.doi.org/10.1137/090760155} {\bibfield  {journal} {\bibinfo
  {journal} {SIAM Journal on Optimization}\ }\textbf {\bibinfo {volume} {20}},\
  \bibinfo {pages} {2157} (\bibinfo {year} {2010})}\BibitemShut {NoStop}%
\bibitem [{\citenamefont {Żukowski}\ and\ \citenamefont
  {Brukner}(2002)}]{GivenQB.NQubit}%
  \BibitemOpen
  \bibfield  {author} {\bibinfo {author} {\bibfnamefont {M.}~\bibnamefont
  {Żukowski}}\ and\ \bibinfo {author} {\bibfnamefont {{\v{C}}.}~\bibnamefont
  {Brukner}},\ }\bibfield  {title} {\enquote {\bibinfo {title} {{Bell’s
  Theorem for General N-Qubit States}},}\ }\href
  {http://link.aps.org/doi/10.1103/PhysRevLett.88.210401} {\bibfield  {journal}
  {\bibinfo  {journal} {Physical Review Letters}\ }\textbf {\bibinfo {volume}
  {88}},\ \bibinfo {pages} {210401} (\bibinfo {year} {2002})}\BibitemShut
  {NoStop}%
\bibitem [{\citenamefont {Cabello}(2002{\natexlab{a}})}]{GivenQB.Cabello}%
  \BibitemOpen
  \bibfield  {author} {\bibinfo {author} {\bibfnamefont {A.}~\bibnamefont
  {Cabello}},\ }\bibfield  {title} {\enquote {\bibinfo {title} {{Bell’s
  Inequality for \(n\) Spin-\(s\) Particles}},}\ }\href
  {http://link.aps.org/doi/10.1103/PhysRevA.65.062105} {\bibfield  {journal}
  {\bibinfo  {journal} {Physical Review A}\ }\textbf {\bibinfo {volume} {65}},\
  \bibinfo {pages} {062105} (\bibinfo {year} {2002}{\natexlab{a}})}\BibitemShut
  {NoStop}%
\bibitem [{\citenamefont {Xiang}(2010)}]{GivenQB.TwoQubit}%
  \BibitemOpen
  \bibfield  {author} {\bibinfo {author} {\bibfnamefont {Y.}~\bibnamefont
  {Xiang}},\ }\bibfield  {title} {\enquote {\bibinfo {title} {{Maximal
  Violation of Bell Inequality for Any Given Two-Qubit Pure State}},}\ }\href
  {http://stacks.iop.org/0256-307X/27/i=12/a=120301} {\bibfield  {journal}
  {\bibinfo  {journal} {Chinese Physics Letters}\ }\textbf {\bibinfo {volume}
  {27}},\ \bibinfo {pages} {120301} (\bibinfo {year} {2010})}\BibitemShut
  {NoStop}%
\bibitem [{\citenamefont {Palazuelos}(2012)}]{GivenQB.arXiv}%
  \BibitemOpen
  \bibfield  {author} {\bibinfo {author} {\bibfnamefont {C.}~\bibnamefont
  {Palazuelos}},\ }\bibfield  {title} {\enquote {\bibinfo {title} {{On the
  Largest Bell Violation Attainable by a Quantum State}},}\ }\href
  {http://arXiv.org/abs/1206.3695} {\bibfield  {journal} {\bibinfo  {journal}
  {arXiv:1206.3695}\ } (\bibinfo {year} {2012})}\BibitemShut {NoStop}%
\bibitem [{\citenamefont {Bennett}\ \emph
  {et~al.}(1999{\natexlab{a}})\citenamefont {Bennett}, \citenamefont
  {DiVincenzo}, \citenamefont {Fuchs}, \citenamefont {Mor}, \citenamefont
  {Rains}, \citenamefont {Shor}, \citenamefont {Smolin},\ and\ \citenamefont
  {Wootters}}]{NonlocalityWithoutEntanglement}%
  \BibitemOpen
  \bibfield  {author} {\bibinfo {author} {\bibfnamefont {C.~H.}\ \bibnamefont
  {Bennett}}, \bibinfo {author} {\bibfnamefont {D.~P.}\ \bibnamefont
  {DiVincenzo}}, \bibinfo {author} {\bibfnamefont {C.~A.}\ \bibnamefont
  {Fuchs}}, \bibinfo {author} {\bibfnamefont {T.}~\bibnamefont {Mor}}, \bibinfo
  {author} {\bibfnamefont {E.}~\bibnamefont {Rains}}, \bibinfo {author}
  {\bibfnamefont {P.~W.}\ \bibnamefont {Shor}}, \bibinfo {author}
  {\bibfnamefont {J.~A.}\ \bibnamefont {Smolin}}, \ and\ \bibinfo {author}
  {\bibfnamefont {W.~K.}\ \bibnamefont {Wootters}},\ }\bibfield  {title}
  {\enquote {\bibinfo {title} {{Quantum Nonlocality Without Entanglement}},}\
  }\href {http://link.aps.org/doi/10.1103/PhysRevA.59.1070} {\bibfield
  {journal} {\bibinfo  {journal} {Physical Review A}\ }\textbf {\bibinfo
  {volume} {59}},\ \bibinfo {pages} {1070} (\bibinfo {year}
  {1999}{\natexlab{a}})}\BibitemShut {NoStop}%
\bibitem [{\citenamefont {Terhal}(2000)}]{NeedEntanglement}%
  \BibitemOpen
  \bibfield  {author} {\bibinfo {author} {\bibfnamefont {B.~M.}\ \bibnamefont
  {Terhal}},\ }\bibfield  {title} {\enquote {\bibinfo {title} {{Bell
  Inequalities and the Separability Criterion}},}\ }\href
  {http://www.sciencedirect.com/science/article/pii/S0375960100004011}
  {\bibfield  {journal} {\bibinfo  {journal} {Physics Letters A}\ }\textbf
  {\bibinfo {volume} {271}},\ \bibinfo {pages} {319} (\bibinfo {year}
  {2000})}\BibitemShut {NoStop}%
\bibitem [{\citenamefont {Scarani}\ \emph {et~al.}(2006)\citenamefont
  {Scarani}, \citenamefont {Gisin}, \citenamefont {Brunner}, \citenamefont
  {Masanes}, \citenamefont {Pino},\ and\ \citenamefont {Acín}}]{CryptoPRA}%
  \BibitemOpen
  \bibfield  {author} {\bibinfo {author} {\bibfnamefont {V.}~\bibnamefont
  {Scarani}}, \bibinfo {author} {\bibfnamefont {N.}~\bibnamefont {Gisin}},
  \bibinfo {author} {\bibfnamefont {N.}~\bibnamefont {Brunner}}, \bibinfo
  {author} {\bibfnamefont {L.}~\bibnamefont {Masanes}}, \bibinfo {author}
  {\bibfnamefont {S.}~\bibnamefont {Pino}}, \ and\ \bibinfo {author}
  {\bibfnamefont {A.}~\bibnamefont {Acín}},\ }\bibfield  {title} {\enquote
  {\bibinfo {title} {{Secrecy Extraction from No-Signaling Correlations}},}\
  }\href {http://link.aps.org/doi/10.1103/PhysRevA.74.042339} {\bibfield
  {journal} {\bibinfo  {journal} {Physical Review A}\ }\textbf {\bibinfo
  {volume} {74}},\ \bibinfo {pages} {042339} (\bibinfo {year}
  {2006})}\BibitemShut {NoStop}%
\bibitem [{\citenamefont {Brunner}\ \emph {et~al.}(2012)\citenamefont
  {Brunner}, \citenamefont {Sharam},\ and\ \citenamefont
  {Vértesi}}]{MultipartiteEntanglement}%
  \BibitemOpen
  \bibfield  {author} {\bibinfo {author} {\bibfnamefont {N.}~\bibnamefont
  {Brunner}}, \bibinfo {author} {\bibfnamefont {J.}~\bibnamefont {Sharam}}, \
  and\ \bibinfo {author} {\bibfnamefont {T.}~\bibnamefont {Vértesi}},\
  }\bibfield  {title} {\enquote {\bibinfo {title} {{Testing the Structure of
  Multipartite Entanglement with Bell Inequalities}},}\ }\href
  {http://link.aps.org/doi/10.1103/PhysRevLett.108.110501} {\bibfield
  {journal} {\bibinfo  {journal} {Physical Review Letters}\ }\textbf {\bibinfo
  {volume} {108}},\ \bibinfo {pages} {110501} (\bibinfo {year}
  {2012})}\BibitemShut {NoStop}%
\bibitem [{\citenamefont {Popescu}(1995)}]{HiddenNonlocality}%
  \BibitemOpen
  \bibfield  {author} {\bibinfo {author} {\bibfnamefont {S.}~\bibnamefont
  {Popescu}},\ }\bibfield  {title} {\enquote {\bibinfo {title} {{Bell's
  Inequalities and Density Matrices: Revealing “Hidden” Nonlocality}},}\
  }\href {http://link.aps.org/doi/10.1103/PhysRevLett.74.2619} {\bibfield
  {journal} {\bibinfo  {journal} {Physical Review Letters}\ }\textbf {\bibinfo
  {volume} {74}},\ \bibinfo {pages} {2619} (\bibinfo {year}
  {1995})}\BibitemShut {NoStop}%
\bibitem [{\citenamefont {Ac{\'i}n}\ \emph {et~al.}(2010)\citenamefont
  {Ac{\'i}n}, \citenamefont {Augusiak}, \citenamefont {Cavalcanti},
  \citenamefont {Hadley}, \citenamefont {Korbicz}, \citenamefont {Lewenstein},
  \citenamefont {Masanes},\ and\ \citenamefont {Piani}}]{UnifiedCorrelations}%
  \BibitemOpen
  \bibfield  {author} {\bibinfo {author} {\bibfnamefont {A.}~\bibnamefont
  {Ac{\'i}n}}, \bibinfo {author} {\bibfnamefont {R.}~\bibnamefont {Augusiak}},
  \bibinfo {author} {\bibfnamefont {D.}~\bibnamefont {Cavalcanti}}, \bibinfo
  {author} {\bibfnamefont {C.}~\bibnamefont {Hadley}}, \bibinfo {author}
  {\bibfnamefont {J.~K.}\ \bibnamefont {Korbicz}}, \bibinfo {author}
  {\bibfnamefont {M.}~\bibnamefont {Lewenstein}}, \bibinfo {author}
  {\bibfnamefont {L.}~\bibnamefont {Masanes}}, \ and\ \bibinfo {author}
  {\bibfnamefont {M.}~\bibnamefont {Piani}},\ }\bibfield  {title} {\enquote
  {\bibinfo {title} {{Unified Framework for Correlations in Terms of Local
  Quantum Observables}},}\ }\href
  {http://link.aps.org/doi/10.1103/PhysRevLett.104.140404} {\bibfield
  {journal} {\bibinfo  {journal} {Physical Review Letters}\ }\textbf {\bibinfo
  {volume} {104}},\ \bibinfo {pages} {140404} (\bibinfo {year}
  {2010})}\BibitemShut {NoStop}%
\bibitem [{\citenamefont {Vértesi}\ and\ \citenamefont
  {Brunner}(2012)}]{PPTNonlocal}%
  \BibitemOpen
  \bibfield  {author} {\bibinfo {author} {\bibfnamefont {T.}~\bibnamefont
  {Vértesi}}\ and\ \bibinfo {author} {\bibfnamefont {N.}~\bibnamefont
  {Brunner}},\ }\bibfield  {title} {\enquote {\bibinfo {title} {{Quantum
  Nonlocality Does Not Imply Entanglement Distillability}},}\ }\href
  {http://link.aps.org/doi/10.1103/PhysRevLett.108.030403} {\bibfield
  {journal} {\bibinfo  {journal} {Physical Review Letters}\ }\textbf {\bibinfo
  {volume} {108}},\ \bibinfo {pages} {030403} (\bibinfo {year}
  {2012})}\BibitemShut {NoStop}%
\bibitem [{\citenamefont {Liang}\ \emph {et~al.}(2012)\citenamefont {Liang},
  \citenamefont {Masanes},\ and\ \citenamefont
  {Rosset}}]{AllEntangledNonlocality}%
  \BibitemOpen
  \bibfield  {author} {\bibinfo {author} {\bibfnamefont {Y.-C.}\ \bibnamefont
  {Liang}}, \bibinfo {author} {\bibfnamefont {L.}~\bibnamefont {Masanes}}, \
  and\ \bibinfo {author} {\bibfnamefont {D.}~\bibnamefont {Rosset}},\
  }\bibfield  {title} {\enquote {\bibinfo {title} {{All Entangled States
  display some Hidden Nonlocality}},}\ }\href
  {http://link.aps.org/doi/10.1103/PhysRevA.86.052115} {\bibfield  {journal}
  {\bibinfo  {journal} {Physical Review A}\ }\textbf {\bibinfo {volume} {86}},\
  \bibinfo {pages} {052115} (\bibinfo {year} {2012})}\BibitemShut {NoStop}%
\bibitem [{\citenamefont {Buscemi}(2012)}]{AllEntangledNonlocal}%
  \BibitemOpen
  \bibfield  {author} {\bibinfo {author} {\bibfnamefont {F.}~\bibnamefont
  {Buscemi}},\ }\bibfield  {title} {\enquote {\bibinfo {title} {{All Entangled
  Quantum States Are Nonlocal}},}\ }\href
  {http://link.aps.org/doi/10.1103/PhysRevLett.108.200401} {\bibfield
  {journal} {\bibinfo  {journal} {Physical Review Letters}\ }\textbf {\bibinfo
  {volume} {108}},\ \bibinfo {pages} {200401} (\bibinfo {year}
  {2012})}\BibitemShut {NoStop}%
\bibitem [{\citenamefont {Massar}\ and\ \citenamefont
  {Pironio}(2012)}]{CloserConnections}%
  \BibitemOpen
  \bibfield  {author} {\bibinfo {author} {\bibfnamefont {S.}~\bibnamefont
  {Massar}}\ and\ \bibinfo {author} {\bibfnamefont {S.}~\bibnamefont
  {Pironio}},\ }\bibfield  {title} {\enquote {\bibinfo {title} {{A Closer
  Connection Between Entanglement and Nonlocality}},}\ }\href
  {http://link.aps.org/doi/10.1103/Physics.5.56} {\bibfield  {journal}
  {\bibinfo  {journal} {Physics}\ }\textbf {\bibinfo {volume} {5}},\ \bibinfo
  {pages} {56} (\bibinfo {year} {2012})}\BibitemShut {NoStop}%
\bibitem [{\citenamefont {{Toner}}\ and\ \citenamefont
  {{Verstraete}}(2006)}]{VerstraeteBellMonogamy}%
  \BibitemOpen
  \bibfield  {author} {\bibinfo {author} {\bibfnamefont {B.}~\bibnamefont
  {{Toner}}}\ and\ \bibinfo {author} {\bibfnamefont {F.}~\bibnamefont
  {{Verstraete}}},\ }\bibfield  {title} {\enquote {\bibinfo {title} {{Monogamy
  of Bell Correlations and Tsirelson's Bound}},}\ }\href
  {http://arxiv.org/abs/quant-ph/0611001} {\bibfield  {journal} {\bibinfo
  {journal} {arXiv:quant-ph/0611001}\ } (\bibinfo {year} {2006})}\BibitemShut
  {NoStop}%
\bibitem [{\citenamefont {Brunner}\ \emph {et~al.}(2008)\citenamefont
  {Brunner}, \citenamefont {Pironio}, \citenamefont {Acin}, \citenamefont
  {Gisin}, \citenamefont {Méthot},\ and\ \citenamefont
  {Scarani}}]{Dimension2008}%
  \BibitemOpen
  \bibfield  {author} {\bibinfo {author} {\bibfnamefont {N.}~\bibnamefont
  {Brunner}}, \bibinfo {author} {\bibfnamefont {S.}~\bibnamefont {Pironio}},
  \bibinfo {author} {\bibfnamefont {A.}~\bibnamefont {Acin}}, \bibinfo {author}
  {\bibfnamefont {N.}~\bibnamefont {Gisin}}, \bibinfo {author} {\bibfnamefont
  {A.~A.}\ \bibnamefont {Méthot}}, \ and\ \bibinfo {author} {\bibfnamefont
  {V.}~\bibnamefont {Scarani}},\ }\bibfield  {title} {\enquote {\bibinfo
  {title} {{Testing the Dimension of Hilbert Spaces}},}\ }\href
  {http://link.aps.org/doi/10.1103/PhysRevLett.100.210503} {\bibfield
  {journal} {\bibinfo  {journal} {Physical Review Letters}\ }\textbf {\bibinfo
  {volume} {100}},\ \bibinfo {pages} {210503} (\bibinfo {year}
  {2008})}\BibitemShut {NoStop}%
\bibitem [{\citenamefont {Gallego}\ \emph {et~al.}(2010)\citenamefont
  {Gallego}, \citenamefont {Brunner}, \citenamefont {Hadley},\ and\
  \citenamefont {Acín}}]{Dimension2010}%
  \BibitemOpen
  \bibfield  {author} {\bibinfo {author} {\bibfnamefont {R.}~\bibnamefont
  {Gallego}}, \bibinfo {author} {\bibfnamefont {N.}~\bibnamefont {Brunner}},
  \bibinfo {author} {\bibfnamefont {C.}~\bibnamefont {Hadley}}, \ and\ \bibinfo
  {author} {\bibfnamefont {A.}~\bibnamefont {Acín}},\ }\bibfield  {title}
  {\enquote {\bibinfo {title} {{Device-Independent Tests of Classical and
  Quantum Dimensions}},}\ }\href
  {http://link.aps.org/doi/10.1103/PhysRevLett.105.230501} {\bibfield
  {journal} {\bibinfo  {journal} {Physical Review Letters}\ }\textbf {\bibinfo
  {volume} {105}},\ \bibinfo {pages} {230501} (\bibinfo {year}
  {2010})}\BibitemShut {NoStop}%
\bibitem [{\citenamefont {Brunner}\ \emph {et~al.}(2013)\citenamefont
  {Brunner}, \citenamefont {Navascués},\ and\ \citenamefont
  {Vértesi}}]{Dimension2013}%
  \BibitemOpen
  \bibfield  {author} {\bibinfo {author} {\bibfnamefont {N.}~\bibnamefont
  {Brunner}}, \bibinfo {author} {\bibfnamefont {M.}~\bibnamefont {Navascués}},
  \ and\ \bibinfo {author} {\bibfnamefont {T.}~\bibnamefont {Vértesi}},\
  }\bibfield  {title} {\enquote {\bibinfo {title} {{Dimension Witnesses and
  Quantum State Discrimination}},}\ }\href
  {http://link.aps.org/doi/10.1103/PhysRevLett.110.150501} {\bibfield
  {journal} {\bibinfo  {journal} {Physical Review Letters}\ }\textbf {\bibinfo
  {volume} {110}},\ \bibinfo {pages} {150501} (\bibinfo {year}
  {2013})}\BibitemShut {NoStop}%
\bibitem [{\citenamefont {Gühne}\ \emph {et~al.}(2014)\citenamefont {Gühne},
  \citenamefont {Budroni}, \citenamefont {Cabello}, \citenamefont {Kleinmann},\
  and\ \citenamefont {Larsson}}]{Dimension2013Cabello}%
  \BibitemOpen
  \bibfield  {author} {\bibinfo {author} {\bibfnamefont {O.}~\bibnamefont
  {Gühne}}, \bibinfo {author} {\bibfnamefont {C.}~\bibnamefont {Budroni}},
  \bibinfo {author} {\bibfnamefont {A.}~\bibnamefont {Cabello}}, \bibinfo
  {author} {\bibfnamefont {M.}~\bibnamefont {Kleinmann}}, \ and\ \bibinfo
  {author} {\bibfnamefont {J.-{\AA}.}\ \bibnamefont {Larsson}},\ }\bibfield
  {title} {\enquote {\bibinfo {title} {{Bounding the Quantum Dimension with
  Contextuality}},}\ }\href
  {http://link.aps.org/doi/10.1103/PhysRevA.89.062107} {\bibfield  {journal}
  {\bibinfo  {journal} {Physical Review A}\ }\textbf {\bibinfo {volume} {89}},\
  \bibinfo {pages} {062107} (\bibinfo {year} {2014})}\BibitemShut {NoStop}%
\bibitem [{\citenamefont {D'Ambrosio}\ \emph {et~al.}(2014)\citenamefont
  {D'Ambrosio}, \citenamefont {Bisesto}, \citenamefont {Sciarrino},
  \citenamefont {Barra}, \citenamefont {Lima},\ and\ \citenamefont
  {Cabello}}]{Dimension2014Cabello}%
  \BibitemOpen
  \bibfield  {author} {\bibinfo {author} {\bibfnamefont {V.}~\bibnamefont
  {D'Ambrosio}}, \bibinfo {author} {\bibfnamefont {F.}~\bibnamefont {Bisesto}},
  \bibinfo {author} {\bibfnamefont {F.}~\bibnamefont {Sciarrino}}, \bibinfo
  {author} {\bibfnamefont {J.~F.}\ \bibnamefont {Barra}}, \bibinfo {author}
  {\bibfnamefont {G.}~\bibnamefont {Lima}}, \ and\ \bibinfo {author}
  {\bibfnamefont {A.}~\bibnamefont {Cabello}},\ }\bibfield  {title} {\enquote
  {\bibinfo {title} {{Device-Independent Certification of High-Dimensional
  Quantum Systems}},}\ }\href
  {http://link.aps.org/doi/10.1103/PhysRevLett.112.140503} {\bibfield
  {journal} {\bibinfo  {journal} {Physical Review Letters}\ }\textbf {\bibinfo
  {volume} {112}},\ \bibinfo {pages} {140503} (\bibinfo {year}
  {2014})}\BibitemShut {NoStop}%
\bibitem [{\citenamefont {Masanes}(2006)}]{MasanesQubits}%
  \BibitemOpen
  \bibfield  {author} {\bibinfo {author} {\bibfnamefont {L.}~\bibnamefont
  {Masanes}},\ }\bibfield  {title} {\enquote {\bibinfo {title} {{Asymptotic
  Violation of Bell Inequalities and Distillability}},}\ }\href
  {http://link.aps.org/doi/10.1103/PhysRevLett.97.050503} {\bibfield  {journal}
  {\bibinfo  {journal} {Physical Review Letters}\ }\textbf {\bibinfo {volume}
  {97}},\ \bibinfo {pages} {050503} (\bibinfo {year} {2006})}\BibitemShut
  {NoStop}%
\bibitem [{\citenamefont {Pironio}(2005)}]{PironioDimension}%
  \BibitemOpen
  \bibfield  {author} {\bibinfo {author} {\bibfnamefont {S.}~\bibnamefont
  {Pironio}},\ }\bibfield  {title} {\enquote {\bibinfo {title} {{Lifting Bell
  Inequalities}},}\ }\href {http://dx.doi.org/10.1063/1.1928727} {\bibfield
  {journal} {\bibinfo  {journal} {Journal of Mathematical Physics}\ }\textbf
  {\bibinfo {volume} {46}},\ \bibinfo {eid} {062112} (\bibinfo {year}
  {2005})}\BibitemShut {NoStop}%
\bibitem [{\citenamefont {Bell}(1966)}]{BellOriginal}%
  \BibitemOpen
  \bibfield  {author} {\bibinfo {author} {\bibfnamefont {J.~S.}\ \bibnamefont
  {Bell}},\ }\bibfield  {title} {\enquote {\bibinfo {title} {{On the Problem of
  Hidden Variables in Quantum Mechanics}},}\ }\href
  {http://link.aps.org/doi/10.1103/RevModPhys.38.447} {\bibfield  {journal}
  {\bibinfo  {journal} {Reviews of Modern Physics}\ }\textbf {\bibinfo {volume}
  {38}},\ \bibinfo {pages} {447} (\bibinfo {year} {1966})}\BibitemShut
  {NoStop}%
\bibitem [{\citenamefont {Sadiq}\ \emph {et~al.}(2013)\citenamefont {Sadiq},
  \citenamefont {Badziąg}, \citenamefont {Bourennane},\ and\ \citenamefont
  {Cabello}}]{VariousCHSHBellInequalities}%
  \BibitemOpen
  \bibfield  {author} {\bibinfo {author} {\bibfnamefont {M.}~\bibnamefont
  {Sadiq}}, \bibinfo {author} {\bibfnamefont {P.}~\bibnamefont {Badziąg}},
  \bibinfo {author} {\bibfnamefont {M.}~\bibnamefont {Bourennane}}, \ and\
  \bibinfo {author} {\bibfnamefont {A.}~\bibnamefont {Cabello}},\ }\bibfield
  {title} {\enquote {\bibinfo {title} {{Bell Inequalities for the Simplest
  Exclusivity Graph}},}\ }\href
  {http://link.aps.org/doi/10.1103/PhysRevA.87.012128} {\bibfield  {journal}
  {\bibinfo  {journal} {Physical Review A}\ }\textbf {\bibinfo {volume} {87}},\
  \bibinfo {pages} {012128} (\bibinfo {year} {2013})}\BibitemShut {NoStop}%
\bibitem [{\citenamefont {Dowker}\ \emph {et~al.}(2014)\citenamefont {Dowker},
  \citenamefont {Henson},\ and\ \citenamefont
  {Wallden}}]{ConsistentHistoriesContextuality}%
  \BibitemOpen
  \bibfield  {author} {\bibinfo {author} {\bibfnamefont {F.}~\bibnamefont
  {Dowker}}, \bibinfo {author} {\bibfnamefont {J.}~\bibnamefont {Henson}}, \
  and\ \bibinfo {author} {\bibfnamefont {P.}~\bibnamefont {Wallden}},\
  }\bibfield  {title} {\enquote {\bibinfo {title} {{A Histories Perspective on
  Characterizing Quantum Non-Locality}},}\ }\href
  {http://stacks.iop.org/1367-2630/16/i=3/a=033033} {\bibfield  {journal}
  {\bibinfo  {journal} {New Journal of Physics}\ }\textbf {\bibinfo {volume}
  {16}},\ \bibinfo {pages} {033033} (\bibinfo {year} {2014})}\BibitemShut
  {NoStop}%
\bibitem [{\citenamefont {Hardy}(1992)}]{Hardy92}%
  \BibitemOpen
  \bibfield  {author} {\bibinfo {author} {\bibfnamefont {L.}~\bibnamefont
  {Hardy}},\ }\bibfield  {title} {\enquote {\bibinfo {title} {{Quantum
  Mechanics, Local Realistic Theories, and Lorentz-Invariant Realistic
  Theories}},}\ }\href {http://link.aps.org/doi/10.1103/PhysRevLett.68.2981}
  {\bibfield  {journal} {\bibinfo  {journal} {Physical Review Letters}\
  }\textbf {\bibinfo {volume} {68}},\ \bibinfo {pages} {2981} (\bibinfo {year}
  {1992})}\BibitemShut {NoStop}%
\bibitem [{\citenamefont {Hardy}(1993)}]{Hardy93}%
  \BibitemOpen
  \bibfield  {author} {\bibinfo {author} {\bibfnamefont {L.}~\bibnamefont
  {Hardy}},\ }\bibfield  {title} {\enquote {\bibinfo {title} {{Nonlocality for
  Two Particles Without Inequalities for Almost All Entangled States}},}\
  }\href {http://link.aps.org/doi/10.1103/PhysRevLett.71.1665} {\bibfield
  {journal} {\bibinfo  {journal} {Physical Review Letters}\ }\textbf {\bibinfo
  {volume} {71}},\ \bibinfo {pages} {1665} (\bibinfo {year}
  {1993})}\BibitemShut {NoStop}%
\bibitem [{\citenamefont {Boschi}\ \emph {et~al.}(1997)\citenamefont {Boschi},
  \citenamefont {Branca}, \citenamefont {De~Martini},\ and\ \citenamefont
  {Hardy}}]{Hardy97}%
  \BibitemOpen
  \bibfield  {author} {\bibinfo {author} {\bibfnamefont {D.}~\bibnamefont
  {Boschi}}, \bibinfo {author} {\bibfnamefont {S.}~\bibnamefont {Branca}},
  \bibinfo {author} {\bibfnamefont {F.}~\bibnamefont {De~Martini}}, \ and\
  \bibinfo {author} {\bibfnamefont {L.}~\bibnamefont {Hardy}},\ }\bibfield
  {title} {\enquote {\bibinfo {title} {{Ladder Proof of Nonlocality without
  Inequalities: Theoretical and Experimental Results}},}\ }\href
  {http://link.aps.org/doi/10.1103/PhysRevLett.79.2755} {\bibfield  {journal}
  {\bibinfo  {journal} {Physical Review Letters}\ }\textbf {\bibinfo {volume}
  {79}},\ \bibinfo {pages} {2755} (\bibinfo {year} {1997})}\BibitemShut
  {NoStop}%
\bibitem [{\citenamefont {Cabello}(2002{\natexlab{b}})}]{CabelloHardy}%
  \BibitemOpen
  \bibfield  {author} {\bibinfo {author} {\bibfnamefont {A.}~\bibnamefont
  {Cabello}},\ }\bibfield  {title} {\enquote {\bibinfo {title} {{Bell's Theorem
  With and Without Inequalities for the Three-Qubit Greenberger-Horne-Zeilinger
  and W States}},}\ }\href {http://link.aps.org/doi/10.1103/PhysRevA.65.032108}
  {\bibfield  {journal} {\bibinfo  {journal} {Physical Review A}\ }\textbf
  {\bibinfo {volume} {65}},\ \bibinfo {pages} {032108} (\bibinfo {year}
  {2002}{\natexlab{b}})}\BibitemShut {NoStop}%
\bibitem [{\citenamefont {{Man{\v c}inska}}\ and\ \citenamefont
  {{Wehner}}(2014)}]{HardyCHSH}%
  \BibitemOpen
  \bibfield  {author} {\bibinfo {author} {\bibfnamefont {L.}~\bibnamefont
  {{Man{\v c}inska}}}\ and\ \bibinfo {author} {\bibfnamefont {S.}~\bibnamefont
  {{Wehner}}},\ }\bibfield  {title} {\enquote {\bibinfo {title} {{A Unified
  View on Hardy's Paradox and the CHSH Inequality}},}\ }\href
  {http://arxiv.org/abs/1407.2320} {\bibfield  {journal} {\bibinfo  {journal}
  {arXiv:1407.2320}\ } (\bibinfo {year} {2014})}\BibitemShut {NoStop}%
\bibitem [{\citenamefont {Kunkri}\ \emph {et~al.}(2006)\citenamefont {Kunkri},
  \citenamefont {Choudhary}, \citenamefont {Ahanj},\ and\ \citenamefont
  {Joag}}]{HardyQubits06}%
  \BibitemOpen
  \bibfield  {author} {\bibinfo {author} {\bibfnamefont {S.}~\bibnamefont
  {Kunkri}}, \bibinfo {author} {\bibfnamefont {S.~K.}\ \bibnamefont
  {Choudhary}}, \bibinfo {author} {\bibfnamefont {A.}~\bibnamefont {Ahanj}}, \
  and\ \bibinfo {author} {\bibfnamefont {P.}~\bibnamefont {Joag}},\ }\bibfield
  {title} {\enquote {\bibinfo {title} {{Nonlocality Without Inequality for
  Almost All Two-Qubit Entangled States based on Cabello's Nonlocality
  Argument}},}\ }\href {http://link.aps.org/doi/10.1103/PhysRevA.73.022346}
  {\bibfield  {journal} {\bibinfo  {journal} {Physical Review A}\ }\textbf
  {\bibinfo {volume} {73}},\ \bibinfo {pages} {022346} (\bibinfo {year}
  {2006})}\BibitemShut {NoStop}%
\bibitem [{\citenamefont {Seshadreesan}\ and\ \citenamefont
  {Ghosh}(2011)}]{HardyQubits11}%
  \BibitemOpen
  \bibfield  {author} {\bibinfo {author} {\bibfnamefont {K.~P.}\ \bibnamefont
  {Seshadreesan}}\ and\ \bibinfo {author} {\bibfnamefont {S.}~\bibnamefont
  {Ghosh}},\ }\bibfield  {title} {\enquote {\bibinfo {title} {{Constancy of
  Maximal Nonlocal Probability in Hardy's Nonlocality Test for Bipartite
  Quantum Systems}},}\ }\href
  {http://stacks.iop.org/1751-8121/44/i=31/a=315305} {\bibfield  {journal}
  {\bibinfo  {journal} {Journal of Physics A: Mathematical and Theoretical}\
  }\textbf {\bibinfo {volume} {44}},\ \bibinfo {pages} {315305} (\bibinfo
  {year} {2011})}\BibitemShut {NoStop}%
\bibitem [{\citenamefont {Rabelo}\ \emph {et~al.}(2012)\citenamefont {Rabelo},
  \citenamefont {Zhi},\ and\ \citenamefont {Scarani}}]{HardyQubits12}%
  \BibitemOpen
  \bibfield  {author} {\bibinfo {author} {\bibfnamefont {R.}~\bibnamefont
  {Rabelo}}, \bibinfo {author} {\bibfnamefont {L.~Y.}\ \bibnamefont {Zhi}}, \
  and\ \bibinfo {author} {\bibfnamefont {V.}~\bibnamefont {Scarani}},\
  }\bibfield  {title} {\enquote {\bibinfo {title} {{Device-Independent Bounds
  for Hardy's Experiment}},}\ }\href
  {http://link.aps.org/doi/10.1103/PhysRevLett.109.180401} {\bibfield
  {journal} {\bibinfo  {journal} {Physical Review Letters}\ }\textbf {\bibinfo
  {volume} {109}},\ \bibinfo {pages} {180401} (\bibinfo {year}
  {2012})}\BibitemShut {NoStop}%
\bibitem [{\citenamefont {{Xu}}\ \emph {et~al.}(2014)\citenamefont {{Xu}},
  \citenamefont {{Su}},\ and\ \citenamefont {{Chen}}}]{HardyVarientDIversion}%
  \BibitemOpen
  \bibfield  {author} {\bibinfo {author} {\bibfnamefont {Z.-P.}\ \bibnamefont
  {{Xu}}}, \bibinfo {author} {\bibfnamefont {H.-Y.}\ \bibnamefont {{Su}}}, \
  and\ \bibinfo {author} {\bibfnamefont {J.-L.}\ \bibnamefont {{Chen}}},\
  }\bibfield  {title} {\enquote {\bibinfo {title} {{Dimension-Independent
  Bounds for Hardy's Experiment}},}\ }\href {http://arxiv.org/abs/1406.5812}
  {\bibfield  {journal} {\bibinfo  {journal} {arXiv:1406.5812}\ } (\bibinfo
  {year} {2014})}\BibitemShut {NoStop}%
\bibitem [{\citenamefont {Güler}(1997)}]{Hyperbolic97}%
  \BibitemOpen
  \bibfield  {author} {\bibinfo {author} {\bibfnamefont {O.}~\bibnamefont
  {Güler}},\ }\bibfield  {title} {\enquote {\bibinfo {title} {{Hyperbolic
  Polynomials and Interior Point Methods for Convex Programming}},}\ }\href
  {http://dx.doi.org/10.1287/moor.22.2.350} {\bibfield  {journal} {\bibinfo
  {journal} {Mathematics of Operations Research}\ }\textbf {\bibinfo {volume}
  {22}},\ \bibinfo {pages} {350} (\bibinfo {year} {1997})}\BibitemShut
  {NoStop}%
\bibitem [{\citenamefont {Bauschke}\ \emph {et~al.}(2001)\citenamefont
  {Bauschke}, \citenamefont {Güler}, \citenamefont {Lewis},\ and\
  \citenamefont {Sendov}}]{Hyperbolic01}%
  \BibitemOpen
  \bibfield  {author} {\bibinfo {author} {\bibfnamefont {H.~H.}\ \bibnamefont
  {Bauschke}}, \bibinfo {author} {\bibfnamefont {O.}~\bibnamefont {Güler}},
  \bibinfo {author} {\bibfnamefont {A.~S.}\ \bibnamefont {Lewis}}, \ and\
  \bibinfo {author} {\bibfnamefont {H.~S.}\ \bibnamefont {Sendov}},\ }\bibfield
   {title} {\enquote {\bibinfo {title} {{Hyperbolic Polynomials and Convex
  Analysis}},}\ }\href {http://dx.doi.org/10.4153/CJM-2001-020-6} {\bibfield
  {journal} {\bibinfo  {journal} {Canadian Journal of Mathematics}\ }\textbf
  {\bibinfo {volume} {53}},\ \bibinfo {pages} {470} (\bibinfo {year}
  {2001})}\BibitemShut {NoStop}%
\bibitem [{\citenamefont {{Navascu{\'e}s}}\ \emph {et~al.}(2014)\citenamefont
  {{Navascu{\'e}s}}, \citenamefont {{Guryanova}}, \citenamefont {{Hoban}},\
  and\ \citenamefont {{Ac{\'{\i}}n}}}]{AlmostQuantum2}%
  \BibitemOpen
  \bibfield  {author} {\bibinfo {author} {\bibfnamefont {M.}~\bibnamefont
  {{Navascu{\'e}s}}}, \bibinfo {author} {\bibfnamefont {Y.}~\bibnamefont
  {{Guryanova}}}, \bibinfo {author} {\bibfnamefont {M.~J.}\ \bibnamefont
  {{Hoban}}}, \ and\ \bibinfo {author} {\bibfnamefont {A.}~\bibnamefont
  {{Ac{\'{\i}}n}}},\ }\bibfield  {title} {\enquote {\bibinfo {title} {{Almost
  Quantum Correlations}},}\ }\href {http://arXiv.org/abs/1403.4621} {\bibfield
  {journal} {\bibinfo  {journal} {arXiv:1403.4621}\ } (\bibinfo {year}
  {2014})}\BibitemShut {NoStop}%
\bibitem [{\citenamefont {Clauser}\ \emph {et~al.}(1969)\citenamefont
  {Clauser}, \citenamefont {Horne}, \citenamefont {Shimony},\ and\
  \citenamefont {Holt}}]{CHSHOriginal}%
  \BibitemOpen
  \bibfield  {author} {\bibinfo {author} {\bibfnamefont {J.~F.}\ \bibnamefont
  {Clauser}}, \bibinfo {author} {\bibfnamefont {M.~A.}\ \bibnamefont {Horne}},
  \bibinfo {author} {\bibfnamefont {A.}~\bibnamefont {Shimony}}, \ and\
  \bibinfo {author} {\bibfnamefont {R.~A.}\ \bibnamefont {Holt}},\ }\bibfield
  {title} {\enquote {\bibinfo {title} {{Proposed Experiment to Test Local
  Hidden-Variable Theories}},}\ }\href
  {http://link.aps.org/doi/10.1103/PhysRevLett.23.880} {\bibfield  {journal}
  {\bibinfo  {journal} {Physical Review Letters}\ }\textbf {\bibinfo {volume}
  {23}},\ \bibinfo {pages} {880} (\bibinfo {year} {1969})}\BibitemShut
  {NoStop}%
\bibitem [{\citenamefont {Scarani}(2012)}]{scarani2012device}%
  \BibitemOpen
  \bibfield  {author} {\bibinfo {author} {\bibfnamefont {V.}~\bibnamefont
  {Scarani}},\ }\bibfield  {title} {\enquote {\bibinfo {title} {{The
  Device-Independent Outlook on Quantum Physics}},}\ }\href
  {http://www.physics.sk/aps/pub.php?y=2012&pub=aps-12-04} {\bibfield
  {journal} {\bibinfo  {journal} {Acta Physica Slovaca}\ }\textbf {\bibinfo
  {volume} {62}},\ \bibinfo {pages} {347} (\bibinfo {year} {2012})}\BibitemShut
  {NoStop}%
\bibitem [{\citenamefont {Uffink}(2002)}]{Uffink}%
  \BibitemOpen
  \bibfield  {author} {\bibinfo {author} {\bibfnamefont {J.}~\bibnamefont
  {Uffink}},\ }\bibfield  {title} {\enquote {\bibinfo {title} {{Quadratic Bell
  Inequalities as Tests for Multipartite Entanglement}},}\ }\href
  {http://link.aps.org/doi/10.1103/PhysRevLett.88.230406} {\bibfield  {journal}
  {\bibinfo  {journal} {Physical Review Letters}\ }\textbf {\bibinfo {volume}
  {88}},\ \bibinfo {pages} {230406} (\bibinfo {year} {2002})}\BibitemShut
  {NoStop}%
\bibitem [{\citenamefont {Ac{\'{i}}n}\ \emph {et~al.}(2006)\citenamefont
  {Ac{\'{i}}n}, \citenamefont {Gisin},\ and\ \citenamefont
  {Masanes}}]{CryptoPRL}%
  \BibitemOpen
  \bibfield  {author} {\bibinfo {author} {\bibfnamefont {A.}~\bibnamefont
  {Ac{\'{i}}n}}, \bibinfo {author} {\bibfnamefont {N.}~\bibnamefont {Gisin}}, \
  and\ \bibinfo {author} {\bibfnamefont {L.}~\bibnamefont {Masanes}},\
  }\bibfield  {title} {\enquote {\bibinfo {title} {{From Bell’s Theorem to
  Secure Quantum Key Distribution}},}\ }\href
  {http://link.aps.org/doi/10.1103/PhysRevLett.97.120405} {\bibfield  {journal}
  {\bibinfo  {journal} {Physical Review Letters}\ }\textbf {\bibinfo {volume}
  {97}},\ \bibinfo {pages} {120405} (\bibinfo {year} {2006})}\BibitemShut
  {NoStop}%
\bibitem [{\citenamefont {Masanes}\ \emph {et~al.}(2011)\citenamefont
  {Masanes}, \citenamefont {Pironio},\ and\ \citenamefont
  {Acín}}]{masanesDIQKD}%
  \BibitemOpen
  \bibfield  {author} {\bibinfo {author} {\bibfnamefont {L.}~\bibnamefont
  {Masanes}}, \bibinfo {author} {\bibfnamefont {S.}~\bibnamefont {Pironio}}, \
  and\ \bibinfo {author} {\bibfnamefont {A.}~\bibnamefont {Acín}},\ }\bibfield
   {title} {\enquote {\bibinfo {title} {{Secure Device-Independent Quantum Key
  Distribution with Causally Independent Measurement Devices}},}\ }\href
  {http://dx.doi.org/10.1038/ncomms1244} {\bibfield  {journal} {\bibinfo
  {journal} {Nature Communications}\ }\textbf {\bibinfo {volume} {2}},\
  \bibinfo {pages} {238+} (\bibinfo {year} {2011})}\BibitemShut {NoStop}%
\bibitem [{\citenamefont {Pironio}\ \emph {et~al.}(2009)\citenamefont
  {Pironio}, \citenamefont {Acín}, \citenamefont {Brunner}, \citenamefont
  {Gisin}, \citenamefont {Massar},\ and\ \citenamefont
  {Scarani}}]{CryptoDIQKD}%
  \BibitemOpen
  \bibfield  {author} {\bibinfo {author} {\bibfnamefont {S.}~\bibnamefont
  {Pironio}}, \bibinfo {author} {\bibfnamefont {A.}~\bibnamefont {Acín}},
  \bibinfo {author} {\bibfnamefont {N.}~\bibnamefont {Brunner}}, \bibinfo
  {author} {\bibfnamefont {N.}~\bibnamefont {Gisin}}, \bibinfo {author}
  {\bibfnamefont {S.}~\bibnamefont {Massar}}, \ and\ \bibinfo {author}
  {\bibfnamefont {V.}~\bibnamefont {Scarani}},\ }\bibfield  {title} {\enquote
  {\bibinfo {title} {{Device-Independent Quantum Key Distribution Secure
  Against Collective Attacks}},}\ }\href
  {http://stacks.iop.org/1367-2630/11/i=4/a=045021} {\bibfield  {journal}
  {\bibinfo  {journal} {New Journal of Physics}\ }\textbf {\bibinfo {volume}
  {11}},\ \bibinfo {pages} {045021} (\bibinfo {year} {2009})}\BibitemShut
  {NoStop}%
\bibitem [{\citenamefont {Ekert}\ and\ \citenamefont
  {Renner}(2014)}]{RenatoQKDNature}%
  \BibitemOpen
  \bibfield  {author} {\bibinfo {author} {\bibfnamefont {A.}~\bibnamefont
  {Ekert}}\ and\ \bibinfo {author} {\bibfnamefont {R.}~\bibnamefont {Renner}},\
  }\bibfield  {title} {\enquote {\bibinfo {title} {{The Ultimate Physical
  Limits of Privacy}},}\ }\href {http://dx.doi.org/10.1038/nature13132}
  {\bibfield  {journal} {\bibinfo  {journal} {Nature}\ }\textbf {\bibinfo
  {volume} {507}},\ \bibinfo {pages} {443} (\bibinfo {year}
  {2014})}\BibitemShut {NoStop}%
\bibitem [{\citenamefont {Nieto-Silleras}\ \emph {et~al.}(2014)\citenamefont
  {Nieto-Silleras}, \citenamefont {Pironio},\ and\ \citenamefont
  {Silman}}]{DIRandomness}%
  \BibitemOpen
  \bibfield  {author} {\bibinfo {author} {\bibfnamefont {O.}~\bibnamefont
  {Nieto-Silleras}}, \bibinfo {author} {\bibfnamefont {S.}~\bibnamefont
  {Pironio}}, \ and\ \bibinfo {author} {\bibfnamefont {J.}~\bibnamefont
  {Silman}},\ }\bibfield  {title} {\enquote {\bibinfo {title} {{Using Complete
  Measurement Statistics for Optimal Device-Independent Randomness
  Evaluation}},}\ }\href {http://stacks.iop.org/1367-2630/16/i=1/a=013035}
  {\bibfield  {journal} {\bibinfo  {journal} {New Journal of Physics}\ }\textbf
  {\bibinfo {volume} {16}},\ \bibinfo {pages} {013035} (\bibinfo {year}
  {2014})}\BibitemShut {NoStop}%
\bibitem [{\citenamefont {{Masanes}}\ \emph {et~al.}(2006)\citenamefont
  {{Masanes}}, \citenamefont {{Renner}}, \citenamefont {{Christandl}},
  \citenamefont {{Winter}},\ and\ \citenamefont
  {{Barrett}}}]{CryptoDIQKDarXiv2006}%
  \BibitemOpen
  \bibfield  {author} {\bibinfo {author} {\bibfnamefont {L.}~\bibnamefont
  {{Masanes}}}, \bibinfo {author} {\bibfnamefont {R.}~\bibnamefont {{Renner}}},
  \bibinfo {author} {\bibfnamefont {M.}~\bibnamefont {{Christandl}}}, \bibinfo
  {author} {\bibfnamefont {A.}~\bibnamefont {{Winter}}}, \ and\ \bibinfo
  {author} {\bibfnamefont {J.}~\bibnamefont {{Barrett}}},\ }\bibfield  {title}
  {\enquote {\bibinfo {title} {{Unconditional Security of Key Distribution from
  Causality Constraints}},}\ }\href {http://arxiv.org/abs/quant-ph/0606049}
  {\bibfield  {journal} {\bibinfo  {journal} {arXiv:quant-ph/0606049}\ }
  (\bibinfo {year} {2006})}\BibitemShut {NoStop}%
\bibitem [{\citenamefont {{H{\"a}nggi}}\ \emph {et~al.}(2009)\citenamefont
  {{H{\"a}nggi}}, \citenamefont {{Renner}},\ and\ \citenamefont
  {{Wolf}}}]{CryptoDIQKDarXiv}%
  \BibitemOpen
  \bibfield  {author} {\bibinfo {author} {\bibfnamefont {E.}~\bibnamefont
  {{H{\"a}nggi}}}, \bibinfo {author} {\bibfnamefont {R.}~\bibnamefont
  {{Renner}}}, \ and\ \bibinfo {author} {\bibfnamefont {S.}~\bibnamefont
  {{Wolf}}},\ }\bibfield  {title} {\enquote {\bibinfo {title} {{Quantum
  Cryptography Based Solely on Bell's Theorem}},}\ }\href
  {http://arxiv.org/abs/0911.4171} {\bibfield  {journal} {\bibinfo  {journal}
  {arXiv:0911.4171}\ } (\bibinfo {year} {2009})}\BibitemShut {NoStop}%
\bibitem [{\citenamefont {Barrett}\ \emph {et~al.}(2005)\citenamefont
  {Barrett}, \citenamefont {Hardy},\ and\ \citenamefont
  {Kent}}]{CryptoDIQKDEarly}%
  \BibitemOpen
  \bibfield  {author} {\bibinfo {author} {\bibfnamefont {J.}~\bibnamefont
  {Barrett}}, \bibinfo {author} {\bibfnamefont {L.}~\bibnamefont {Hardy}}, \
  and\ \bibinfo {author} {\bibfnamefont {A.}~\bibnamefont {Kent}},\ }\bibfield
  {title} {\enquote {\bibinfo {title} {{No Signaling and Quantum Key
  Distribution}},}\ }\href
  {http://link.aps.org/doi/10.1103/PhysRevLett.95.010503} {\bibfield  {journal}
  {\bibinfo  {journal} {Physical Review Letters}\ }\textbf {\bibinfo {volume}
  {95}},\ \bibinfo {pages} {010503} (\bibinfo {year} {2005})}\BibitemShut
  {NoStop}%
\bibitem [{\citenamefont {{Horodecki}}\ \emph {et~al.}(2010)\citenamefont
  {{Horodecki}}, \citenamefont {{Horodecki}}, \citenamefont {{Horodecki}},
  \citenamefont {{Horodecki}}, \citenamefont {{Pawłowski}},\ and\
  \citenamefont {{Bourennane}}}]{CryptoDIQKD2010}%
  \BibitemOpen
  \bibfield  {author} {\bibinfo {author} {\bibfnamefont {K.}~\bibnamefont
  {{Horodecki}}}, \bibinfo {author} {\bibfnamefont {M.}~\bibnamefont
  {{Horodecki}}}, \bibinfo {author} {\bibfnamefont {P.}~\bibnamefont
  {{Horodecki}}}, \bibinfo {author} {\bibfnamefont {R.}~\bibnamefont
  {{Horodecki}}}, \bibinfo {author} {\bibfnamefont {M.}~\bibnamefont
  {{Pawłowski}}}, \ and\ \bibinfo {author} {\bibfnamefont {M.}~\bibnamefont
  {{Bourennane}}},\ }\bibfield  {title} {\enquote {\bibinfo {title}
  {{Contextuality offers Device-Independent Security}},}\ }\href
  {http://arxiv.org/abs/1006.0468} {\bibfield  {journal} {\bibinfo  {journal}
  {arXiv:1006.0468}\ } (\bibinfo {year} {2010})},\ \Eprint
  {http://arxiv.org/abs/1006.0468} {arXiv:1006.0468 [quant-ph]} \BibitemShut
  {NoStop}%
\bibitem [{\citenamefont {Acín}\ \emph {et~al.}(2006)\citenamefont {Acín},
  \citenamefont {Massar},\ and\ \citenamefont {Pironio}}]{CryptoDIQKDNJP}%
  \BibitemOpen
  \bibfield  {author} {\bibinfo {author} {\bibfnamefont {A.}~\bibnamefont
  {Acín}}, \bibinfo {author} {\bibfnamefont {S.}~\bibnamefont {Massar}}, \
  and\ \bibinfo {author} {\bibfnamefont {S.}~\bibnamefont {Pironio}},\
  }\bibfield  {title} {\enquote {\bibinfo {title} {{Efficient Quantum Key
  Distribution Secure against No-Signalling Eavesdroppers}},}\ }\href
  {http://stacks.iop.org/1367-2630/8/i=8/a=126} {\bibfield  {journal} {\bibinfo
   {journal} {New Journal of Physics}\ }\textbf {\bibinfo {volume} {8}},\
  \bibinfo {pages} {126} (\bibinfo {year} {2006})}\BibitemShut {NoStop}%
\bibitem [{\citenamefont {{Haur Yang}}\ \emph {et~al.}(2013)\citenamefont
  {{Haur Yang}}, \citenamefont {{V{\'e}rtesi}}, \citenamefont {{Bancal}},
  \citenamefont {{Scarani}},\ and\ \citenamefont
  {{Navascu{\'e}s}}}]{DITomagraphy2013}%
  \BibitemOpen
  \bibfield  {author} {\bibinfo {author} {\bibfnamefont {T.}~\bibnamefont
  {{Haur Yang}}}, \bibinfo {author} {\bibfnamefont {T.}~\bibnamefont
  {{V{\'e}rtesi}}}, \bibinfo {author} {\bibfnamefont {J.-D.}\ \bibnamefont
  {{Bancal}}}, \bibinfo {author} {\bibfnamefont {V.}~\bibnamefont {{Scarani}}},
  \ and\ \bibinfo {author} {\bibfnamefont {M.}~\bibnamefont
  {{Navascu{\'e}s}}},\ }\bibfield  {title} {\enquote {\bibinfo {title}
  {{Opening the Black Box: How to Estimate Physical Properties from Non-Local
  Correlations}},}\ }\href {http://arxiv.org/abs/1307.7053} {\bibfield
  {journal} {\bibinfo  {journal} {arXiv:1307.7053}\ } (\bibinfo {year}
  {2013})}\BibitemShut {NoStop}%
\bibitem [{\citenamefont {F.~Pal}\ \emph {et~al.}(2014)\citenamefont {F.~Pal},
  \citenamefont {{V{\'e}rtesi}},\ and\ \citenamefont
  {{Navascu{\'e}s}}}]{DITomagraphy2014}%
  \BibitemOpen
  \bibfield  {author} {\bibinfo {author} {\bibfnamefont {K.}~\bibnamefont
  {F.~Pal}}, \bibinfo {author} {\bibfnamefont {T.}~\bibnamefont
  {{V{\'e}rtesi}}}, \ and\ \bibinfo {author} {\bibfnamefont {M.}~\bibnamefont
  {{Navascu{\'e}s}}},\ }\bibfield  {title} {\enquote {\bibinfo {title}
  {{Device-Independent Tomography of Multipartite Quantum States}},}\ }\href
  {http://arxiv.org/abs/1407.5911} {\bibfield  {journal} {\bibinfo  {journal}
  {arXiv:1407.5911}\ } (\bibinfo {year} {2014})}\BibitemShut {NoStop}%
\bibitem [{\citenamefont {Bruß}(2002)}]{characterizingentanglement}%
  \BibitemOpen
  \bibfield  {author} {\bibinfo {author} {\bibfnamefont {D.}~\bibnamefont
  {Bruß}},\ }\bibfield  {title} {\enquote {\bibinfo {title} {{Characterizing
  Entanglement}},}\ }\href {http://dx.doi.org/10.1063/1.1494474} {\bibfield
  {journal} {\bibinfo  {journal} {Journal of Mathematical Physics}\ }\textbf
  {\bibinfo {volume} {43}},\ \bibinfo {pages} {4237} (\bibinfo {year}
  {2002})}\BibitemShut {NoStop}%
\bibitem [{\citenamefont {Knill}\ and\ \citenamefont
  {Laflamme}(1998)}]{DQC1Original}%
  \BibitemOpen
  \bibfield  {author} {\bibinfo {author} {\bibfnamefont {E.}~\bibnamefont
  {Knill}}\ and\ \bibinfo {author} {\bibfnamefont {R.}~\bibnamefont
  {Laflamme}},\ }\bibfield  {title} {\enquote {\bibinfo {title} {{Power of One
  Bit of Quantum Information}},}\ }\href
  {http://link.aps.org/doi/10.1103/PhysRevLett.81.5672} {\bibfield  {journal}
  {\bibinfo  {journal} {Physical Review Letters}\ }\textbf {\bibinfo {volume}
  {81}},\ \bibinfo {pages} {5672} (\bibinfo {year} {1998})}\BibitemShut
  {NoStop}%
\bibitem [{\citenamefont {Ollivier}\ and\ \citenamefont
  {Zurek}(2001)}]{Discord2001}%
  \BibitemOpen
  \bibfield  {author} {\bibinfo {author} {\bibfnamefont {H.}~\bibnamefont
  {Ollivier}}\ and\ \bibinfo {author} {\bibfnamefont {W.~H.}\ \bibnamefont
  {Zurek}},\ }\bibfield  {title} {\enquote {\bibinfo {title} {{Quantum Discord:
  A Measure of the Quantumness of Correlations}},}\ }\href
  {http://link.aps.org/doi/10.1103/PhysRevLett.88.017901} {\bibfield  {journal}
  {\bibinfo  {journal} {Physical Review Letters}\ }\textbf {\bibinfo {volume}
  {88}},\ \bibinfo {pages} {017901} (\bibinfo {year} {2001})}\BibitemShut
  {NoStop}%
\bibitem [{\citenamefont {Datta}\ \emph {et~al.}(2008)\citenamefont {Datta},
  \citenamefont {Shaji},\ and\ \citenamefont {Caves}}]{DQC1}%
  \BibitemOpen
  \bibfield  {author} {\bibinfo {author} {\bibfnamefont {A.}~\bibnamefont
  {Datta}}, \bibinfo {author} {\bibfnamefont {A.}~\bibnamefont {Shaji}}, \ and\
  \bibinfo {author} {\bibfnamefont {C.~M.}\ \bibnamefont {Caves}},\ }\bibfield
  {title} {\enquote {\bibinfo {title} {{Quantum Discord and the Power of One
  Qubit}},}\ }\href {http://link.aps.org/doi/10.1103/PhysRevLett.100.050502}
  {\bibfield  {journal} {\bibinfo  {journal} {Physical Review Letters}\
  }\textbf {\bibinfo {volume} {100}},\ \bibinfo {pages} {050502} (\bibinfo
  {year} {2008})}\BibitemShut {NoStop}%
\bibitem [{\citenamefont {Dakić}\ \emph {et~al.}(2010)\citenamefont {Dakić},
  \citenamefont {Vedral},\ and\ \citenamefont {Brukner}}]{Discord2010}%
  \BibitemOpen
  \bibfield  {author} {\bibinfo {author} {\bibfnamefont {B.}~\bibnamefont
  {Dakić}}, \bibinfo {author} {\bibfnamefont {V.}~\bibnamefont {Vedral}}, \
  and\ \bibinfo {author} {\bibfnamefont {{\v{C}}.}~\bibnamefont {Brukner}},\
  }\bibfield  {title} {\enquote {\bibinfo {title} {{Necessary and Sufficient
  Condition for Nonzero Quantum Discord}},}\ }\href
  {http://link.aps.org/doi/10.1103/PhysRevLett.105.190502} {\bibfield
  {journal} {\bibinfo  {journal} {Physical Review Letters}\ }\textbf {\bibinfo
  {volume} {105}},\ \bibinfo {pages} {190502} (\bibinfo {year}
  {2010})}\BibitemShut {NoStop}%
\bibitem [{\citenamefont {Cavalcanti}\ \emph {et~al.}(2011)\citenamefont
  {Cavalcanti}, \citenamefont {Aolita}, \citenamefont {Boixo}, \citenamefont
  {Modi}, \citenamefont {Piani},\ and\ \citenamefont {Winter}}]{Discord2011}%
  \BibitemOpen
  \bibfield  {author} {\bibinfo {author} {\bibfnamefont {D.}~\bibnamefont
  {Cavalcanti}}, \bibinfo {author} {\bibfnamefont {L.}~\bibnamefont {Aolita}},
  \bibinfo {author} {\bibfnamefont {S.}~\bibnamefont {Boixo}}, \bibinfo
  {author} {\bibfnamefont {K.}~\bibnamefont {Modi}}, \bibinfo {author}
  {\bibfnamefont {M.}~\bibnamefont {Piani}}, \ and\ \bibinfo {author}
  {\bibfnamefont {A.}~\bibnamefont {Winter}},\ }\bibfield  {title} {\enquote
  {\bibinfo {title} {{Operational Interpretations of Quantum Discord}},}\
  }\href {http://link.aps.org/doi/10.1103/PhysRevA.83.032324} {\bibfield
  {journal} {\bibinfo  {journal} {Physical Review A}\ }\textbf {\bibinfo
  {volume} {83}},\ \bibinfo {pages} {032324} (\bibinfo {year}
  {2011})}\BibitemShut {NoStop}%
\bibitem [{\citenamefont {Ferrie}\ \emph {et~al.}(2010)\citenamefont {Ferrie},
  \citenamefont {Morris},\ and\ \citenamefont
  {Emerson}}]{NegativityPostDiscord}%
  \BibitemOpen
  \bibfield  {author} {\bibinfo {author} {\bibfnamefont {C.}~\bibnamefont
  {Ferrie}}, \bibinfo {author} {\bibfnamefont {R.}~\bibnamefont {Morris}}, \
  and\ \bibinfo {author} {\bibfnamefont {J.}~\bibnamefont {Emerson}},\
  }\bibfield  {title} {\enquote {\bibinfo {title} {{Necessity of Negativity in
  Quantum Theory}},}\ }\href
  {http://link.aps.org/doi/10.1103/PhysRevA.82.044103} {\bibfield  {journal}
  {\bibinfo  {journal} {Physical Review A}\ }\textbf {\bibinfo {volume} {82}},\
  \bibinfo {pages} {044103} (\bibinfo {year} {2010})}\BibitemShut {NoStop}%
\bibitem [{\citenamefont {Van~den Nest}(2013)}]{MinEntangComputing}%
  \BibitemOpen
  \bibfield  {author} {\bibinfo {author} {\bibfnamefont {M.}~\bibnamefont
  {Van~den Nest}},\ }\bibfield  {title} {\enquote {\bibinfo {title} {{Universal
  Quantum Computation with Little Entanglement}},}\ }\href
  {http://link.aps.org/doi/10.1103/PhysRevLett.110.060504} {\bibfield
  {journal} {\bibinfo  {journal} {Physical Review Letters}\ }\textbf {\bibinfo
  {volume} {110}},\ \bibinfo {pages} {060504} (\bibinfo {year}
  {2013})}\BibitemShut {NoStop}%
\bibitem [{\citenamefont {Bennett}\ \emph
  {et~al.}(1999{\natexlab{b}})\citenamefont {Bennett}, \citenamefont
  {DiVincenzo}, \citenamefont {Mor}, \citenamefont {Shor}, \citenamefont
  {Smolin},\ and\ \citenamefont {Terhal}}]{UPBOriginal}%
  \BibitemOpen
  \bibfield  {author} {\bibinfo {author} {\bibfnamefont {C.~H.}\ \bibnamefont
  {Bennett}}, \bibinfo {author} {\bibfnamefont {D.~P.}\ \bibnamefont
  {DiVincenzo}}, \bibinfo {author} {\bibfnamefont {T.}~\bibnamefont {Mor}},
  \bibinfo {author} {\bibfnamefont {P.~W.}\ \bibnamefont {Shor}}, \bibinfo
  {author} {\bibfnamefont {J.~A.}\ \bibnamefont {Smolin}}, \ and\ \bibinfo
  {author} {\bibfnamefont {B.~M.}\ \bibnamefont {Terhal}},\ }\bibfield  {title}
  {\enquote {\bibinfo {title} {{Unextendible Product Bases and Bound
  Entanglement}},}\ }\href
  {http://link.aps.org/doi/10.1103/PhysRevLett.82.5385} {\bibfield  {journal}
  {\bibinfo  {journal} {Physical Review Letters}\ }\textbf {\bibinfo {volume}
  {82}},\ \bibinfo {pages} {5385} (\bibinfo {year}
  {1999}{\natexlab{b}})}\BibitemShut {NoStop}%
\bibitem [{\citenamefont {{Ac{\'{\i}}n}}\ \emph {et~al.}(2005)\citenamefont
  {{Ac{\'{\i}}n}}, \citenamefont {Gill},\ and\ \citenamefont
  {Gisin}}]{EntanglementSurprise2005}%
  \BibitemOpen
  \bibfield  {author} {\bibinfo {author} {\bibfnamefont {A.}~\bibnamefont
  {{Ac{\'{\i}}n}}}, \bibinfo {author} {\bibfnamefont {R.}~\bibnamefont {Gill}},
  \ and\ \bibinfo {author} {\bibfnamefont {N.}~\bibnamefont {Gisin}},\
  }\bibfield  {title} {\enquote {\bibinfo {title} {{Optimal Bell Tests Do Not
  Require Maximally Entangled States}},}\ }\href
  {http://link.aps.org/doi/10.1103/PhysRevLett.95.210402} {\bibfield  {journal}
  {\bibinfo  {journal} {Physical Review Letters}\ }\textbf {\bibinfo {volume}
  {95}},\ \bibinfo {pages} {210402} (\bibinfo {year} {2005})}\BibitemShut
  {NoStop}%
\bibitem [{\citenamefont {Brunner}\ \emph {et~al.}(2005)\citenamefont
  {Brunner}, \citenamefont {Gisin},\ and\ \citenamefont
  {Scarani}}]{EntanglementSurprise2005b}%
  \BibitemOpen
  \bibfield  {author} {\bibinfo {author} {\bibfnamefont {N.}~\bibnamefont
  {Brunner}}, \bibinfo {author} {\bibfnamefont {N.}~\bibnamefont {Gisin}}, \
  and\ \bibinfo {author} {\bibfnamefont {V.}~\bibnamefont {Scarani}},\
  }\bibfield  {title} {\enquote {\bibinfo {title} {{Entanglement and
  Non-Locality are Different Resources}},}\ }\href
  {http://stacks.iop.org/1367-2630/7/i=1/a=088} {\bibfield  {journal} {\bibinfo
   {journal} {New Journal of Physics}\ }\textbf {\bibinfo {volume} {7}},\
  \bibinfo {pages} {88} (\bibinfo {year} {2005})}\BibitemShut {NoStop}%
\bibitem [{\citenamefont {M{\'e}thot}\ and\ \citenamefont
  {Scarani}(2007)}]{EntanglementSurprise2007}%
  \BibitemOpen
  \bibfield  {author} {\bibinfo {author} {\bibfnamefont {A.~A.}\ \bibnamefont
  {M{\'e}thot}}\ and\ \bibinfo {author} {\bibfnamefont {V.}~\bibnamefont
  {Scarani}},\ }\bibfield  {title} {\enquote {\bibinfo {title} {{An Anomaly of
  Non-Locality}},}\ }\href {http://arxiv.org/abs/quant-ph/0601210} {\bibfield
  {journal} {\bibinfo  {journal} {Quantum Information {\&} Computation}\
  }\textbf {\bibinfo {volume} {7}},\ \bibinfo {pages} {157} (\bibinfo {year}
  {2007})}\BibitemShut {NoStop}%
\bibitem [{\citenamefont {Vidick}\ and\ \citenamefont
  {Wehner}(2011)}]{EntanglementSurprise2011}%
  \BibitemOpen
  \bibfield  {author} {\bibinfo {author} {\bibfnamefont {T.}~\bibnamefont
  {Vidick}}\ and\ \bibinfo {author} {\bibfnamefont {S.}~\bibnamefont
  {Wehner}},\ }\bibfield  {title} {\enquote {\bibinfo {title} {{More
  Nonlocality with Less Entanglement}},}\ }\href
  {http://link.aps.org/doi/10.1103/PhysRevA.83.052310} {\bibfield  {journal}
  {\bibinfo  {journal} {Physical Review A}\ }\textbf {\bibinfo {volume} {83}},\
  \bibinfo {pages} {052310} (\bibinfo {year} {2011})}\BibitemShut {NoStop}%
\bibitem [{\citenamefont {Vallone}\ \emph {et~al.}(2014)\citenamefont
  {Vallone}, \citenamefont {Lima}, \citenamefont {G\'omez}, \citenamefont
  {Ca{\~n}as}, \citenamefont {Larsson}, \citenamefont {Mataloni},\ and\
  \citenamefont {Cabello}}]{EntanglementSurprise2014}%
  \BibitemOpen
  \bibfield  {author} {\bibinfo {author} {\bibfnamefont {G.}~\bibnamefont
  {Vallone}}, \bibinfo {author} {\bibfnamefont {G.}~\bibnamefont {Lima}},
  \bibinfo {author} {\bibfnamefont {E.~S.}\ \bibnamefont {G\'omez}}, \bibinfo
  {author} {\bibfnamefont {G.}~\bibnamefont {Ca{\~n}as}}, \bibinfo {author}
  {\bibfnamefont {J.-{\AA}.}\ \bibnamefont {Larsson}}, \bibinfo {author}
  {\bibfnamefont {P.}~\bibnamefont {Mataloni}}, \ and\ \bibinfo {author}
  {\bibfnamefont {A.}~\bibnamefont {Cabello}},\ }\bibfield  {title} {\enquote
  {\bibinfo {title} {{Bell Scenarios in which Nonlocality and Entanglement are
  Inversely Related}},}\ }\href
  {http://link.aps.org/doi/10.1103/PhysRevA.89.012102} {\bibfield  {journal}
  {\bibinfo  {journal} {Physical Review A}\ }\textbf {\bibinfo {volume} {89}},\
  \bibinfo {pages} {012102} (\bibinfo {year} {2014})}\BibitemShut {NoStop}%
\bibitem [{\citenamefont {Acín}\ \emph {et~al.}(2012)\citenamefont {Acín},
  \citenamefont {Massar},\ and\ \citenamefont
  {Pironio}}]{RandomnessFromNonlocality}%
  \BibitemOpen
  \bibfield  {author} {\bibinfo {author} {\bibfnamefont {A.}~\bibnamefont
  {Acín}}, \bibinfo {author} {\bibfnamefont {S.}~\bibnamefont {Massar}}, \
  and\ \bibinfo {author} {\bibfnamefont {S.}~\bibnamefont {Pironio}},\
  }\bibfield  {title} {\enquote {\bibinfo {title} {{Randomness versus
  Nonlocality and Entanglement}},}\ }\href
  {http://link.aps.org/doi/10.1103/PhysRevLett.108.100402} {\bibfield
  {journal} {\bibinfo  {journal} {Physical Review Letters}\ }\textbf {\bibinfo
  {volume} {108}},\ \bibinfo {pages} {100402} (\bibinfo {year}
  {2012})}\BibitemShut {NoStop}%
\bibitem [{\citenamefont {{T{\'u}lio Quintino}}\ \emph
  {et~al.}(2014)\citenamefont {{T{\'u}lio Quintino}}, \citenamefont
  {{V{\'e}rtesi}},\ and\ \citenamefont
  {{Brunner}}}]{SteeringNonlocalityBrunner}%
  \BibitemOpen
  \bibfield  {author} {\bibinfo {author} {\bibfnamefont {M.}~\bibnamefont
  {{T{\'u}lio Quintino}}}, \bibinfo {author} {\bibfnamefont {T.}~\bibnamefont
  {{V{\'e}rtesi}}}, \ and\ \bibinfo {author} {\bibfnamefont {N.}~\bibnamefont
  {{Brunner}}},\ }\bibfield  {title} {\enquote {\bibinfo {title} {{Joint
  Measurability, EPR Steering, and Bell Nonlocality}},}\ }\href
  {http://arxiv.org/abs/1406.6976} {\bibfield  {journal} {\bibinfo  {journal}
  {arXiv:1406.6976}\ } (\bibinfo {year} {2014})}\BibitemShut {NoStop}%
\bibitem [{\citenamefont {{Uola}}\ \emph {et~al.}(2014)\citenamefont {{Uola}},
  \citenamefont {{Moroder}},\ and\ \citenamefont
  {{G{\"u}hne}}}]{SteeringNonlocalityArXiv}%
  \BibitemOpen
  \bibfield  {author} {\bibinfo {author} {\bibfnamefont {R.}~\bibnamefont
  {{Uola}}}, \bibinfo {author} {\bibfnamefont {T.}~\bibnamefont {{Moroder}}}, \
  and\ \bibinfo {author} {\bibfnamefont {O.}~\bibnamefont {{G{\"u}hne}}},\
  }\bibfield  {title} {\enquote {\bibinfo {title} {{Joint Measurability of
  Generalized Measurements Implies Classicality}},}\ }\href
  {http://arxiv.org/abs/1407.2224} {\bibfield  {journal} {\bibinfo  {journal}
  {arXiv:1407.2224}\ } (\bibinfo {year} {2014})}\BibitemShut {NoStop}%
\bibitem [{\citenamefont {{Augusiak}}\ \emph {et~al.}(2014)\citenamefont
  {{Augusiak}}, \citenamefont {{Demianowicz}}, \citenamefont {{Tura}},\ and\
  \citenamefont {{Ac{\'{\i}}n}}}]{EntanglementInequivalent2014}%
  \BibitemOpen
  \bibfield  {author} {\bibinfo {author} {\bibfnamefont {R.}~\bibnamefont
  {{Augusiak}}}, \bibinfo {author} {\bibfnamefont {M.}~\bibnamefont
  {{Demianowicz}}}, \bibinfo {author} {\bibfnamefont {J.}~\bibnamefont
  {{Tura}}}, \ and\ \bibinfo {author} {\bibfnamefont {A.}~\bibnamefont
  {{Ac{\'{\i}}n}}},\ }\bibfield  {title} {\enquote {\bibinfo {title}
  {{Entanglement and Nonlocality are Inequivalent for Any Number of
  Particles}},}\ }\href {http://arxiv.org/abs/1407.3114} {\bibfield  {journal}
  {\bibinfo  {journal} {arXiv:1407.3114}\ } (\bibinfo {year}
  {2014})}\BibitemShut {NoStop}%
\bibitem [{\citenamefont {Gisin}(1991)}]{Gisin1991}%
  \BibitemOpen
  \bibfield  {author} {\bibinfo {author} {\bibfnamefont {N.}~\bibnamefont
  {Gisin}},\ }\bibfield  {title} {\enquote {\bibinfo {title} {{Bell's
  Inequality Holds for All Non-Product States}},}\ }\href
  {http://www.sciencedirect.com/science/article/pii/037596019190805I}
  {\bibfield  {journal} {\bibinfo  {journal} {Physics Letters A}\ }\textbf
  {\bibinfo {volume} {154}},\ \bibinfo {pages} {201 } (\bibinfo {year}
  {1991})}\BibitemShut {NoStop}%
\bibitem [{\citenamefont {Yu}\ \emph {et~al.}(2012)\citenamefont {Yu},
  \citenamefont {Chen}, \citenamefont {Zhang}, \citenamefont {Lai},\ and\
  \citenamefont {Oh}}]{GisinPure2012}%
  \BibitemOpen
  \bibfield  {author} {\bibinfo {author} {\bibfnamefont {S.}~\bibnamefont
  {Yu}}, \bibinfo {author} {\bibfnamefont {Q.}~\bibnamefont {Chen}}, \bibinfo
  {author} {\bibfnamefont {C.}~\bibnamefont {Zhang}}, \bibinfo {author}
  {\bibfnamefont {C.~H.}\ \bibnamefont {Lai}}, \ and\ \bibinfo {author}
  {\bibfnamefont {C.~H.}\ \bibnamefont {Oh}},\ }\bibfield  {title} {\enquote
  {\bibinfo {title} {{All Entangled Pure States Violate a Single Bell's
  Inequality}},}\ }\href
  {http://link.aps.org/doi/10.1103/PhysRevLett.109.120402} {\bibfield
  {journal} {\bibinfo  {journal} {Physical Review Letters}\ }\textbf {\bibinfo
  {volume} {109}},\ \bibinfo {pages} {120402} (\bibinfo {year}
  {2012})}\BibitemShut {NoStop}%
\bibitem [{\citenamefont {{Chen}}\ \emph {et~al.}(2014)\citenamefont {{Chen}},
  \citenamefont {{Su}}, \citenamefont {{Xu}}, \citenamefont {{Wu}},
  \citenamefont {{Wu}},\ and\ \citenamefont {{Kwek}}}]{EnhancedGisinSteering}%
  \BibitemOpen
  \bibfield  {author} {\bibinfo {author} {\bibfnamefont {J.-L.}\ \bibnamefont
  {{Chen}}}, \bibinfo {author} {\bibfnamefont {H.-Y.}\ \bibnamefont {{Su}}},
  \bibinfo {author} {\bibfnamefont {Z.-P.}\ \bibnamefont {{Xu}}}, \bibinfo
  {author} {\bibfnamefont {Y.-C.}\ \bibnamefont {{Wu}}}, \bibinfo {author}
  {\bibfnamefont {C.}~\bibnamefont {{Wu}}}, \ and\ \bibinfo {author}
  {\bibfnamefont {L.~C.}\ \bibnamefont {{Kwek}}},\ }\bibfield  {title}
  {\enquote {\bibinfo {title} {{Enhanced Gisin's Theorem: Bridging Quantum
  Entanglement and Bell Nonlocality by Einstein-Podolsky-Rosen Steering}},}\
  }\href {http://arxiv.org/abs/1404.2675} {\bibfield  {journal} {\bibinfo
  {journal} {arXiv:1404.2675}\ } (\bibinfo {year} {2014})}\BibitemShut
  {NoStop}%
\bibitem [{\citenamefont {Shimony}\ \emph {et~al.}(2009)\citenamefont
  {Shimony}, \citenamefont {Myrvold},\ and\ \citenamefont
  {Christian}}]{shimony2009quantum}%
  \BibitemOpen
  \bibfield  {author} {\bibinfo {author} {\bibfnamefont {A.}~\bibnamefont
  {Shimony}}, \bibinfo {author} {\bibfnamefont {W.}~\bibnamefont {Myrvold}}, \
  and\ \bibinfo {author} {\bibfnamefont {J.}~\bibnamefont {Christian}},\ }\href
  {http://books.google.com/books?id=TaM1Nv09hJAC} {\emph {\bibinfo {title}
  {{Quantum Reality, Relativistic Causality, and Closing the Epistemic Circle:
  Essays in Honour of Abner Shimony}}}},\ The Western Ontario Series in
  Philosophy of Science\ (\bibinfo  {publisher} {Springer},\ \bibinfo {year}
  {2009})\BibitemShut {NoStop}%
\bibitem [{\citenamefont {Ac{\'i}n}(2001)}]{acin2001bound}%
  \BibitemOpen
  \bibfield  {author} {\bibinfo {author} {\bibfnamefont {A.}~\bibnamefont
  {Ac{\'i}n}},\ }\bibfield  {title} {\enquote {\bibinfo {title}
  {{Distillability, Bell Inequalities, and Multiparticle Bound
  Entanglement}},}\ }\href {http://dx.doi.org/10.1103/physrevlett.88.027901}
  {\bibfield  {journal} {\bibinfo  {journal} {Physical Review Letters}\
  }\textbf {\bibinfo {volume} {88}},\ \bibinfo {pages} {027901} (\bibinfo
  {year} {2001})}\BibitemShut {NoStop}%
\bibitem [{\citenamefont {Jevtic}\ \emph {et~al.}(2014)\citenamefont {Jevtic},
  \citenamefont {Pusey}, \citenamefont {Jennings},\ and\ \citenamefont
  {Rudolph}}]{SteeringEllipsoids}%
  \BibitemOpen
  \bibfield  {author} {\bibinfo {author} {\bibfnamefont {S.}~\bibnamefont
  {Jevtic}}, \bibinfo {author} {\bibfnamefont {M.}~\bibnamefont {Pusey}},
  \bibinfo {author} {\bibfnamefont {D.}~\bibnamefont {Jennings}}, \ and\
  \bibinfo {author} {\bibfnamefont {T.}~\bibnamefont {Rudolph}},\ }\bibfield
  {title} {\enquote {\bibinfo {title} {{Quantum Steering Ellipsoids}},}\ }\href
  {http://link.aps.org/doi/10.1103/PhysRevLett.113.020402} {\bibfield
  {journal} {\bibinfo  {journal} {Physical Review Letters}\ }\textbf {\bibinfo
  {volume} {113}},\ \bibinfo {pages} {020402} (\bibinfo {year}
  {2014})}\BibitemShut {NoStop}%
\bibitem [{\citenamefont {Kwiat}\ \emph {et~al.}(1995)\citenamefont {Kwiat},
  \citenamefont {Mattle}, \citenamefont {Weinfurter}, \citenamefont
  {Zeilinger}, \citenamefont {Sergienko},\ and\ \citenamefont
  {Shih}}]{DownConversion}%
  \BibitemOpen
  \bibfield  {author} {\bibinfo {author} {\bibfnamefont {P.~G.}\ \bibnamefont
  {Kwiat}}, \bibinfo {author} {\bibfnamefont {K.}~\bibnamefont {Mattle}},
  \bibinfo {author} {\bibfnamefont {H.}~\bibnamefont {Weinfurter}}, \bibinfo
  {author} {\bibfnamefont {A.}~\bibnamefont {Zeilinger}}, \bibinfo {author}
  {\bibfnamefont {A.~V.}\ \bibnamefont {Sergienko}}, \ and\ \bibinfo {author}
  {\bibfnamefont {Y.}~\bibnamefont {Shih}},\ }\bibfield  {title} {\enquote
  {\bibinfo {title} {{New High-Intensity Source of Polarization-Entangled
  Photon Pairs}},}\ }\href
  {http://link.aps.org/doi/10.1103/PhysRevLett.75.4337} {\bibfield  {journal}
  {\bibinfo  {journal} {Physical Review Letters}\ }\textbf {\bibinfo {volume}
  {75}},\ \bibinfo {pages} {4337} (\bibinfo {year} {1995})}\BibitemShut
  {NoStop}%
\bibitem [{\citenamefont {Cohen}\ and\ \citenamefont
  {Sham}(2013)}]{ResonantFluorescence}%
  \BibitemOpen
  \bibfield  {author} {\bibinfo {author} {\bibfnamefont {G.~Z.}\ \bibnamefont
  {Cohen}}\ and\ \bibinfo {author} {\bibfnamefont {L.~J.}\ \bibnamefont
  {Sham}},\ }\bibfield  {title} {\enquote {\bibinfo {title} {{Rapid Creation of
  Distant Entanglement by Multiphoton Resonant Fluorescence}},}\ }\href
  {http://link.aps.org/doi/10.1103/PhysRevB.88.245306} {\bibfield  {journal}
  {\bibinfo  {journal} {Physical Review B}\ }\textbf {\bibinfo {volume} {88}},\
  \bibinfo {pages} {245306} (\bibinfo {year} {2013})}\BibitemShut {NoStop}%
\bibitem [{\citenamefont {Vanderbruggen}\ \emph {et~al.}(2011)\citenamefont
  {Vanderbruggen}, \citenamefont {Bernon}, \citenamefont {Bertoldi},
  \citenamefont {Landragin},\ and\ \citenamefont {Bouyer}}]{DickePrepration}%
  \BibitemOpen
  \bibfield  {author} {\bibinfo {author} {\bibfnamefont {T.}~\bibnamefont
  {Vanderbruggen}}, \bibinfo {author} {\bibfnamefont {S.}~\bibnamefont
  {Bernon}}, \bibinfo {author} {\bibfnamefont {A.}~\bibnamefont {Bertoldi}},
  \bibinfo {author} {\bibfnamefont {A.}~\bibnamefont {Landragin}}, \ and\
  \bibinfo {author} {\bibfnamefont {P.}~\bibnamefont {Bouyer}},\ }\bibfield
  {title} {\enquote {\bibinfo {title} {{Spin-Squeezing and Dicke-State
  Preparation by Heterodyne Measurement}},}\ }\href
  {http://link.aps.org/doi/10.1103/PhysRevA.83.013821} {\bibfield  {journal}
  {\bibinfo  {journal} {Physical Review A}\ }\textbf {\bibinfo {volume} {83}},\
  \bibinfo {pages} {013821} (\bibinfo {year} {2011})}\BibitemShut {NoStop}%
\bibitem [{\citenamefont {Morrison}\ and\ \citenamefont
  {Parkins}(2008)}]{SpinEntanglementMorrison}%
  \BibitemOpen
  \bibfield  {author} {\bibinfo {author} {\bibfnamefont {S.}~\bibnamefont
  {Morrison}}\ and\ \bibinfo {author} {\bibfnamefont {A.~S.}\ \bibnamefont
  {Parkins}},\ }\bibfield  {title} {\enquote {\bibinfo {title}
  {{Dissipation-Driven Quantum Phase Transitions in Collective Spin
  Systems}},}\ }\href {http://stacks.iop.org/0953-4075/41/i=19/a=195502}
  {\bibfield  {journal} {\bibinfo  {journal} {Journal of Physics B: Atomic,
  Molecular and Optical Physics}\ }\textbf {\bibinfo {volume} {41}},\ \bibinfo
  {pages} {195502} (\bibinfo {year} {2008})}\BibitemShut {NoStop}%
\bibitem [{\citenamefont {L\"ucke}\ \emph {et~al.}(2014)\citenamefont
  {L\"ucke}, \citenamefont {Peise}, \citenamefont {Vitagliano}, \citenamefont
  {Arlt}, \citenamefont {Santos}, \citenamefont {T\'oth},\ and\ \citenamefont
  {Klempt}}]{TothDickePRL}%
  \BibitemOpen
  \bibfield  {author} {\bibinfo {author} {\bibfnamefont {B.}~\bibnamefont
  {L\"ucke}}, \bibinfo {author} {\bibfnamefont {J.}~\bibnamefont {Peise}},
  \bibinfo {author} {\bibfnamefont {G.}~\bibnamefont {Vitagliano}}, \bibinfo
  {author} {\bibfnamefont {J.}~\bibnamefont {Arlt}}, \bibinfo {author}
  {\bibfnamefont {L.}~\bibnamefont {Santos}}, \bibinfo {author} {\bibfnamefont
  {G.}~\bibnamefont {T\'oth}}, \ and\ \bibinfo {author} {\bibfnamefont
  {C.}~\bibnamefont {Klempt}},\ }\bibfield  {title} {\enquote {\bibinfo {title}
  {{Detecting Multiparticle Entanglement of Dicke States}},}\ }\href
  {http://link.aps.org/doi/10.1103/PhysRevLett.112.155304} {\bibfield
  {journal} {\bibinfo  {journal} {Physical Review Letters}\ }\textbf {\bibinfo
  {volume} {112}},\ \bibinfo {pages} {155304} (\bibinfo {year}
  {2014})}\BibitemShut {NoStop}%
\bibitem [{\citenamefont {Chiuri}\ \emph {et~al.}(2012)\citenamefont {Chiuri},
  \citenamefont {Greganti}, \citenamefont {Paternostro}, \citenamefont
  {Vallone},\ and\ \citenamefont {Mataloni}}]{DickeNetworking}%
  \BibitemOpen
  \bibfield  {author} {\bibinfo {author} {\bibfnamefont {A.}~\bibnamefont
  {Chiuri}}, \bibinfo {author} {\bibfnamefont {C.}~\bibnamefont {Greganti}},
  \bibinfo {author} {\bibfnamefont {M.}~\bibnamefont {Paternostro}}, \bibinfo
  {author} {\bibfnamefont {G.}~\bibnamefont {Vallone}}, \ and\ \bibinfo
  {author} {\bibfnamefont {P.}~\bibnamefont {Mataloni}},\ }\bibfield  {title}
  {\enquote {\bibinfo {title} {{Experimental Quantum Networking Protocols via
  Four-Qubit Hyperentangled Dicke States}},}\ }\href
  {http://link.aps.org/doi/10.1103/PhysRevLett.109.173604} {\bibfield
  {journal} {\bibinfo  {journal} {Physical Review Letters}\ }\textbf {\bibinfo
  {volume} {109}},\ \bibinfo {pages} {173604} (\bibinfo {year}
  {2012})}\BibitemShut {NoStop}%
\bibitem [{\citenamefont {Cleve}\ \emph {et~al.}(1998)\citenamefont {Cleve},
  \citenamefont {Ekert}, \citenamefont {Macchiavello},\ and\ \citenamefont
  {Mosca}}]{mosca_algorithms_revisited}%
  \BibitemOpen
  \bibfield  {author} {\bibinfo {author} {\bibfnamefont {R.}~\bibnamefont
  {Cleve}}, \bibinfo {author} {\bibfnamefont {A.}~\bibnamefont {Ekert}},
  \bibinfo {author} {\bibfnamefont {C.}~\bibnamefont {Macchiavello}}, \ and\
  \bibinfo {author} {\bibfnamefont {M.}~\bibnamefont {Mosca}},\ }\bibfield
  {title} {\enquote {\bibinfo {title} {{Quantum Algorithms Revisited}},}\
  }\href {http://dx.doi.org/10.1098/rspa.1998.0164} {\bibfield  {journal}
  {\bibinfo  {journal} {Proceedings of the Royal Society of London. Series A:
  Mathematical, Physical and Engineering Sciences}\ }\textbf {\bibinfo {volume}
  {454}},\ \bibinfo {pages} {339} (\bibinfo {year} {1998})}\BibitemShut
  {NoStop}%
\bibitem [{\citenamefont {Van~den Nest}\ \emph {et~al.}(2007)\citenamefont
  {Van~den Nest}, \citenamefont {Dür}, \citenamefont {Miyake},\ and\
  \citenamefont {Briegel}}]{MBQC}%
  \BibitemOpen
  \bibfield  {author} {\bibinfo {author} {\bibfnamefont {M.}~\bibnamefont
  {Van~den Nest}}, \bibinfo {author} {\bibfnamefont {W.}~\bibnamefont {Dür}},
  \bibinfo {author} {\bibfnamefont {A.}~\bibnamefont {Miyake}}, \ and\ \bibinfo
  {author} {\bibfnamefont {H.~J.}\ \bibnamefont {Briegel}},\ }\bibfield
  {title} {\enquote {\bibinfo {title} {{Fundamentals of Universality in One-Way
  Quantum Computation}},}\ }\href {http://stacks.iop.org/1367-2630/9/i=6/a=204}
  {\bibfield  {journal} {\bibinfo  {journal} {New Journal of Physics}\ }\textbf
  {\bibinfo {volume} {9}},\ \bibinfo {pages} {204} (\bibinfo {year}
  {2007})}\BibitemShut {NoStop}%
\bibitem [{\citenamefont {Van~den Nest}\ \emph {et~al.}(2006)\citenamefont
  {Van~den Nest}, \citenamefont {Miyake}, \citenamefont {D\"ur},\ and\
  \citenamefont {Briegel}}]{MBQC2}%
  \BibitemOpen
  \bibfield  {author} {\bibinfo {author} {\bibfnamefont {M.}~\bibnamefont
  {Van~den Nest}}, \bibinfo {author} {\bibfnamefont {A.}~\bibnamefont
  {Miyake}}, \bibinfo {author} {\bibfnamefont {W.}~\bibnamefont {D\"ur}}, \
  and\ \bibinfo {author} {\bibfnamefont {H.~J.}\ \bibnamefont {Briegel}},\
  }\bibfield  {title} {\enquote {\bibinfo {title} {{Universal Resources for
  Measurement-Based Quantum Computation}},}\ }\href
  {http://link.aps.org/doi/10.1103/PhysRevLett.97.150504} {\bibfield  {journal}
  {\bibinfo  {journal} {Physical Review Letters}\ }\textbf {\bibinfo {volume}
  {97}},\ \bibinfo {pages} {150504} (\bibinfo {year} {2006})}\BibitemShut
  {NoStop}%
\bibitem [{\citenamefont {Kok}\ and\ \citenamefont
  {Lovett}(2010)}]{QuantOptBook}%
  \BibitemOpen
  \bibfield  {author} {\bibinfo {author} {\bibfnamefont {P.}~\bibnamefont
  {Kok}}\ and\ \bibinfo {author} {\bibfnamefont {B.}~\bibnamefont {Lovett}},\
  }\href {http://books.google.com/books?id=G2zKNooOeKcC} {\emph {\bibinfo
  {title} {Introduction to Optical Quantum Information Processing}}}\ (\bibinfo
   {publisher} {Cambridge University Press},\ \bibinfo {year}
  {2010})\BibitemShut {NoStop}%
\bibitem [{\citenamefont {Schumacher}\ and\ \citenamefont
  {Westmoreland}(2010)}]{Schumacher2010Quantum}%
  \BibitemOpen
  \bibfield  {author} {\bibinfo {author} {\bibfnamefont {B.}~\bibnamefont
  {Schumacher}}\ and\ \bibinfo {author} {\bibfnamefont {M.}~\bibnamefont
  {Westmoreland}},\ }\href
  {http://www.amazon.com/exec/obidos/redirect?tag=citeulike07-20\&path=ASIN/052187534X}
  {\emph {\bibinfo {title} {{Quantum Processes Systems, and Information}}}}\
  (\bibinfo  {publisher} {Cambridge University Press},\ \bibinfo {year}
  {2010})\BibitemShut {NoStop}%
\bibitem [{\citenamefont {Nielsen}\ and\ \citenamefont
  {Chuang}(2011)}]{Nielsen2011Quantum}%
  \BibitemOpen
  \bibfield  {author} {\bibinfo {author} {\bibfnamefont {M.~A.}\ \bibnamefont
  {Nielsen}}\ and\ \bibinfo {author} {\bibfnamefont {I.~L.}\ \bibnamefont
  {Chuang}},\ }\href
  {http://www.amazon.com/exec/obidos/redirect?tag=citeulike07-20\&path=ASIN/1107002176}
  {\emph {\bibinfo {title} {{Quantum Computation and Quantum Information: 10th
  Anniversary Edition}}}},\ \bibinfo {edition} {10th}\ ed.\ (\bibinfo
  {publisher} {Cambridge University Press},\ \bibinfo {year}
  {2011})\BibitemShut {NoStop}%
\bibitem [{\citenamefont {Wilde}(2013)}]{Wilde2013Quantum}%
  \BibitemOpen
  \bibfield  {author} {\bibinfo {author} {\bibfnamefont {M.~M.}\ \bibnamefont
  {Wilde}},\ }\href
  {http://www.amazon.com/exec/obidos/redirect?tag=citeulike07-20\&path=ASIN/1107034256}
  {\emph {\bibinfo {title} {{Quantum Information Theory}}}},\ \bibinfo
  {edition} {1st}\ ed.\ (\bibinfo  {publisher} {Cambridge University Press},\
  \bibinfo {year} {2013})\BibitemShut {NoStop}%
\bibitem [{\citenamefont {S\o{}rensen}\ and\ \citenamefont
  {M\o{}lmer}(2001)}]{SpinSqueezing2001}%
  \BibitemOpen
  \bibfield  {author} {\bibinfo {author} {\bibfnamefont {A.~S.}\ \bibnamefont
  {S\o{}rensen}}\ and\ \bibinfo {author} {\bibfnamefont {K.}~\bibnamefont
  {M\o{}lmer}},\ }\bibfield  {title} {\enquote {\bibinfo {title} {{Entanglement
  and Extreme Spin Squeezing}},}\ }\href
  {http://link.aps.org/doi/10.1103/PhysRevLett.86.4431} {\bibfield  {journal}
  {\bibinfo  {journal} {Physical Review Letters}\ }\textbf {\bibinfo {volume}
  {86}},\ \bibinfo {pages} {4431} (\bibinfo {year} {2001})}\BibitemShut
  {NoStop}%
\bibitem [{\citenamefont {T\'oth}\ \emph {et~al.}(2009)\citenamefont {T\'oth},
  \citenamefont {Knapp}, \citenamefont {G\"uhne},\ and\ \citenamefont
  {Briegel}}]{TothSpinSqueezing}%
  \BibitemOpen
  \bibfield  {author} {\bibinfo {author} {\bibfnamefont {G.}~\bibnamefont
  {T\'oth}}, \bibinfo {author} {\bibfnamefont {C.}~\bibnamefont {Knapp}},
  \bibinfo {author} {\bibfnamefont {O.}~\bibnamefont {G\"uhne}}, \ and\
  \bibinfo {author} {\bibfnamefont {H.~J.}\ \bibnamefont {Briegel}},\
  }\bibfield  {title} {\enquote {\bibinfo {title} {{Spin Squeezing and
  Entanglement}},}\ }\href {http://link.aps.org/doi/10.1103/PhysRevA.79.042334}
  {\bibfield  {journal} {\bibinfo  {journal} {Physical Review A}\ }\textbf
  {\bibinfo {volume} {79}},\ \bibinfo {pages} {042334} (\bibinfo {year}
  {2009})}\BibitemShut {NoStop}%
\bibitem [{\citenamefont {Ma}\ \emph {et~al.}(2011)\citenamefont {Ma},
  \citenamefont {Wang}, \citenamefont {Sun},\ and\ \citenamefont
  {Nori}}]{NoriSpinSqueezing}%
  \BibitemOpen
  \bibfield  {author} {\bibinfo {author} {\bibfnamefont {J.}~\bibnamefont
  {Ma}}, \bibinfo {author} {\bibfnamefont {X.}~\bibnamefont {Wang}}, \bibinfo
  {author} {\bibfnamefont {C.}~\bibnamefont {Sun}}, \ and\ \bibinfo {author}
  {\bibfnamefont {F.}~\bibnamefont {Nori}},\ }\bibfield  {title} {\enquote
  {\bibinfo {title} {{Quantum Spin Squeezing}},}\ }\href
  {http://www.sciencedirect.com/science/article/pii/S0370157311002201}
  {\bibfield  {journal} {\bibinfo  {journal} {Physics Reports}\ }\textbf
  {\bibinfo {volume} {509}},\ \bibinfo {pages} {89 } (\bibinfo {year}
  {2011})}\BibitemShut {NoStop}%
\bibitem [{\citenamefont {Gross}\ \emph {et~al.}(2010)\citenamefont {Gross},
  \citenamefont {Zibold}, \citenamefont {Nicklas}, \citenamefont {Estève},\
  and\ \citenamefont {Oberthaler}}]{Gross2010Nature}%
  \BibitemOpen
  \bibfield  {author} {\bibinfo {author} {\bibfnamefont {C.}~\bibnamefont
  {Gross}}, \bibinfo {author} {\bibfnamefont {T.}~\bibnamefont {Zibold}},
  \bibinfo {author} {\bibfnamefont {E.}~\bibnamefont {Nicklas}}, \bibinfo
  {author} {\bibfnamefont {J.}~\bibnamefont {Estève}}, \ and\ \bibinfo
  {author} {\bibfnamefont {M.~K.}\ \bibnamefont {Oberthaler}},\ }\bibfield
  {title} {\enquote {\bibinfo {title} {{Nonlinear Atom Interferometer Surpasses
  Classical Precision Limit}},}\ }\href {http://dx.doi.org/10.1038/nature08919}
  {\bibfield  {journal} {\bibinfo  {journal} {Nature}\ }\textbf {\bibinfo
  {volume} {464}},\ \bibinfo {pages} {1165} (\bibinfo {year}
  {2010})}\BibitemShut {NoStop}%
\bibitem [{\citenamefont {T\'oth}(2012)}]{PrecisionToth}%
  \BibitemOpen
  \bibfield  {author} {\bibinfo {author} {\bibfnamefont {G.}~\bibnamefont
  {T\'oth}},\ }\bibfield  {title} {\enquote {\bibinfo {title} {{Multipartite
  Entanglement and High-Precision Metrology}},}\ }\href
  {http://link.aps.org/doi/10.1103/PhysRevA.85.022322} {\bibfield  {journal}
  {\bibinfo  {journal} {Physical Review A}\ }\textbf {\bibinfo {volume} {85}},\
  \bibinfo {pages} {022322} (\bibinfo {year} {2012})}\BibitemShut {NoStop}%
\bibitem [{\citenamefont {Gross}(2012)}]{Gross2012IOP}%
  \BibitemOpen
  \bibfield  {author} {\bibinfo {author} {\bibfnamefont {C.}~\bibnamefont
  {Gross}},\ }\bibfield  {title} {\enquote {\bibinfo {title} {{Spin Squeezing,
  Entanglement and Quantum Metrology with Bose–Einstein Condensates}},}\
  }\href {http://stacks.iop.org/0953-4075/45/i=10/a=103001} {\bibfield
  {journal} {\bibinfo  {journal} {Journal of Physics B: Atomic, Molecular and
  Optical Physics}\ }\textbf {\bibinfo {volume} {45}},\ \bibinfo {pages}
  {103001} (\bibinfo {year} {2012})}\BibitemShut {NoStop}%
\bibitem [{\citenamefont {Hyllus}\ \emph {et~al.}(2012)\citenamefont {Hyllus},
  \citenamefont {Laskowski}, \citenamefont {Krischek}, \citenamefont
  {Schwemmer}, \citenamefont {Wieczorek}, \citenamefont {Weinfurter},
  \citenamefont {Pezz\'e},\ and\ \citenamefont
  {Smerzi}}]{FisherPrecisionMeasurement}%
  \BibitemOpen
  \bibfield  {author} {\bibinfo {author} {\bibfnamefont {P.}~\bibnamefont
  {Hyllus}}, \bibinfo {author} {\bibfnamefont {W.}~\bibnamefont {Laskowski}},
  \bibinfo {author} {\bibfnamefont {R.}~\bibnamefont {Krischek}}, \bibinfo
  {author} {\bibfnamefont {C.}~\bibnamefont {Schwemmer}}, \bibinfo {author}
  {\bibfnamefont {W.}~\bibnamefont {Wieczorek}}, \bibinfo {author}
  {\bibfnamefont {H.}~\bibnamefont {Weinfurter}}, \bibinfo {author}
  {\bibfnamefont {L.}~\bibnamefont {Pezz\'e}}, \ and\ \bibinfo {author}
  {\bibfnamefont {A.}~\bibnamefont {Smerzi}},\ }\bibfield  {title} {\enquote
  {\bibinfo {title} {{Fisher Information and Multiparticle Entanglement}},}\
  }\href {http://link.aps.org/doi/10.1103/PhysRevA.85.022321} {\bibfield
  {journal} {\bibinfo  {journal} {Physical Review A}\ }\textbf {\bibinfo
  {volume} {85}},\ \bibinfo {pages} {022321} (\bibinfo {year}
  {2012})}\BibitemShut {NoStop}%
\bibitem [{\citenamefont {Bergmann}\ and\ \citenamefont
  {Gühne}(2013)}]{DickeEntanglement}%
  \BibitemOpen
  \bibfield  {author} {\bibinfo {author} {\bibfnamefont {M.}~\bibnamefont
  {Bergmann}}\ and\ \bibinfo {author} {\bibfnamefont {O.}~\bibnamefont
  {Gühne}},\ }\bibfield  {title} {\enquote {\bibinfo {title} {{Entanglement
  Criteria for Dicke States}},}\ }\href
  {http://stacks.iop.org/1751-8121/46/i=38/a=385304} {\bibfield  {journal}
  {\bibinfo  {journal} {Journal of Physics A: Mathematical and Theoretical}\
  }\textbf {\bibinfo {volume} {46}},\ \bibinfo {pages} {385304} (\bibinfo
  {year} {2013})}\BibitemShut {NoStop}%
\bibitem [{\citenamefont {Verstraete}\ \emph {et~al.}(2001)\citenamefont
  {Verstraete}, \citenamefont {Audenaert}, \citenamefont {Dehaene},\ and\
  \citenamefont {Moor}}]{OneNegativeEigenvalue}%
  \BibitemOpen
  \bibfield  {author} {\bibinfo {author} {\bibfnamefont {F.}~\bibnamefont
  {Verstraete}}, \bibinfo {author} {\bibfnamefont {K.}~\bibnamefont
  {Audenaert}}, \bibinfo {author} {\bibfnamefont {J.}~\bibnamefont {Dehaene}},
  \ and\ \bibinfo {author} {\bibfnamefont {B.~D.}\ \bibnamefont {Moor}},\
  }\bibfield  {title} {\enquote {\bibinfo {title} {{A Comparison of the
  Entanglement Measures Negativity and Concurrence}},}\ }\href
  {http://stacks.iop.org/0305-4470/34/i=47/a=329} {\bibfield  {journal}
  {\bibinfo  {journal} {Journal of Physics A: Mathematical and General}\
  }\textbf {\bibinfo {volume} {34}},\ \bibinfo {pages} {10327} (\bibinfo {year}
  {2001})}\BibitemShut {NoStop}%
\bibitem [{\citenamefont {Vidal}\ and\ \citenamefont
  {Werner}(2002)}]{NegativityIntro}%
  \BibitemOpen
  \bibfield  {author} {\bibinfo {author} {\bibfnamefont {G.}~\bibnamefont
  {Vidal}}\ and\ \bibinfo {author} {\bibfnamefont {R.~F.}\ \bibnamefont
  {Werner}},\ }\bibfield  {title} {\enquote {\bibinfo {title} {{Computable
  Measure of Entanglement}},}\ }\href
  {http://link.aps.org/doi/10.1103/PhysRevA.65.032314} {\bibfield  {journal}
  {\bibinfo  {journal} {Physical Review A}\ }\textbf {\bibinfo {volume} {65}},\
  \bibinfo {pages} {032314} (\bibinfo {year} {2002})}\BibitemShut {NoStop}%
\bibitem [{\citenamefont {Mandilara}\ \emph {et~al.}(2006)\citenamefont
  {Mandilara}, \citenamefont {Akulin}, \citenamefont {Smilga},\ and\
  \citenamefont {Viola}}]{NilpotentReview}%
  \BibitemOpen
  \bibfield  {author} {\bibinfo {author} {\bibfnamefont {A.}~\bibnamefont
  {Mandilara}}, \bibinfo {author} {\bibfnamefont {V.~M.}\ \bibnamefont
  {Akulin}}, \bibinfo {author} {\bibfnamefont {A.~V.}\ \bibnamefont {Smilga}},
  \ and\ \bibinfo {author} {\bibfnamefont {L.}~\bibnamefont {Viola}},\
  }\bibfield  {title} {\enquote {\bibinfo {title} {{Quantum Entanglement via
  Nilpotent Polynomials}},}\ }\href
  {http://link.aps.org/doi/10.1103/PhysRevA.74.022331} {\bibfield  {journal}
  {\bibinfo  {journal} {Physical Review A}\ }\textbf {\bibinfo {volume} {74}},\
  \bibinfo {pages} {022331} (\bibinfo {year} {2006})}\BibitemShut {NoStop}%
\bibitem [{\citenamefont {Bengtsson}\ and\ \citenamefont
  {Życzkowski.}(2006)}]{GeomQuantStatesBook}%
  \BibitemOpen
  \bibfield  {author} {\bibinfo {author} {\bibfnamefont {I.}~\bibnamefont
  {Bengtsson}}\ and\ \bibinfo {author} {\bibfnamefont {K.}~\bibnamefont
  {Życzkowski.}},\ }\href {http://dx.doi.org/10.1017/CBO9780511535048} {\emph
  {\bibinfo {title} {Geometry of Quantum States}}}\ (\bibinfo  {publisher}
  {Cambridge University Press},\ \bibinfo {year} {2006})\BibitemShut {NoStop}%
\bibitem [{\citenamefont {Plbnio}\ and\ \citenamefont
  {Virmani}(2007)}]{PlenioEntanglementMeasures}%
  \BibitemOpen
  \bibfield  {author} {\bibinfo {author} {\bibfnamefont {M.~B.}\ \bibnamefont
  {Plbnio}}\ and\ \bibinfo {author} {\bibfnamefont {S.}~\bibnamefont
  {Virmani}},\ }\bibfield  {title} {\enquote {\bibinfo {title} {{An
  Introduction to Entanglement Measures}},}\ }\href
  {http://www.rintonpress.com/journals/qiconline.html#v7n12} {\bibfield
  {journal} {\bibinfo  {journal} {Quantum Information and Computation}\
  }\textbf {\bibinfo {volume} {7}},\ \bibinfo {pages} {1} (\bibinfo {year}
  {2007})}\BibitemShut {NoStop}%
\bibitem [{\citenamefont {Amico}\ \emph {et~al.}(2008)\citenamefont {Amico},
  \citenamefont {Osterloh},\ and\ \citenamefont {Vedral}}]{multireview}%
  \BibitemOpen
  \bibfield  {author} {\bibinfo {author} {\bibfnamefont {L.}~\bibnamefont
  {Amico}}, \bibinfo {author} {\bibfnamefont {A.}~\bibnamefont {Osterloh}}, \
  and\ \bibinfo {author} {\bibfnamefont {V.}~\bibnamefont {Vedral}},\
  }\bibfield  {title} {\enquote {\bibinfo {title} {{Entanglement in Many-Body
  Systems}},}\ }\href {http://dx.doi.org/10.1103/revmodphys.80.517} {\bibfield
  {journal} {\bibinfo  {journal} {Reviews of Modern Physics}\ }\textbf
  {\bibinfo {volume} {80}},\ \bibinfo {pages} {517} (\bibinfo {year}
  {2008})}\BibitemShut {NoStop}%
\bibitem [{\citenamefont {G\"{u}hne}\ and\ \citenamefont
  {T\'{o}th}(2009)}]{entang.review.toth}%
  \BibitemOpen
  \bibfield  {author} {\bibinfo {author} {\bibfnamefont {O.}~\bibnamefont
  {G\"{u}hne}}\ and\ \bibinfo {author} {\bibfnamefont {G.}~\bibnamefont
  {T\'{o}th}},\ }\bibfield  {title} {\enquote {\bibinfo {title} {{Entanglement
  Detection}},}\ }\href {http://dx.doi.org/10.1016/j.physrep.2009.02.004}
  {\bibfield  {journal} {\bibinfo  {journal} {Physics Reports}\ }\textbf
  {\bibinfo {volume} {474}},\ \bibinfo {pages} {1} (\bibinfo {year}
  {2009})}\BibitemShut {NoStop}%
\bibitem [{\citenamefont {T\'{o}th}\ and\ \citenamefont
  {G\"{u}hne}(2010)}]{SymmetricEquivalents}%
  \BibitemOpen
  \bibfield  {author} {\bibinfo {author} {\bibfnamefont {G.}~\bibnamefont
  {T\'{o}th}}\ and\ \bibinfo {author} {\bibfnamefont {O.}~\bibnamefont
  {G\"{u}hne}},\ }\bibfield  {title} {\enquote {\bibinfo {title} {{Separability
  Criteria and Entanglement Witnesses for Symmetric Quantum States}},}\ }\href
  {http://dx.doi.org/10.1007/s00340-009-3839-7} {\bibfield  {journal} {\bibinfo
   {journal} {Applied Physics B}\ }\textbf {\bibinfo {volume} {98}},\ \bibinfo
  {pages} {617} (\bibinfo {year} {2010})}\BibitemShut {NoStop}%
\bibitem [{\citenamefont {Mintert}\ \emph {et~al.}(2005)\citenamefont
  {Mintert}, \citenamefont {Ku\ifmmode~\acute{s}\else \'{s}\fi{}},\ and\
  \citenamefont {Buchleitner}}]{MultipartiteConc}%
  \BibitemOpen
  \bibfield  {author} {\bibinfo {author} {\bibfnamefont {F.}~\bibnamefont
  {Mintert}}, \bibinfo {author} {\bibfnamefont {M.}~\bibnamefont
  {Ku\ifmmode~\acute{s}\else \'{s}\fi{}}}, \ and\ \bibinfo {author}
  {\bibfnamefont {A.}~\bibnamefont {Buchleitner}},\ }\bibfield  {title}
  {\enquote {\bibinfo {title} {{Concurrence of Mixed Multipartite Quantum
  States}},}\ }\href {http://link.aps.org/doi/10.1103/PhysRevLett.95.260502}
  {\bibfield  {journal} {\bibinfo  {journal} {Physical Review Letters}\
  }\textbf {\bibinfo {volume} {95}},\ \bibinfo {pages} {260502} (\bibinfo
  {year} {2005})}\BibitemShut {NoStop}%
\bibitem [{\citenamefont {Wei}\ \emph {et~al.}(2004)\citenamefont {Wei},
  \citenamefont {Altepeter}, \citenamefont {Goldbart},\ and\ \citenamefont
  {Munro}}]{ConvVsNeg}%
  \BibitemOpen
  \bibfield  {author} {\bibinfo {author} {\bibfnamefont {T.-C.}\ \bibnamefont
  {Wei}}, \bibinfo {author} {\bibfnamefont {J.~B.}\ \bibnamefont {Altepeter}},
  \bibinfo {author} {\bibfnamefont {P.~M.}\ \bibnamefont {Goldbart}}, \ and\
  \bibinfo {author} {\bibfnamefont {W.~J.}\ \bibnamefont {Munro}},\ }\bibfield
  {title} {\enquote {\bibinfo {title} {{Measures of Entanglement in
  Multipartite Bound Entangled States}},}\ }\href
  {http://link.aps.org/doi/10.1103/PhysRevA.70.022322} {\bibfield  {journal}
  {\bibinfo  {journal} {Physical Review A}\ }\textbf {\bibinfo {volume} {70}},\
  \bibinfo {pages} {022322} (\bibinfo {year} {2004})}\BibitemShut {NoStop}%
\bibitem [{\citenamefont {Branciard}\ \emph {et~al.}(2010)\citenamefont
  {Branciard}, \citenamefont {Zhu}, \citenamefont {Chen},\ and\ \citenamefont
  {Scarani}}]{ConcVsGeom}%
  \BibitemOpen
  \bibfield  {author} {\bibinfo {author} {\bibfnamefont {C.}~\bibnamefont
  {Branciard}}, \bibinfo {author} {\bibfnamefont {H.}~\bibnamefont {Zhu}},
  \bibinfo {author} {\bibfnamefont {L.}~\bibnamefont {Chen}}, \ and\ \bibinfo
  {author} {\bibfnamefont {V.}~\bibnamefont {Scarani}},\ }\bibfield  {title}
  {\enquote {\bibinfo {title} {{Evaluation of Two Different Entanglement
  Measures on a Bound Entangled State}},}\ }\href
  {http://link.aps.org/doi/10.1103/PhysRevA.82.012327} {\bibfield  {journal}
  {\bibinfo  {journal} {Physical Review A}\ }\textbf {\bibinfo {volume} {82}},\
  \bibinfo {pages} {012327} (\bibinfo {year} {2010})}\BibitemShut {NoStop}%
\bibitem [{\citenamefont {Tamaryan}\ \emph {et~al.}(2009)\citenamefont
  {Tamaryan}, \citenamefont {Wei},\ and\ \citenamefont
  {Park}}]{Geometric3Qubits}%
  \BibitemOpen
  \bibfield  {author} {\bibinfo {author} {\bibfnamefont {S.}~\bibnamefont
  {Tamaryan}}, \bibinfo {author} {\bibfnamefont {T.-C.}\ \bibnamefont {Wei}}, \
  and\ \bibinfo {author} {\bibfnamefont {D.}~\bibnamefont {Park}},\ }\bibfield
  {title} {\enquote {\bibinfo {title} {{Maximally Entangled Three-Qubit States
  via Geometric Measure of Entanglement}},}\ }\href
  {http://link.aps.org/doi/10.1103/PhysRevA.80.052315} {\bibfield  {journal}
  {\bibinfo  {journal} {Physical Review A}\ }\textbf {\bibinfo {volume} {80}},\
  \bibinfo {pages} {052315} (\bibinfo {year} {2009})}\BibitemShut {NoStop}%
\bibitem [{\citenamefont {Chen}\ \emph {et~al.}(2010)\citenamefont {Chen},
  \citenamefont {Xu},\ and\ \citenamefont {Zhu}}]{GeometricMultiQubit}%
  \BibitemOpen
  \bibfield  {author} {\bibinfo {author} {\bibfnamefont {L.}~\bibnamefont
  {Chen}}, \bibinfo {author} {\bibfnamefont {A.}~\bibnamefont {Xu}}, \ and\
  \bibinfo {author} {\bibfnamefont {H.}~\bibnamefont {Zhu}},\ }\bibfield
  {title} {\enquote {\bibinfo {title} {{Computation of the Geometric Measure of
  Entanglement for Pure Multiqubit States}},}\ }\href
  {http://link.aps.org/doi/10.1103/PhysRevA.82.032301} {\bibfield  {journal}
  {\bibinfo  {journal} {Physical Review A}\ }\textbf {\bibinfo {volume} {82}},\
  \bibinfo {pages} {032301} (\bibinfo {year} {2010})}\BibitemShut {NoStop}%
\bibitem [{\citenamefont {{Moroder}}\ \emph {et~al.}(2014)\citenamefont
  {{Moroder}}, \citenamefont {{Gittsovich}}, \citenamefont {{Huber}},\ and\
  \citenamefont {{G{\"u}hne}}}]{SteeringEntanglement}%
  \BibitemOpen
  \bibfield  {author} {\bibinfo {author} {\bibfnamefont {T.}~\bibnamefont
  {{Moroder}}}, \bibinfo {author} {\bibfnamefont {O.}~\bibnamefont
  {{Gittsovich}}}, \bibinfo {author} {\bibfnamefont {M.}~\bibnamefont
  {{Huber}}}, \ and\ \bibinfo {author} {\bibfnamefont {O.}~\bibnamefont
  {{G{\"u}hne}}},\ }\bibfield  {title} {\enquote {\bibinfo {title} {{Steering
  Bound Entangled States: A Counterexample to the Stronger Peres
  Conjecture}},}\ }\href {http://arxiv.org/abs/1405.0262} {\bibfield  {journal}
  {\bibinfo  {journal} {arXiv:1405.0262}\ } (\bibinfo {year}
  {2014})}\BibitemShut {NoStop}%
\bibitem [{\citenamefont {D\"{u}r}\ \emph {et~al.}(2000)\citenamefont
  {D\"{u}r}, \citenamefont {Vidal},\ and\ \citenamefont
  {Cirac}}]{3Qubits2Ways}%
  \BibitemOpen
  \bibfield  {author} {\bibinfo {author} {\bibfnamefont {W.}~\bibnamefont
  {D\"{u}r}}, \bibinfo {author} {\bibfnamefont {G.}~\bibnamefont {Vidal}}, \
  and\ \bibinfo {author} {\bibfnamefont {J.~I.}\ \bibnamefont {Cirac}},\
  }\bibfield  {title} {\enquote {\bibinfo {title} {{Three Qubits can be
  Entangled in Two Inequivalent Ways}},}\ }\href
  {http://dx.doi.org/10.1103/physreva.62.062314} {\bibfield  {journal}
  {\bibinfo  {journal} {Physical Review A}\ }\textbf {\bibinfo {volume} {62}},\
  \bibinfo {pages} {062314} (\bibinfo {year} {2000})}\BibitemShut {NoStop}%
\bibitem [{\citenamefont {Ac{\'i}n}\ \emph {et~al.}(2001)\citenamefont
  {Ac{\'i}n}, \citenamefont {Bru{\ss}}, \citenamefont {Lewenstein},\ and\
  \citenamefont {Sanpera}}]{ThreeQubitClassification04}%
  \BibitemOpen
  \bibfield  {author} {\bibinfo {author} {\bibfnamefont {A.}~\bibnamefont
  {Ac{\'i}n}}, \bibinfo {author} {\bibfnamefont {D.}~\bibnamefont {Bru{\ss}}},
  \bibinfo {author} {\bibfnamefont {M.}~\bibnamefont {Lewenstein}}, \ and\
  \bibinfo {author} {\bibfnamefont {A.}~\bibnamefont {Sanpera}},\ }\bibfield
  {title} {\enquote {\bibinfo {title} {{Classification of Mixed Three-Qubit
  States}},}\ }\href {http://link.aps.org/doi/10.1103/PhysRevLett.87.040401}
  {\bibfield  {journal} {\bibinfo  {journal} {Physical Review Letters}\
  }\textbf {\bibinfo {volume} {87}},\ \bibinfo {pages} {040401} (\bibinfo
  {year} {2001})}\BibitemShut {NoStop}%
\bibitem [{\citenamefont {Chen}\ and\ \citenamefont
  {Chen}(2006)}]{ThreeQubitClassification06}%
  \BibitemOpen
  \bibfield  {author} {\bibinfo {author} {\bibfnamefont {L.}~\bibnamefont
  {Chen}}\ and\ \bibinfo {author} {\bibfnamefont {Y.-X.}\ \bibnamefont
  {Chen}},\ }\bibfield  {title} {\enquote {\bibinfo {title} {{Classification of
  GHZ-type, W-type, and GHZ-W-type Multiqubit Entanglement}},}\ }\href
  {http://link.aps.org/doi/10.1103/PhysRevA.74.062310} {\bibfield  {journal}
  {\bibinfo  {journal} {Physical Review A}\ }\textbf {\bibinfo {volume} {74}},\
  \bibinfo {pages} {062310} (\bibinfo {year} {2006})}\BibitemShut {NoStop}%
\bibitem [{\citenamefont {Miyake}(2003)}]{ClassificationHyperdeterminants}%
  \BibitemOpen
  \bibfield  {author} {\bibinfo {author} {\bibfnamefont {A.}~\bibnamefont
  {Miyake}},\ }\bibfield  {title} {\enquote {\bibinfo {title} {{Classification
  of Multipartite Entangled States by Multidimensional Determinants}},}\ }\href
  {http://link.aps.org/doi/10.1103/PhysRevA.67.012108} {\bibfield  {journal}
  {\bibinfo  {journal} {Physical Review A}\ }\textbf {\bibinfo {volume} {67}},\
  \bibinfo {pages} {012108} (\bibinfo {year} {2003})}\BibitemShut {NoStop}%
\bibitem [{\citenamefont {Mathonet}\ \emph {et~al.}(2010)\citenamefont
  {Mathonet}, \citenamefont {Krins}, \citenamefont {Godefroid}, \citenamefont
  {Lamata}, \citenamefont {Solano},\ and\ \citenamefont
  {Bastin}}]{NQubitEquivalences}%
  \BibitemOpen
  \bibfield  {author} {\bibinfo {author} {\bibfnamefont {P.}~\bibnamefont
  {Mathonet}}, \bibinfo {author} {\bibfnamefont {S.}~\bibnamefont {Krins}},
  \bibinfo {author} {\bibfnamefont {M.}~\bibnamefont {Godefroid}}, \bibinfo
  {author} {\bibfnamefont {L.}~\bibnamefont {Lamata}}, \bibinfo {author}
  {\bibfnamefont {E.}~\bibnamefont {Solano}}, \ and\ \bibinfo {author}
  {\bibfnamefont {T.}~\bibnamefont {Bastin}},\ }\bibfield  {title} {\enquote
  {\bibinfo {title} {{Entanglement Equivalence of N-Qubit Symmetric States}},}\
  }\href {http://link.aps.org/doi/10.1103/PhysRevA.81.052315} {\bibfield
  {journal} {\bibinfo  {journal} {Physical Review A}\ }\textbf {\bibinfo
  {volume} {81}},\ \bibinfo {pages} {052315} (\bibinfo {year}
  {2010})}\BibitemShut {NoStop}%
\bibitem [{\citenamefont {Peres}(1996)}]{PPTAsher}%
  \BibitemOpen
  \bibfield  {author} {\bibinfo {author} {\bibfnamefont {A.}~\bibnamefont
  {Peres}},\ }\bibfield  {title} {\enquote {\bibinfo {title} {{Separability
  Criterion for Density Matrices}},}\ }\href
  {http://link.aps.org/doi/10.1103/PhysRevLett.77.1413} {\bibfield  {journal}
  {\bibinfo  {journal} {Physical Review Letters}\ }\textbf {\bibinfo {volume}
  {77}},\ \bibinfo {pages} {1413} (\bibinfo {year} {1996})}\BibitemShut
  {NoStop}%
\bibitem [{\citenamefont {Horodecki}\ \emph {et~al.}(1996)\citenamefont
  {Horodecki}, \citenamefont {Horodecki},\ and\ \citenamefont
  {Horodecki}}]{PPTHorodecki}%
  \BibitemOpen
  \bibfield  {author} {\bibinfo {author} {\bibfnamefont {M.}~\bibnamefont
  {Horodecki}}, \bibinfo {author} {\bibfnamefont {P.}~\bibnamefont
  {Horodecki}}, \ and\ \bibinfo {author} {\bibfnamefont {R.}~\bibnamefont
  {Horodecki}},\ }\bibfield  {title} {\enquote {\bibinfo {title} {{Separability
  of Mixed States: Necessary and Sufficient Conditions}},}\ }\href
  {http://www.sciencedirect.com/science/article/pii/S0375960196007062}
  {\bibfield  {journal} {\bibinfo  {journal} {Physics Letters A}\ }\textbf
  {\bibinfo {volume} {223}},\ \bibinfo {pages} {1} (\bibinfo {year}
  {1996})}\BibitemShut {NoStop}%
\bibitem [{\citenamefont {Tura}\ \emph {et~al.}(2012)\citenamefont {Tura},
  \citenamefont {Augusiak}, \citenamefont {Hyllus}, \citenamefont {Ku\'{s}},
  \citenamefont {Samsonowicz},\ and\ \citenamefont {Lewenstein}}]{ppt4qubit}%
  \BibitemOpen
  \bibfield  {author} {\bibinfo {author} {\bibfnamefont {J.}~\bibnamefont
  {Tura}}, \bibinfo {author} {\bibfnamefont {R.}~\bibnamefont {Augusiak}},
  \bibinfo {author} {\bibfnamefont {P.}~\bibnamefont {Hyllus}}, \bibinfo
  {author} {\bibfnamefont {M.}~\bibnamefont {Ku\'{s}}}, \bibinfo {author}
  {\bibfnamefont {J.}~\bibnamefont {Samsonowicz}}, \ and\ \bibinfo {author}
  {\bibfnamefont {M.}~\bibnamefont {Lewenstein}},\ }\bibfield  {title}
  {\enquote {\bibinfo {title} {{Four-Qubit Entangled Symmetric States with
  Positive Partial Transpositions}},}\ }\href
  {http://dx.doi.org/10.1103/physreva.85.060302} {\bibfield  {journal}
  {\bibinfo  {journal} {Physical Review A}\ }\textbf {\bibinfo {volume} {85}},\
  \bibinfo {pages} {060302} (\bibinfo {year} {2012})}\BibitemShut {NoStop}%
\bibitem [{\citenamefont {Augusiak}\ \emph {et~al.}(2012)\citenamefont
  {Augusiak}, \citenamefont {Tura}, \citenamefont {Samsonowicz},\ and\
  \citenamefont {Lewenstein}}]{pptNqubit}%
  \BibitemOpen
  \bibfield  {author} {\bibinfo {author} {\bibfnamefont {R.}~\bibnamefont
  {Augusiak}}, \bibinfo {author} {\bibfnamefont {J.}~\bibnamefont {Tura}},
  \bibinfo {author} {\bibfnamefont {J.}~\bibnamefont {Samsonowicz}}, \ and\
  \bibinfo {author} {\bibfnamefont {M.}~\bibnamefont {Lewenstein}},\ }\bibfield
   {title} {\enquote {\bibinfo {title} {{Entangled Symmetric States of \(N\)
  Qubits with All Positive Partial Transpositions}},}\ }\href
  {http://link.aps.org/doi/10.1103/PhysRevA.86.042316} {\bibfield  {journal}
  {\bibinfo  {journal} {Physical Review A}\ }\textbf {\bibinfo {volume} {86}},\
  \bibinfo {pages} {042316} (\bibinfo {year} {2012})}\BibitemShut {NoStop}%
\bibitem [{\citenamefont {{Wang}}\ \emph {et~al.}(2014)\citenamefont {{Wang}},
  \citenamefont {{Xu}},\ and\ \citenamefont
  {{Severini}}}]{EntangWitnessExtend}%
  \BibitemOpen
  \bibfield  {author} {\bibinfo {author} {\bibfnamefont {B.-H.}\ \bibnamefont
  {{Wang}}}, \bibinfo {author} {\bibfnamefont {H.-R.}\ \bibnamefont {{Xu}}}, \
  and\ \bibinfo {author} {\bibfnamefont {S.}~\bibnamefont {{Severini}}},\
  }\bibfield  {title} {\enquote {\bibinfo {title} {{Universal Methods for
  Extending any Entanglement Witness from the Bipartite to the Multipartite
  Case}},}\ }\href {http://arxiv.org/abs/1405.2021} {\bibfield  {journal}
  {\bibinfo  {journal} {arXiv:1405.2021}\ } (\bibinfo {year}
  {2014})}\BibitemShut {NoStop}%
\bibitem [{\citenamefont {Doherty}\ \emph {et~al.}(2004)\citenamefont
  {Doherty}, \citenamefont {Parrilo},\ and\ \citenamefont
  {Spedalieri}}]{EntanglementHierarchy}%
  \BibitemOpen
  \bibfield  {author} {\bibinfo {author} {\bibfnamefont {A.~C.}\ \bibnamefont
  {Doherty}}, \bibinfo {author} {\bibfnamefont {P.~A.}\ \bibnamefont
  {Parrilo}}, \ and\ \bibinfo {author} {\bibfnamefont {F.~M.}\ \bibnamefont
  {Spedalieri}},\ }\bibfield  {title} {\enquote {\bibinfo {title} {{Complete
  Family of Separability Criteria}},}\ }\href
  {http://link.aps.org/doi/10.1103/PhysRevA.69.022308} {\bibfield  {journal}
  {\bibinfo  {journal} {Physical Review A}\ }\textbf {\bibinfo {volume} {69}},\
  \bibinfo {pages} {022308} (\bibinfo {year} {2004})}\BibitemShut {NoStop}%
\bibitem [{\citenamefont {Doherty}\ \emph {et~al.}(2005)\citenamefont
  {Doherty}, \citenamefont {Parrilo},\ and\ \citenamefont
  {Spedalieri}}]{EntanglementHierarchyMultipartite}%
  \BibitemOpen
  \bibfield  {author} {\bibinfo {author} {\bibfnamefont {A.~C.}\ \bibnamefont
  {Doherty}}, \bibinfo {author} {\bibfnamefont {P.~A.}\ \bibnamefont
  {Parrilo}}, \ and\ \bibinfo {author} {\bibfnamefont {F.~M.}\ \bibnamefont
  {Spedalieri}},\ }\bibfield  {title} {\enquote {\bibinfo {title} {{Detecting
  Multipartite Entanglement}},}\ }\href
  {http://link.aps.org/doi/10.1103/PhysRevA.71.032333} {\bibfield  {journal}
  {\bibinfo  {journal} {Physical Review A}\ }\textbf {\bibinfo {volume} {71}},\
  \bibinfo {pages} {032333} (\bibinfo {year} {2005})}\BibitemShut {NoStop}%
\bibitem [{\citenamefont {Kraus}\ \emph {et~al.}(2000)\citenamefont {Kraus},
  \citenamefont {Cirac}, \citenamefont {Karnas},\ and\ \citenamefont
  {Lewenstein}}]{KrausCiracKarnasLewenstein2000}%
  \BibitemOpen
  \bibfield  {author} {\bibinfo {author} {\bibfnamefont {B.}~\bibnamefont
  {Kraus}}, \bibinfo {author} {\bibfnamefont {J.~I.}\ \bibnamefont {Cirac}},
  \bibinfo {author} {\bibfnamefont {S.}~\bibnamefont {Karnas}}, \ and\ \bibinfo
  {author} {\bibfnamefont {M.}~\bibnamefont {Lewenstein}},\ }\bibfield  {title}
  {\enquote {\bibinfo {title} {{Separability in \(2 \times N\) Composite
  Quantum Systems}},}\ }\href
  {http://link.aps.org/doi/10.1103/PhysRevA.61.062302} {\bibfield  {journal}
  {\bibinfo  {journal} {Physical Review A}\ }\textbf {\bibinfo {volume} {61}},\
  \bibinfo {pages} {062302} (\bibinfo {year} {2000})}\BibitemShut {NoStop}%
\bibitem [{\citenamefont {Karnas}\ and\ \citenamefont
  {Lewenstein}(2001)}]{KarnasLewenstein2001}%
  \BibitemOpen
  \bibfield  {author} {\bibinfo {author} {\bibfnamefont {S.}~\bibnamefont
  {Karnas}}\ and\ \bibinfo {author} {\bibfnamefont {M.}~\bibnamefont
  {Lewenstein}},\ }\bibfield  {title} {\enquote {\bibinfo {title}
  {{Separability and Entanglement in \(\mathbb{C}^2\otimes \mathbb{C}^2 \otimes
  \mathbb{C}^N\) Composite Quantum Systems}},}\ }\href
  {http://link.aps.org/doi/10.1103/PhysRevA.64.042313} {\bibfield  {journal}
  {\bibinfo  {journal} {Physical Review A}\ }\textbf {\bibinfo {volume} {64}},\
  \bibinfo {pages} {042313} (\bibinfo {year} {2001})}\BibitemShut {NoStop}%
\bibitem [{\citenamefont {Szalay}\ and\ \citenamefont
  {K\"{o}k\'{e}nyesi}(2012)}]{Szalay2012Partial}%
  \BibitemOpen
  \bibfield  {author} {\bibinfo {author} {\bibfnamefont {S.}~\bibnamefont
  {Szalay}}\ and\ \bibinfo {author} {\bibfnamefont {Z.}~\bibnamefont
  {K\"{o}k\'{e}nyesi}},\ }\bibfield  {title} {\enquote {\bibinfo {title}
  {{Partial Separability Revisited: Necessary and Sufficient Criteria}},}\
  }\href {http://dx.doi.org/10.1103/physreva.86.032341} {\bibfield  {journal}
  {\bibinfo  {journal} {Physical Review A}\ }\textbf {\bibinfo {volume} {86}},\
  \bibinfo {pages} {032341} (\bibinfo {year} {2012})}\BibitemShut {NoStop}%
\bibitem [{\citenamefont {Wolfe}\ and\ \citenamefont
  {Yelin}(2014)}]{SuperradSeparable}%
  \BibitemOpen
  \bibfield  {author} {\bibinfo {author} {\bibfnamefont {E.}~\bibnamefont
  {Wolfe}}\ and\ \bibinfo {author} {\bibfnamefont {S.}~\bibnamefont {Yelin}},\
  }\bibfield  {title} {\enquote {\bibinfo {title} {{Certifying Separability in
  Symmetric Mixed States of \(N\) Qubits, and Superradiance}},}\ }\href
  {http://link.aps.org/doi/10.1103/PhysRevLett.112.140402} {\bibfield
  {journal} {\bibinfo  {journal} {Physical Review Letters}\ }\textbf {\bibinfo
  {volume} {112}},\ \bibinfo {pages} {140402} (\bibinfo {year}
  {2014})}\BibitemShut {NoStop}%
\bibitem [{\citenamefont {Korbicz}\ \emph {et~al.}(2005)\citenamefont
  {Korbicz}, \citenamefont {Cirac},\ and\ \citenamefont
  {Lewenstein}}]{CiracSpinSqueezing}%
  \BibitemOpen
  \bibfield  {author} {\bibinfo {author} {\bibfnamefont {J.~K.}\ \bibnamefont
  {Korbicz}}, \bibinfo {author} {\bibfnamefont {J.~I.}\ \bibnamefont {Cirac}},
  \ and\ \bibinfo {author} {\bibfnamefont {M.}~\bibnamefont {Lewenstein}},\
  }\bibfield  {title} {\enquote {\bibinfo {title} {{Spin Squeezing Inequalities
  and Entanglement of \(N\) Qubit States}},}\ }\href
  {http://link.aps.org/doi/10.1103/PhysRevLett.95.120502} {\bibfield  {journal}
  {\bibinfo  {journal} {Physical Review Letters}\ }\textbf {\bibinfo {volume}
  {95}},\ \bibinfo {pages} {120502} (\bibinfo {year} {2005})}\BibitemShut
  {NoStop}%
\bibitem [{\citenamefont {Chen}\ \emph
  {et~al.}(2014{\natexlab{a}})\citenamefont {Chen}, \citenamefont {Ma},\ and\
  \citenamefont {Fei}}]{BipartiteMUBEntanglement}%
  \BibitemOpen
  \bibfield  {author} {\bibinfo {author} {\bibfnamefont {B.}~\bibnamefont
  {Chen}}, \bibinfo {author} {\bibfnamefont {T.}~\bibnamefont {Ma}}, \ and\
  \bibinfo {author} {\bibfnamefont {S.-M.}\ \bibnamefont {Fei}},\ }\bibfield
  {title} {\enquote {\bibinfo {title} {{Entanglement Detection using Mutually
  Unbiased Measurements}},}\ }\href
  {http://link.aps.org/doi/10.1103/PhysRevA.89.064302} {\bibfield  {journal}
  {\bibinfo  {journal} {Physical Review A}\ }\textbf {\bibinfo {volume} {89}},\
  \bibinfo {pages} {064302} (\bibinfo {year} {2014}{\natexlab{a}})}\BibitemShut
  {NoStop}%
\bibitem [{\citenamefont {Chen}\ \emph
  {et~al.}(2014{\natexlab{b}})\citenamefont {Chen}, \citenamefont {Ma},\ and\
  \citenamefont {Fei}}]{BipartiteSICEntanglement}%
  \BibitemOpen
  \bibfield  {author} {\bibinfo {author} {\bibfnamefont {B.}~\bibnamefont
  {Chen}}, \bibinfo {author} {\bibfnamefont {T.}~\bibnamefont {Ma}}, \ and\
  \bibinfo {author} {\bibfnamefont {S.-M.}\ \bibnamefont {Fei}},\ }\bibfield
  {title} {\enquote {\bibinfo {title} {{General SIC-Measurement Based
  Entanglement Detection}},}\ }\href {http://arxiv.org/abs/1406.7820}
  {\bibfield  {journal} {\bibinfo  {journal} {arXiv:1406.7820}\ } (\bibinfo
  {year} {2014}{\natexlab{b}})}\BibitemShut {NoStop}%
\bibitem [{\citenamefont {Bandyopadhyay}\ \emph {et~al.}(2002)\citenamefont
  {Bandyopadhyay}, \citenamefont {Boykin}, \citenamefont {Roychowdhury},\ and\
  \citenamefont {Vatan}}]{MUBConstruction}%
  \BibitemOpen
  \bibfield  {author} {\bibinfo {author} {\bibnamefont {Bandyopadhyay}},
  \bibinfo {author} {\bibnamefont {Boykin}}, \bibinfo {author} {\bibnamefont
  {Roychowdhury}}, \ and\ \bibinfo {author} {\bibnamefont {Vatan}},\ }\bibfield
   {title} {\enquote {\bibinfo {title} {{A New Proof for the Existence of
  Mutually Unbiased Bases}},}\ }\href
  {http://dx.doi.org/10.1007/s00453-002-0980-7} {\bibfield  {journal} {\bibinfo
   {journal} {Algorithmica}\ }\textbf {\bibinfo {volume} {34}},\ \bibinfo
  {pages} {512} (\bibinfo {year} {2002})}\BibitemShut {NoStop}%
\bibitem [{\citenamefont {Durt}\ \emph {et~al.}(2010)\citenamefont {Durt},
  \citenamefont {Englert}, \citenamefont {Bengtsson},\ and\ \citenamefont
  {Życzkowski}}]{MUBReview}%
  \BibitemOpen
  \bibfield  {author} {\bibinfo {author} {\bibfnamefont {T.}~\bibnamefont
  {Durt}}, \bibinfo {author} {\bibfnamefont {B.-G.}\ \bibnamefont {Englert}},
  \bibinfo {author} {\bibfnamefont {I.}~\bibnamefont {Bengtsson}}, \ and\
  \bibinfo {author} {\bibfnamefont {K.}~\bibnamefont {Życzkowski}},\
  }\bibfield  {title} {\enquote {\bibinfo {title} {{On Mutually Unbiased
  Bases}},}\ }\href
  {http://www.worldscientific.com/doi/abs/10.1142/S0219749910006502} {\bibfield
   {journal} {\bibinfo  {journal} {International Journal of Quantum
  Information}\ }\textbf {\bibinfo {volume} {08}},\ \bibinfo {pages} {535}
  (\bibinfo {year} {2010})}\BibitemShut {NoStop}%
\bibitem [{\citenamefont {Bengtsson}(2010)}]{MUBSICPOVM}%
  \BibitemOpen
  \bibfield  {author} {\bibinfo {author} {\bibfnamefont {I.}~\bibnamefont
  {Bengtsson}},\ }\bibfield  {title} {\enquote {\bibinfo {title} {{From SICs
  and MUBs to Eddington}},}\ }\href
  {http://stacks.iop.org/1742-6596/254/i=1/a=012007} {\bibfield  {journal}
  {\bibinfo  {journal} {Journal of Physics: Conference Series}\ }\textbf
  {\bibinfo {volume} {254}},\ \bibinfo {pages} {012007} (\bibinfo {year}
  {2010})}\BibitemShut {NoStop}%
\bibitem [{\citenamefont {Caves}\ \emph {et~al.}(2002)\citenamefont {Caves},
  \citenamefont {Fuchs},\ and\ \citenamefont {Schack}}]{SICPOVM2002}%
  \BibitemOpen
  \bibfield  {author} {\bibinfo {author} {\bibfnamefont {C.~M.}\ \bibnamefont
  {Caves}}, \bibinfo {author} {\bibfnamefont {C.~A.}\ \bibnamefont {Fuchs}}, \
  and\ \bibinfo {author} {\bibfnamefont {R.}~\bibnamefont {Schack}},\
  }\bibfield  {title} {\enquote {\bibinfo {title} {{Unknown Quantum States: The
  Quantum de Finetti Representation}},}\ }\href
  {http://scitation.aip.org/content/aip/journal/jmp/43/9/10.1063/1.1494475}
  {\bibfield  {journal} {\bibinfo  {journal} {Journal of Mathematical Physics}\
  }\textbf {\bibinfo {volume} {43}},\ \bibinfo {pages} {4537} (\bibinfo {year}
  {2002})}\BibitemShut {NoStop}%
\bibitem [{\citenamefont {Renes}\ \emph {et~al.}(2004)\citenamefont {Renes},
  \citenamefont {Blume-Kohout}, \citenamefont {Scott},\ and\ \citenamefont
  {Caves}}]{SICPOVM2004}%
  \BibitemOpen
  \bibfield  {author} {\bibinfo {author} {\bibfnamefont {J.~M.}\ \bibnamefont
  {Renes}}, \bibinfo {author} {\bibfnamefont {R.}~\bibnamefont {Blume-Kohout}},
  \bibinfo {author} {\bibfnamefont {A.~J.}\ \bibnamefont {Scott}}, \ and\
  \bibinfo {author} {\bibfnamefont {C.~M.}\ \bibnamefont {Caves}},\ }\bibfield
  {title} {\enquote {\bibinfo {title} {{Symmetric Informationally Complete
  Quantum Measurements}},}\ }\href
  {http://scitation.aip.org/content/aip/journal/jmp/45/6/10.1063/1.1737053}
  {\bibfield  {journal} {\bibinfo  {journal} {Journal of Mathematical Physics}\
  }\textbf {\bibinfo {volume} {45}},\ \bibinfo {pages} {2171} (\bibinfo {year}
  {2004})}\BibitemShut {NoStop}%
\bibitem [{\citenamefont {Eckert}\ \emph {et~al.}(2002)\citenamefont {Eckert},
  \citenamefont {Schliemann}, \citenamefont {Bru{\ss}},\ and\ \citenamefont
  {Lewenstein}}]{eckert2002quantum}%
  \BibitemOpen
  \bibfield  {author} {\bibinfo {author} {\bibfnamefont {K.}~\bibnamefont
  {Eckert}}, \bibinfo {author} {\bibfnamefont {J.}~\bibnamefont {Schliemann}},
  \bibinfo {author} {\bibfnamefont {D.}~\bibnamefont {Bru{\ss}}}, \ and\
  \bibinfo {author} {\bibfnamefont {M.}~\bibnamefont {Lewenstein}},\ }\bibfield
   {title} {\enquote {\bibinfo {title} {{Quantum Correlations in Systems of
  Indistinguishable Particles}},}\ }\href@noop {} {\bibfield  {journal}
  {\bibinfo  {journal} {Annals of Physics}\ }\textbf {\bibinfo {volume}
  {299}},\ \bibinfo {pages} {88} (\bibinfo {year} {2002})}\BibitemShut
  {NoStop}%
\bibitem [{\citenamefont {{Jeong}}\ \emph {et~al.}(2014)\citenamefont
  {{Jeong}}, \citenamefont {{Kang}},\ and\ \citenamefont
  {{Kwon}}}]{MacroscropicQuantumness}%
  \BibitemOpen
  \bibfield  {author} {\bibinfo {author} {\bibfnamefont {H.}~\bibnamefont
  {{Jeong}}}, \bibinfo {author} {\bibfnamefont {M.}~\bibnamefont {{Kang}}}, \
  and\ \bibinfo {author} {\bibfnamefont {H.}~\bibnamefont {{Kwon}}},\
  }\bibfield  {title} {\enquote {\bibinfo {title} {{Characterizations and
  Quantifications of Macroscopic Quantumness and Its Implementations using
  Optical Fields}},}\ }\href {http://arxiv.org/abs/1407.0126} {\bibfield
  {journal} {\bibinfo  {journal} {arXiv:1407.0126}\ } (\bibinfo {year}
  {2014})}\BibitemShut {NoStop}%
\bibitem [{\citenamefont {{Yadin}}\ and\ \citenamefont
  {{Vedral}}(2014)}]{MacroscopicArXiv}%
  \BibitemOpen
  \bibfield  {author} {\bibinfo {author} {\bibfnamefont {B.}~\bibnamefont
  {{Yadin}}}\ and\ \bibinfo {author} {\bibfnamefont {V.}~\bibnamefont
  {{Vedral}}},\ }\bibfield  {title} {\enquote {\bibinfo {title} {{A New
  Criterion for Macroscopic Quantum States}},}\ }\href
  {http://arxiv.org/abs/1407.2442} {\bibfield  {journal} {\bibinfo  {journal}
  {arXiv:1407.2442}\ } (\bibinfo {year} {2014})}\BibitemShut {NoStop}%
\bibitem [{\citenamefont {Tura}\ \emph {et~al.}(2014)\citenamefont {Tura},
  \citenamefont {Augusiak}, \citenamefont {Sainz}, \citenamefont {Vértesi},
  \citenamefont {Lewenstein},\ and\ \citenamefont {Acín}}]{AcinDetecting}%
  \BibitemOpen
  \bibfield  {author} {\bibinfo {author} {\bibfnamefont {J.}~\bibnamefont
  {Tura}}, \bibinfo {author} {\bibfnamefont {R.}~\bibnamefont {Augusiak}},
  \bibinfo {author} {\bibfnamefont {A.~B.}\ \bibnamefont {Sainz}}, \bibinfo
  {author} {\bibfnamefont {T.}~\bibnamefont {Vértesi}}, \bibinfo {author}
  {\bibfnamefont {M.}~\bibnamefont {Lewenstein}}, \ and\ \bibinfo {author}
  {\bibfnamefont {A.}~\bibnamefont {Acín}},\ }\bibfield  {title} {\enquote
  {\bibinfo {title} {{Detecting Nonlocality in Many-Body Quantum States}},}\
  }\href {http://www.sciencemag.org/content/344/6189/1256.abstract} {\bibfield
  {journal} {\bibinfo  {journal} {Science}\ }\textbf {\bibinfo {volume}
  {344}},\ \bibinfo {pages} {1256} (\bibinfo {year} {2014})}\BibitemShut
  {NoStop}%
\bibitem [{\citenamefont {{Wolfe}}\ and\ \citenamefont
  {{Yelin}}(2014)}]{DrivenSuperrad}%
  \BibitemOpen
  \bibfield  {author} {\bibinfo {author} {\bibfnamefont {E.}~\bibnamefont
  {{Wolfe}}}\ and\ \bibinfo {author} {\bibfnamefont {S.~F.}\ \bibnamefont
  {{Yelin}}},\ }\bibfield  {title} {\enquote {\bibinfo {title} {{Spin Squeezing
  by means of Driven Superradiance}},}\ }\href {http://arxiv.org/abs/1405.5288}
  {\bibfield  {journal} {\bibinfo  {journal} {arXiv:1405.5288}\ } (\bibinfo
  {year} {2014})}\BibitemShut {NoStop}%
\bibitem [{\citenamefont {Quesada}\ and\ \citenamefont
  {Sanpera}(2014)}]{BestSeparableQubits}%
  \BibitemOpen
  \bibfield  {author} {\bibinfo {author} {\bibfnamefont {R.}~\bibnamefont
  {Quesada}}\ and\ \bibinfo {author} {\bibfnamefont {A.}~\bibnamefont
  {Sanpera}},\ }\bibfield  {title} {\enquote {\bibinfo {title} {{Best Separable
  Approximation of Multipartite Diagonal Symmetric States}},}\ }\href
  {http://link.aps.org/doi/10.1103/PhysRevA.89.052319} {\bibfield  {journal}
  {\bibinfo  {journal} {Physical Review A}\ }\textbf {\bibinfo {volume} {89}},\
  \bibinfo {pages} {052319} (\bibinfo {year} {2014})}\BibitemShut {NoStop}%
\bibitem [{\citenamefont {Życzkowski}\ \emph {et~al.}(1998)\citenamefont
  {Życzkowski}, \citenamefont {Horodecki}, \citenamefont {Sanpera},\ and\
  \citenamefont {Lewenstein}}]{Zyczkowski98}%
  \BibitemOpen
  \bibfield  {author} {\bibinfo {author} {\bibfnamefont {K.}~\bibnamefont
  {Życzkowski}}, \bibinfo {author} {\bibfnamefont {P.}~\bibnamefont
  {Horodecki}}, \bibinfo {author} {\bibfnamefont {A.}~\bibnamefont {Sanpera}},
  \ and\ \bibinfo {author} {\bibfnamefont {M.}~\bibnamefont {Lewenstein}},\
  }\bibfield  {title} {\enquote {\bibinfo {title} {{Volume of the Set of
  Separable States}},}\ }\href
  {http://link.aps.org/doi/10.1103/PhysRevA.58.883} {\bibfield  {journal}
  {\bibinfo  {journal} {Physical Review A}\ }\textbf {\bibinfo {volume} {58}},\
  \bibinfo {pages} {883} (\bibinfo {year} {1998})}\BibitemShut {NoStop}%
\bibitem [{\citenamefont {Życzkowski}(1999)}]{Zyczkowski99}%
  \BibitemOpen
  \bibfield  {author} {\bibinfo {author} {\bibfnamefont {K.}~\bibnamefont
  {Życzkowski}},\ }\bibfield  {title} {\enquote {\bibinfo {title} {{Volume of
  the Set of Separable States. II}},}\ }\href
  {http://link.aps.org/doi/10.1103/PhysRevA.60.3496} {\bibfield  {journal}
  {\bibinfo  {journal} {Physical Review A}\ }\textbf {\bibinfo {volume} {60}},\
  \bibinfo {pages} {3496} (\bibinfo {year} {1999})}\BibitemShut {NoStop}%
\bibitem [{\citenamefont {Życzkowski}\ and\ \citenamefont
  {Sommers}()}]{Zyczkowski01}%
  \BibitemOpen
  \bibfield  {author} {\bibinfo {author} {\bibfnamefont {K.}~\bibnamefont
  {Życzkowski}}\ and\ \bibinfo {author} {\bibfnamefont {H.-J.}\ \bibnamefont
  {Sommers}},\ }\bibfield  {title} {\enquote {\bibinfo {title} {{Induced
  Measures in the Space of Mixed Quantum States}},}\ }\href
  {http://stacks.iop.org/0305-4470/34/i=35/a=335} {\bibfield  {journal}
  {\bibinfo  {journal} {Journal of Physics A: Mathematical and General}\
  }\textbf {\bibinfo {volume} {34}},\ \bibinfo {pages} {7111}}\BibitemShut
  {NoStop}%
\bibitem [{\citenamefont {Życzkowski}\ and\ \citenamefont
  {Sommers}(2003)}]{ZyczkowskiHS}%
  \BibitemOpen
  \bibfield  {author} {\bibinfo {author} {\bibfnamefont {K.}~\bibnamefont
  {Życzkowski}}\ and\ \bibinfo {author} {\bibfnamefont {H.-J.}\ \bibnamefont
  {Sommers}},\ }\bibfield  {title} {\enquote {\bibinfo {title}
  {{Hilbert–Schmidt Volume of the Set of Mixed Quantum States}},}\ }\href
  {http://dx.doi.org/10.1088/0305-4470/36/39/310} {\bibfield  {journal}
  {\bibinfo  {journal} {Journal of Physics A: Mathematical and General}\
  }\textbf {\bibinfo {volume} {36}},\ \bibinfo {pages} {10115} (\bibinfo {year}
  {2003})}\BibitemShut {NoStop}%
\bibitem [{\citenamefont {Sommers}\ and\ \citenamefont
  {Życzkowski}()}]{ZyczkowskiBures}%
  \BibitemOpen
  \bibfield  {author} {\bibinfo {author} {\bibfnamefont {H.-J.}\ \bibnamefont
  {Sommers}}\ and\ \bibinfo {author} {\bibfnamefont {K.}~\bibnamefont
  {Życzkowski}},\ }\bibfield  {title} {\enquote {\bibinfo {title} {{Bures
  Volume of the Set of Mixed Quantum States}},}\ }\href
  {http://stacks.iop.org/0305-4470/36/i=39/a=308} {\bibfield  {journal}
  {\bibinfo  {journal} {Journal of Physics A: Mathematical and General}\
  }\textbf {\bibinfo {volume} {36}},\ \bibinfo {pages} {10083}}\BibitemShut
  {NoStop}%
\bibitem [{\citenamefont {Novo}\ \emph {et~al.}(2013)\citenamefont {Novo},
  \citenamefont {Moroder},\ and\ \citenamefont
  {G\"uhne}}]{symmetricMultiparticleEntanglement}%
  \BibitemOpen
  \bibfield  {author} {\bibinfo {author} {\bibfnamefont {L.}~\bibnamefont
  {Novo}}, \bibinfo {author} {\bibfnamefont {T.}~\bibnamefont {Moroder}}, \
  and\ \bibinfo {author} {\bibfnamefont {O.}~\bibnamefont {G\"uhne}},\
  }\bibfield  {title} {\enquote {\bibinfo {title} {{Genuine Multiparticle
  Entanglement of Permutationally Invariant States}},}\ }\href
  {http://link.aps.org/doi/10.1103/PhysRevA.88.012305} {\bibfield  {journal}
  {\bibinfo  {journal} {Physical Review A}\ }\textbf {\bibinfo {volume} {88}},\
  \bibinfo {pages} {012305} (\bibinfo {year} {2013})}\BibitemShut {NoStop}%
\bibitem [{\citenamefont {Dicke}(1954)}]{Dicke54}%
  \BibitemOpen
  \bibfield  {author} {\bibinfo {author} {\bibfnamefont {R.~H.}\ \bibnamefont
  {Dicke}},\ }\bibfield  {title} {\enquote {\bibinfo {title} {{Coherence in
  Spontaneous Radiation Processes}},}\ }\href
  {http://link.aps.org/doi/10.1103/PhysRev.93.99} {\bibfield  {journal}
  {\bibinfo  {journal} {Physical Review}\ }\textbf {\bibinfo {volume} {93}},\
  \bibinfo {pages} {99} (\bibinfo {year} {1954})}\BibitemShut {NoStop}%
\bibitem [{\citenamefont {Gross}\ and\ \citenamefont
  {Haroche}(1982)}]{superrad.original}%
  \BibitemOpen
  \bibfield  {author} {\bibinfo {author} {\bibfnamefont {M.}~\bibnamefont
  {Gross}}\ and\ \bibinfo {author} {\bibfnamefont {S.}~\bibnamefont
  {Haroche}},\ }\bibfield  {title} {\enquote {\bibinfo {title} {{Superradiance:
  An Essay on the Theory of Collective Spontaneous Emission}},}\ }\href
  {http://dx.doi.org/10.1016/0370-1573(82)90102-8} {\bibfield  {journal}
  {\bibinfo  {journal} {Physics Reports}\ }\textbf {\bibinfo {volume} {93}},\
  \bibinfo {pages} {301} (\bibinfo {year} {1982})}\BibitemShut {NoStop}%
\bibitem [{\citenamefont {Lin}\ and\ \citenamefont
  {Yelin}(2012{\natexlab{a}})}]{superrad.yelinPRA}%
  \BibitemOpen
  \bibfield  {author} {\bibinfo {author} {\bibfnamefont {G.-D.}\ \bibnamefont
  {Lin}}\ and\ \bibinfo {author} {\bibfnamefont {S.~F.}\ \bibnamefont
  {Yelin}},\ }\bibfield  {title} {\enquote {\bibinfo {title} {{Superradiance in
  Spin-\(j\) Particles: Effects of Multiple Levels}},}\ }\href
  {http://link.aps.org/doi/10.1103/PhysRevA.85.033831} {\bibfield  {journal}
  {\bibinfo  {journal} {Physical Review A}\ }\textbf {\bibinfo {volume} {85}},\
  \bibinfo {pages} {033831} (\bibinfo {year} {2012}{\natexlab{a}})}\BibitemShut
  {NoStop}%
\bibitem [{\citenamefont {Lin}\ and\ \citenamefont
  {Yelin}(2012{\natexlab{b}})}]{superrad.yelinBook}%
  \BibitemOpen
  \bibfield  {author} {\bibinfo {author} {\bibfnamefont {G.-D.}\ \bibnamefont
  {Lin}}\ and\ \bibinfo {author} {\bibfnamefont {S.~F.}\ \bibnamefont
  {Yelin}},\ }\href
  {http://www.sciencedirect.com/science/article/pii/B9780123964823000065}
  {\emph {\bibinfo {title} {{Chapter 6 - Superradiance: An Integrated Approach
  to Cooperative Effects in Various Systems}, editor = {Paul Berman, Ennio
  Arimondo and Chun Lin}, booktitle.hide = {Advances in Atomic, Molecular, and
  Optical Physics}}}},\ \bibinfo {series} {Advances In Atomic, Molecular, and
  Optical Physics}, Vol.~\bibinfo {volume} {61}\ (\bibinfo  {publisher}
  {Academic Press},\ \bibinfo {year} {2012})\ pp.\ \bibinfo {pages} {295 --
  329}\BibitemShut {NoStop}%
\bibitem [{\citenamefont {Breuer}(2007)}]{Breuer2007Theory}%
  \BibitemOpen
  \bibfield  {author} {\bibinfo {author} {\bibfnamefont {H.-P.}\ \bibnamefont
  {Breuer}},\ }\href {http://www.worldcat.org/isbn/9780199213900} {\emph
  {\bibinfo {title} {{The Theory of Open Quantum Systems}}}}\ (\bibinfo
  {publisher} {Oxford University Press},\ \bibinfo {year} {2007})\BibitemShut
  {NoStop}%
\bibitem [{\citenamefont {Scully}\ and\ \citenamefont
  {Zubairy}(1997)}]{scully1997quantum}%
  \BibitemOpen
  \bibfield  {author} {\bibinfo {author} {\bibfnamefont {M.}~\bibnamefont
  {Scully}}\ and\ \bibinfo {author} {\bibfnamefont {S.}~\bibnamefont
  {Zubairy}},\ }\href {http://books.google.com/books?id=20ISsQCKKmQC} {\emph
  {\bibinfo {title} {{Quantum Optics}}}}\ (\bibinfo  {publisher} {Cambridge
  University Press},\ \bibinfo {year} {1997})\BibitemShut {NoStop}%
\bibitem [{\citenamefont {Hall}\ \emph {et~al.}(2014)\citenamefont {Hall},
  \citenamefont {Cresser}, \citenamefont {Li},\ and\ \citenamefont
  {Andersson}}]{NonMarkovianity}%
  \BibitemOpen
  \bibfield  {author} {\bibinfo {author} {\bibfnamefont {M.~J.~W.}\
  \bibnamefont {Hall}}, \bibinfo {author} {\bibfnamefont {J.~D.}\ \bibnamefont
  {Cresser}}, \bibinfo {author} {\bibfnamefont {L.}~\bibnamefont {Li}}, \ and\
  \bibinfo {author} {\bibfnamefont {E.}~\bibnamefont {Andersson}},\ }\bibfield
  {title} {\enquote {\bibinfo {title} {{Canonical Form of Master Equations and
  Characterization of Non-Markovianity}},}\ }\href
  {http://link.aps.org/doi/10.1103/PhysRevA.89.042120} {\bibfield  {journal}
  {\bibinfo  {journal} {Physical Review A}\ }\textbf {\bibinfo {volume} {89}},\
  \bibinfo {pages} {042120} (\bibinfo {year} {2014})}\BibitemShut {NoStop}%
\bibitem [{\citenamefont {Cetnar}(2006)}]{Bateman2006}%
  \BibitemOpen
  \bibfield  {author} {\bibinfo {author} {\bibfnamefont {J.}~\bibnamefont
  {Cetnar}},\ }\bibfield  {title} {\enquote {\bibinfo {title} {{General
  Solution of Bateman Equations for Nuclear Transmutations}},}\ }\href
  {http://www.sciencedirect.com/science/article/pii/S0306454906000284}
  {\bibfield  {journal} {\bibinfo  {journal} {Annals of Nuclear Energy}\
  }\textbf {\bibinfo {volume} {33}},\ \bibinfo {pages} {640 } (\bibinfo {year}
  {2006})}\BibitemShut {NoStop}%
\bibitem [{\citenamefont {Drummond}\ and\ \citenamefont
  {Carmichael}(1978)}]{Drummund1978}%
  \BibitemOpen
  \bibfield  {author} {\bibinfo {author} {\bibfnamefont {P.}~\bibnamefont
  {Drummond}}\ and\ \bibinfo {author} {\bibfnamefont {H.}~\bibnamefont
  {Carmichael}},\ }\bibfield  {title} {\enquote {\bibinfo {title} {{Volterra
  Cycles and the Cooperative Fluorescence Critical Point}},}\ }\href
  {http://www.sciencedirect.com/science/article/pii/0030401878901980}
  {\bibfield  {journal} {\bibinfo  {journal} {Optics Communications}\ }\textbf
  {\bibinfo {volume} {27}},\ \bibinfo {pages} {160 } (\bibinfo {year}
  {1978})}\BibitemShut {NoStop}%
\bibitem [{\citenamefont {Drummond}(1980)}]{Drummond1980}%
  \BibitemOpen
  \bibfield  {author} {\bibinfo {author} {\bibfnamefont {P.~D.}\ \bibnamefont
  {Drummond}},\ }\bibfield  {title} {\enquote {\bibinfo {title} {{Observables
  and Moments of Cooperative Resonance Fluorescence}},}\ }\href
  {http://link.aps.org/doi/10.1103/PhysRevA.22.1179} {\bibfield  {journal}
  {\bibinfo  {journal} {Physical Review A}\ }\textbf {\bibinfo {volume} {22}},\
  \bibinfo {pages} {1179} (\bibinfo {year} {1980})}\BibitemShut {NoStop}%
\bibitem [{\citenamefont {{Aparicio Alcalde, M. and Cardenas, A. H. and
  Svaiter, N. F. and Bezerra, V. B.}}(2010)}]{superrad2010}%
  \BibitemOpen
  \bibfield  {author} {\bibinfo {author} {\bibnamefont {{Aparicio Alcalde, M.
  and Cardenas, A. H. and Svaiter, N. F. and Bezerra, V. B.}}},\ }\bibfield
  {title} {\enquote {\bibinfo {title} {{Entangled States and Superradiant Phase
  Transitions}},}\ }\href {http://link.aps.org/doi/10.1103/PhysRevA.81.032335}
  {\bibfield  {journal} {\bibinfo  {journal} {Physical Review A}\ }\textbf
  {\bibinfo {volume} {81}},\ \bibinfo {pages} {032335} (\bibinfo {year}
  {2010})}\BibitemShut {NoStop}%
\bibitem [{\citenamefont {Schneider}\ and\ \citenamefont
  {Milburn}(2002)}]{DrivenSuperradAlmost}%
  \BibitemOpen
  \bibfield  {author} {\bibinfo {author} {\bibfnamefont {S.}~\bibnamefont
  {Schneider}}\ and\ \bibinfo {author} {\bibfnamefont {G.~J.}\ \bibnamefont
  {Milburn}},\ }\bibfield  {title} {\enquote {\bibinfo {title} {{Entanglement
  in the Steady State of a Collective-Angular-Momentum (Dicke) Model}},}\
  }\href {http://link.aps.org/doi/10.1103/PhysRevA.65.042107} {\bibfield
  {journal} {\bibinfo  {journal} {Physical Review A}\ }\textbf {\bibinfo
  {volume} {65}},\ \bibinfo {pages} {042107} (\bibinfo {year}
  {2002})}\BibitemShut {NoStop}%
\bibitem [{\citenamefont {Gonz\'alez-Tudela}\ and\ \citenamefont
  {Porras}(2013)}]{DrivenSuperradPorras}%
  \BibitemOpen
  \bibfield  {author} {\bibinfo {author} {\bibfnamefont {A.}~\bibnamefont
  {Gonz\'alez-Tudela}}\ and\ \bibinfo {author} {\bibfnamefont {D.}~\bibnamefont
  {Porras}},\ }\bibfield  {title} {\enquote {\bibinfo {title} {{Mesoscopic
  Entanglement Induced by Spontaneous Emission in Solid-State Quantum
  Optics}},}\ }\href {http://link.aps.org/doi/10.1103/PhysRevLett.110.080502}
  {\bibfield  {journal} {\bibinfo  {journal} {Physical Review Letters}\
  }\textbf {\bibinfo {volume} {110}},\ \bibinfo {pages} {080502} (\bibinfo
  {year} {2013})}\BibitemShut {NoStop}%
\bibitem [{\citenamefont {Irish}\ \emph {et~al.}(2005)\citenamefont {Irish},
  \citenamefont {Gea-Banacloche}, \citenamefont {Martin},\ and\ \citenamefont
  {Schwab}}]{RWADerivation1}%
  \BibitemOpen
  \bibfield  {author} {\bibinfo {author} {\bibfnamefont {E.~K.}\ \bibnamefont
  {Irish}}, \bibinfo {author} {\bibfnamefont {J.}~\bibnamefont
  {Gea-Banacloche}}, \bibinfo {author} {\bibfnamefont {I.}~\bibnamefont
  {Martin}}, \ and\ \bibinfo {author} {\bibfnamefont {K.~C.}\ \bibnamefont
  {Schwab}},\ }\bibfield  {title} {\enquote {\bibinfo {title} {{Dynamics of a
  Two-Level System Strongly Coupled to a High-Frequency Quantum Oscillator}},}\
  }\href {http://link.aps.org/doi/10.1103/PhysRevB.72.195410} {\bibfield
  {journal} {\bibinfo  {journal} {Physical Review B}\ }\textbf {\bibinfo
  {volume} {72}},\ \bibinfo {pages} {195410} (\bibinfo {year}
  {2005})}\BibitemShut {NoStop}%
\bibitem [{\citenamefont {Frasca}(2003)}]{RWADerivation2}%
  \BibitemOpen
  \bibfield  {author} {\bibinfo {author} {\bibfnamefont {M.}~\bibnamefont
  {Frasca}},\ }\bibfield  {title} {\enquote {\bibinfo {title} {{A Modern Review
  of the Two-Level Approximation }},}\ }\href
  {http://www.sciencedirect.com/science/article/pii/S0003491603000782}
  {\bibfield  {journal} {\bibinfo  {journal} {Annals of Physics}\ }\textbf
  {\bibinfo {volume} {306}},\ \bibinfo {pages} {193 } (\bibinfo {year}
  {2003})}\BibitemShut {NoStop}%
\bibitem [{\citenamefont {T\'oth}\ \emph {et~al.}(2007)\citenamefont {T\'oth},
  \citenamefont {Knapp}, \citenamefont {G\"uhne},\ and\ \citenamefont
  {Briegel}}]{OptimalSpinSqueezingParameterEarly}%
  \BibitemOpen
  \bibfield  {author} {\bibinfo {author} {\bibfnamefont {G.}~\bibnamefont
  {T\'oth}}, \bibinfo {author} {\bibfnamefont {C.}~\bibnamefont {Knapp}},
  \bibinfo {author} {\bibfnamefont {O.}~\bibnamefont {G\"uhne}}, \ and\
  \bibinfo {author} {\bibfnamefont {H.~J.}\ \bibnamefont {Briegel}},\
  }\bibfield  {title} {\enquote {\bibinfo {title} {{Optimal Spin Squeezing
  Inequalities Detect Bound Entanglement in Spin Models}},}\ }\href
  {http://link.aps.org/doi/10.1103/PhysRevLett.99.250405} {\bibfield  {journal}
  {\bibinfo  {journal} {Physical Review Letters}\ }\textbf {\bibinfo {volume}
  {99}},\ \bibinfo {pages} {250405} (\bibinfo {year} {2007})}\BibitemShut
  {NoStop}%
\bibitem [{\citenamefont {Vitagliano}\ \emph {et~al.}(2014)\citenamefont
  {Vitagliano}, \citenamefont {Apellaniz}, \citenamefont {Egusquiza},\ and\
  \citenamefont {Tóth}}]{OptimalSpinSqueezingParameter}%
  \BibitemOpen
  \bibfield  {author} {\bibinfo {author} {\bibfnamefont {G.}~\bibnamefont
  {Vitagliano}}, \bibinfo {author} {\bibfnamefont {I.}~\bibnamefont
  {Apellaniz}}, \bibinfo {author} {\bibfnamefont {I.~L.}\ \bibnamefont
  {Egusquiza}}, \ and\ \bibinfo {author} {\bibfnamefont {G.}~\bibnamefont
  {Tóth}},\ }\bibfield  {title} {\enquote {\bibinfo {title} {{Spin Squeezing
  and Entanglement for an Arbitrary Spin}},}\ }\href
  {http://link.aps.org/doi/10.1103/PhysRevA.89.032307} {\bibfield  {journal}
  {\bibinfo  {journal} {Physical Review A}\ }\textbf {\bibinfo {volume} {89}},\
  \bibinfo {pages} {032307} (\bibinfo {year} {2014})}\BibitemShut {NoStop}%
\bibitem [{\citenamefont {Lee}\ and\ \citenamefont
  {Chan}(2013)}]{SpinSqueeze2013}%
  \BibitemOpen
  \bibfield  {author} {\bibinfo {author} {\bibfnamefont {T.~E.}\ \bibnamefont
  {Lee}}\ and\ \bibinfo {author} {\bibfnamefont {C.-K.}\ \bibnamefont {Chan}},\
  }\bibfield  {title} {\enquote {\bibinfo {title} {{Dissipative
  Transverse-Field Ising model: Steady-State Correlations and Spin
  Squeezing}},}\ }\href {http://link.aps.org/doi/10.1103/PhysRevA.88.063811}
  {\bibfield  {journal} {\bibinfo  {journal} {Physical Review A}\ }\textbf
  {\bibinfo {volume} {88}},\ \bibinfo {pages} {063811} (\bibinfo {year}
  {2013})}\BibitemShut {NoStop}%
\bibitem [{\citenamefont {Kitagawa}\ and\ \citenamefont
  {Ueda}(1993)}]{KitagawaUeda}%
  \BibitemOpen
  \bibfield  {author} {\bibinfo {author} {\bibfnamefont {M.}~\bibnamefont
  {Kitagawa}}\ and\ \bibinfo {author} {\bibfnamefont {M.}~\bibnamefont
  {Ueda}},\ }\bibfield  {title} {\enquote {\bibinfo {title} {{Squeezed Spin
  States}},}\ }\href {http://link.aps.org/doi/10.1103/PhysRevA.47.5138}
  {\bibfield  {journal} {\bibinfo  {journal} {Physical Review A}\ }\textbf
  {\bibinfo {volume} {47}},\ \bibinfo {pages} {5138} (\bibinfo {year}
  {1993})}\BibitemShut {NoStop}%
\bibitem [{\citenamefont {Wang}\ and\ \citenamefont
  {Sanders}(2003)}]{SymmetricSpinSqueezing}%
  \BibitemOpen
  \bibfield  {author} {\bibinfo {author} {\bibfnamefont {X.}~\bibnamefont
  {Wang}}\ and\ \bibinfo {author} {\bibfnamefont {B.~C.}\ \bibnamefont
  {Sanders}},\ }\bibfield  {title} {\enquote {\bibinfo {title} {{Spin Squeezing
  and Pairwise Entanglement for Symmetric Multiqubit States}},}\ }\href
  {http://link.aps.org/doi/10.1103/PhysRevA.68.012101} {\bibfield  {journal}
  {\bibinfo  {journal} {Physical Review A}\ }\textbf {\bibinfo {volume} {68}},\
  \bibinfo {pages} {012101} (\bibinfo {year} {2003})}\BibitemShut {NoStop}%
\bibitem [{\citenamefont {Ulam-Orgikh}\ and\ \citenamefont
  {Kitagawa}(2001)}]{RamseySpinSqueezing}%
  \BibitemOpen
  \bibfield  {author} {\bibinfo {author} {\bibfnamefont {D.}~\bibnamefont
  {Ulam-Orgikh}}\ and\ \bibinfo {author} {\bibfnamefont {M.}~\bibnamefont
  {Kitagawa}},\ }\bibfield  {title} {\enquote {\bibinfo {title} {{Spin
  Squeezing and Decoherence Limit in Ramsey Spectroscopy}},}\ }\href
  {http://link.aps.org/doi/10.1103/PhysRevA.64.052106} {\bibfield  {journal}
  {\bibinfo  {journal} {Physical Review A}\ }\textbf {\bibinfo {volume} {64}},\
  \bibinfo {pages} {052106} (\bibinfo {year} {2001})}\BibitemShut {NoStop}%
\bibitem [{\citenamefont {T{\'o}th}\ and\ \citenamefont
  {G{\"u}hne}(2009)}]{SymmetricEquivalentsPRL}%
  \BibitemOpen
  \bibfield  {author} {\bibinfo {author} {\bibfnamefont {G.}~\bibnamefont
  {T{\'o}th}}\ and\ \bibinfo {author} {\bibfnamefont {O.}~\bibnamefont
  {G{\"u}hne}},\ }\bibfield  {title} {\enquote {\bibinfo {title} {{Entanglement
  and Permutational Symmetry}},}\ }\href
  {http://link.aps.org/doi/10.1103/PhysRevLett.102.170503} {\bibfield
  {journal} {\bibinfo  {journal} {Physical Review Letters}\ }\textbf {\bibinfo
  {volume} {102}},\ \bibinfo {pages} {170503} (\bibinfo {year}
  {2009})}\BibitemShut {NoStop}%
\bibitem [{\citenamefont {Sharp}(1960)}]{Sharp1960}%
  \BibitemOpen
  \bibfield  {author} {\bibinfo {author} {\bibfnamefont {R.~T.}\ \bibnamefont
  {Sharp}},\ }\bibfield  {title} {\enquote {\bibinfo {title} {{Simple
  Derivation of the Clebsch-Gordan Coefficients}},}\ }\href
  {http://dx.doi.org/10.1119/1.1935073} {\bibfield  {journal} {\bibinfo
  {journal} {American Journal of Physics}\ }\textbf {\bibinfo {volume} {28}},\
  \bibinfo {pages} {116+} (\bibinfo {year} {1960})}\BibitemShut {NoStop}%
\bibitem [{\citenamefont {Schertler}\ and\ \citenamefont
  {Thoma}(1996)}]{Schertler1996Combinatorial}%
  \BibitemOpen
  \bibfield  {author} {\bibinfo {author} {\bibfnamefont {K.}~\bibnamefont
  {Schertler}}\ and\ \bibinfo {author} {\bibfnamefont {M.~H.}\ \bibnamefont
  {Thoma}},\ }\bibfield  {title} {\enquote {\bibinfo {title} {{Combinatorial
  Computation of Clebsch-Gordan Coefficients}},}\ }\href
  {http://dx.doi.org/10.1002/andp.2065080108} {\bibfield  {journal} {\bibinfo
  {journal} {Annals of Physics}\ }\textbf {\bibinfo {volume} {508}},\ \bibinfo
  {pages} {103} (\bibinfo {year} {1996})}\BibitemShut {NoStop}%
\bibitem [{\citenamefont {Biedenharn}\ and\ \citenamefont
  {Louck}(1981)}]{Biedenharn1981}%
  \BibitemOpen
  \bibfield  {author} {\bibinfo {author} {\bibfnamefont {L.~C.}\ \bibnamefont
  {Biedenharn}}\ and\ \bibinfo {author} {\bibfnamefont {J.~D.}\ \bibnamefont
  {Louck}},\ }\href {http://www.worldcat.org/isbn/0201135078} {\emph {\bibinfo
  {title} {{Angular Momentum in Quantum Physics: Theory and Application}}}}\
  (\bibinfo  {publisher} {Addison-Wesley Publishing Company},\ \bibinfo {year}
  {1981})\BibitemShut {NoStop}%
\bibitem [{\citenamefont {Caola}(2010)}]{Coala2010}%
  \BibitemOpen
  \bibfield  {author} {\bibinfo {author} {\bibfnamefont {M.~J.}\ \bibnamefont
  {Caola}},\ }\bibfield  {title} {\enquote {\bibinfo {title} {{A Simple Short
  Derivation of the Clebsch-Gordan Coefficients}},}\ }\href
  {http://www.lajpe.org/jan10/15_Michael_Caola.pdf} {\bibfield  {journal}
  {\bibinfo  {journal} {Latin-American Journal of Physics Education}\ }\textbf
  {\bibinfo {volume} {4}},\ \bibinfo {pages} {84} (\bibinfo {year}
  {2010})}\BibitemShut {NoStop}%
\bibitem [{\citenamefont {O'Hara}(2001)}]{OHara2001}%
  \BibitemOpen
  \bibfield  {author} {\bibinfo {author} {\bibfnamefont {P.}~\bibnamefont
  {O'Hara}},\ }\bibfield  {title} {\enquote {\bibinfo {title} {{Clebsch-Gordan
  Coefficients and the Binomial Distribution}},}\ }\href
  {http://arxiv.org/abs/quant-ph/0112096} {\bibfield  {journal} {\bibinfo
  {journal} {arXiv:quant-ph/0112096}\ } (\bibinfo {year} {2001})}\BibitemShut
  {NoStop}%
\bibitem [{\citenamefont {Guseinov}\ \emph {et~al.}(2009)\citenamefont
  {Guseinov}, \citenamefont {Mamedov},\ and\ \citenamefont
  {\c{C}opuro\u{g}lu}}]{Guseinov2009}%
  \BibitemOpen
  \bibfield  {author} {\bibinfo {author} {\bibfnamefont {I.~I.}\ \bibnamefont
  {Guseinov}}, \bibinfo {author} {\bibfnamefont {B.~A.}\ \bibnamefont
  {Mamedov}}, \ and\ \bibinfo {author} {\bibfnamefont {E.}~\bibnamefont
  {\c{C}opuro\u{g}lu}},\ }\bibfield  {title} {\enquote {\bibinfo {title} {{Use
  of Binomial Coefficients in Fast and Accurate Calculation of Clebsch–Gordan
  and Gaunt Coefficients, and Wigner n - j Symbols}},}\ }\href
  {http://dx.doi.org/10.1142/S0219633609004782} {\bibfield  {journal} {\bibinfo
   {journal} {Journal of Theoretical and Computational Chemistry}\ }\textbf
  {\bibinfo {volume} {08}},\ \bibinfo {pages} {251} (\bibinfo {year}
  {2009})}\BibitemShut {NoStop}%
\bibitem [{\citenamefont {Louck}(2008)}]{Louck2008}%
  \BibitemOpen
  \bibfield  {author} {\bibinfo {author} {\bibfnamefont {J.~D.}\ \bibnamefont
  {Louck}},\ }\bibfield  {title} {\enquote {\bibinfo {title} {{Properties of
  Clebsch–Gordan Numbers}},}\ }\href
  {http://dx.doi.org/10.1088/1742-6596/104/1/012015} {\bibfield  {journal}
  {\bibinfo  {journal} {Journal of Physics: Conference Series}\ }\textbf
  {\bibinfo {volume} {104}},\ \bibinfo {pages} {012015+} (\bibinfo {year}
  {2008})}\BibitemShut {NoStop}%
\bibitem [{\citenamefont {Ulfbeck}(2007)}]{Ulfbeck2007}%
  \BibitemOpen
  \bibfield  {author} {\bibinfo {author} {\bibfnamefont {O.}~\bibnamefont
  {Ulfbeck}},\ }\bibfield  {title} {\enquote {\bibinfo {title} {{Simple
  Derivation of the Vector Addition Coefficients}},}\ }\href
  {http://dx.doi.org/10.1063/1.2768437} {\bibfield  {journal} {\bibinfo
  {journal} {Journal of Mathematical Physics}\ }\textbf {\bibinfo {volume}
  {48}},\ \bibinfo {pages} {082105+} (\bibinfo {year} {2007})}\BibitemShut
  {NoStop}%
\bibitem [{\citenamefont {{Wolfe}}(2014)}]{WolfeCG}%
  \BibitemOpen
  \bibfield  {author} {\bibinfo {author} {\bibfnamefont {E.}~\bibnamefont
  {{Wolfe}}},\ }\bibfield  {title} {\enquote {\bibinfo {title} {{Combinatorial
  Derivation of the Clebsch-Gordan Coefficients}},}\ }\href
  {https://www.researchgate.net/publication/259465309} {\bibfield  {journal}
  {\bibinfo  {journal} {researchgate.net/publication/259465309}\ } (\bibinfo
  {year} {2014})}\BibitemShut {NoStop}%
\bibitem [{\citenamefont {Tóth}\ and\ \citenamefont
  {Gühne}(2006)}]{EntanglementOnlyTwoBodyToth}%
  \BibitemOpen
  \bibfield  {author} {\bibinfo {author} {\bibfnamefont {G.}~\bibnamefont
  {Tóth}}\ and\ \bibinfo {author} {\bibfnamefont {O.}~\bibnamefont {Gühne}},\
  }\bibfield  {title} {\enquote {\bibinfo {title} {Detection of multipartite
  entanglement with two-body correlations},}\ }\href
  {http://dx.doi.org/10.1007/s00340-005-2057-1} {\bibfield  {journal} {\bibinfo
   {journal} {Applied Physics B}\ }\textbf {\bibinfo {volume} {82}},\ \bibinfo
  {pages} {237} (\bibinfo {year} {2006})}\BibitemShut {NoStop}%
\bibitem [{\citenamefont {{Tura}}\ \emph {et~al.}(2013)\citenamefont {{Tura}},
  \citenamefont {{Sainz}}, \citenamefont {{V{\'e}rtesi}}, \citenamefont
  {{Ac{\'{\i}}n}}, \citenamefont {{Lewenstein}},\ and\ \citenamefont
  {{Augusiak}}}]{BellTwoBodyOnlyArXiv}%
  \BibitemOpen
  \bibfield  {author} {\bibinfo {author} {\bibfnamefont {J.}~\bibnamefont
  {{Tura}}}, \bibinfo {author} {\bibfnamefont {A.~B.}\ \bibnamefont {{Sainz}}},
  \bibinfo {author} {\bibfnamefont {T.}~\bibnamefont {{V{\'e}rtesi}}}, \bibinfo
  {author} {\bibfnamefont {A.}~\bibnamefont {{Ac{\'{\i}}n}}}, \bibinfo {author}
  {\bibfnamefont {M.}~\bibnamefont {{Lewenstein}}}, \ and\ \bibinfo {author}
  {\bibfnamefont {R.}~\bibnamefont {{Augusiak}}},\ }\bibfield  {title}
  {\enquote {\bibinfo {title} {{Translationally Invariant Multipartite Bell
  Inequalities Involving only Two-Body Correlators}},}\ }\href
  {http://arxiv.org/abs/1312.0265} {\bibfield  {journal} {\bibinfo  {journal}
  {arXiv:1312.0265}\ } (\bibinfo {year} {2013})}\BibitemShut {NoStop}%
\bibitem [{\citenamefont {{Chen}}\ \emph {et~al.}(2014)\citenamefont {{Chen}},
  \citenamefont {{Su}}, \citenamefont {{Xu}}, \citenamefont {{Wu}},\ and\
  \citenamefont {{Chen}}}]{BellInequalityHomogenization}%
  \BibitemOpen
  \bibfield  {author} {\bibinfo {author} {\bibfnamefont {X.}~\bibnamefont
  {{Chen}}}, \bibinfo {author} {\bibfnamefont {H.-Y.}\ \bibnamefont {{Su}}},
  \bibinfo {author} {\bibfnamefont {Z.-P.}\ \bibnamefont {{Xu}}}, \bibinfo
  {author} {\bibfnamefont {Y.-C.}\ \bibnamefont {{Wu}}}, \ and\ \bibinfo
  {author} {\bibfnamefont {J.-L.}\ \bibnamefont {{Chen}}},\ }\bibfield  {title}
  {\enquote {\bibinfo {title} {{Quantum Nonlocality Enhanced by
  Homogenization}},}\ }\href {http://arxiv.org/abs/1406.4581} {\bibfield
  {journal} {\bibinfo  {journal} {arXiv:1406.4581}\ } (\bibinfo {year}
  {2014})}\BibitemShut {NoStop}%
\bibitem [{\citenamefont {{Bera}}(2014)}]{PrecisionViaNonlocality}%
  \BibitemOpen
  \bibfield  {author} {\bibinfo {author} {\bibfnamefont {M.~N.}\ \bibnamefont
  {{Bera}}},\ }\bibfield  {title} {\enquote {\bibinfo {title} {{Role of Quantum
  Correlation in Metrology Beyond Standard Quantum Limit}},}\ }\href
  {http://arxiv.org/abs/1405.5357} {\bibfield  {journal} {\bibinfo  {journal}
  {arXiv:1405.5357}\ } (\bibinfo {year} {2014})}\BibitemShut {NoStop}%
\bibitem [{\citenamefont {Cabello}(2013{\natexlab{a}})}]{EPSpeckerVid}%
  \BibitemOpen
  \bibfield  {author} {\bibinfo {author} {\bibfnamefont {A.}~\bibnamefont
  {Cabello}},\ }\bibfield  {title} {\enquote {\bibinfo {title} {{Ernst Specker
  and the Fundamental Theorem of Quantum Mechanics}},}\ }\href
  {http://vimeo.com/52923835} {\bibfield  {journal} {\bibinfo  {journal}
  {vimeo.com/52923835}\ } (\bibinfo {year} {2013}{\natexlab{a}})}\BibitemShut
  {NoStop}%
\bibitem [{\citenamefont {Mermin}(1993)}]{MerminContextuality}%
  \BibitemOpen
  \bibfield  {author} {\bibinfo {author} {\bibfnamefont {N.~D.}\ \bibnamefont
  {Mermin}},\ }\bibfield  {title} {\enquote {\bibinfo {title} {{Hidden
  Variables and the Two Theorems of John Bell}},}\ }\href
  {http://link.aps.org/doi/10.1103/RevModPhys.65.803} {\bibfield  {journal}
  {\bibinfo  {journal} {Reviews of Modern Physics}\ }\textbf {\bibinfo {volume}
  {65}},\ \bibinfo {pages} {803} (\bibinfo {year} {1993})}\BibitemShut
  {NoStop}%
\bibitem [{\citenamefont {Spekkens}(2005)}]{PreperationContextuality}%
  \BibitemOpen
  \bibfield  {author} {\bibinfo {author} {\bibfnamefont {R.~W.}\ \bibnamefont
  {Spekkens}},\ }\bibfield  {title} {\enquote {\bibinfo {title} {{Contextuality
  for Preparations, Transformations, and Unsharp Measurements}},}\ }\href
  {http://link.aps.org/doi/10.1103/PhysRevA.71.052108} {\bibfield  {journal}
  {\bibinfo  {journal} {Physical Review A}\ }\textbf {\bibinfo {volume} {71}},\
  \bibinfo {pages} {052108} (\bibinfo {year} {2005})}\BibitemShut {NoStop}%
\bibitem [{\citenamefont {Klyachko}\ \emph {et~al.}(2008)\citenamefont
  {Klyachko}, \citenamefont {Can}, \citenamefont
  {Binicio\ifmmode~\breve{g}\else \u{g}\fi{}lu},\ and\ \citenamefont
  {Shumovsky}}]{KCBS}%
  \BibitemOpen
  \bibfield  {author} {\bibinfo {author} {\bibfnamefont {A.~A.}\ \bibnamefont
  {Klyachko}}, \bibinfo {author} {\bibfnamefont {M.~A.}\ \bibnamefont {Can}},
  \bibinfo {author} {\bibfnamefont {S.}~\bibnamefont
  {Binicio\ifmmode~\breve{g}\else \u{g}\fi{}lu}}, \ and\ \bibinfo {author}
  {\bibfnamefont {A.~S.}\ \bibnamefont {Shumovsky}},\ }\bibfield  {title}
  {\enquote {\bibinfo {title} {{Simple Test for Hidden Variables in Spin-1
  Systems}},}\ }\href {http://link.aps.org/doi/10.1103/PhysRevLett.101.020403}
  {\bibfield  {journal} {\bibinfo  {journal} {Physical Review Letters}\
  }\textbf {\bibinfo {volume} {101}},\ \bibinfo {pages} {020403} (\bibinfo
  {year} {2008})}\BibitemShut {NoStop}%
\bibitem [{\citenamefont {Bub}\ and\ \citenamefont
  {Stairs}(2009)}]{contextualitycircle}%
  \BibitemOpen
  \bibfield  {author} {\bibinfo {author} {\bibfnamefont {J.}~\bibnamefont
  {Bub}}\ and\ \bibinfo {author} {\bibfnamefont {A.}~\bibnamefont {Stairs}},\
  }\bibfield  {title} {\enquote {\bibinfo {title} {{Contextuality and
  Nonlocality in ‘No Signaling’ Theories}},}\ }\href
  {http://dx.doi.org/10.1007/s10701-009-9307-8} {\bibfield  {journal} {\bibinfo
   {journal} {Foundations of Physics}\ }\textbf {\bibinfo {volume} {39}},\
  \bibinfo {pages} {690} (\bibinfo {year} {2009})}\BibitemShut {NoStop}%
\bibitem [{\citenamefont {{Man'ko}}\ and\ \citenamefont
  {{Markovich}}(2014)}]{TwoQubitsSeparabilityContextuality}%
  \BibitemOpen
  \bibfield  {author} {\bibinfo {author} {\bibfnamefont {V.~I.}\ \bibnamefont
  {{Man'ko}}}\ and\ \bibinfo {author} {\bibfnamefont {L.~A.}\ \bibnamefont
  {{Markovich}}},\ }\bibfield  {title} {\enquote {\bibinfo {title}
  {{Separability and Entanglement of Spin \(1\) Particle Composed from Two Spin
  \(1/2\) Particles}},}\ }\href {http://arxiv.org/abs/1406.7118} {\bibfield
  {journal} {\bibinfo  {journal} {arXiv:1406.7118}\ } (\bibinfo {year}
  {2014})}\BibitemShut {NoStop}%
\bibitem [{\citenamefont {Bartlett}(2014)}]{naturecompeditorial}%
  \BibitemOpen
  \bibfield  {author} {\bibinfo {author} {\bibfnamefont {S.~D.}\ \bibnamefont
  {Bartlett}},\ }\bibfield  {title} {\enquote {\bibinfo {title} {{Quantum
  Computing: Powered by Magic}},}\ }\href
  {http://dx.doi.org/10.1038/nature13504} {\bibfield  {journal} {\bibinfo
  {journal} {Nature}\ }\textbf {\bibinfo {volume} {510}},\ \bibinfo {pages}
  {345} (\bibinfo {year} {2014})}\BibitemShut {NoStop}%
\bibitem [{\citenamefont {Howard}\ \emph {et~al.}(2014)\citenamefont {Howard},
  \citenamefont {Wallman}, \citenamefont {Veitch},\ and\ \citenamefont
  {Emerson}}]{emersoncontextuality}%
  \BibitemOpen
  \bibfield  {author} {\bibinfo {author} {\bibfnamefont {M.}~\bibnamefont
  {Howard}}, \bibinfo {author} {\bibfnamefont {J.}~\bibnamefont {Wallman}},
  \bibinfo {author} {\bibfnamefont {V.}~\bibnamefont {Veitch}}, \ and\ \bibinfo
  {author} {\bibfnamefont {J.}~\bibnamefont {Emerson}},\ }\bibfield  {title}
  {\enquote {\bibinfo {title} {{Contextuality Supplies the `Magic' for Quantum
  Computation}},}\ }\href {http://dx.doi.org/10.1038/nature13460} {\bibfield
  {journal} {\bibinfo  {journal} {Nature}\ }\textbf {\bibinfo {volume} {510}},\
  \bibinfo {pages} {351} (\bibinfo {year} {2014})}\BibitemShut {NoStop}%
\bibitem [{\citenamefont {{Cabello}}\ \emph {et~al.}(2010)\citenamefont
  {{Cabello}}, \citenamefont {{Severini}},\ and\ \citenamefont
  {{Winter}}}]{CSWOld}%
  \BibitemOpen
  \bibfield  {author} {\bibinfo {author} {\bibfnamefont {A.}~\bibnamefont
  {{Cabello}}}, \bibinfo {author} {\bibfnamefont {S.}~\bibnamefont
  {{Severini}}}, \ and\ \bibinfo {author} {\bibfnamefont {A.}~\bibnamefont
  {{Winter}}},\ }\bibfield  {title} {\enquote {\bibinfo {title}
  {{(Non-)Contextuality of Physical Theories as an Axiom}},}\ }\href
  {http://arxiv.org/abs/1010.2163} {\bibfield  {journal} {\bibinfo  {journal}
  {arXiv:1010.2163}\ } (\bibinfo {year} {2010})}\BibitemShut {NoStop}%
\bibitem [{\citenamefont {Cabello}(2013{\natexlab{b}})}]{EPOriginal}%
  \BibitemOpen
  \bibfield  {author} {\bibinfo {author} {\bibfnamefont {A.}~\bibnamefont
  {Cabello}},\ }\bibfield  {title} {\enquote {\bibinfo {title} {{Simple
  Explanation of the Quantum Violation of a Fundamental Inequality}},}\ }\href
  {http://www.ncbi.nlm.nih.gov/pubmed/23432221} {\bibfield  {journal} {\bibinfo
   {journal} {Physical Review Letters}\ }\textbf {\bibinfo {volume} {110}},\
  \bibinfo {pages} {060402} (\bibinfo {year} {2013}{\natexlab{b}})}\BibitemShut
  {NoStop}%
\bibitem [{\citenamefont {Rabelo}\ \emph {et~al.}(2014)\citenamefont {Rabelo},
  \citenamefont {Duarte}, \citenamefont {López-Tarrida}, \citenamefont
  {Cunha},\ and\ \citenamefont {Cabello}}]{CabelloMultigraph}%
  \BibitemOpen
  \bibfield  {author} {\bibinfo {author} {\bibfnamefont {R.}~\bibnamefont
  {Rabelo}}, \bibinfo {author} {\bibfnamefont {C.}~\bibnamefont {Duarte}},
  \bibinfo {author} {\bibfnamefont {A.~J.}\ \bibnamefont {López-Tarrida}},
  \bibinfo {author} {\bibfnamefont {M.~T.}\ \bibnamefont {Cunha}}, \ and\
  \bibinfo {author} {\bibfnamefont {A.}~\bibnamefont {Cabello}},\ }\bibfield
  {title} {\enquote {\bibinfo {title} {{Multigraph Approach to Quantum
  Nonlocality}},}\ }\href {http://arxiv.org/abs/1407.5340} {\bibfield
  {journal} {\bibinfo  {journal} {arXiv:1407.5340}\ } (\bibinfo {year}
  {2014})}\BibitemShut {NoStop}%
\bibitem [{\citenamefont {Knuth}(1993)}]{knuth1993sandwich}%
  \BibitemOpen
  \bibfield  {author} {\bibinfo {author} {\bibfnamefont {D.~E.}\ \bibnamefont
  {Knuth}},\ }\bibfield  {title} {\enquote {\bibinfo {title} {{The Sandwich
  Theorem}},}\ }\href
  {http://www.combinatorics.org/ojs/index.php/eljc/article/view/v1i1a1}
  {\bibfield  {journal} {\bibinfo  {journal} {Electronic Journal of
  Combinatorics}\ }\textbf {\bibinfo {volume} {1}},\ \bibinfo {pages} {A1}
  (\bibinfo {year} {1993})}\BibitemShut {NoStop}%
\bibitem [{\citenamefont {Lovász}(1979)}]{LovaszHandle}%
  \BibitemOpen
  \bibfield  {author} {\bibinfo {author} {\bibfnamefont {L.}~\bibnamefont
  {Lovász}},\ }\bibfield  {title} {\enquote {\bibinfo {title} {{On the Shannon
  Capacity of a Graph}},}\ }\href
  {http://ieeexplore.ieee.org/xpl/articleDetails.jsp?arnumber=1055985}
  {\bibfield  {journal} {\bibinfo  {journal} {IEEE Transactions on Information
  Theory}\ }\textbf {\bibinfo {volume} {25}},\ \bibinfo {pages} {1} (\bibinfo
  {year} {1979})}\BibitemShut {NoStop}%
\bibitem [{\citenamefont {Hardy}(2001)}]{MultiPrinciples2001}%
  \BibitemOpen
  \bibfield  {author} {\bibinfo {author} {\bibfnamefont {L.}~\bibnamefont
  {Hardy}},\ }\bibfield  {title} {\enquote {\bibinfo {title} {{Quantum Theory
  from Five Reasonable Axioms}},}\ }\href
  {http://arxiv.org/abs/quant-ph/0101012} {\bibfield  {journal} {\bibinfo
  {journal} {arXiv:quant-ph/0101012}\ } (\bibinfo {year} {2001})}\BibitemShut
  {NoStop}%
\bibitem [{\citenamefont {Masanes}\ and\ \citenamefont
  {Müller}(2011)}]{MultiPrinciples2011Masanes}%
  \BibitemOpen
  \bibfield  {author} {\bibinfo {author} {\bibfnamefont {L.}~\bibnamefont
  {Masanes}}\ and\ \bibinfo {author} {\bibfnamefont {M.~P.}\ \bibnamefont
  {Müller}},\ }\bibfield  {title} {\enquote {\bibinfo {title} {{A Derivation
  of Quantum Theory from Physical Requirements}},}\ }\href
  {http://stacks.iop.org/1367-2630/13/i=6/a=063001} {\bibfield  {journal}
  {\bibinfo  {journal} {New Journal of Physics}\ }\textbf {\bibinfo {volume}
  {13}},\ \bibinfo {pages} {063001} (\bibinfo {year} {2011})}\BibitemShut
  {NoStop}%
\bibitem [{\citenamefont {Chiribella}\ \emph {et~al.}(2011)\citenamefont
  {Chiribella}, \citenamefont {D’Ariano},\ and\ \citenamefont
  {Perinotti}}]{MultiPrinciples2011Chiribella}%
  \BibitemOpen
  \bibfield  {author} {\bibinfo {author} {\bibfnamefont {G.}~\bibnamefont
  {Chiribella}}, \bibinfo {author} {\bibfnamefont {G.~M.}\ \bibnamefont
  {D’Ariano}}, \ and\ \bibinfo {author} {\bibfnamefont {P.}~\bibnamefont
  {Perinotti}},\ }\bibfield  {title} {\enquote {\bibinfo {title}
  {{Informational Derivation of Quantum Theory}},}\ }\href
  {http://link.aps.org/doi/10.1103/PhysRevA.84.012311} {\bibfield  {journal}
  {\bibinfo  {journal} {Physical Review A}\ }\textbf {\bibinfo {volume} {84}},\
  \bibinfo {pages} {012311} (\bibinfo {year} {2011})}\BibitemShut {NoStop}%
\bibitem [{\citenamefont {de~la Torre}\ \emph {et~al.}(2012)\citenamefont
  {de~la Torre}, \citenamefont {Masanes}, \citenamefont {Short},\ and\
  \citenamefont {Müller}}]{MultiPrinciples2012}%
  \BibitemOpen
  \bibfield  {author} {\bibinfo {author} {\bibfnamefont {G.}~\bibnamefont
  {de~la Torre}}, \bibinfo {author} {\bibfnamefont {L.}~\bibnamefont
  {Masanes}}, \bibinfo {author} {\bibfnamefont {A.~J.}\ \bibnamefont {Short}},
  \ and\ \bibinfo {author} {\bibfnamefont {M.~P.}\ \bibnamefont {Müller}},\
  }\bibfield  {title} {\enquote {\bibinfo {title} {{Deriving Quantum Theory
  from Its Local Structure and Reversibility}},}\ }\href
  {http://link.aps.org/doi/10.1103/PhysRevLett.109.090403} {\bibfield
  {journal} {\bibinfo  {journal} {Physical Review Letters}\ }\textbf {\bibinfo
  {volume} {109}},\ \bibinfo {pages} {090403} (\bibinfo {year}
  {2012})}\BibitemShut {NoStop}%
\bibitem [{\citenamefont {{Cabello}}(2012)}]{EPSpecker}%
  \BibitemOpen
  \bibfield  {author} {\bibinfo {author} {\bibfnamefont {A.}~\bibnamefont
  {{Cabello}}},\ }\bibfield  {title} {\enquote {\bibinfo {title} {{Specker's
  Fundamental Principle of Quantum Mechanics}},}\ }\href
  {http://arxiv.org/abs/1212.1756} {\bibfield  {journal} {\bibinfo  {journal}
  {arXiv:1212.1756}\ } (\bibinfo {year} {2012})}\BibitemShut {NoStop}%
\bibitem [{\citenamefont {Cabello}\ \emph {et~al.}(2013)\citenamefont
  {Cabello}, \citenamefont {Danielsen}, \citenamefont {López-Tarrida},\ and\
  \citenamefont {Portillo}}]{EPExampleGraphs}%
  \BibitemOpen
  \bibfield  {author} {\bibinfo {author} {\bibfnamefont {A.}~\bibnamefont
  {Cabello}}, \bibinfo {author} {\bibfnamefont {L.~E.}\ \bibnamefont
  {Danielsen}}, \bibinfo {author} {\bibfnamefont {A.~J.}\ \bibnamefont
  {López-Tarrida}}, \ and\ \bibinfo {author} {\bibfnamefont {J.~R.}\
  \bibnamefont {Portillo}},\ }\bibfield  {title} {\enquote {\bibinfo {title}
  {{Basic Exclusivity Graphs in Quantum Correlations}},}\ }\href
  {http://link.aps.org/doi/10.1103/PhysRevA.88.032104} {\bibfield  {journal}
  {\bibinfo  {journal} {Physical Review A}\ }\textbf {\bibinfo {volume} {88}},\
  \bibinfo {pages} {032104} (\bibinfo {year} {2013})}\BibitemShut {NoStop}%
\bibitem [{\citenamefont {{Henson}}(2012)}]{EPHenson}%
  \BibitemOpen
  \bibfield  {author} {\bibinfo {author} {\bibfnamefont {J.}~\bibnamefont
  {{Henson}}},\ }\bibfield  {title} {\enquote {\bibinfo {title} {{Quantum
  Contextuality from a Simple Principle?}}}\ }\href
  {http://arxiv.org/abs/1210.5978} {\bibfield  {journal} {\bibinfo  {journal}
  {arXiv:1210.5978}\ } (\bibinfo {year} {2012})}\BibitemShut {NoStop}%
\bibitem [{\citenamefont {Yan}(2013)}]{EPYan}%
  \BibitemOpen
  \bibfield  {author} {\bibinfo {author} {\bibfnamefont {B.}~\bibnamefont
  {Yan}},\ }\bibfield  {title} {\enquote {\bibinfo {title} {{Quantum
  Correlations are Tightly Bound by the Exclusivity Principle}},}\ }\href
  {http://link.aps.org/doi/10.1103/PhysRevLett.110.260406} {\bibfield
  {journal} {\bibinfo  {journal} {Physical Review Letters}\ }\textbf {\bibinfo
  {volume} {110}},\ \bibinfo {pages} {260406} (\bibinfo {year}
  {2013})}\BibitemShut {NoStop}%
\bibitem [{\citenamefont {Amaral}\ \emph {et~al.}(2014)\citenamefont {Amaral},
  \citenamefont {Cunha},\ and\ \citenamefont {Cabello}}]{EP2013}%
  \BibitemOpen
  \bibfield  {author} {\bibinfo {author} {\bibfnamefont {B.}~\bibnamefont
  {Amaral}}, \bibinfo {author} {\bibfnamefont {M.~T.}\ \bibnamefont {Cunha}}, \
  and\ \bibinfo {author} {\bibfnamefont {A.}~\bibnamefont {Cabello}},\
  }\bibfield  {title} {\enquote {\bibinfo {title} {{Exclusivity Principle
  Forbids Sets of Correlations Larger than the Quantum Set}},}\ }\href
  {http://link.aps.org/doi/10.1103/PhysRevA.89.030101} {\bibfield  {journal}
  {\bibinfo  {journal} {Physical Review A}\ }\textbf {\bibinfo {volume} {89}},\
  \bibinfo {pages} {030101} (\bibinfo {year} {2014})}\BibitemShut {NoStop}%
\bibitem [{\citenamefont {Nawareg}\ \emph {et~al.}(2013)\citenamefont
  {Nawareg}, \citenamefont {Bisesto}, \citenamefont {D'Ambrosio}, \citenamefont
  {Amselem}, \citenamefont {Sciarrino}, \citenamefont {Bourennane},\ and\
  \citenamefont {Cabello}}]{EPTwoCities}%
  \BibitemOpen
  \bibfield  {author} {\bibinfo {author} {\bibfnamefont {M.}~\bibnamefont
  {Nawareg}}, \bibinfo {author} {\bibfnamefont {F.}~\bibnamefont {Bisesto}},
  \bibinfo {author} {\bibfnamefont {V.}~\bibnamefont {D'Ambrosio}}, \bibinfo
  {author} {\bibfnamefont {E.}~\bibnamefont {Amselem}}, \bibinfo {author}
  {\bibfnamefont {F.}~\bibnamefont {Sciarrino}}, \bibinfo {author}
  {\bibfnamefont {M.}~\bibnamefont {Bourennane}}, \ and\ \bibinfo {author}
  {\bibfnamefont {A.}~\bibnamefont {Cabello}},\ }\bibfield  {title} {\enquote
  {\bibinfo {title} {{Bounding Quantum Theory with the Exclusivity Principle in
  a Two-City Experiment}},}\ }\href {http://arXiv.org/abs/1311.3495} {\bibfield
   {journal} {\bibinfo  {journal} {arXiv:1311.3495}\ } (\bibinfo {year}
  {2013})}\BibitemShut {NoStop}%
\bibitem [{\citenamefont {{Cabello}}(2014)}]{EPTsirelson}%
  \BibitemOpen
  \bibfield  {author} {\bibinfo {author} {\bibfnamefont {A.}~\bibnamefont
  {{Cabello}}},\ }\bibfield  {title} {\enquote {\bibinfo {title} {{The
  Exclusivity Principle Singles out the Quantum Violation of the Bell
  Inequality}},}\ }\href {http://arxiv.org/abs/1406.5656} {\bibfield  {journal}
  {\bibinfo  {journal} {arXiv:1406.5656}\ } (\bibinfo {year}
  {2014})}\BibitemShut {NoStop}%
\bibitem [{\citenamefont {Sorkin}(1994{\natexlab{a}})}]{SorkinOriginal}%
  \BibitemOpen
  \bibfield  {author} {\bibinfo {author} {\bibfnamefont {R.~D.}\ \bibnamefont
  {Sorkin}},\ }\bibfield  {title} {\enquote {\bibinfo {title} {{Quantum
  Mechanics as Quantum Measure Theory}},}\ }\href
  {http://www.worldscientific.com/doi/abs/10.1142/S021773239400294X} {\bibfield
   {journal} {\bibinfo  {journal} {Modern Physics Letters A}\ }\textbf
  {\bibinfo {volume} {09}},\ \bibinfo {pages} {3119} (\bibinfo {year}
  {1994}{\natexlab{a}})}\BibitemShut {NoStop}%
\bibitem [{\citenamefont
  {Sorkin}(1994{\natexlab{b}})}]{Sorkin94quantummeasure}%
  \BibitemOpen
  \bibfield  {author} {\bibinfo {author} {\bibfnamefont {R.~D.}\ \bibnamefont
  {Sorkin}},\ }\bibfield  {title} {\enquote {\bibinfo {title} {{Quantum Measure
  Theory and its Interpretation}},}\ }in\ \href
  {http://arxiv.org/abs/gr-qc/9507057} {\emph {\bibinfo {booktitle} {Quantum
  Classical Correspondence: Proceedings of the 4 th Drexel Symposium on Quantum
  Nonintegrability, held Philadelphia, September 8-11}}}\ (\bibinfo
  {publisher} {International Press},\ \bibinfo {year} {1994})\ pp.\ \bibinfo
  {pages} {229--251}\BibitemShut {NoStop}%
\bibitem [{\citenamefont {Dowker}\ and\ \citenamefont
  {Kent}(1996)}]{HistoriesReviewDowker}%
  \BibitemOpen
  \bibfield  {author} {\bibinfo {author} {\bibfnamefont {F.}~\bibnamefont
  {Dowker}}\ and\ \bibinfo {author} {\bibfnamefont {A.}~\bibnamefont {Kent}},\
  }\bibfield  {title} {\enquote {\bibinfo {title} {{On the Consistent Histories
  Approach to Quantum Mechanics}},}\ }\href
  {http://dx.doi.org/10.1007/BF02183396} {\bibfield  {journal} {\bibinfo
  {journal} {Journal of Statistical Physics}\ }\textbf {\bibinfo {volume}
  {82}},\ \bibinfo {pages} {1575} (\bibinfo {year} {1996})}\BibitemShut
  {NoStop}%
\bibitem [{\citenamefont {Niestegge}(2014)}]{AltSorkingQ1a}%
  \BibitemOpen
  \bibfield  {author} {\bibinfo {author} {\bibfnamefont {G.}~\bibnamefont
  {Niestegge}},\ }\bibfield  {title} {\enquote {\bibinfo {title} {{Super
  Quantum Probabilities and Three-Slit Experiments — Wright's Pentagon State
  and the Popescu–Rohrlich Box require Third-Order Interference}},}\ }\href
  {http://stacks.iop.org/1402-4896/2014/i=T160/a=014034} {\bibfield  {journal}
  {\bibinfo  {journal} {Physica Scripta}\ }\textbf {\bibinfo {volume} {2014}},\
  \bibinfo {pages} {014034} (\bibinfo {year} {2014})}\BibitemShut {NoStop}%
\bibitem [{\citenamefont {Niestegge}(2013)}]{AltSorkingQ1b}%
  \BibitemOpen
  \bibfield  {author} {\bibinfo {author} {\bibfnamefont {G.}~\bibnamefont
  {Niestegge}},\ }\bibfield  {title} {\enquote {\bibinfo {title} {{Three-Slit
  Experiments and Quantum Nonlocality}},}\ }\href
  {http://dx.doi.org/10.1007/s10701-013-9719-3} {\bibfield  {journal} {\bibinfo
   {journal} {Foundations of Physics}\ }\textbf {\bibinfo {volume} {43}},\
  \bibinfo {pages} {805} (\bibinfo {year} {2013})}\BibitemShut {NoStop}%
\bibitem [{\citenamefont {{Henson}}(2014)}]{SorkinQ1}%
  \BibitemOpen
  \bibfield  {author} {\bibinfo {author} {\bibfnamefont {J.}~\bibnamefont
  {{Henson}}},\ }\bibfield  {title} {\enquote {\bibinfo {title} {{Bounding
  Quantum Contextuality with Lack of Third-Order Interference}},}\ }\href
  {http://arxiv.org/abs/1406.3281} {\bibfield  {journal} {\bibinfo  {journal}
  {arXiv:1406.3281}\ } (\bibinfo {year} {2014})}\BibitemShut {NoStop}%
\bibitem [{\citenamefont {Bell}(2003)}]{BellInteractionInformation}%
  \BibitemOpen
  \bibfield  {author} {\bibinfo {author} {\bibfnamefont {A.~J.}\ \bibnamefont
  {Bell}},\ }\bibfield  {title} {\enquote {\bibinfo {title} {{The
  Co-Information Lattice}},}\ }in\ \href {http://www.rni.org/bell/nara4.pdf}
  {\emph {\bibinfo {booktitle} {in Proc. 4th Int. Symp. Independent Component
  Analysis and Blind Source Separation}}}\ (\bibinfo {year} {2003})\ pp.\
  \bibinfo {pages} {921--926}\BibitemShut {NoStop}%
\bibitem [{\citenamefont {Kochen}\ and\ \citenamefont
  {Specker}(1975)}]{KSOriginal}%
  \BibitemOpen
  \bibfield  {author} {\bibinfo {author} {\bibfnamefont {S.}~\bibnamefont
  {Kochen}}\ and\ \bibinfo {author} {\bibfnamefont {E.}~\bibnamefont
  {Specker}},\ }\bibfield  {title} {\enquote {\bibinfo {title} {{The Problem of
  Hidden Variables in Quantum Mechanics}},}\ }in\ \href
  {http://dx.doi.org/10.1007/978-94-010-1795-4_17} {\emph {\bibinfo {booktitle}
  {{The Logico-Algebraic Approach to Quantum Mechanics}}}},\ \bibinfo {series}
  {{The University of Western Ontario Series in Philosophy of Science}},
  Vol.~\bibinfo {volume} {5a},\ \bibinfo {editor} {edited by\ \bibinfo {editor}
  {\bibfnamefont {C.}~\bibnamefont {Hooker}}}\ (\bibinfo  {publisher} {Springer
  Netherlands},\ \bibinfo {year} {1975})\ pp.\ \bibinfo {pages}
  {293--328}\BibitemShut {NoStop}%
\bibitem [{\citenamefont {{Seevinck}}(2011)}]{SeevinckSpeckerTranslation}%
  \BibitemOpen
  \bibfield  {author} {\bibinfo {author} {\bibfnamefont {M.~P.}\ \bibnamefont
  {{Seevinck}}},\ }\bibfield  {title} {\enquote {\bibinfo {title} {{E. Specker:
  ''The logic of non-simultaneously decidable propositions'' (1960)}},}\ }\href
  {http://arxiv.org/abs/1103.4537} {\bibfield  {journal} {\bibinfo  {journal}
  {arXiv:1103.4537}\ } (\bibinfo {year} {2011})}\BibitemShut {NoStop}%
\bibitem [{\citenamefont
  {{Stoica}}(2014)}]{NonlocalityContextualityCommutativity}%
  \BibitemOpen
  \bibfield  {author} {\bibinfo {author} {\bibfnamefont {O.-C.}\ \bibnamefont
  {{Stoica}}},\ }\bibfield  {title} {\enquote {\bibinfo {title} {{A Principle
  of Quantumness}},}\ }\href {http://arxiv.org/abs/1402.2252} {\bibfield
  {journal} {\bibinfo  {journal} {arXiv:1402.2252}\ } (\bibinfo {year}
  {2014})}\BibitemShut {NoStop}%
\bibitem [{\citenamefont {Dowker}\ and\ \citenamefont
  {Kent}(1995)}]{HistoriesShort}%
  \BibitemOpen
  \bibfield  {author} {\bibinfo {author} {\bibfnamefont {F.}~\bibnamefont
  {Dowker}}\ and\ \bibinfo {author} {\bibfnamefont {A.}~\bibnamefont {Kent}},\
  }\bibfield  {title} {\enquote {\bibinfo {title} {{Properties of Consistent
  Histories}},}\ }\href {http://link.aps.org/doi/10.1103/PhysRevLett.75.3038}
  {\bibfield  {journal} {\bibinfo  {journal} {Physical Review Letters}\
  }\textbf {\bibinfo {volume} {75}},\ \bibinfo {pages} {3038} (\bibinfo {year}
  {1995})}\BibitemShut {NoStop}%
\bibitem [{\citenamefont {Griffiths}(2014)}]{HistoriesReview}%
  \BibitemOpen
  \bibfield  {author} {\bibinfo {author} {\bibfnamefont {R.~B.}\ \bibnamefont
  {Griffiths}},\ }\bibfield  {title} {\enquote {\bibinfo {title} {{The New
  Quantum Logic}},}\ }\href {http://dx.doi.org/10.1007/s10701-014-9802-4}
  {\bibfield  {journal} {\bibinfo  {journal} {Foundations of Physics}\ }\textbf
  {\bibinfo {volume} {44}},\ \bibinfo {pages} {610} (\bibinfo {year}
  {2014})}\BibitemShut {NoStop}%
\bibitem [{\citenamefont {Collins}\ and\ \citenamefont
  {Gisin}(2004)}]{I3322Original}%
  \BibitemOpen
  \bibfield  {author} {\bibinfo {author} {\bibfnamefont {D.}~\bibnamefont
  {Collins}}\ and\ \bibinfo {author} {\bibfnamefont {N.}~\bibnamefont
  {Gisin}},\ }\bibfield  {title} {\enquote {\bibinfo {title} {{A Relevant Two
  Qubit Bell Inequality Inequivalent to the CHSH Inequality}},}\ }\href
  {http://stacks.iop.org/0305-4470/37/i=5/a=021} {\bibfield  {journal}
  {\bibinfo  {journal} {Journal of Physics A: Mathematical and General}\
  }\textbf {\bibinfo {volume} {37}},\ \bibinfo {pages} {1775} (\bibinfo {year}
  {2004})}\BibitemShut {NoStop}%
\bibitem [{\citenamefont {P\'al}\ and\ \citenamefont
  {V\'ertesi}(2009)}]{I3322NPA1}%
  \BibitemOpen
  \bibfield  {author} {\bibinfo {author} {\bibfnamefont {K.~F.}\ \bibnamefont
  {P\'al}}\ and\ \bibinfo {author} {\bibfnamefont {T.}~\bibnamefont
  {V\'ertesi}},\ }\bibfield  {title} {\enquote {\bibinfo {title} {{Quantum
  Bounds on Bell Inequalities}},}\ }\href
  {http://link.aps.org/doi/10.1103/PhysRevA.79.022120} {\bibfield  {journal}
  {\bibinfo  {journal} {Physical Review A}\ }\textbf {\bibinfo {volume} {79}},\
  \bibinfo {pages} {022120} (\bibinfo {year} {2009})}\BibitemShut {NoStop}%
\bibitem [{\citenamefont {P\'al}\ and\ \citenamefont
  {V\'ertesi}(2010)}]{I3322NPA2}%
  \BibitemOpen
  \bibfield  {author} {\bibinfo {author} {\bibfnamefont {K.~F.}\ \bibnamefont
  {P\'al}}\ and\ \bibinfo {author} {\bibfnamefont {T.}~\bibnamefont
  {V\'ertesi}},\ }\bibfield  {title} {\enquote {\bibinfo {title} {{Maximal
  Violation of a Bipartite Three-Setting, Two-Outcome Bell Inequality using
  Infinite-Dimensional Quantum Systems}},}\ }\href
  {http://link.aps.org/doi/10.1103/PhysRevA.82.022116} {\bibfield  {journal}
  {\bibinfo  {journal} {Physical Review A}\ }\textbf {\bibinfo {volume} {82}},\
  \bibinfo {pages} {022116} (\bibinfo {year} {2010})}\BibitemShut {NoStop}%
\bibitem [{\citenamefont {Collins}\ \emph {et~al.}(2002)\citenamefont
  {Collins}, \citenamefont {Gisin}, \citenamefont {Linden}, \citenamefont
  {Massar},\ and\ \citenamefont {Popescu}}]{CGLMP02}%
  \BibitemOpen
  \bibfield  {author} {\bibinfo {author} {\bibfnamefont {D.}~\bibnamefont
  {Collins}}, \bibinfo {author} {\bibfnamefont {N.}~\bibnamefont {Gisin}},
  \bibinfo {author} {\bibfnamefont {N.}~\bibnamefont {Linden}}, \bibinfo
  {author} {\bibfnamefont {S.}~\bibnamefont {Massar}}, \ and\ \bibinfo {author}
  {\bibfnamefont {S.}~\bibnamefont {Popescu}},\ }\bibfield  {title} {\enquote
  {\bibinfo {title} {{Bell Inequalities for Arbitrarily High-Dimensional
  Systems}},}\ }\href {http://link.aps.org/doi/10.1103/PhysRevLett.88.040404}
  {\bibfield  {journal} {\bibinfo  {journal} {Physical Review Letters}\
  }\textbf {\bibinfo {volume} {88}},\ \bibinfo {pages} {040404} (\bibinfo
  {year} {2002})}\BibitemShut {NoStop}%
\bibitem [{\citenamefont {Zohren}\ and\ \citenamefont {Gill}(2008)}]{CGLMP08}%
  \BibitemOpen
  \bibfield  {author} {\bibinfo {author} {\bibfnamefont {S.}~\bibnamefont
  {Zohren}}\ and\ \bibinfo {author} {\bibfnamefont {R.~D.}\ \bibnamefont
  {Gill}},\ }\bibfield  {title} {\enquote {\bibinfo {title} {{Maximal Violation
  of the Collins-Gisin-Linden-Massar-Popescu Inequality for Infinite
  Dimensional States}},}\ }\href
  {http://link.aps.org/doi/10.1103/PhysRevLett.100.120406} {\bibfield
  {journal} {\bibinfo  {journal} {Physical Review Letters}\ }\textbf {\bibinfo
  {volume} {100}},\ \bibinfo {pages} {120406} (\bibinfo {year}
  {2008})}\BibitemShut {NoStop}%
\bibitem [{\citenamefont {Bell}(2004)}]{bell2004speakable}%
  \BibitemOpen
  \bibfield  {author} {\bibinfo {author} {\bibfnamefont {J.}~\bibnamefont
  {Bell}},\ }\href {http://books.google.com/books?id=FGnnHxh2YtQC} {\emph
  {\bibinfo {title} {{Speakable and Unspeakable in Quantum Mechanics: Collected
  Papers on Quantum Philosophy}}}},\ Collected papers on quantum philosophy\
  (\bibinfo  {publisher} {Cambridge University Press},\ \bibinfo {year}
  {2004})\BibitemShut {NoStop}%
\bibitem [{\citenamefont {{Dowling}}\ and\ \citenamefont
  {{Milburn}}(2002)}]{QuantumRevArXiv}%
  \BibitemOpen
  \bibfield  {author} {\bibinfo {author} {\bibfnamefont {J.~P.}\ \bibnamefont
  {{Dowling}}}\ and\ \bibinfo {author} {\bibfnamefont {G.~J.}\ \bibnamefont
  {{Milburn}}},\ }\bibfield  {title} {\enquote {\bibinfo {title} {{Quantum
  Technology: The Second Quantum Revolution}},}\ }\href
  {http://arxiv.org/abs/quant-ph/0206091} {\bibfield  {journal} {\bibinfo
  {journal} {arXiv:quant-ph/0206091}\ } (\bibinfo {year} {2002})}\BibitemShut
  {NoStop}%
\bibitem [{\citenamefont {Zhang}(2013)}]{YanbaoThesis}%
  \BibitemOpen
  \bibfield  {author} {\bibinfo {author} {\bibfnamefont {Y.}~\bibnamefont
  {Zhang}},\ }\emph {\bibinfo {title} {{Analysis of Tests of Local Realism}}},\
  \href {http://search.proquest.com/docview/1493001786} {Ph.D. thesis}
  (\bibinfo {year} {2013})\BibitemShut {NoStop}%
\bibitem [{\citenamefont {Zhang}\ \emph {et~al.}(2010)\citenamefont {Zhang},
  \citenamefont {Knill},\ and\ \citenamefont {Glancy}}]{PhysRevA.81.032117}%
  \BibitemOpen
  \bibfield  {author} {\bibinfo {author} {\bibfnamefont {Y.}~\bibnamefont
  {Zhang}}, \bibinfo {author} {\bibfnamefont {E.}~\bibnamefont {Knill}}, \ and\
  \bibinfo {author} {\bibfnamefont {S.}~\bibnamefont {Glancy}},\ }\bibfield
  {title} {\enquote {\bibinfo {title} {{Statistical Strength of Experiments to
  Reject Local Realism with Photon Pairs and Inefficient Detectors}},}\ }\href
  {http://link.aps.org/doi/10.1103/PhysRevA.81.032117} {\bibfield  {journal}
  {\bibinfo  {journal} {Physical Review A}\ }\textbf {\bibinfo {volume} {81}},\
  \bibinfo {pages} {032117} (\bibinfo {year} {2010})}\BibitemShut {NoStop}%
\bibitem [{\citenamefont {Zhang}\ \emph {et~al.}(2011)\citenamefont {Zhang},
  \citenamefont {Glancy},\ and\ \citenamefont {Knill}}]{PhysRevA.84.062118}%
  \BibitemOpen
  \bibfield  {author} {\bibinfo {author} {\bibfnamefont {Y.}~\bibnamefont
  {Zhang}}, \bibinfo {author} {\bibfnamefont {S.}~\bibnamefont {Glancy}}, \
  and\ \bibinfo {author} {\bibfnamefont {E.}~\bibnamefont {Knill}},\ }\bibfield
   {title} {\enquote {\bibinfo {title} {{Asymptotically Optimal Data Analysis
  for Rejecting Local Realism}},}\ }\href
  {http://link.aps.org/doi/10.1103/PhysRevA.84.062118} {\bibfield  {journal}
  {\bibinfo  {journal} {Physical Review A}\ }\textbf {\bibinfo {volume} {84}},\
  \bibinfo {pages} {062118} (\bibinfo {year} {2011})}\BibitemShut {NoStop}%
\bibitem [{\citenamefont {Zhang}\ \emph {et~al.}(2013)\citenamefont {Zhang},
  \citenamefont {Glancy},\ and\ \citenamefont {Knill}}]{PhysRevA.88.052119}%
  \BibitemOpen
  \bibfield  {author} {\bibinfo {author} {\bibfnamefont {Y.}~\bibnamefont
  {Zhang}}, \bibinfo {author} {\bibfnamefont {S.}~\bibnamefont {Glancy}}, \
  and\ \bibinfo {author} {\bibfnamefont {E.}~\bibnamefont {Knill}},\ }\bibfield
   {title} {\enquote {\bibinfo {title} {{Efficient Quantification of
  Experimental Evidence Against Local Realism}},}\ }\href
  {http://link.aps.org/doi/10.1103/PhysRevA.88.052119} {\bibfield  {journal}
  {\bibinfo  {journal} {Physical Review A}\ }\textbf {\bibinfo {volume} {88}},\
  \bibinfo {pages} {052119} (\bibinfo {year} {2013})}\BibitemShut {NoStop}%
\end{thebibliography}%
\end{spacing}


\setglossarystyle{altlist}
\glsaddall
\printglossary[title=Glossary,toctitle=Glossary]

\end{document}